\documentclass[aps,pra, superscriptaddress, floatfix, 12pt, longbibliography]{revtex4-2}

\makeatletter
\def\tocdepth@munge{%
  \let\l@section@saved\l@section
  \let\l@section\@gobble@tw@
  \let\l@subsection@saved\l@subsection
  \let\l@subsection\@gobble@tw@
}
\def\tocdepth@restore{%
  \let\l@section\l@section@saved
  \let\l@subsection\l@subsection@saved
}
\makeatother

\usepackage{siunitx}
\usepackage{hyperref}
\usepackage{comment}

\usepackage{amsmath}
\usepackage{amsfonts}
\usepackage{graphicx}

\usepackage{comment}

\newcommand{\scref}[1]{\cref{#1}}

\def\cmntsoff{} 

\usepackage{amsmath,amsthm,amssymb}
\usepackage{stmaryrd} 
\usepackage{dsfont}   
\usepackage{bm}
\usepackage{comment}
\usepackage{multirow}
\usepackage{tabularx}
\usepackage{array}
\usepackage{graphicx}
\usepackage{booktabs}
\usepackage[dvipsnames]{xcolor}
\usepackage{tikz}
\usepackage[normalem]{ulem} 
\usepackage{enumitem}
\usepackage[ruled,lined]{algorithm2e}
\usepackage{lineno}
\usepackage{hyperref}
\hypersetup{pdfstartview=}
\usepackage{url}
\usepackage{cleveref}

\crefname{figure}{Fig.}{Figs.}
\Crefname{figure}{Fig.}{Figs.}
\crefname{equation}{Eq.}{Eqs.}
\Crefname{equation}{equation}{Eqs.}
\crefname{section}{Sect.}{Sects.}
\Crefname{section}{Sect.}{Sects.}
\crefname{theorem}{Thm.}{Thms.}
\Crefname{theorem}{Thm.}{Thms.}

\SetAlgorithmName{Protocol}{List of Protocols}

\setlist{nosep} 
\newlist{enuma}{enumerate}{10}
\setlist[enuma]{label*=\arabic*.}
\setlistdepth{10} 

\numberwithin{equation}{section}
\newtheorem{theorem}{Theorem}[section]
\newtheorem*{theorem*}{Theorem}
\newtheorem{result}[theorem]{Result}
\newtheorem*{result*}{Result}

\newtheorem*{lemma*}{Lemma}

\usetikzlibrary{calc, intersections}

\usepackage{physics}
\NewCommandCopy{\pbr}{\qty} 
\usepackage{siunitx}
\let\siunitxqty\qty 
\let\pqty\pbr     
\let\qty\siunitxqty 
\AtBeginDocument{\RenewCommandCopy\qty\SI}

\newif\ifcmnt
\cmnttrue 
\ifdefined\cmntsoff\cmntfalse\fi

\ifcmnt
    \providecommand{\aucmnt}[1]{#1}
    \providecommand{\MKcolor}{\color{teal}}

    \providecommand{\Pcolor}{\color{gray}}
    
\else
    \providecommand{\aucmnt}[1]{}
    \providecommand{\MKcolor}{}

    \providecommand{\Pcolor}{}
    
\fi

\providecommand{\ignore}[1]{}
\providecommand{\auedit}[1]{#1}

\newcommand{\MK}[1]{\auedit{{\MKcolor #1}}}
\newcommand{\MKc}[1]{\aucmnt{{\MKcolor [MK: {\color{darkgray}#1}]}}}
\newcommand{\MKs}[1]{\aucmnt{{\MKcolor \sout{#1}}}}

\newcommand{\Ac}[1]{\aucmnt{{\Pcolor \textbf{[}Authors' record: {\color{gray}#1}\textbf{]}}}}
\newcommand\bluesout{\bgroup\markoverwith{\textcolor{Cerulean}{\rule[0.5ex]{2pt}{1pt}}}\ULon}

\newcommand{\one}{\mathds{1}}

\newcommand{\rls}{\mathds{R}}

\providecommand{\Prob}{\mathbb{P}}

\newcommand{\cA}{\mathcal{A}} \newcommand{\cB}{\mathcal{B}} \newcommand{\cC}{\mathcal{C}}
\newcommand{\cD}{\mathcal{D}} \newcommand{\cE}{\mathcal{E}} 
 \newcommand{\cH}{\mathcal{H}} 
 \newcommand{\cM}{\mathcal{M}} \newcommand{\cN}{\mathcal{N}}
\newcommand{\cO}{\mathcal{O}} \newcommand{\cP}{\mathcal{P}} \newcommand{\cQ}{\mathcal{Q}}
 \newcommand{\cS}{\mathcal{S}} \newcommand{\cT}{\mathcal{T}}
 \newcommand{\cV}{\mathcal{V}} 
  \newcommand{\cZ}{\mathcal{Z}}

\newcommand{\Id}{\mathds{1}}

\newcommand{\LR}{{\rm LR}}

\newcommand{\TV}{\textrm{TV}}
\newcommand{\Exp}{\textrm{Exp}}

\DeclareMathOperator{\argmax}{arg\,max}

\newcommand{\Va}{{\sf A}'}   
\newcommand{\Vap}{{\sf A}}    
\newcommand{\Vb}{{\sf B}}     
\newcommand{\Vs}{{\sf A_{\text{source}}}} 
\newcommand{\Vp}{{\sf P}}     
\newcommand{\Ef}{{\sf F}}     
\newcommand{\Qa}{{\sf Q}_{\Va}} 
\newcommand{\Qp}{{\sf Q}_{\Vp}} 
\newcommand{\tR}{E}          
\newcommand{\cba}{u_{\Vap}{}}\newcommand{\Cba}{U_{\Vap}{}} 
\newcommand{\cbb}{u_{\Vb}{}}\newcommand{\Cbb}{U_{\Vb}{}}  
\newcommand{\cnu}{\nu_U}     
\newcommand{\sfun}{f}        
\newcommand{\mqa}{m_{\Va}}\newcommand{\Mqa}{M_{\Va}} 
\newcommand{\mmua}{\mu_{\Va}} 
\newcommand{\mset}{{\rm set}}\newcommand{\mmuap}{\mu_{\mset}} 
\newcommand{\mqp}{m_{\Vp}}\newcommand{\Mqp}{M_{\Vp}} 
\newcommand{\oqa}{o_{\Va}{}}\newcommand{\Oqa}{O_{\Va}{}} 
\newcommand{\oqp}{o_{\Vp}{}}\newcommand{\Oqp}{O_{\Vp}{}} 
\newcommand{\zqa}{z_{\Vap}{}}\newcommand{\Zqa}{Z_{\Vap}{}} 
\newcommand{\zqb}{z_{\Vb}{}}\newcommand{\Zqb}{Z_{\Vb}{}} 
\newcommand{\Trec}{R_{{\rm trial}}{}}\newcommand{\trec}{r_{{\rm trial}}{}}
\newcommand{\ttrial}{t_{{\rm trial}}}
\newcommand{\In}{{{\rm in}}}\newcommand{\Out}{{{\rm out}}}
\newcommand{\Src}{{{\rm src}}}\newcommand{\Tgt}{{{\rm tgt}}}

\newcommand{\Ain}{\cA_{{\rm in}}} \newcommand{\Aout}{\cA_{{\rm out}}}
\newcommand{\Aastart}{\cA_{\Vap,{\rm start}}}
\newcommand{\Abstart}{\cA_{{\Vb, \rm start}}}
\newcommand{\Aaend}{\cA_{{\Vap, \rm end}}}
\newcommand{\Abend}{\cA_{{\Vb, \rm end}}} \newcommand{\Aa}{\cA_{\Vap}}
\newcommand{\Ab}{\cA_{\Vb}}

\newcommand{\valVerifierSep}{\qty{195.1 \pm 0.3}{\meter}}
\newcommand{\valCircleOneRad}{\qty{157.3 \pm 0.2}{\meter}}
\newcommand{\valCircleTwoRad}{\qty{116.7 \pm 0.2}{\meter}}
\newcommand{\valEllipseOne}{\qty{273.1 \pm 0.2}{\meter}}
\newcommand{\valEllipseTwo}{\qty{274.8 \pm 0.2}{\meter}}

\newcommand{\valDistVtwoP}{\qty{92.8 \pm 0.1}{\meter}}
\newcommand{\valDistVtwoBttag}{\qty{192.2 \pm 0.3}{\meter}}
\newcommand{\valDistVtwoBrng}{\qty{192.2 \pm 0.2}{\meter}}

\newcommand{\valMeanInvArea}{\qty{4.02 \pm 0.03}{}}
\newcommand{\valMeanInvVol}{\qty{4.53 \pm 0.05}{}}
\newcommand{\valMeanInvLineInt}{\qty{4.48 \pm 0.02}{}}
\newcommand{\valMeanInvIdealLine}{\qty{2.47 \pm 0.02}{}}

\newcommand{\valSound}{\ensuremath{2^{-64}}}
\newcommand{\valComplete}{0.97725}
\newcommand{\valThresh}{\ensuremath{8\times 10^{-6}}}
\newcommand{\valBasicInstances}{232}
\newcommand{\valBasicSuccess}{224}
\newcommand{\valBasicFail}{8}
\newcommand{\valEntInstances}{103}
\newcommand{\valEntSuccess}{102}

\newcommand{\valTotInstances}{335}

\newcommand{\Sp}[2]{\mathsf{S}\!\left(#1,#2\right)} 
\newcommand{\El}[1]{\mathsf{E}\!\left({\sf A},{\sf B};#1\right)}

\newcommand{\timeCbaOut}{\qty{1291.0 \pm 0.5}{\nano\second}}
\newcommand{\timeCbbOut}{\qty{1429.1 \pm 0.6}{\nano\second}}
\newcommand{\timeQpLas}{\qty{1023.3 \pm 0.3}{\nano\second}}
\newcommand{\timeQpEm}{\qty{1133 \pm 2}{\nano\second}}
\newcommand{\timeSetMqa}{\qty{1738.8 \pm 0.4}{\nano\second}}
\newcommand{\timeOutOqa}{\qty{1874.5 \pm 0.3}{\nano\second}}
\newcommand{\timeCbaIn}{\qty{1766 \pm 1}{\nano\second}}
\newcommand{\timeCbbIn}{\qty{1769 \pm 1}{\nano\second}}
\newcommand{\timeSigPc}{\qty{1798 \pm 1}{\nano\second}}
\newcommand{\timePhotPc}{\qty{1831 \pm 1}{\nano\second}}
\newcommand{\timePhotSnspd}{\qty{1858 \pm 2}{\nano\second}}
\newcommand{\timeZCreate}{\qty{1868 \pm 2}{\nano\second}}
\newcommand{\timeZqbIn}{\qty{2207.7 \pm 0.6}{\nano\second}}
\newcommand{\timeZqaIn}{\qty{2340.3 \pm 0.5}{\nano\second}}

\newcommand{\gexp}{2.3223\times 10^{-6}}
\newcommand{\Deltamaxexp}{0.3592}
\newcommand{\Deltabarexp}{0.3188}
\newcommand{\gchsh}{0.032075}
\newcommand{\Deltamaxchsh}{0.5523}
\newcommand{\Deltabarchsh}{0.5523}
\newcommand{\Deltafiveexp}{1.6748}
\newcommand{\Deltafivechsh}{2.7615}
\newcommand{\Epsilonexp}{1.1555\times 10^{-6}}
\newcommand{\Epsilonexpg}{6.0704\times 10^{-6}}
\newcommand{\lexptwo}{127}
\newcommand{\lexpthree}{229}
\newcommand{\lexpfour}{375} 
\newcommand{\lchshtwo}{49} 
\newcommand{\lchshfour}{141}
\newcommand{\lexpgl}{999}
\newcommand{\lexpgq}{8}
\newcommand{\lexpgbnd}{120}
\ignore{%
g=2.3223e-6; Deltamax=0.3592; Deltabar=0.3188;
Delta5=3*Deltabar+2*Deltamax; eps=(e^(5*g/6)-1)/Delta5;
f = @(eps, d, l) -(l - 3) / 2 + (d^2 / 2) * log2(1 + 2 / eps) - log2(eps);
f(eps,2,127)
f(eps,3,229)
f(eps,4,375)
epsg=(e^(5*g/6)-1)/Deltabar;
q=8; l=999;
G=((l-3)/2+log2(epsg))/(log2(q)+1)

g=0.032075; Deltamax=0.5523; Deltabar=0.5523;
Delta5=3*Deltabar+2*Deltamax; eps=(e^(5*g/6)-1)/Delta5;
f(eps,2,49)
f(eps,4,141)
}%

\makeatletter
\@ifundefined{linenomath}{%
  \newenvironment{linenomath}{}{}
  \newenvironment{linenomath*}{}{}
}{}
\makeatother

\begin{document}

\addtocontents{toc}{\string\tocdepth@munge} 

\title{Quantum Position Verification with Remote Untrusted Devices}
\author{Gautam A. Kavuri}\email{gautamk@duck.com}
\affiliation{Department of Physics, University of Colorado, Boulder, CO, 80309, USA}
\affiliation{Associate of the National Institute of Standards and Technology, Boulder, CO, 80305, USA}
\author{Yanbao Zhang}\email{zhangy2@ornl.gov}
\affiliation{Quantum Information Science Section, Computational Sciences and Engineering Division, Oak Ridge National Laboratory, Oak Ridge, TN, 37831, USA}
\author{Abigail R. Gookin}
\affiliation{Department of Electrical, Computer, and Energy Engineering, University of Colorado, Boulder, CO, 80309, USA}
\affiliation{Associate of the National Institute of Standards and Technology, Boulder, CO, 80305, USA}
\author{Soumyadip Patra}
\affiliation{Department of Mathematics, University of New Orleans, New Orleans, LA, 70148, USA}
\author{Joshua C. Bienfang}
\affiliation{Joint Quantum Institute, National Institute of Standards and Technology and University of Maryland, 100 Bureau Drive,
Gaithersburg, MD, 20899, USA.}
\author{Honghao Fu}
\affiliation{Concordia Institute for Information Systems Engineering, Concordia University, Montreal, QC, H3G 1M8, Canada}
\author{Yusuf Alnawakhtha}
\affiliation{Joint Center for Quantum Information and Computer Science, University of Maryland, College Park, MD, 20742, USA}
\author{Dileep V. Reddy}
\affiliation{Department of Physics, University of Colorado, Boulder, CO, 80309, USA}
\affiliation{Associate of the National Institute of Standards and Technology, Boulder, CO, 80305, USA}
\author{Michael D. Mazurek}
\affiliation{Department of Physics, University of Colorado, Boulder, CO, 80309, USA}
\affiliation{Associate of the National Institute of Standards and Technology, Boulder, CO, 80305, USA}
\author{Carlos Abell\'an}
\author{Waldimar Amaya}
\affiliation{Quside Technologies S.L., Castelldefels (Barcelona), Spain}
\author{Morgan W. Mitchell}
\affiliation{ICFO-Institut de Ciencies Fotoniques, The Barcelona Institute of Science and Technology, 08860 Castelldefels (Barcelona), Spain.}
\affiliation{ICREA - Instituci\'{o} Catalana de Recerca i Estudis Avan{\c{c}}ats, 08010, Barcelona, Spain}
\author{Carl A. Miller}
\affiliation{Computer Security Division, National Institute of Standards and Technology, Gaithersburg, MD, 20899, USA}
\affiliation{Joint Center for Quantum Information and Computer Science, University of Maryland, College Park, MD, 20742, USA}
\author{Richard P. Mirin}
\affiliation{Physical Measurement Laboratory, National Institute of Standards and Technology, Boulder, CO, 80305, USA}
\author{Martin J. Stevens}
\affiliation{Physical Measurement Laboratory, National Institute of Standards and Technology, Boulder, CO, 80305, USA}
\author{Scott Glancy}
\affiliation{Applied and Computational Mathematics Division, National Institute of Standards and Technology, Boulder, CO, 80305, USA}
\affiliation{Department of Physics, University of Colorado, Boulder, CO, 80309, USA}
\author{Emanuel Knill}
\affiliation{Applied and Computational Mathematics Division, National Institute of Standards and Technology, Boulder, CO, 80305, USA}
\affiliation{Department of Physics, University of Colorado, Boulder, CO, 80309, USA}
\affiliation{Center for Theory of Quantum Matter, University of Colorado, Boulder, CO, 80305, USA}
\author{Lynden K. Shalm}\email{krister.shalm@nist.gov}
\affiliation{Department of Physics, University of Colorado, Boulder, CO, 80309, USA}
\affiliation{Physical Measurement Laboratory, National Institute of Standards and Technology, Boulder, CO, 80305, USA}
\affiliation{Quantum Engineering Initiative, Department of Electrical, Computer, and Energy Engineering, University of Colorado, Boulder, CO, 80309, USA}

\date{\today}

\maketitle

\textbf{Position information underpins many modern technologies, from
  navigation and timing to authentication and critical
  infrastructure~\cite{chandran_positionbased_2014,bhamidipati_gps_2018,ardizzon_authenticated_2022}. However,
  classical methods of proving that information originates from a
  particular position are vulnerable to
  spoofing~\cite{tippenhauer_attacks_2009,brands_distancebounding_1994,drimer_keep_2007}. This
  limitation can be overcome with quantum
  technologies~\cite{Malaney_LocVer2010, Kent_TagVer2011,
    buhrman2013garden, buhrman_positionbased_2014,
    lim_losstolerant_2016, bluhm_singlequbit_2022,
    allerstorfer_practical_2022, escola-farras_singlequbit_2023,
    asadi2025linear, allerstorfer2025making} but current protocols
  rely on trust in quantum hardware that can be
  undermined~\cite{lydersen_hacking_2010, weier_quantum_2011}, or
  require quantum computers and bounds on adversarial \MKs{classical}\MKc{Both quantum and classical computation and complexity assumptions, I thought it best to not specify in this abstract.}%
  computation~\cite{kaleoglu_equivalence_2025}.  Nevertheless, there
  has been significant interest in experimental demonstrations, and
  aspects of these protocols have been
  implemented~\cite{kanneworff_experimental_2025,liu_certified_2025a,fan2026relativistic}.
  Here we introduce and experimentally demonstrate the Bell-test
  quantum position verification protocol for device-independent
  quantum position verification that guarantees security with only
  observed correlations from a loophole-free Bell
  test~\cite{shalm_strong_2015, giustina_significantloopholefree_2015,
    Bell_nonlocality_review2014} across a quantum network. We
  experimentally implement a version of this device-independent
  protocol against adversaries who, before each trial, are weakly
  entangled. Our demonstration achieves a one-dimensional localization
  2.47(2) times smaller than the best, necessarily non-remote,
  classical localization protocol. Compared to classical protocols
  with identical latencies, the localization volume is 4.53(5) times
  smaller, and represents a certifiable quantum advantage. The general
  Bell-test protocol is loss tolerant and secure against adversaries
  with significant quantum resources. This work allows digital
  security to be anchored to physically trusted locations, enabling
  new position-based authentication protocols for applications such as
  financial transactions, legal agreements, and securing critical
  infrastructure~\cite{alabdulatif_novel_2023,denning_locationbased_1996,zhang_locationbased_2012,alawami_locauth_2020}.}

\section*{Introduction}

Position is a fundamental aspect of
security~\cite{denning_locationbased_1996}. Sensitive systems and
information are often stored in secure locations that can only be
accessed in person, and permissions for certain actions are granted to
entities based on where they are. Physical access controls, such as
air-gapped systems~\cite{langner_stuxnet_2011} or analog locks,
mitigate remote attacks~\cite{whitman_principles_2009,
  anderson_security_2020b}. A central challenge is extending the
security of a trusted location into the digital world---if a party can
prove they are communicating from a particular place, they can be
remotely granted comparable levels of access and trust. Such
guarantees would underpin a wide range of applications, from
authenticating that commands to a power grid originate from the
utility's control center rather than a hostile actor
abroad~\cite{bhamidipati_gps_2018}, to confirming that a GPS signal
genuinely originates from a satellite at a known orbital
position~\cite{fernandez-hernandez_navigation_2016} rather than a
nearby software-defined radio, to ensuring that a transaction request
to a bank's servers comes from a customer at a verified
branch~\cite{alabdulatif_novel_2023,
  zhang_locationbased_2012}. Securely verifying position is the first
step to building many of these advanced position-based authentication
applications~\cite{chandran_positionbased_2014,
  buhrman_positionbased_2014}.

Accurately localizing a position can be achieved using ranging methods
that measure the round-trip time of signals sent at the speed of light
to a target region. Several trusted parties, known as verifiers, can
then work together to triangulate position. However, the signals used
in such ranging protocols can in principle always be intercepted and
manipulated by adversaries to spoof a location they do not
occupy~\cite{buhrman_positionbased_2014, chandran_positionbased_2014,
  tippenhauer_attacks_2009}. Cryptographic authentication can mitigate
this~\cite{brands_distancebounding_1994, drimer_keep_2007,
  hancke_rfid_2005, cowperthwaite_practical_2023}, but only at the
cost of trusted hardware and pre-shared keys or secrets. If those keys
are leaked, hardware devices cloned, or if cryptographic hardness
assumptions fail, a wide range of sensitive systems that depend on
position, navigation, and timing become vulnerable to spoofing.

An alternative approach to avoiding spoofing in position verification
is to use quantum states that cannot be cloned in combination with
classical ranging \cite{Kent_TagVer2011,Malaney_LocVer2010}.  In the
simplest case of BB84 quantum position verification
(QPV)~\cite{Kent_TagVer2011,buhrman_positionbased_2014} two trusted
verifiers work together to determine if a third party, the prover, is
located within a target location. One verifier randomly prepares a
qubit in one of several bases and sends it to the prover. The
verifiers also send classical challenge bits to the prover who uses
them to perform a predefined computation. The result of this
computation sets the measurement basis for the prover. The prover
performs this measurement and returns the result to the verifiers, who
check that the result is correct and use the round-trip time of the
signals to localize the position of the prover.  While unconditional
security against entangled attackers is
impossible~\cite{buhrman_positionbased_2014}, it is possible to design
protocols where the resources that the adversaries require to
successfully attack the protocol increase with the number of classical
challenge bits. Ideally, a protocol also would require an honest
prover to manipulate only a single qubit while requiring more
resources from successful attackers~\cite{bluhm_singlequbit_2022,
  asadi2025linear}.  It is also desirable to allow the quantum signals
in the protocol to be arbitrarily slow and rely only on the timing of
classical communication for ranging.  Protocols with this feature are
more practical for deployment over large
distances~\cite{bluhm_singlequbit_2022, allerstorfer2025making,
  escola-farras_singlequbit_2023}.

Besides BB84 QPV a range of other QPV protocols have been proposed,
including extensions of the original BB84
QPV~\cite{bluhm_singlequbit_2022, escola-farras_singlequbit_2023,
  allerstorfer2025making}, methods based on ideas from
measurement-device-independent quantum key distribution (QKD) and
entanglement swapping~\cite{lim_losstolerant_2016}, qubit
routing~\cite{Kent_TagVer2011,bluhm_singlequbit_2022} and SWAP
  protocols~\cite{allerstorfer_practical_2022}.  Importantly, the security of these protocols requires
that the quantum devices used to perform specific state preparations
and measurements are trusted and operating correctly.  Such protocols
are said to be device-dependent, and this dependence introduces
opportunities for attackers to subvert the quantum devices or exploit
side-channels. Such exploits have been demonstrated for the closely
related QKD
protocols~\cite{lydersen_hacking_2010,weier_quantum_2011,tippenhauer_attacks_2009,brands_distancebounding_1994,drimer_keep_2007}.
Despite much
theoretical development in device-dependent QPV protocols, their
complete experimental demonstration has been lagging. Experiments
include demonstrations of the basic
principles~\cite{kanneworff_experimental_2025}, with distance
simulated with fiber-spools~\cite{fan2026relativistic}, and of the
main components of a high-complexity scheme based on advanced quantum
computers~\cite{liu_certified_2025a}.

Here we introduce the Bell-test QPV protocol, a device-independent
protocol that is based on a loophole-free Bell test. Bell-test QPV
uses the violation of a Bell inequality by a verifier and the prover
to certify that the prover is operating a quantum device in a
specified target region in spacetime, where the quantum device must be
capable of non-trivial quantum actions conditional on classical input.
No assumptions are made about the internal workings of the required
quantum devices, including the quantum measurement devices and the entanglement source. Device-independent security is achieved based only on
analysis of the inputs to and outputs from the experimental
devices. Like all device-independent protocols such as those
  implemented in Refs.~\cite{shalm_deviceindependent_2021,
  kavuri_traceable_2025, liu_deviceindependent_2021,
  li_experimental_2021, nadlinger_experimental_2022, lu_deviceindependent_2026, Kulikov2026},  Bell-test QPV
requires trust in the classical devices used for recording results,
measuring timing, and performing distance measurements.
The protocol performance
does not depend on the speed of the qubits provided by the entanglement source, which is treated as fully untrusted, but only on the round-trip
time of the classical signals sent. 
As long as a verifier and the
prover share an entangled state, be it entangled photons or
matter-based qubits entangled via a quantum repeater, the protocol can
proceed. The compatibility of the protocol with heralded sources of
entanglement such as those reported in the recent
work~\cite{lu_deviceindependent_2026, Liu_LongLivedHeraledEnt_2026}
means that the protocol is scalable with distance over a quantum
network.  For precise localization, it suffices to use
near-speed-of-light classical communication such as wireless
communication or hollow-core fiber~\cite{fan2026relativistic}.

We implemented the Bell-test QPV protocol with two verifiers separated
by \valVerifierSep, performing \(\valTotInstances\) experimental
protocol instances, each lasting two or four minutes depending on the
class of adversaries considered, over two days. Each protocol instance
consists of a sequence of trials. Measurement outcomes and messages
from the trials are recorded and determine whether the instance
successfully verified position. While we do not limit the computation
or communication power of adversaries during each experimental trial,
our security analysis requires a limit on the amount of entanglement
shared between adversaries prior to the trial. Throughout this work,
we refer to such pre-shared entanglement as prior entanglement. A
basic version of the protocol is secure against adversaries who share
no prior entanglement. An extended protocol implemented in the
experiment demonstrates security against adversaries possessing
limited prior entanglement, bounded from above by an average
robustness of entanglement~\cite{vidal_robustness_1999} of
\(\valThresh\) over the trials of a protocol instance. A coarse such
bound was previously obtained for general QPV protocols in
Ref.~\cite{beigi2011simplified}. To protect against adversaries with significant quantum
resources including prior entanglement, we prove that it suffices to
increase the number of bits in the challenges.  The security of our protocol is
derived from the impossibility of faster-than-light signaling and
constraints on the shareability of three-party non-signaling
correlations~\cite{masanes_general_2006a}.  These constraints can be
effectively falsified by a loophole-free Bell violation in the
configuration we experimentally implemented.  Each trial requires the
prover to compute a function of challenges received from the
verifiers. We experimentally demonstrate the simplest nontrivial instance of such
a function. 

Conditional on successfully passing the Bell-test protocol, we can be
confident that a prover must have performed an input-conditional
quantum operation in a localized target spacetime region \(\tR\).
Ideally, the localization could approach arbitrary precision,
independent of verifier separation. In practice, it is constrained by
communication and processing latencies. Using low-latency processing
and communication, we are able to certify the position of a prover to
a volume \valMeanInvVol\ smaller than that achievable by a
classical protocol that is constrained by the same
latencies. Our results establish a new cryptographic primitive for
quantum networks, allowing access to a secure physical location to be
used to authenticate a network user. This bridges physical and network
security and opens new directions for position-based cryptography and
computation~\cite{chandran_positionbased_2014,
  buhrman_positionbased_2014}.

\section*{Position verification}

\begin{figure}
    \centering
    \includegraphics[width=0.5\linewidth]{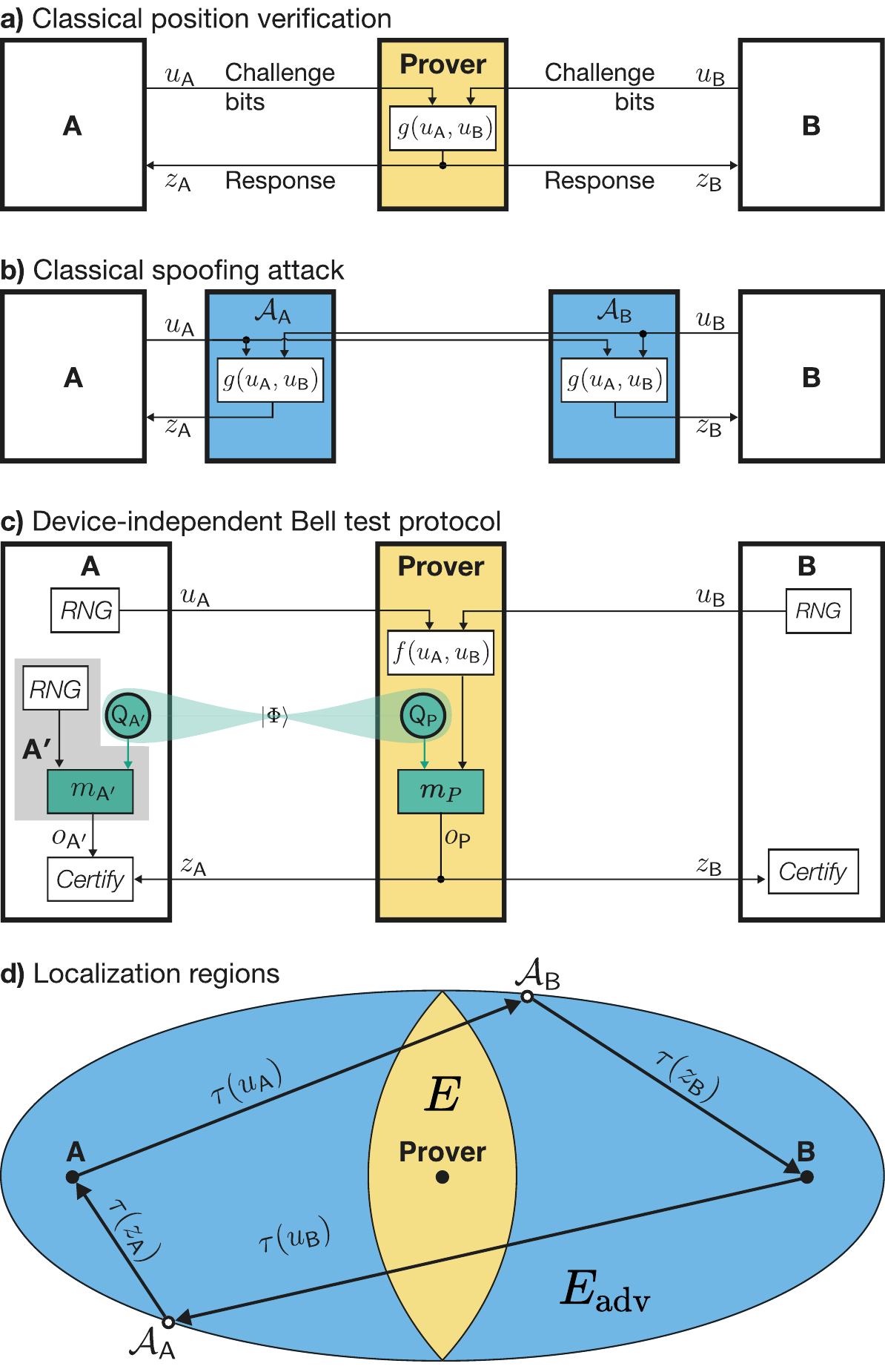}
    \caption{Schematic representations of position verification protocols and possible attacks. Subfigure a) shows a single-prover classical protocol for position verification that enforces the target region \(E\) in yellow with classical challenges from the verifiers \(\Vap\) and \(\Vb\). The verifiers desire that the computation \(g(\cba,\cbb)\) is localized to the target region. Subfigure b) sketches a specific attack on the classical protocol, with colluding adversaries \(\Aa\) and \(\Ab\), situated in the blue regions outside the target regions, each performing the computation locally. By communicating the challenges to one another, they can spoof a successful pass of the classical protocol from anywhere inside \(E_{\mathrm{adv}}\). To ensure security against general attacks perpetrated by unentangled or weakly entangled adversaries, we introduce our quantum protocol, sketched in c). Here, the honest prover \(\Vp\) and a quantum measurement station \(\Va\) located at \(\Vap\) share an entangled state \(|\Phi\rangle\). The state is measured using untrusted quantum hardware, with the measurement basis at \(\Vap\) decided by a local random number generator and the basis at \(\Vp\) decided by the output of \(f(\cba,\cbb)\).  Subfigure d) is a schematic of the spatial extents of the  target regions enforceable by the classical (\(E_{\mathrm{adv}}\)) and quantum (\(E\)) protocols, projected into two dimensions for  a realistic two-verifier protocol. }
    \label{fig:protocol}
\end{figure}

Formally, position verification is a cryptographic protocol that
involves two or more verifiers working together to ensure that only
provers \(\Vp\) with access to a specified target region \(\tR\) can
pass the protocol. In this work, we consider the case of two verifiers
\(\Vap\) and \(\Vb\) who use time-of-flight ranging to establish
presence of a prover within the target region. We consider first a
classical protocol where we assume that the verifiers are only
interacting with a single prover, as shown in~\cref{fig:protocol} a),
the verifiers \(\Vap\) and \(\Vb\) send classical challenges
\(\cba\) and \(\cbb\) timed to arrive simultaneously at the
prover. The prover performs a computation using the challenges from both
verifiers and a predetermined, publicly known function
\(g(\cba,\cbb)\), and then returns the result.  Conditioned on the
prover computing the correct answer, each verifier can use the time
difference between when the challenge was sent and when the answer
was received to bound the maximum distance to the prover. The
intersection of the regions within the maximum distances from each
verifier localizes the position of the prover. If this intersection is
contained in the region to be secured, the protocol passes. 
With light-speed signals and zero processing
latency, the verifiers can localize the prover to a single
point. However, any latencies in the classical signals or the
processing increases the size of the smallest possible localization
region, \(E\). Projected to two dimensions, this forms a
lenticular region, as shown by the yellow region
in~\cref{fig:protocol} d).

However, a pair of adversaries working together can spoof responses
from inside \(E\), as shown in Figure \ref{fig:protocol} b). In the
attack, each adversary stores a copy of the challenge bits sent from
the verifier closest to them. When the challenge bits from the other
verifier arrive, they compute the response and return the answer to
the verifier closest to them. In this way, the adversaries can
convince the verifiers they are inside the target region \(E\). In the
ideal case where there is no excess latency, the spoofing adversaries
must both be stationed on the line joining the verifiers, or the
verifier axis. However, if there is processing or other excess latency
in the system, then adversaries with negligible processing latency can
exploit this to move away from the verifier axis in three dimensions
and still respond in time. As described in the methods section, the
region that successful adversaries are required to occupy is bounded
by a set of geometric and timing constraints. In the symmetric case,
where the target region is centered at the midpoint between the
verifiers, and any timings are symmetric, the region that the
adversaries are required to occupy is an ellipsoid with foci at the
verifier positions, and whose major axis is twice the distance light
travels in the time from when a challenge is emitted to when a
response is received, at either verifier. The surface of the ellipsoid
corresponds to the set of positions from which an adversary can just
barely receive a challenge from one verifier, compute a response, and
deliver it to the other verifier in time, as shown
in~\cref{fig:protocol} d). Any adversaries outside this ellipsoidal
region \(E_{\rm adv}\) are unable to spoof the protocol. This means
the verifiers can achieve a degree of localization with a classical
protocol, but this region contains the verifier positions and has
diameter substantially larger than the verifier separation.

The device-independent Bell-test QPV protocol is based on the prover
and verifiers jointly violating a loophole-free Bell inequality, as
shown in~\cref{fig:protocol} c). In a Bell test, two parties perform
independent, space-like separated measurements on an entangled
state. These measurements are repeated many times in a series of
trials. If the joint measurement statistics violate a Bell inequality,
it is possible to rule out the presence of local hidden variable
models as an explanation for the results. In our protocol, in each
trial, verifier \(\Vap\) and prover \(\Vp\) share an entangled Bell
state. At the start of the trial, the verifiers send challenges
\(\cba\) and \(\cbb\) to the prover.  When they arrive, the prover
computes a measurement settings function \(f(\cba,\cbb)\) whose value
is used to choose the prover's measurement basis.  Simultaneously,
verifier \(\Vap\) uses a random number generator to set their
measurement basis. \(\Vap\) and \(\Vp\) then measure their respective
entangled qubits in a space-like separated manner, and \(\Vp\)
responds to the challenges by sending their measurement outcome
separately to \(\Vap\) and \(\Vb\). For each trial, a Bell-like test
statistic is evaluated on the trial record, which consists of the
measurement basis choice and measurement outcome of \(\Vap\), and the
prover measurement basis \(f(\cba,\cbb)\) and the two prover responses
to \(\Vap\) and \(\Vb\). The two responses may not be identical due to
transmission error or other sources of noise.  Intuitively, empirical
violation of the inequality associated with the test statistic can be
used to rule out the presence of spoofing adversaries. Security of the
Bell-test QPV protocol places additional demands on the test statistic
and its interpretation, see the next section. The round-trip time
between the challenges being sent and the prover's response with
the outcome of their measurement allows the verifiers to localize the
prover. The target region \(\tR\) for the quantum protocol is
determined in the Supplementary
Information~\scref{subsect:spacetimeana} and is the same as the
single-prover target region \(E\) in the classical
protocol above, while being secure against colluding adversaries. The
ratio between the achievable localization region \(E\) by the Bell-test QPV
protocol and the localization region \(E_{\rm adv}\) of the classical protocol 
represents a quantum advantage in
localization. In principle, the quantum protocol's target region could
be made arbitrarily small with instantaneous processing and direct,
speed-of-light classical communication. The only requirement is that the
verifier and prover have established entanglement before the
measurements need to be performed. This can be done, for instance,
through direct transmission over a quantum channel or by using a
quantum repeater.

While the preceding discussion has focused on the target region in the
three spatial dimensions, it can be specified in all four dimensions
including time, see~\scref{subsect:spacetimeana} in the Supplementary
Information. Given that our experiment is performed as a series of
trials, the localization has a periodic structure in time.

\section*{Theory}
We establish security of Bell-test QPV against several classes of
adversaries, where colluding adversaries attempt to pass the Bell-test
protocol without having any presence in the target region. The classes
are distinguished by the resources available to the adversaries in a
trial. The resources considered include prior entanglement in a trial,
and quantum computation and communication.  We do not restrict
adversaries ability to move provided that they do not enter the target
region's spacetime extent. Security is established under standard
assumptions for device-independent protocols and allows for complete
control over the source by the adversaries. The assumptions are
summarized in the Methods section and a detailed list of assumptions
is given in Supplementary
Information~\scref{subsect:protocol_assumptions}. Bell-test QPV
depends on choosing measurement bases and challenge distributions, and
the settings function \(f(\cba,\cbb)\) of the challenges. Requirements
on these distributions and their dependencies are included in the list
of assumptions.

Security of QPV protocols  requires soundness and completeness.
Soundness with parameter \(\delta\) requires that before a protocol
starts, the maximum probability that adversaries in a given class can
pass the protocol is at most \(\delta\). For high security, the
soundness parameter must be extremely small.  Completeness with
parameter \(\epsilon\) requires that the actual prover has probability
at least \(\epsilon\) of passing.  It is desirable for completeness to
be close to \(1\).  For the experimental demonstration we choose
\(\delta=2^{-64}\).  For a summary of Bell-test QPV soundness and
completeness results, see the Methods.

Completeness of Bell-test QPV is witnessed by any verifier-prover pair
able to violate local realism in a Bell test with two measurement
bases for the prover, such as we experimentally demonstrated.
Scalability of the protocol requires that the number of trials grows
sufficiently slowly as the soundness is reduced and the completeness
is increased.  Like most useful protocols, Bell-test protocols require
only \(O(\log(1/\delta))+O(\log(1/(1-\epsilon)))\) trials for soundness
\(\delta\) and completeness \(\epsilon\). The suppressed constants
depend on the quality of the honest prover's performance.

Bell-test protocols require a departure from the conventional approach
of classifying trials by whether the prover responses are correct or
not. Instead we use test factors originally introduced for test of
local realism~\cite{zhang_y:qc2011a}, which are special
test statistics \(W\) applied to each trial and calculated from the
trial record. Test factors have recently been found to have broad
applications in statistics for constructing so-called \(e\)-values,
\MKs{which are chainable \(p\)-values}\MK{which are a chainable
  alternative to \(p\)-values}~\cite{vovk2021values,grunwald2024safe}.
For position verification, these test statistics satisfy \(W\geq 0\)
and the trial expectation of \(W\) is at most \(1\) for all
adversaries in a given class regardless of what happened in the past.
To determine whether the protocol passed, the test factor \(W\) is
applied to each trial's record and the resulting values are multiplied
to obtain the test-factor product \(P\). A key consequence of the
properties of test factors is that the probability that
\(P\geq 1/\delta\) for adversaries is at most \(\delta\).  Soundness
is therefore assured if the protocol is passed only when
\(P\geq 1/\delta\).

The use of Bell tests with test factors and the need to establish
target regions in \(3\)+\(1\)-dimensional spacetime require a detailed
analysis of the spacetime capabilities of adversaries. We adapt the
spacetime circuit method used for a parallel, one-round QPV protocol
by Unruh~\cite{unruh2014quantum} to determine the trial target
region from the timing information for the verifiers sending out the
challenges and receiving the responses. We then partition the
adversary accessible regions to reduce adversary attacks to those that
can be implemented by two adversaries who can communicate only once
during a trial. See Supplementary
Information~\scref{subsect:spacetimeana}.  Such reductions to two
adversaries are typical for QPV protocols but are usually only
implemented in one spatial dimension, which omits several spacetime
regions that could be exploited by adversaries and does not address
the effects of realistic latencies and transmission delays.  Because our treatment is
fully relativistic, it is also necessary to determine the specific
trial boundary at which we quantify the adversary entanglement
resources.  The trial boundary obtained is non-trivial and can be
described as the boundary of a spacetime region constructed from lightcones, see
Supplementary Information~\scref{fig:1+1d_regions}.

The basis for our soundness results against different classes of
adversaries is an analysis of the power of adversaries with no prior
entanglement, but unlimited classical and quantum computing and
communication resources. Such an analysis forms the basis of the
security of many other QPV protocols and was introduced in
Ref.~\cite{buhrman_positionbased_2014}. Subsequent improved analyses
connected security of a class of existing QPV protocols to monogamy of
entanglement games~\cite{tomamichel2013monogamy}.  For Bell-test QPV,
the analysis requires a different approach for simplifying adversary
strategies that takes advantage of test-factor properties. See
Supplementary Information~\scref{sect:reduction}.  The simplification
reduces the problem of test-factor design for Bell-test QPV to
constructing test factors for three-party quantum non-signaling
scenarios.   For provers
with two measurement basis choices, the problem can be further reduced
to that of modifying test factors against local realism. 

An advantage of test factors is their flexibility: they can be adapted
during a protocol, enable heralding and early stopping, and can be
rescaled against adversaries with prior entanglement. Heralded trials
can be used to mitigate the problem of loss in photon transmission. In
particular, previous proofs of loss tolerance by trial commitment by
the prover~\cite{allerstorfer2025making} generalize to Bell-test QPV,
see Supplementary Information~\scref{sect:heraldetc}.  Rescaling can
be used directly for the security against weakly entangled adversaries
demonstrated in our experiment, see below.  This is a quantified
generalization of an observation in Ref.~\cite{beigi2011simplified},
see Supplementary Information~\scref{sec:smallentanglement}.  We
quantify the entanglement by robustness of entanglement, which
measures how far an entangled state is from the set of separable
states. We show that the increase in the expectation of the test
factor \(W\) due to the adversaries' prior entanglement can be bounded
in terms of robustness of this prior entanglement, and this increase
determines the required test-factor rescaling.

Rescaled test factors are also used against adversaries with
significant entanglement or other quantum resources. This requires
increasing the number \(l\) of bits in each of the challenges \(\cba\)
and \(\cbb\) and a suitable choice of settings function
\(f(\cba,\cbb)\).  That increasing the complexity of \(f(\cba,\cbb)\)
is sufficient for security against adversaries with significant
quantum resources has been established for some previous
protocols~\cite{bluhm_singlequbit_2022,asadi2025linear} by reduction
to a communication-complexity problem. Test factors enable a tighter
such reduction to a different communication-complexity problem, see
Supplementary Information~\scref{sect:entangled_adversaries}.  We can
protect against adversaries with prior entanglement in
\((\log(l)/2-O(1))\) qubits and unlimited classical communication
and local quantum computation, or adversaries with
\(O(l/(\log(q)+1))\) quantum operations on \(q\) qubits and
  unlimited classical communication.  With our experimental
parameters and an increase of the number of trials required by 
a factor of \(6\), if \(l=\lexpthree\), then adversaries, in addition to
having to occupy intercepting paths on two opposite sides of the
target region and to intercept and control multiple signals, must use
at least \(1\) qutrit for entanglement or quantum communication per
trial to have a reasonable chance of passing the protocol. In the
experiment, we used one weakly entangled qubit per trial.  If
\(l=\lexpgl\), then adversaries with fewer than \(q=\lexpgq\) qubits
require more than \(\lexpgbnd\) operations in a standard universal
gate and measurement set to break security.  There is a simple
function \(f(\cba,\cbb)\) that achieves this security and that is
suitable for low-latency implementation. The function is described in
the Methods.

\section*{Experiment}
\begin{figure}
    \centering
    \includegraphics[width=1.\linewidth]{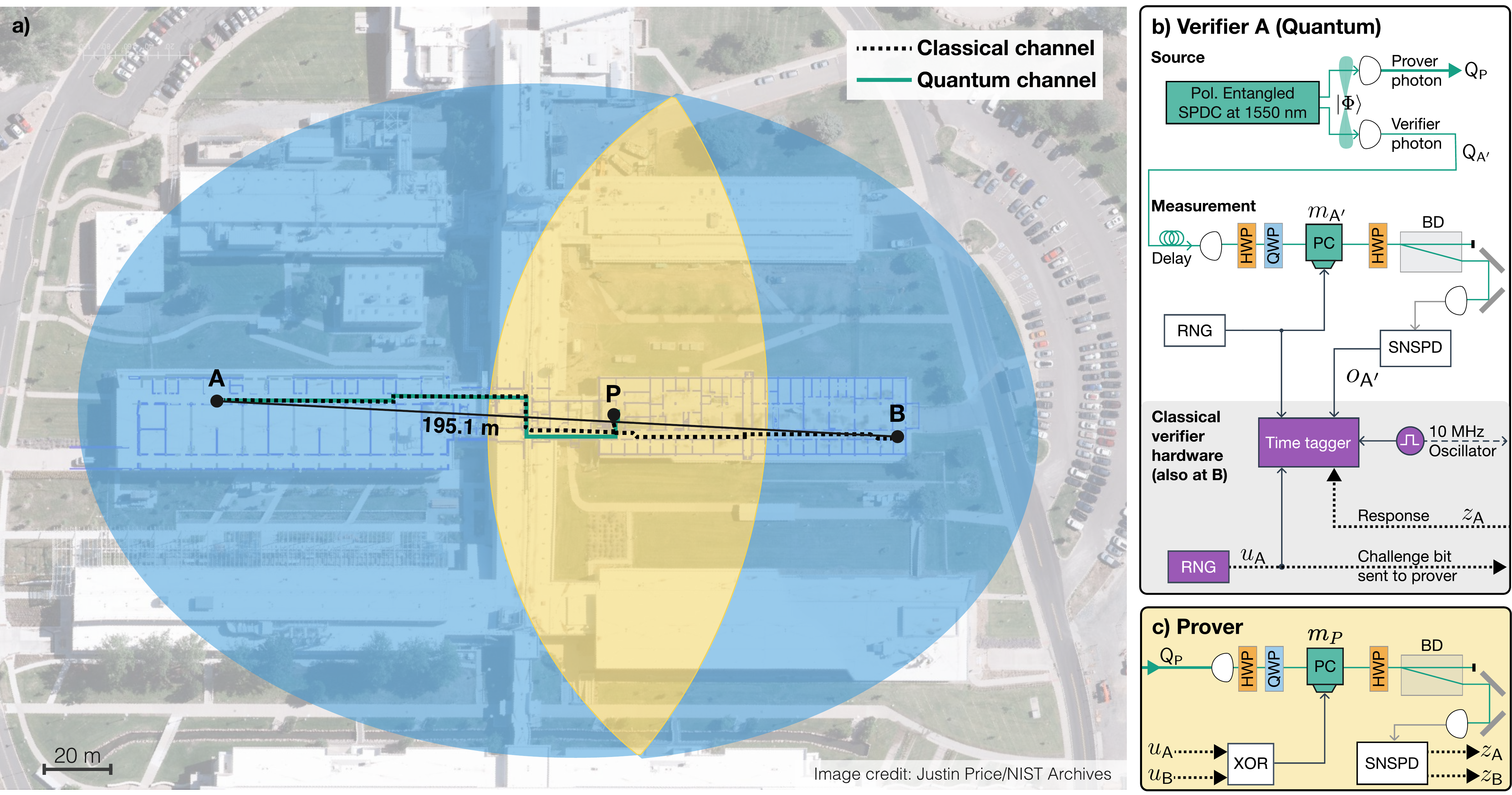}
    \caption{Figure depicting the experimental setup and verifier and
      prover stations used for our demonstration. a) The quantum
      target region \(\tR\) in yellow, and the classical target region
      of a comparable classical protocol in blue overlaid on an aerial
      photograph of the NIST building, where the experiment is
      housed. b) Schematic of the hardware at verifier \(\Vap\). From
      an untrusted entangled photon pair source, one photon of the
      pair is sent to prover \(\Vp\), and the other to the measurement
      station \(\Va\) at verifier \(\Vap\). At measurement station
      \(\Va\), the incoming photon is delayed in a fiber loop and
      measured by a superconducting nanowire detector in a
      polarization basis determined by a fast Pockels cell and
      waveplates. The setting is chosen via a random number generator
      (RNG)~\cite{AbellanPRL2015}, and the outcome is recorded on a
      time-tagger. c) Measurement station at the prover \(\Vp\). The
      choice of measurement basis here is set via an XOR of the bits
      \(u_{\Vap}\) and \(u_{\Vb}\) from the verifiers. The outcome of
      this measurement is sent back to both verifiers, who record it
      on their timetaggers. SNSPD: Superconducting nanowire single
      photon detector; XOR: Exclusive OR; PC: Pockels cell; HWP:
      Half-wave plate; QWP: Quarter-wave plate.}
    \label{fig:apparatus}
\end{figure}

In our experiment, we demonstrate successful instances of the Bell-test protocol using two verifier stations, \(\Vap\) and \(\Vb\), and a prover station, \(\Vp\). Each verifier is equipped with a random number generator to produce challenge bits and a time-tagging device to record detection events. Verifier \(\Vap\) also houses an entangled photon source and a polarization measurement apparatus \(\Va\).
As shown in~\cref{fig:apparatus}, verifier stations \(\Vap\) and \(\Vb\) are separated by \valVerifierSep. The prover station is \valDistVtwoP\ from verifier \(\Vb\). All three stations are arranged approximately collinearly, and much of our hardware is similar to past device-independent experiments~\cite{kavuri_traceable_2025, shalm_deviceindependent_2021}.

The protocol requires the accumulation of statistics over many
trials. At the beginning of each trial, the quantum verifier \(\Vap\)
attempts to generate a pair of polarization-entangled photons in a
state close to \(0.383 \ket{HH} + 0.924 \ket{VV}\), where $\ket{H}$
and $\ket{V}$ represent horizontal and vertical polarizations
respectively. The target state was chosen based on a simulation
numerically maximizing Bell violation, taking measured background
counts and photon efficiencies into account. One photon is delayed
locally via a fiber loop, while the other is transmitted to the prover
in fiber. Each verifier then sends a randomly generated challenge bit
to the prover. The challenge bits are timed to arrive at the prover
station within \(\qty{3\pm 2}{\nano \second}\) of each other and are
transmitted using specialized coaxial cables with signal transmission
speeds about \(86\%\) the speed of light. We note that we allow
adversaries to transmit signals at the speed of light. The implemented
\(86\%\) transmission speed makes our localization region larger than
it would be in an implementation with speed-of-light classical signal
transmission.

\begin{figure}
    \centering
    \includegraphics[width=\linewidth]{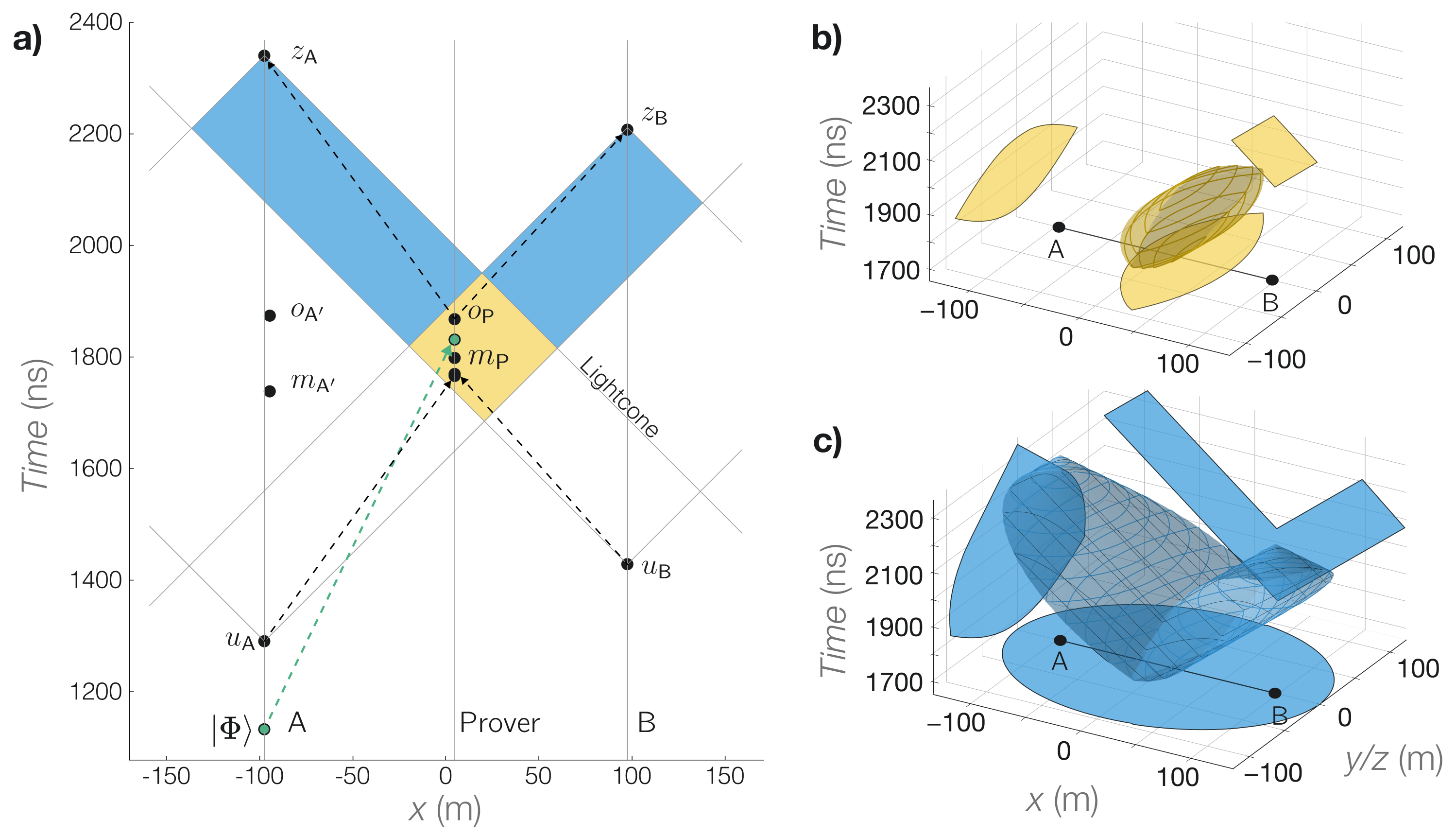}
    \caption{Figure depicting the spacetime target regions in our
      demonstration. a) The two-dimensional spacetime slice, where the
      spatial dimension is along the line joining the provers, here
      assigned the \(x\)-axis. Both the quantum (yellow), and
      comparable classical (blue) regions are completely defined by
      forward and backward lightcones from the spacetime events
      corresponding to the release of \(\cba\) and \(\cbb\) and the
      last moments when \(\zqa\) and \(\zqb\) are accepted. Events
      \(\oqa\) and \(\mqa\) occur at \(\Va\), which is slightly
      displaced from \(\Vap\). The region in blue is also the region
      from which adversaries with sufficient quantum resources can
      successfully attack the quantum protocol. Subfigures b) and c)
      show three-dimensional projections of the quantum and classical
      target regions, with the time dimension vertically up on the
      page, and the spatial dimensions parallel and perpendicular to
      the verifier axis depicted in the other two directions. The
      regions are all rotationally symmetric about the verifier axis,
      and so are identical in the spatial dimensions (\(y\) and \(z\))
      perpendicular to the verifier axis. At the displayed scale, error
      bars indicating uncertainty on the positions and timings are too
      small to be discernible. A full accounting of
      uncertainties and their effect on our results is detailed in the
      Supplementary Information.}
    \label{fig:minkowski}
\end{figure}

When the challenge bits \(\cba, \cbb\) arrive at the prover station,
an XOR circuit is used to compute a settings bit
\(\mqp = \cba \oplus \cbb\). The prover selects their polarization
measurement bases, defined as
$\cos(\theta_{P})\ket{H} + \sin(\theta_{P})\ket{V}$, by using this bit
to choose between two angles,
$\theta_{P} \in \{6.7^{\circ}, -29.5^{\circ}\}$. The selection is
implemented using a Pockels cell to rotate the polarization of the
incoming photon, followed by a polarizer to perform a polarization
measurement, see~\cref{fig:apparatus}. The output of the polarizer is
coupled into a single-mode fiber and incident on a superconducting
nanowire detector. The result of the single-photon measurement (click
or no-click) is sent back to the verifier stations via another set of
coaxial cables, where the arrival times are recorded on time tagging
devices. As depicted in~\cref{fig:minkowski} a), the incoming photon
arrives just in time for the measurement.  At nearly the same time that the prover performs its quantum measurement (see~\cref{fig:minkowski} a)), verifier \(\Vap\) measures the polarization of the stored local photon \(\Qa\) using a similar setup. The verifier uses a local random number generator to select \(\mqa\), which sets the polarization measurement  basis $\cos(\theta_{A'})\ket{H} + \sin(\theta_{A'})\ket{V}$ to the angle $\theta_{A'}$ to either \(\qty{-6.7}{\degree}\) or \(\qty{29.26}{\degree}\). The setting choice and outcome are recorded on a local time tagger. The quantum verifier and prover locations are sufficiently separated to ensure space-like separation of measurement settings and measurement outcomes at the other station, see~\cref{fig:minkowski}.

  

The system operates at a rate of approximately 250,000 trials per
second, limited by the operating rate of the Pockels cells. Data was
recorded in \qty{60}{\second} intervals over two days. At the start of
each day, an automated calibration routine was performed to correct
for polarization drifts in the optical fiber and to ensure high
coupling efficiency for the entangled photons. The measured system
detection efficiency was \(>81 \%\) for both verifier \(\Vap\) and the
prover station \(\Vp\), sufficient for an analysis that does not rely
on fair-sampling assumptions. Specifically, this means we do
not post-select or perform  background subtraction. Loss or
background counts beyond the acceptable threshold would result in
repeated protocol failure. The efficiency threshold for our protocol
is \(2/3\), the same as a two-party Bell
test~\cite{rowe_experimental_2001, eberhard_background_1993}, see
Supplementary Information~\scref{sect:reduction}. Loophole-free Bell
tests have been successfully implemented on many
platforms~\cite{hensen_loopholefree_2015, shalm_strong_2015, giustina_significantloopholefree_2015,  
  rosenfeld_eventready_2017, loopholefree_shanghai2018, storz_loopholefree_2023}, making the
Bell-test QPV protocol amenable to a wide range of quantum hardware.

\section*{Analysis and Results}

We first collected calibration data to set key parameters in our
analysis.  We then performed a series of QPV experiments on two
separate days, with data collected in \qty{60}{\second} chunks with
each chunk containing about \(15\) million experimental
trials. Approximately eight hours of data was collected on September
20, 2024 (day 1) and seven hours of data on October 7, 2024 (day
2). The day 1 data was reserved for testing the basic protocol against
adversaries without prior entanglement, while the day 2 data was used
to test the extended protocol against adversaries with prior
entanglement whose robustness averaged over trials is limited by a
threshold \(r_{\text{th}}\).  The soundness parameter for both
protocols was chosen to be \(\delta = \valSound\), and the targeted
completeness, or success rate, was set to \(\epsilon = 0.97725\),
corresponding to a Gaussian \(2\sigma\) probability of a particular
protocol run successfully passing.  After evaluating the calibration
data to determine the anticipated performance of the experiment, for
the extended protocol the robustness threshold was fixed at
\(r_{\text{th}} = \valThresh\).  We determined that for these
parameters, \(30\) million trials (two minutes of data) were required
per run of the basic protocol, and \(60\) million trials (four minutes
of data) were needed per run of the extended protocol.

The robustness threshold of \(r_{\text{th}} = \valThresh\) can be
compared to the robustness per trial \(2\times 10^{-3}\) of the
entangled state designed to be emitted by the source. The latter
robustness accounts for the probability of about \(1/350\) that a
photon pair is emitted in a trial (see Supplementary
Information~\scref{subsec:exp_prot_imp}). The robustness of
entanglement of a Bell state is \(1\). With a version of the
``garden-hose'' attack~\cite{buhrman2013garden} given in Supplementary
Information~\scref{sect:entangled_adversaries}, it is possible to
break the security of the experimentally implemented version of
Bell-test QPV with adversaries sharing two pairs of qubits, each with
robustness of entanglement \(2\times 10^{-3}\).  As explained above,
for security against adversaries with more quantum resources than used
in the experiment, it suffices to increase the number of classical
challenge bits in each trial.

\begin{figure}
    \centering
    \includegraphics[width=0.8\linewidth]{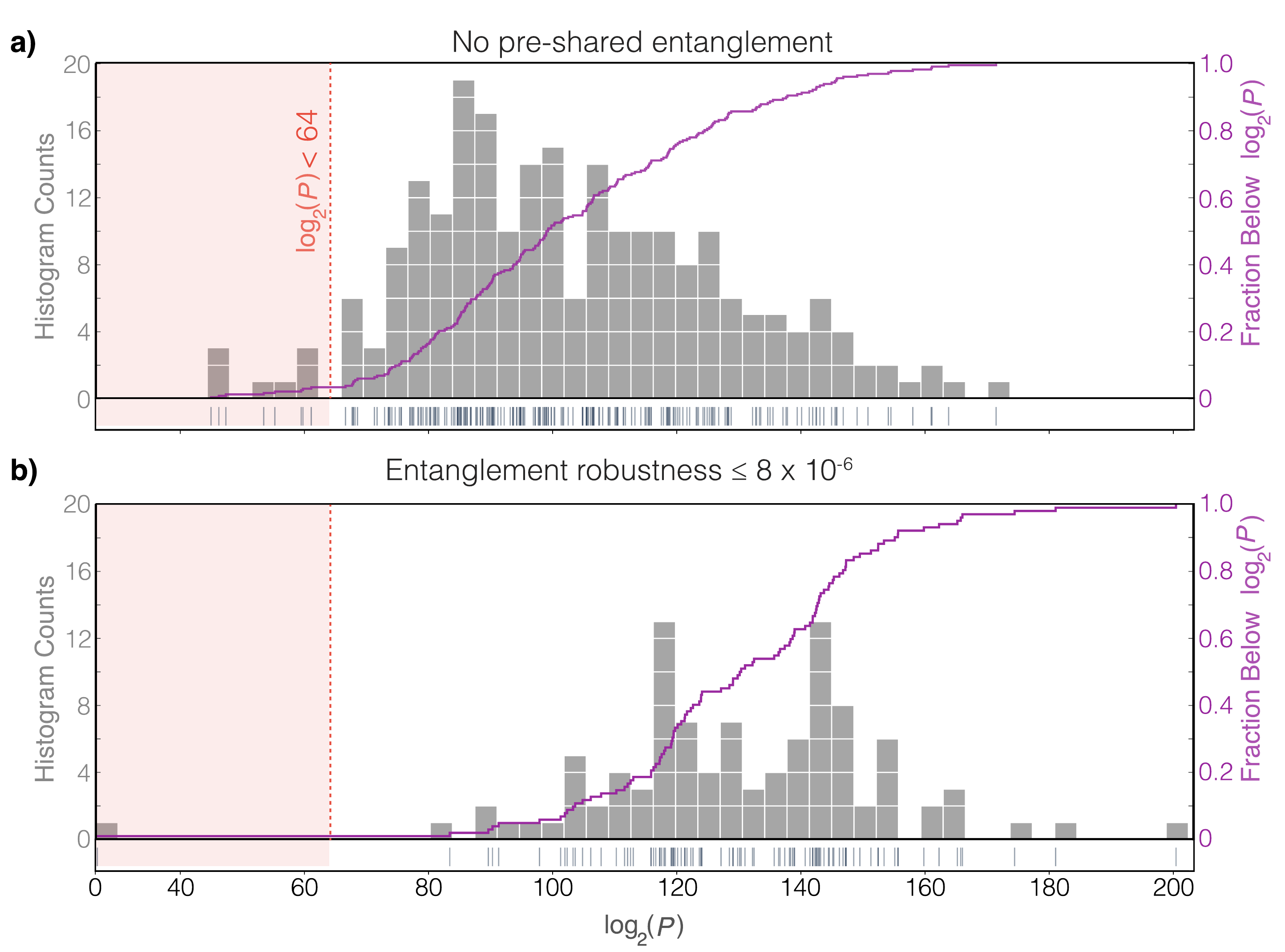}
    \caption{Plots of results from our experimental runs. Two sets of
      protocol instances were run as part of our demonstration: a) is
      day 1, analyzed for security against adversaries pre-sharing no
      entanglement; and b) is day 2, analyzed against adversaries
      pre-sharing entanglement such that the average robustness over
      an instance's trials is upper bounded by
      \(r_{\text{th}} = \valThresh\). The soundness target for all
      instances was \(\delta = \valSound\). Plotted are histograms of
      the data binned based on the calculated \(\log_{2}(P)\), where
      \(P\) is the product of the test-factor values over the trials
      in a protocol instance.  The \(\log_2(P)\) value for each
      instance is indicated with a tick directly above the \(x\)-axis
      and below the histograms on both plots. Any instances with a
      \(\log_2(P)\) less than \(-\log_2(\delta) = 64\) are declared
      failures, and correspond to ticks inside the red regions. Also
      plotted in purple is the experimental fraction of instances
      below \(\log_2(P)\) as a function of \(\log_2(P)\). At a higher
      \(\log_2(P)\), which corresponds to a stricter choice for the
      soundness, a larger fraction of the data would be considered a
      failure, reducing success probability. The plots both share the
      same \(x\)-axis.}
    \label{fig:histogram}
\end{figure}

After running the two series of QPV experiments, the collected data
were compressed and placed in blind storage prior to
analysis. Occasional errors where our single-photon nanowire detectors
failed to properly reset, and therefore were not light sensitive for
one or more of the trials during a chunk, were flagged. On both day 1
and day 2, the first 10 minutes of data free of detector errors were
used to construct the initial test factor \(W\). The trials from the
subsequent two or four minutes, depending on whether running the basic
or extended protocol, and irrespective of detector errors, were then
un-blinded and analyzed to determine the product \(P\) of the values
of \(W\) witnessed by each trial record.  A product \(P \geq 2^{64}\) results in a
pass; otherwise, that instance is considered a fail. Subsequent
instances were processed sequentially using the next two or four
minutes of data if available. To track long-term drifts in the
experiment, we updated the test factor \(W\) before each subsequent
instance, using the preceding 10 minutes of processed data without
detector errors. This yielded \(\valBasicInstances\) runs of the basic
protocol on day 1 and \(\valEntInstances\) runs of the extended
protocol on day 2. Of the \(\valBasicInstances\) evaluated for the
basic analysis, \(\valBasicSuccess\) passed (\(96.5\%\)), and of the
\(\valEntInstances\) evaluated for the extended analysis,
\(\valEntSuccess\) succeeded (\(99.0\%)\). Both of these agree well
with the expected completeness of \(\epsilon = 0.97725\), and are
shown in~\cref{fig:histogram}.

In each of the successful instances, the target three-dimensional
volume in which a prover must perform a quantum operation is
\(\valMeanInvVol\) times smaller than the volume they would need to be
in to pass a comparable classical protocol. This is a measure of our
quantum position verification advantage (\({\rm QA}\) =
\(E/E_{\rm adv}\), see above). This comparison can also be made in
two-dimensions, \({\rm QA_{\rm area} = \valMeanInvArea}\), and in
one-dimension, \({\rm QA_{\rm length} = \valMeanInvLineInt}\). For
intuition on our metric, consider that in \cref{fig:protocol} for
example, the (\({\rm QA}\)) is the ratio of the total volume of the
blue and yellow regions to that of the yellow region, where the figure
shows the two-dimensional projections of these regions. Finally, we
can also compare with the ideal classical protocol in one dimension
where the target region is the line segment between the two
verifiers. In this case the advantage is
\({\rm QA_{\rm length-ideal} =
  \valMeanInvIdealLine}\). See~\scref{subsec:target_region} in the
Supplementary Information for more details. Prior theory
works~\cite{bluhm_singlequbit_2022, chandran_positionbased_2014} often
consider an ideal version of this protocol. In this case, the quantum
target region reduces to a point, the best classical target region
reduces to the line joining the verifiers, and the QA tends to
infinity.

\section*{Conclusion}

This work presents the first complete demonstration of a quantum
position verification protocol---a major step forward in the
development of secure quantum cryptographic primitives.  Our
implementation achieves strong localization guarantees and rules out
powerful classes of adversaries without relying on assumptions about
the internal functioning of the quantum devices involved. We have
shown that Bell-test QPV is scalable and robust. Because the source is
untrusted with no restriction on speed of transmission of qubits,
transmission loss can be tolerated provided that trials are heralded
via entanglement swapping or commitment. The effect of all noise
sources including detector inefficiencies on
completeness is the same as that on Bell tests: It suffices for the
Bell test to be successful for the honest prover to be able to pass
the protocol~\cite{eberhard_background_1993}.  For future scalability, entangled photon pairs can be
provided by quantum repeaters.  No source of noise can affect
  soundness, and neither do adversary actions such as interference
  with communication and calibrations, including calibrations used to
  choose the test factors. Such actions at worst result in denial of
  service, an issue that is faced by all communication protocols.

While the protocol as demonstrated already offers robust security within minutes, there are several areas in which future experiments can be improved. At present, the Pockels cells used to perform the measurements can operate only at \qty{250}{kHz} due to the large switching voltages required, whereas all other components in the system can support trial rates of up to \qty{100}{MHz}. Faster active measurement devices could reduce run times by two orders of magnitude, leading to position-verification times on the order of one second while maintaining the same security parameters.

Our localization accuracy is currently limited by transmission latencies as well as measurement processing time at the prover station. For larger distances on the metropolitan scale, low-latency line-of-sight classical communication links will be required, using either laser communications or wireless technologies. Both technologies are mature and should be readily adaptable to larger distances. The line-of-sight requirement for ranging means that transmitting signals through buried cables, whether high-velocity-factor cables or hollow-core fibers, will not be feasible in most situations. Buried cables rarely follow line-of-sight paths, leading to significant geometry-induced latencies. Using custom low-latency circuits and FPGAs, it should be possible with existing classical technology to reduce the combined transmission and measurement latency to below \qty{30}{\nano \second}, even over distances of many kilometers. This would lead to prover regions that are well contained within a large room or a single building along the verifier axis. Such low-latency classical communications will also be valuable for future quantum repeater networks, as they can significantly reduce the feed-forward time between repeater nodes, thereby reducing the storage times required by quantum memories.

Finally, to exclude adversaries possessing more entanglement than that prepared by the source, we have shown that it suffices to increase the number of bits used for the challenges \(\cba,\cbb\). Recent work in~\cite{fan2026relativistic} has demonstrated the feasibility of 20-bit challenges for each of \(\cba\) and \(\cbb\), together with random functions \(f(\cba,\cbb)\). Adapting these classical processing techniques will enable systems based on our protocol to safeguard against more powerful quantum adversaries.

While there is still much work to be done before QPV systems can be widely deployed in real-world settings, our work makes significant progress toward demonstrating the requirements for a complete protocol capable of achieving a quantum advantage in a nontrivial distributed environment. 

\section*{Methods}

\subsection*{Assumptions}

Completeness and soundness of Bell-test QPV are established under the
assumptions detailed in Supplementary
Information~\scref{subsect:protocol_assumptions}.  The assumptions are
standard for device-independent
protocols~\cite{shalm_deviceindependent_2021,
  liu_deviceindependent_2021, kavuri_traceable_2025} and include
secure classical computing, known verifier laboratory locations, and
sufficiently accurate event timing. The verifier laboratories must be physically
  secured against unauthorized access. According to
  device-independence, the quantum measurement apparatus and the
  entangled photon source are untrusted: no assumptions are made about
  their internal operation. Like other device-independent
  protocols~\cite{nadlinger_experimental_2022}, we require shielding
  of the measurement station's external interface.  In the
  no-prior-entanglement setting, it is desirable to add shielding
  to the source to prevent emissions outside defined time
    windows. For significant prior entanglement, source contributions are accounted
    for by the adversary entanglement bounds, which exceed the entanglement
  required by the verifiers and honest prover.

Protocol specific assumptions include conditions on measurement
settings and challenge distributions, defined target regions, 
trial boundaries, and specified classes of adversaries.
Verifier settings and challenges are drawn from
fixed distributions satisfying two independence conditions: (1) the
prover setting $\Mqp$ and verifier setting $\Mqa$ are independent of
$\Vap$'s challenge $\Cba$; and (2) $\Mqa$ is conditionally independent
of both challenges given $\Mqp$. These hold, for example, when the
challenges are uniform and independent of $\Mqa$, and the settings
function $f(\cba,\cbb)$ is chosen appropriately.

Depending on the settings function and test-factor construction,
  Bell-test QPV is sound against different classes of adversaries.
For all classes, adversaries can perform instantaneous
  computation and speed-of-light communication.  Each class is
  defined by a set of assumptions on adversaries required for
  soundness.  All classes of adversaries allow for unlimited classical
  communication.  For the simplest class, we assume no prior
  entanglement but allow unlimited quantum communication. For classes
  of adversaries with significant prior entanglement quantum
  communication is replaced by quantum teleportation and contributes
  to the prior entanglement.  The maximum amount of prior entanglement
  allowed while ensuring protocol soundness is determined by the
  number of challenge bits and the settings function used in each
  trial. See the completeness and soundness results below for details.

\subsection*{Completeness and Soundness Results}

\noindent\textbf{Protocol completeness:}
\emph{For all
  \(\epsilon\in (0,1)\), Bell-test QPV  is
  \(\epsilon\)-complete. To witness completeness it suffices for the
  verifiers and honest prover to violate local realism for some
  Bell test configuration with two measurement settings for the
  prover. For every class of adversaries under consideration and a
  given violation of local realism, the number of trials required
  for completeness \(\epsilon\) and soundness \(\delta\) is
  \(O(\log(1/\delta))+ O(\log(1/(1-\epsilon)))\). }

The proof of this result is in Supplementary
Information~\scref{sect:reduction}. Because there is no restriction
on how the entangled state for the Bell-test is obtained by the
verifiers and prover before each trial, completeness is preserved if
a slow quantum network is used to prepare these states. The quantum
link may be heralded and can support prover commitment
strategies~\cite{allerstorfer2025making}.  Therefore loss in
photonic links is well tolerated. See Supplementary
Information~\scref{sect:heraldetc} for how the protocol supports
heralded sources and commitment.

The soundness results depend on the class of adversaries under
consideration. In each case, soundness holds for all spatial
dimensions, and the source is untrusted and may be controlled by one
of the adversaries. Classical communication by the adversaries is
always unconstrained.  The first soundness result involves the class of
adversaries with no prior entanglement.

\noindent
\textbf{Protocol soundness against adversaries with no prior entanglement:}
\emph{For all \(\delta>0\), Bell-test QPV is \(\delta\)-sound
  against adversaries who have no prior entanglement for each
  trial. The adversaries can freely use classical and quantum computation and
  communication during each trial.}

The proof is contained in Supplementary
Information~\scref{subsect:spacetimeana},~\scref{sect:reduction}
and~\scref{sec:test_factors}. The first step of the proof  follows the
traditional path of reducing the adversary attacks to those involving
two adversaries on opposite sides of the target
region~\cite{buhrman_positionbased_2014}.  Bell-test QPV
requires an analysis that is independent of spatial dimension, takes
into account the time required by the honest prover to perform its
measurement task, and supports device-independence with test factors.
For this we adapt the spacetime circuit analysis introduced in
Ref.~\cite{unruh2014quantum} and fully formalize the reduction to two
adversaries. The subsequent analysis of adversary attacks deviates
from the traditional path in order to reduce the statistics of
adversarial trial records to a three-party non-signaling scenario for
which we can construct test factors. 

The absence of prior entanglement for the above soundness result
implies a constraint on the location of the source entanglement at the
trial boundary: The source entanglement must be located on the side
closer to verifier \(\Va\), which requires that we prevent the quantum
system destined for the prover from leaking to the other side. This
constraint on the location of the source entanglement is mitigated by
the next soundness result and absent in the last two soundness
results.

The second soundness result allows for small amounts of prior entanglement.

\noindent \textbf{Protocol soundness against adversaries with small amounts
  of prior entanglement:} \emph{For all \(\delta>0\), the Bell-test
  protocol of Sect.~\scref{subsect:protocol_assumptions} is
  \(\delta\)-sound against adversaries with small amounts of prior
  entanglement for each trial or on average over the trials.  The
  adversaries can freely use classical and quantum computation and communication
  during each trial.}

This is proven in Supplementary
Information~\scref{sec:smallentanglement}. The amount of entanglement
allowed depends on the quality of the Bell test used by the verifiers
and prover. The amount of prior entanglement is quantified either as
maximum robustness of entanglement over trials, or as average
robustness of entanglement over trials.  For the experiment, the upper
bounds on allowed prior entanglement are smaller than the entanglement
created by the source, limiting the strength of the result.  The
technique used here can be viewed as a quantified strengthening of
Lemma~5.3 of Ref.~\cite{beigi2011simplified}.  It is also related to
the entanglement forging method of Ref.~\cite{eddins2022doubling}.
Neither our above bounds nor those based on
Ref.~\cite{beigi2011simplified} have so far led to bounds on prior
entanglement of adversaries that are higher than the resources used by
the verifiers and honest prover when quantum computation and
communication during a trial is unrestricted.

To prove soundness against adversaries with significant amounts of
prior entanglement requires increasing the complexity of the settings
function \(f(\cba,\cbb)\).  For the next two results, we set
$f(\cba,\cbb)=g_0(\cba\oplus\cbb)$, where \(g_{0}\) is the semi-bent
function
\( g_{0}(z) = z_{0} + \sum_{i=1}^{l-1} z_{i}\,z_{i+l-1} \mod 2 \) and
the bit-length \(l\) of \(\cba\) and \(\cbb\) is odd. The semi-bent
function is explicit, efficiently computable, and therefore scalable.

\noindent \textbf{Protocol soundness against adversaries with logarithmic
  prior entanglement:} \emph{There are efficiently implementable
  settings functions \(f(\cba,\cbb)\) such that for all \(\delta>0\),
  Bell-test QPV is \(\delta\)-sound against adversaries for whom the
  sum of the number of qubits quantum communicated and the number of
  qubits used for prior entanglement is bounded by
  \(\log(l)/2 - O(1)\).  The adversaries may use arbitrary local
  quantum computation and classical communication during each trial.
}

The proof is in Supplementary
Information~\scref{sect:entangled_adversaries}. Since quantum
communication can be replaced by prior entanglement and classical
communication, the soundness can be interpreted as being against
adversaries with \(\log(l)/2-O(1)\) qubits of prior entanglement and
no quantum communication. To support practical applications, we fully
quantify the allowed prior-entanglement bound. To maintain a given
soundness and completeness, it suffices to increase the number of
trials by a constant factor. We choose a factor of \(6\) in the
examples mentioned in the main text.  The suppressed constant in the
bound depends on the test factor used and its gain.  See the
paragraphs starting at~\scref{eq:epsilon_n_d} in Supplementary
Information~\scref{sect:entangled_adversaries}.  The proof of this
soundness result is inspired by the proof of Prop. 6.1 in
Ref.~\cite{bluhm_singlequbit_2022}.  A specific analysis of the
relationship between adversarial test-factor expectations and guessing
probabilities for the settings function is required to efficiently
preserve completeness under all circumstances.  A benefit of our
result is that there is no constraint on classical communication, and
there are no added constraints on honest-prover performance.
Refs.~\cite{bluhm_singlequbit_2022,escola-farras_singlequbit_2023}
show non-constructively that there are device-dependent protocols
where the verifiers and prover require one qubit, but adversaries need
at least a linear total amount of quantum memory, classical memory and
classical communication to pass the protocol. However, it is an open
problem to have a better than logarithmic gap between the qubit
resource required by the verifiers and honest prover, and the
adversary prior entanglement in qubits, when the classical memory and
communication accessible by adversaries are unrestricted. For
Bell-test QPV, there are similar open questions.

Instead of bounding prior entanglement, we can bound qubits and
gates used by the adversary nearer \(\Vb\).

\noindent \textbf{Protocol soundness against adversaries with linear quantum
  operations:} \emph{There are efficiently implementable settings
  functions \(f(\cba,\cbb)\) such that for all \(\delta>0\), 
  Bell-test QPV is \(\delta\)-sound against adversaries with
  arbitrary classical communication but no quantum communication,
  and where the adversary nearer \(\Vap\) can use arbitrary local
  quantum resources, but the adversary nearer \(\Vb\) is restricted to
  \(O(l/(\log(q)+1))\) quantum operations in a basic universal set,
  where \(q\) is the number of qubits used by this adversary.}

The proof of this result is based on the proof of the previous
soundness result and given at the end of Supplementary
Information~\scref{sect:entangled_adversaries}.  Again, to support
practical applications, we fully quantify the bound on quantum
resources allowed by adversaries.  The proof is motivated by the
argument given in Ref.~\cite{asadi2025linear} to prove a similar
result for BB84 and routing QPV protocols. The suppressed constant in
the bound \(O(l/(\log(q)+1))\) depends on the size of the basic
universal set of quantum operations, which includes qubit
measurements.

\subsection*{Target-region comparison}

To benchmark our demonstration's performance, we compare the size of
the target region in our demonstration to the smallest achievable
target region with a classical protocol having the same latencies. In
the classical protocol illustrated in~\cref{fig:protocol} a),
verifiers \(\Vap\) and \(\Vb\) start by broadcasting challenge bits
\(\cba\) and \(\cbb\) at times \(s_{\Vap}\) and \(s_{\Vb}\), timed to
arrive simultaneously at some position between them. Appropriately
positioned provers are assumed to have access to these bits as soon as
they are released. An honest prover \(\Vp\), or adversaries, can
receive the challenge bits and use them as inputs to a predetermined
function \(g(\cba,\cbb )\) that is publicly known. The value of this
function determines the specific measurement to be performed. The
measurement outcome is then sent back to $\Vap$ and $\Vb$, who receive
$\zqa$ and $\zqb$, respectively, and jointly validate the results.

The verifiers choose the target region by deciding when they send
challenges and accept responses. There are four relevant times:
\(s_{\Vap}\) and \(s_{\Vb}\) are the times when \(\Vap\) and \(\Vb\)
send their challenge bits, and \(r_{\Vap}\) and \(r_{\Vb}\) are the
latest times when the verifiers accept the responses. This procedure
is repeated multiple times to build confidence in the honest running
of the protocol. Repeated late responses, or trials where
\(\zqa \neq \zqb\) will result in failure with high likelihood. We
show in the Supplementary Information~\scref{subsect:spacetimeana}
that target regions for the classical protocol and our quantum
protocol are determined by the speed of light \(c\) and the four
time intervals
\renewcommand{\theequation}{\arabic{equation}}
\begin{linenomath}
  \begin{align}
    \Delta\tau_1 &= (r_{\Vap} - s_{\Vap})/2, \label{eq:space_constraints1}\\
    \Delta\tau_2  &= (r_{\Vb} - s_{\Vb})/2,\label{eq:space_constraints2} \\
    \Delta\tau_3  &= r_{\Vap} - s_{\Vb},\label{eq:space_constraints3}\\
    \Delta\tau_4  &= r_{\Vb} - s_{\Vap}.\label{eq:space_constraints4}
  \end{align}%
\end{linenomath}  

The impossibility of superluminal signaling constrains the locations from which a stationary prover can receive \(\cba\) and reply to verifier \(\Vap\) to points in a sphere \(\Sp{\Vap}{c\Delta\tau_1}\) of radius \(c\times\Delta\tau_1\) centered on \(\Vap\). Similarly, \(c\times\Delta\tau_2\) is the radius of an analogous sphere \(\Sp{\Vb}{c\Delta\tau_2}\) centered on \(\Vb\). The set of all locations from which a stationary prover can receive \(\cbb\) (released at \(s_{\Vb}\)) and respond to verifier \(\Vap\) is a prolate ellipsoid \(\El{c\Delta\tau_3}\) with the verifier \(\Vap\) and \(\Vb\) locations as foci, and major axis  \(c\times\Delta\tau_3\). Prolate ellipsoids are the set of all points such that the sum of the distances from the verifiers is less than or equal to the major axis. This means any prover outside this shape could not receive \(\cbb\) and reply to \(\Vap\) within \(r_{\Vap}\). An analogous ellipsoid \(\El{c\Delta\tau_4}\) constrains receiving \(\cba\) and replying to \(\Vb\). 

If the verifiers can ensure that they only ever interact with a single
prover, the target region is the intersection of all the regions
listed above,
\(E\ = \Sp{\Vap}{c\Delta\tau_1} \cap \Sp{\Vb}{c\Delta\tau_2} \cap
\El{c\Delta\tau_3} \cap \El{c\Delta\tau_4}\). A projection of this
region to a 2-D plane for a particular choice of release and receive
times is indicated in yellow in~\cref{fig:protocol} d). This
single-prover target region could be made arbitrarily small if one
could approach zero-latency processing and direct speed-of-light
transmission. However, the verifiers have no way to enforce the
single-prover assumption in practice, and must allow for multiple
interacting provers. As we show in the Supplementary
Information~\scref{subsect:spacetimeana}, multiple adversaries can use
the ``classical spoofing'' protocol, shown in~\cref{fig:protocol} b),
to pass the protocol, provided they have access to the larger region
\(E_{\rm adv} = \Big(\Sp{\Vap}{c\Delta\tau_1}
\cap\El{c\Delta\tau_3}\Big) \cup \Big(\Sp{\Vb}{c\Delta\tau_2} \cap
\El{c\Delta\tau_4}\Big) \supseteq E\). This spoofing is
indistinguishable from an honest single-prover strategy from the
verifiers' point of view. A 2-D projection of \(E_{\rm adv}\) is
indicated with blue in~\cref{fig:protocol} c).  The target region for
the classical protocol \(E_{\rm adv}\) precludes remote position
verification because the target region includes the verifier positions
themselves in addition to the space between the verifiers. While
arbitrarily small target regions that include the verifier positions
are always possible with straightforward
ranging~\cite{sastry_secure_2003}, no classical protocols admit remote
position verification~\cite{chandran_positionbased_2014}

\subsection*{Note on Uncertainties}
All of the uncertainties reported in the main and supplementary texts are standard uncertainties (\(1\)-sigma). When we report a quantity with uncertainty as (for example) \qty{774.3(2)}{nm}, the number in parentheses is the numerical value of the combined standard uncertainty referred to the corresponding last digits of the quoted result.

\section*{Use of AI Tools}

AI tools were used in editing the text to help improve readability, to
assist in producing schematics, and as coding assistants for
implementing analysis code. The authors thoroughly reviewed text,
schematics and code produced by the AI tools and take full
responsibility for the final results.

\section*{Data Availability}
All experimental data from the quantum position verification protocol shown in this manuscript are available upon request from the corresponding authors.

\section*{Acknowledgment}
The authors would like to thank Peter Bierhorst and Michael Grayson for useful discussions, Yi-Kai Liu, Adam McCaughan and Aneesh Ramaswamy for close reading of the manuscript, and Daniel Sorensen for help with cable routing.  The authors acknowledge support from the National Science Foundation QLCI OMA-2016244, National Science Foundation NQVL:QSTD:Pilot 2435378, the National Science Foundation ECCS 2330228, the National Science Foundation OSI 2328800, the University of Colorado Quantum Engineering Initiative, the National Institute of Standards and Technology, and the U.S. Department of Energy, Office of Science, Advanced
Scientific Computing Research (Field Work Proposal ERKJ381). Yusuf Alnawakhtha acknowledges support from the Crown Prince International Scholarship Program and the Arabian Gulf University.

\section*{Disclaimer}
Any mention of commercial products within the manuscript is for information only; it does not imply recommendation or endorsement by NIST. This work includes contributions of the National Institute of Standards and Technology, which are not subject to U.S. copyright. The U.S. Government is authorized to reproduce and distribute reprints for governmental purposes notwithstanding any copyright annotation thereon.

Additionally, this manuscript has been co-authored by UT-Battelle, LLC, under contract DE-AC05-00OR22725 with the US Department of Energy (DOE). The US government retains and the publisher, by accepting the article for publication, acknowledges that the US government retains a nonexclusive, paid-up, irrevocable, worldwide license to publish or reproduce the published form of this manuscript, or allow others to do so, for US government purposes. DOE will provide public access to these results of federally sponsored research in accordance with the DOE Public Access Plan (http://energy.gov/downloads/doe-public-access-plan).

\section*{Disclosures}
The authors declare no conflicts of interest.

\section*{Author Contributions}
Conceptualization: G.A.K., L.K.S., E.K., Y.Z., S.G., M.J.S., H.F., C.A.M., Y.A., A.R.G., S.P., J.C.B.; Investigation: G.A.K., A.R.G., L.K.S., M.J.S., M.D.M; Methodology: Y.Z., G.A.K., L.K.S, E.K., S.G.; Software: G.A.K.,  Y.Z., L.K.S.; Formal analysis: Y.Z., S.P., E.K., G.A.K., S.G.C., L.K.S.; Security analysis: E.K., Y.Z., C.A.M., Y.A.; Funding acquisition: L.K.S., R.P.M., M.J.S.;  Project administration: L.K.S., E.K., M.J.S.; Resources: D.V.R., J.C.B., C.A., W.A., M.W.M.; Supervision: L.K.S., E.K., S.G., M.J.S., R.P.M.; Visualization: G.A.K., L.K.S., E.K., Y.Z.; Writing – original draft: G.A.K, E.K., L.K.S., Y.Z., S.G.;
Writing – review \& editing: G.A.K, E.K., Y.Z., S.G, L.K.S., S.P., A.R.G., H.F., Y.A., M.W.M., C.A.M.


\ignore{
Position information underpins many modern technologies, from navigation and timing to authentication and critical infrastructure. However, classical methods of proving that information originates from a particular position are vulnerable to spoofing. This  limitation can be overcome with quantum technologies but current protocols rely on trust in quantum hardware that can be undermined, or require quantum computers and bounds on adversarial computation.  Nevertheless, there has been significant interest in experimental demonstrations, and aspects of these protocols have been implemented. Here we introduce and experimentally demonstrate the Bell-test quantum position verification protocol for device-independent quantum position verification that guarantees security with only observed correlations from a loophole-free Bell test across a quantum network. We experimentally implement a version of this device-independent protocol against adversaries who, before each trial, are weakly entangled. Our demonstration achieves a one-dimensional localization 2.47(2) times smaller than the best, necessarily non-remote, classical localization protocol. Compared to classical protocols with identical latencies, the localization volume is 4.53(5) times smaller, and represents a certifiable quantum advantage. The general Bell-test protocol is loss tolerant and secure against adversaries with significant quantum resources. This work allows digital security to be anchored to physically trusted locations, enabling new position-based authentication protocols for applications such as financial transactions, legal agreements, and securing critical infrastructure.
  }

\bibliography{theory, experiment}

@article{alabdulatif_novel_2023,
  title = {A {{Novel Robust Geolocation-Based Multi-Factor Authentication Method}} for {{Securing ATM Payment Transactions}}},
  author = {Alabdulatif, Abdullah and Samarasinghe, Rohan and Thilakarathne, Navod Neranjan},
  year = 2023,
  month = sep,
  journal = {Applied Sciences},
  volume = {13},
  number = {19},
  pages = {10743},
  issn = {2076-3417},
  doi = {10.3390/app131910743},
  urldate = {2025-11-24},
  langid = {english}
}

@article{alawami_locauth_2020,
  title = {{{LocAuth}}: {{A}} Fine-Grained Indoor Location-Based Authentication System Using Wireless Networks Characteristics},
  shorttitle = {{{LocAuth}}},
  author = {Alawami, Mohsen A. and Kim, Hyoungshick},
  year = 2020,
  month = feb,
  journal = {Computers \& Security},
  volume = {89},
  pages = {101683},
  issn = {01674048},
  doi = {10.1016/j.cose.2019.101683},
  urldate = {2025-11-24},
  langid = {english}
}

@misc{allerstorfer_practical_2022,
  title = {Towards Practical and Error-Robust Quantum Position Verification},
  author = {Allerstorfer, Rene and Buhrman, Harry and Speelman, Florian and Lunel, Philip Verduyn},
  year = 2022,
  month = aug,
  number = {arXiv:2106.12911},
  eprint = {2106.12911},
  primaryclass = {quant-ph},
  publisher = {arXiv},
  doi = {10.48550/arXiv.2106.12911},
  urldate = {2025-05-30},
  archiveprefix = {arXiv}
}

@article{bluhm_singlequbit_2022,
  title = {A Single-Qubit Position Verification Protocol That Is Secure against Multi-Qubit Attacks},
  author = {Bluhm, Andreas and Christandl, Matthias and Speelman, Florian},
  year = 2022,
  month = jun,
  journal = {Nature Physics},
  volume = {18},
  number = {6},
  pages = {623--626},
  issn = {1745-2473, 1745-2481},
  doi = {10.1038/s41567-022-01577-0},
  urldate = {2025-11-24},
  langid = {english}
}

@inproceedings{brands_distancebounding_1994,
  title = {Distance-{{Bounding Protocols}}},
  booktitle = {Advances in {{Cryptology}} --- {{EUROCRYPT}} '93},
  author = {Brands, Stefan and Chaum, David},
  editor = {Helleseth, Tor},
  year = 1994,
  pages = {344--359},
  publisher = {Springer},
  address = {Berlin, Heidelberg},
  doi = {10.1007/3-540-48285-7_30},
  isbn = {978-3-540-48285-7},
  langid = {english}
}

@article{buhrman_positionbased_2014,
  title = {Position-{{Based Quantum Cryptography}}: {{Impossibility}} and {{Constructions}}},
  shorttitle = {Position-{{Based Quantum Cryptography}}},
  author = {Buhrman, Harry and Chandran, Nishanth and Fehr, Serge and Gelles, Ran and Goyal, Vipul and Ostrovsky, Rafail and Schaffner, Christian},
  year = 2014,
  month = jan,
  journal = {SIAM Journal on Computing},
  volume = {43},
  number = {1},
  pages = {150--178},
  issn = {0097-5397, 1095-7111},
  doi = {10.1137/130913687},
  urldate = {2025-01-13},
  langid = {english}
}

@article{chandran_positionbased_2014,
  title = {Position-{{Based Cryptography}}},
  author = {Chandran, Nishanth and Goyal, Vipul and Moriarty, Ryan and Ostrovsky, Rafail},
  year = 2014,
  month = jan,
  journal = {SIAM Journal on Computing},
  volume = {43},
  number = {4},
  pages = {1291--1341},
  issn = {0097-5397, 1095-7111},
  doi = {10.1137/100805005},
  urldate = {2025-01-13},
  langid = {english}
}

@incollection{cooke_coaxial_2021,
  title = {Coaxial {{Cables}}},
  booktitle = {The {{Global Cable Industry}}},
  author = {Cooke, Timothy},
  editor = {Beyer, G{\"u}nter},
  year = 2021,
  month = may,
  edition = {1},
  pages = {331--349},
  publisher = {Wiley},
  doi = {10.1002/9783527822263.ch12},
  urldate = {2025-11-10},
  copyright = {http://doi.wiley.com/10.1002/tdm\_license\_1.1},
  isbn = {978-3-527-34627-1 978-3-527-82226-3},
  langid = {english}
}

@article{das_practically_2021,
  title = {Practically Secure Quantum Position Verification},
  author = {Das, Siddhartha and Siopsis, George},
  year = 2021,
  month = jun,
  journal = {New Journal of Physics},
  volume = {23},
  number = {6},
  pages = {063069},
  issn = {1367-2630},
  doi = {10.1088/1367-2630/ac0755},
  urldate = {2025-11-24}
}

@article{denning_locationbased_1996,
  title = {Location-Based Authentication: {{Grounding}} Cyberspace for Better Security},
  shorttitle = {Location-Based Authentication},
  author = {Denning, Dorothy E. and MacDoran, Peter F.},
  year = 1996,
  month = feb,
  journal = {Computer Fraud \& Security},
  volume = {1996},
  number = {2},
  pages = {12--16},
  issn = {13613723},
  doi = {10.1016/S1361-3723(97)82613-9},
  urldate = {2025-11-24},
  copyright = {https://www.elsevier.com/tdm/userlicense/1.0/},
  langid = {english}
}

@inproceedings{drimer_keep_2007,
  title = {Keep Your Enemies Close: Distance Bounding against Smartcard Relay Attacks},
  shorttitle = {Keep Your Enemies Close},
  booktitle = {Proceedings of 16th {{USENIX Security Symposium}} on {{USENIX Security Symposium}}},
  author = {Drimer, Saar and Murdoch, Steven J.},
  year = 2007,
  month = aug,
  series = {{{SS}}'07},
  pages = {1--16},
  publisher = {USENIX Association},
  address = {USA},
  urldate = {2025-11-26}
}

@article{eberhard_background_1993,
  title = {Background Level and Counter Efficiencies Required for a Loophole-Free {{Einstein-Podolsky-Rosen}} Experiment},
  author = {Eberhard, Philippe H.},
  year = 1993,
  month = feb,
  journal = {Physical Review A},
  volume = {47},
  number = {2},
  pages = {R747-R750},
  issn = {1050-2947, 1094-1622},
  doi = {10.1103/PhysRevA.47.R747},
  urldate = {2022-05-19},
  langid = {english}
}

@article{escola-farras_singlequbit_2023,
  title = {Single-{{Qubit Loss-Tolerant Quantum Position Verification Protocol Secure}} against {{Entangled Attackers}}},
  author = {{Escol{\`a}-Farr{\`a}s}, Lloren{\c c} and Speelman, Florian},
  year = 2023,
  month = oct,
  journal = {Physical Review Letters},
  volume = {131},
  number = {14},
  pages = {140802},
  issn = {0031-9007, 1079-7114},
  doi = {10.1103/PhysRevLett.131.140802},
  urldate = {2025-01-09},
  langid = {english}
}

@article{giustina_significantloopholefree_2015,
  title = {Significant-{{Loophole-Free Test}} of {{Bell}}'s {{Theorem}} with {{Entangled Photons}}},
  author = {Giustina, Marissa and Versteegh, Marijn A. M. and Wengerowsky, S{\"o}ren and Handsteiner, Johannes and Hochrainer, Armin and Phelan, Kevin and Steinlechner, Fabian and Kofler, Johannes and Larsson, Jan-{\AA}ke and Abell{\'a}n, Carlos and Amaya, Waldimar and Pruneri, Valerio and Mitchell, Morgan W. and Beyer, J{\"o}rn and Gerrits, Thomas and Lita, Adriana E. and Shalm, Lynden K. and Nam, Sae Woo and Scheidl, Thomas and Ursin, Rupert and Wittmann, Bernhard and Zeilinger, Anton},
  year = 2015,
  month = dec,
  journal = {Physical Review Letters},
  volume = {115},
  number = {25},
  pages = {250401},
  issn = {0031-9007, 1079-7114},
  doi = {10.1103/PhysRevLett.115.250401},
  urldate = {2019-12-11},
  langid = {english}
}

@inproceedings{hancke_rfid_2005,
  title = {An {{RFID Distance Bounding Protocol}}},
  booktitle = {First {{International Conference}} on {{Security}} and {{Privacy}} for {{Emerging Areas}} in {{Communications Networks}} ({{SECURECOMM}}'05)},
  author = {Hancke, G.P. and Kuhn, M.G.},
  year = 2005,
  pages = {67--73},
  publisher = {IEEE},
  address = {Athens, Greece},
  doi = {10.1109/SECURECOMM.2005.56},
  urldate = {2025-11-25},
  isbn = {978-0-7695-2369-9}
}

@misc{kaleoglu_equivalence_2025,
  title = {On the {{Equivalence}} between {{Classical Position Verification}} and {{Certified Randomness}}},
  author = {Kaleoglu, Fatih and Liu, Minzhao and Chakraborty, Kaushik and Cui, David and Amer, Omar and Pistoia, Marco and Lim, Charles},
  year = 2025,
  month = jun,
  number = {arXiv:2410.03982},
  eprint = {2410.03982},
  primaryclass = {quant-ph},
  publisher = {arXiv},
  doi = {10.48550/arXiv.2410.03982},
  urldate = {2025-11-24},
  archiveprefix = {arXiv}
}

@misc{kanneworff_experimental_2025,
  title = {Towards Experimental Demonstration of Quantum Position Verification Using True Single Photons},
  author = {Kanneworff, Kirsten and Poortvliet, Mio and Bouwmeester, Dirk and Allerstorfer, Rene and Lunel, Philip Verduyn and Speelman, Florian and Buhrman, Harry and Steindl, Petr and L{\"o}ffler, Wolfgang},
  year = 2025,
  month = feb,
  number = {arXiv:2502.04125},
  eprint = {2502.04125},
  primaryclass = {quant-ph},
  publisher = {arXiv},
  doi = {10.48550/arXiv.2502.04125},
  urldate = {2025-05-30},
  archiveprefix = {arXiv}
}

@article{kavuri_traceable_2025,
  title = {Traceable Random Numbers from a Non-Local Quantum Advantage},
  author = {Kavuri, Gautam A. and Palfree, Jasper and Reddy, Dileep V. and Zhang, Yanbao and Bienfang, Joshua C. and Mazurek, Michael D. and Alhejji, Mohammad A. and Siddiqui, Aliza U. and Cavanagh, Joseph M. and Dalal, Aagam and Abell{\'a}n, Carlos and Amaya, Waldimar and Mitchell, Morgan W. and Stange, Katherine E. and Beale, Paul D. and Brand{\~a}o, Lu{\'i}s T. A. N. and Booth, Harold and Peralta, Ren{\'e} and Nam, Sae Woo and Mirin, Richard P. and Stevens, Martin J. and Knill, Emanuel and Shalm, Lynden K.},
  year = 2025,
  month = jun,
  journal = {Nature},
  volume = {642},
  number = {8069},
  pages = {916--921},
  issn = {0028-0836, 1476-4687},
  doi = {10.1038/s41586-025-09054-3},
  urldate = {2025-11-09},
  langid = {english}
}

@article{lim_losstolerant_2016,
  title = {Loss-Tolerant Quantum Secure Positioning with Weak Laser Sources},
  author = {Lim, Charles Ci Wen and Xu, Feihu and Siopsis, George and Chitambar, Eric and Evans, Philip G. and Qi, Bing},
  year = 2016,
  month = sep,
  journal = {Physical Review A},
  volume = {94},
  number = {3},
  pages = {032315},
  issn = {2469-9926, 2469-9934},
  doi = {10.1103/PhysRevA.94.032315},
  urldate = {2025-01-09},
  copyright = {http://link.aps.org/licenses/aps-default-license},
  langid = {english}
}

@misc{liu_certified_2025a,
  title = {Certified Randomness Amplification by Dynamically Probing Remote Random Quantum States},
  author = {Liu, Minzhao and Niroula, Pradeep and DeCross, Matthew and Foreman, Cameron and Kon, Wen Yu and Primaatmaja, Ignatius William and Allman, M. S. and Campora, J. P. and Isanaka, Akhil and Singhal, Kartik and Amer, Omar and Chakrabarti, Shouvanik and Chakraborty, Kaushik and Cooper, Samuel F. and Delaney, Robert D. and Dreiling, Joan M. and Estey, Brian and Figgatt, Caroline and Foltz, Cameron and Gaebler, John P. and Hall, Alex and He, Zichang and Holliman, Craig A. and Humble, Travis S. and Hung, Shih-Han and Husain, Ali A. and Jin, Yuwei and Kaleoglu, Fatih and Kennedy, Colin J. and Kotibhaskar, Nikhil and Lysne, Nathan K. and Madjarov, Ivaylo S. and Mills, Michael and Milne, Alistair R. and Milner, Kevin and Narmour, Louis and Omanakuttan, Sivaprasad and Park, Annie J. and Perlin, Michael A. and Reed, Adam P. and Self, Chris N. and Steinberg, Matthew and Stephen, David T. and Sullivan, Joseph and Chernoguzov, Alex and Curchod, Florian J. and Ransford, Anthony and Bohnet, Justin G. and Neyenhuis, Brian and {Foss-Feig}, Michael and Otter, Rob and Shaydulin, Ruslan},
  year = 2025,
  month = nov,
  number = {arXiv:2511.03686},
  eprint = {2511.03686},
  primaryclass = {quant-ph},
  publisher = {arXiv},
  doi = {10.48550/arXiv.2511.03686},
  urldate = {2025-11-24},
  archiveprefix = {arXiv}
}

@article{liu_deviceindependent_2021,
  title = {Device-Independent Randomness Expansion against Quantum Side Information},
  author = {Liu, Wen-Zhao and Li, Ming-Han and Ragy, Sammy and Zhao, Si-Ran and Bai, Bing and Liu, Yang and Brown, Peter J. and Zhang, Jun and Colbeck, Roger and Fan, Jingyun and Zhang, Qiang and Pan, Jian-Wei},
  year = 2021,
  month = apr,
  journal = {Nature Physics},
  volume = {17},
  number = {4},
  pages = {448--451},
  issn = {1745-2473, 1745-2481},
  doi = {10.1038/s41567-020-01147-2},
  urldate = {2025-02-04},
  langid = {english}
}

@article{li_experimental_2021,
  title = {Experimental {{Realization}} of {{Device-Independent Quantum Randomness Expansion}}},
  author = {Li, Ming-Han and Zhang, Xingjian and Liu, Wen-Zhao and Zhao, Si-Ran and Bai, Bing and Liu, Yang and Zhao, Qi and Peng, Yuxiang and Zhang, Jun and Zhang, Yanbao and Munro, W. J. and Ma, Xiongfeng and Zhang, Qiang and Fan, Jingyun and Pan, Jian-Wei},
  year = {2021},
  month = feb,
  journal = {Physical Review Letters},
  volume = {126},
  number = {5},
  pages = {050503},
  issn = {0031-9007, 1079-7114},
  doi = {10.1103/PhysRevLett.126.050503},
  urldate = {2022-11-19},
  langid = {english}
}

@article{lydersen_hacking_2010,
  title = {Hacking Commercial Quantum Cryptography Systems by Tailored Bright Illumination},
  author = {Lydersen, Lars and Wiechers, Carlos and Wittmann, Christoffer and Elser, Dominique and Skaar, Johannes and Makarov, Vadim},
  year = 2010,
  month = oct,
  journal = {Nature Photonics},
  volume = {4},
  number = {10},
  pages = {686--689},
  issn = {1749-4885, 1749-4893},
  doi = {10.1038/nphoton.2010.214},
  urldate = {2025-11-30},
  copyright = {http://www.springer.com/tdm},
  langid = {english}
}

@article{masanes_general_2006a,
  title = {General Properties of Nonsignaling Theories},
  author = {Masanes, {\relax Ll}. and Acin, A. and Gisin, N.},
  year = 2006,
  month = jan,
  journal = {Physical Review A},
  volume = {73},
  number = {1},
  pages = {012112},
  issn = {1050-2947, 1094-1622},
  doi = {10.1103/PhysRevA.73.012112},
  urldate = {2025-11-24},
  copyright = {http://link.aps.org/licenses/aps-default-license},
  langid = {english}
}

@article{nadlinger_experimental_2022,
  title = {Experimental Quantum Key Distribution Certified by {{Bell}}'s Theorem},
  author = {Nadlinger, D. P. and Drmota, P. and Nichol, B. C. and Araneda, G. and Main, D. and Srinivas, R. and Lucas, D. M. and Ballance, C. J. and Ivanov, K. and Tan, E. Y.-Z. and Sekatski, P. and Urbanke, R. L. and Renner, R. and Sangouard, N. and Bancal, J.-D.},
  year = 2022,
  month = jul,
  journal = {Nature},
  volume = {607},
  number = {7920},
  pages = {682--686},
  publisher = {Nature Publishing Group},
  issn = {1476-4687},
  doi = {10.1038/s41586-022-04941-5},
  urldate = {2025-05-14},
  copyright = {2022 The Author(s), under exclusive licence to Springer Nature Limited},
  langid = {english}
}

@article{rowe_experimental_2001,
  title = {Experimental Violation of a {{Bell}}'s Inequality with Efficient Detection},
  author = {Rowe, M. A. and Kielpinski, D. and Meyer, V. and Sackett, C. A. and Itano, W. M. and Monroe, C. and Wineland, D. J.},
  year = 2001,
  month = feb,
  journal = {Nature},
  volume = {409},
  number = {6822},
  pages = {791--794},
  publisher = {Nature Publishing Group},
  issn = {1476-4687},
  doi = {10.1038/35057215},
  urldate = {2025-06-03},
  copyright = {2001 Macmillan Magazines Ltd.},
  langid = {english}
}

@inproceedings{sastry_secure_2003,
  title = {Secure Verification of Location Claims},
  booktitle = {Proceedings of the 2nd {{ACM}} Workshop on {{Wireless}} Security},
  author = {Sastry, Naveen and Shankar, Umesh and Wagner, David},
  year = 2003,
  month = sep,
  pages = {1--10},
  publisher = {ACM},
  address = {San Diego CA USA},
  doi = {10.1145/941311.941313},
  urldate = {2025-11-25},
  isbn = {978-1-58113-769-9},
  langid = {english}
}

@article{shalm_deviceindependent_2021,
  title = {Device-Independent Randomness Expansion with Entangled Photons},
  author = {Shalm, Lynden K. and Zhang, Yanbao and Bienfang, Joshua C. and Schlager, Collin and Stevens, Martin J. and Mazurek, Michael D. and Abell{\'a}n, Carlos and Amaya, Waldimar and Mitchell, Morgan W. and Alhejji, Mohammad A. and Fu, Honghao and Ornstein, Joel and Mirin, Richard P. and Nam, Sae Woo and Knill, Emanuel},
  year = 2021,
  month = apr,
  journal = {Nature Physics},
  volume = {17},
  number = {4},
  pages = {452--456},
  issn = {1745-2473, 1745-2481},
  doi = {10.1038/s41567-020-01153-4},
  urldate = {2022-05-19},
  langid = {english}
}

@article{shalm_strong_2015,
  title = {Strong {{Loophole-Free Test}} of {{Local Realism}}},
  author = {Shalm, Lynden K. and {Meyer-Scott}, Evan and Christensen, Bradley G. and Bierhorst, Peter and Wayne, Michael A. and Stevens, Martin J. and Gerrits, Thomas and Glancy, Scott and Hamel, Deny R. and Allman, Michael S. and Coakley, Kevin J. and Dyer, Shellee D. and Hodge, Carson and Lita, Adriana E. and Verma, Varun B. and Lambrocco, Camilla and Tortorici, Edward and Migdall, Alan L. and Zhang, Yanbao and Kumor, Daniel R. and Farr, William H. and Marsili, Francesco and Shaw, Matthew D. and Stern, Jeffrey A. and Abell{\'a}n, Carlos and Amaya, Waldimar and Pruneri, Valerio and Jennewein, Thomas and Mitchell, Morgan W. and Kwiat, Paul G. and Bienfang, Joshua C. and Mirin, Richard P. and Knill, Emanuel and Nam, Sae Woo},
  year = 2015,
  month = dec,
  journal = {Physical Review Letters},
  volume = {115},
  number = {25},
  pages = {250402},
  issn = {0031-9007, 1079-7114},
  doi = {10.1103/PhysRevLett.115.250402},
  urldate = {2019-07-11},
  langid = {english}
}

@inproceedings{tippenhauer_attacks_2009,
  title = {Attacks on Public {{WLAN-based}} Positioning Systems},
  booktitle = {Proceedings of the 7th International Conference on {{Mobile}} Systems, Applications, and Services},
  author = {Tippenhauer, Nils Ole and Rasmussen, Kasper Bonne and P{\"o}pper, Christina and {\v C}apkun, Srdjan},
  year = 2009,
  month = jun,
  pages = {29--40},
  publisher = {ACM},
  address = {Krak\'ow Poland},
  doi = {10.1145/1555816.1555820},
  urldate = {2025-11-24},
  isbn = {978-1-60558-566-6},
  langid = {english}
}

@article{vidal_robustness_1999,
  title = {Robustness of Entanglement},
  author = {Vidal, Guifr{\'e} and Tarrach, Rolf},
  year = 1999,
  month = jan,
  journal = {Physical Review A},
  volume = {59},
  number = {1},
  pages = {141--155},
  issn = {1050-2947, 1094-1622},
  doi = {10.1103/PhysRevA.59.141},
  urldate = {2025-11-24},
  copyright = {http://link.aps.org/licenses/aps-default-license},
  langid = {english}
}

@article{weier_quantum_2011,
  title = {Quantum Eavesdropping without Interception: An Attack Exploiting the Dead Time of Single-Photon Detectors},
  shorttitle = {Quantum Eavesdropping without Interception},
  author = {Weier, Henning and Krauss, Harald and Rau, Markus and F{\"u}rst, Martin and Nauerth, Sebastian and Weinfurter, Harald},
  year = 2011,
  month = jul,
  journal = {New Journal of Physics},
  volume = {13},
  number = {7},
  pages = {073024},
  issn = {1367-2630},
  doi = {10.1088/1367-2630/13/7/073024},
  urldate = {2025-11-30}
}

@article{xu_deviceindependent_2022,
  title = {Device-{{Independent Quantum Key Distribution}} with {{Random Postselection}}},
  author = {Xu, Feihu and Zhang, Yu-Zhe and Zhang, Qiang and Pan, Jian-Wei},
  year = 2022,
  month = mar,
  journal = {Physical Review Letters},
  volume = {128},
  number = {11},
  pages = {110506},
  issn = {0031-9007, 1079-7114},
  doi = {10.1103/PhysRevLett.128.110506},
  urldate = {2025-11-09},
  langid = {english}
}

@article{Bell_nonlocality_review2014,
  title = {Bell nonlocality},
  author = {Brunner, Nicolas and Cavalcanti, Daniel and Pironio, Stefano and Scarani, Valerio and Wehner, Stephanie},
  journal = {Rev. Mod. Phys.},
  volume = {86},
  issue = {2},
  pages = {419--478},
  numpages = {60},
  year = {2014},
  month = {Apr},
  publisher = {American Physical Society},
  doi = {10.1103/RevModPhys.86.419},
  url = {https://link.aps.org/doi/10.1103/RevModPhys.86.419}
}

@article{loopholefree_shanghai2018,
  title = {Test of Local Realism into the Past without Detection and Locality Loopholes},
  author = {Li, Ming-Han and Wu, Cheng and Zhang, Yanbao and Liu, Wen-Zhao and Bai, Bing and Liu, Yang and Zhang, Weijun and Zhao, Qi and Li, Hao and Wang, Zhen and You, Lixing and Munro, W. J. and Yin, Juan and Zhang, Jun and Peng, Cheng-Zhi and Ma, Xiongfeng and Zhang, Qiang and Fan, Jingyun and Pan, Jian-Wei},
  journal = {Phys. Rev. Lett.},
  volume = {121},
  issue = {8},
  pages = {080404},
  numpages = {7},
  year = {2018},
  month = {Aug},
  publisher = {American Physical Society},
  doi = {10.1103/PhysRevLett.121.080404},
  url = {https://link.aps.org/doi/10.1103/PhysRevLett.121.080404}
}

@article{zhang_deviceindependent_2022,
  title = {A Device-Independent Quantum Key Distribution System for Distant Users},
  author = {Zhang, Wei and {van Leent}, Tim and Redeker, Kai and Garthoff, Robert and Schwonnek, Ren{\'e} and Fertig, Florian and Eppelt, Sebastian and Rosenfeld, Wenjamin and Scarani, Valerio and Lim, Charles C.-W. and Weinfurter, Harald},
  year = 2022,
  month = jul,
  journal = {Nature},
  volume = {607},
  number = {7920},
  pages = {687--691},
  publisher = {Nature Publishing Group},
  issn = {1476-4687},
  doi = {10.1038/s41586-022-04891-y},
  urldate = {2025-05-14},
  copyright = {2022 The Author(s)},
  langid = {english}
}

@inproceedings{zhang_locationbased_2012,
  title = {Location-{{Based Authentication}} and {{Authorization Using Smart Phones}}},
  booktitle = {2012 {{IEEE}} 11th {{International Conference}} on {{Trust}}, {{Security}} and {{Privacy}} in {{Computing}} and {{Communications}}},
  author = {Zhang, Feng and Kondoro, Aron and Muftic, Sead},
  year = 2012,
  month = jun,
  pages = {1285--1292},
  publisher = {IEEE},
  address = {Liverpool, United Kingdom},
  doi = {10.1109/TrustCom.2012.198},
  urldate = {2025-11-24},
  isbn = {978-1-4673-2172-3 978-0-7695-4745-9}
}

@article{Malaney_LocVer2010,
  title = {Location-dependent communications using quantum entanglement},
  author = {Malaney, Robert A.},
  journal = {Phys. Rev. A},
  volume = {81},
  issue = {4},
  pages = {042319},
  numpages = {4},
  year = {2010},
  month = {Apr},
  publisher = {American Physical Society},
  doi = {10.1103/PhysRevA.81.042319},
  url = {https://link.aps.org/doi/10.1103/PhysRevA.81.042319}
}

@article{Kent_TagVer2011,
  title = {Quantum tagging: Authenticating location via quantum information and relativistic signaling constraints},
  author = {Kent, Adrian and Munro, William J. and Spiller, Timothy P.},
  journal = {Phys. Rev. A},
  volume = {84},
  issue = {1},
  pages = {012326},
  numpages = {7},
  year = {2011},
  month = {Jul},
  publisher = {American Physical Society},
  doi = {10.1103/PhysRevA.84.012326},
  url = {https://link.aps.org/doi/10.1103/PhysRevA.84.012326}
}

@unpublished{fan2026relativistic,
  title={Relativistic Position Verification with Coherent States},
  author={Fan-Yuan, Guan-Jie and Shan, Yang-Guang and Zhang, Cong and Wang, Yu-Long and Fan, Yu-Xuan and Xie, Wei-Xin and He, De-Yong and Wang, Shuang and Yin, Zhen-Qiang and Chen, Wei and others},
  note={arXiv preprint arXiv:2602.01787},
  year={2026},
  doi={10.48550/arXiv.2602.01787}
}

@inproceedings{bhamidipati_gps_2018,
  title = {{{GPS}} Time Authentication against Spoofing via a Network of Receivers for Power Systems},
  booktitle = {2018 {{IEEE}}/{{ION Position}}, {{Location}} and {{Navigation Symposium}} ({{PLANS}})},
  author = {Bhamidipati, Sriramya and Mina, Tara Yasmin and Gao, Grace Xingxin},
  year = 2018,
  month = apr,
  pages = {1485--1491},
  publisher = {IEEE},
  address = {Monterey, CA},
  doi = {10.1109/PLANS.2018.8373542},
  urldate = {2026-05-24},
  isbn = {978-1-5386-1647-5}
}

@article{ardizzon_authenticated_2022,
  title = {Authenticated {{Timing Protocol Based}} on {{Galileo ACAS}}},
  author = {Ardizzon, Francesco and Crosara, Laura and Laurenti, Nicola and Tomasin, Stefano and Montini, Nicola},
  year = 2022,
  month = aug,
  journal = {Sensors},
  volume = {22},
  number = {16},
  pages = {6298},
  issn = {1424-8220},
  doi = {10.3390/s22166298},
  urldate = {2026-05-24},
  langid = {english}
}

@article{AbellanPRL2015,
author = {Abell\'an, Carlos and Amaya, Waldimar and Mitrani, Daniel and Pruneri, Valerio and Mitchell, Morgan W.},
doi = {10.1103/PhysRevLett.115.250403},
issue = {25},
journal = {Physical Review Letters},
journal-iso = {Phys. Rev. Lett.},
month = {Dec},
numpages = {5},
pages = {250403},
publisher = {American Physical Society},
title = {Generation of Fresh and Pure Random Numbers for Loophole-Free {Bell} Tests},
url = {http://link.aps.org/doi/10.1103/PhysRevLett.115.250403},
volume = {115},
year = {2015}}

@article{langner_stuxnet_2011,
  title = {Stuxnet: {{Dissecting}} a {{Cyberwarfare Weapon}}},
  shorttitle = {Stuxnet},
  author = {Langner, Ralph},
  year = 2011,
  month = may,
  journal = {IEEE Security \& Privacy Magazine},
  volume = {9},
  number = {3},
  pages = {49--51},
  issn = {1540-7993},
  doi = {10.1109/MSP.2011.67},
  urldate = {2026-05-25},
  copyright = {https://ieeexplore.ieee.org/Xplorehelp/downloads/license-information/IEEE.html}
}

@book{anderson_security_2020b,
  title = {Security Engineering: A Guide to Building Dependable Distributed Systems},
  shorttitle = {Security Engineering},
  author = {Anderson, Ross},
  year = 2020,
  edition = {Third edition},
  publisher = {John Wiley \& Sons},
  address = {Indianapolis},
  isbn = {978-1-119-64278-7 978-1-119-64468-2 978-1-119-64283-1 978-1-119-64281-7},
  langid = {english}
}

@book{whitman_principles_2009,
  title = {Principles of Information Security},
  author = {Whitman, Michael E. and Mattord, Herbert J.},
  year = 2009,
  edition = {3. ed},
  publisher = {Thomson Course Technology},
  address = {Boston, Mass},
  isbn = {978-1-4239-0177-8},
  langid = {english}
}

@article{fernandez-hernandez_navigation_2016,
  title = {A {{Navigation Message Authentication Proposal}} for the {{Galileo Open Service}}: {{NMA}} Proposal for {{Galileo OS}}},
  shorttitle = {A {{Navigation Message Authentication Proposal}} for the {{Galileo Open Service}}},
  author = {{Fern{\'a}ndez-Hern{\'a}ndez}, Ignacio and Rijmen, Vincent and {Seco-Granados}, Gonzalo and Simon, Javier and Rodr{\'i}guez, Irma and Calle, J. David},
  year = 2016,
  month = mar,
  journal = {Navigation},
  volume = {63},
  number = {1},
  pages = {85--102},
  issn = {00281522},
  doi = {10.1002/navi.125},
  urldate = {2026-05-25},
  copyright = {http://doi.wiley.com/10.1002/tdm\_license\_1.1},
  langid = {english}
}

@misc{cowperthwaite_practical_2023,
  title = {Towards Practical Quantum Position Verification},
  author = {Cowperthwaite, George and Kent, Adrian and {Pitalua-Garcia}, Damian},
  year = 2023,
  month = nov,
  number = {arXiv:2309.10070},
  eprint = {2309.10070},
  primaryclass = {quant-ph},
  publisher = {arXiv},
  doi = {10.48550/arXiv.2309.10070},
  urldate = {2026-05-31},
  archiveprefix = {arXiv}
}

@article{storz_loopholefree_2023,
  title = {Loophole-Free {{Bell}} Inequality Violation with Superconducting Circuits},
  author = {Storz, Simon and Sch{\"a}r, Josua and Kulikov, Anatoly and Magnard, Paul and Kurpiers, Philipp and L{\"u}tolf, Janis and Walter, Theo and Copetudo, Adrian and Reuer, Kevin and Akin, Abdulkadir and Besse, Jean-Claude and Gabureac, Mihai and Norris, Graham J. and Rosario, Andr{\'e}s and Martin, Ferran and Martinez, Jos{\'e} and Amaya, Waldimar and Mitchell, Morgan W. and Abellan, Carlos and Bancal, Jean-Daniel and Sangouard, Nicolas and Royer, Baptiste and Blais, Alexandre and Wallraff, Andreas},
  year = 2023,
  month = may,
  journal = {Nature},
  volume = {617},
  number = {7960},
  pages = {265--270},
  publisher = {Nature Publishing Group},
  issn = {1476-4687},
  doi = {10.1038/s41586-023-05885-0},
  urldate = {2024-05-15},
  copyright = {2023 The Author(s)},
  langid = {english}
}

@article{rosenfeld_eventready_2017,
  title = {Event-{{Ready Bell Test Using Entangled Atoms Simultaneously Closing Detection}} and {{Locality Loopholes}}},
  author = {Rosenfeld, Wenjamin and Burchardt, Daniel and Garthoff, Robert and Redeker, Kai and Ortegel, Norbert and Rau, Markus and Weinfurter, Harald},
  year = 2017,
  month = jul,
  journal = {Physical Review Letters},
  volume = {119},
  number = {1},
  pages = {010402},
  publisher = {American Physical Society},
  doi = {10.1103/PhysRevLett.119.010402},
  urldate = {2025-05-14}
}

@article{hensen_loopholefree_2015,
  title = {Loophole-Free {{Bell}} Inequality Violation Using Electron Spins Separated by 1.3 Kilometres},
  author = {Hensen, B. and Bernien, H. and Dr{\'e}au, A. E. and Reiserer, A. and Kalb, N. and Blok, M. S. and Ruitenberg, J. and Vermeulen, R. F. L. and Schouten, R. N. and Abell{\'a}n, C. and Amaya, W. and Pruneri, V. and Mitchell, M. W. and Markham, M. and Twitchen, D. J. and Elkouss, D. and Wehner, S. and Taminiau, T. H. and Hanson, R.},
  year = 2015,
  month = oct,
  journal = {Nature},
  volume = {526},
  number = {7575},
  pages = {682--686},
  issn = {0028-0836, 1476-4687},
  doi = {10.1038/nature15759},
  urldate = {2019-12-11},
  langid = {english}
}

@article{Kulikov2026,
	author = {Kulikov, Anatoly and Storz, Simon and Sch{\"a}r, Josua D. and Sandfuchs, Martin and Wolf, Ramona and Berterotti{\`e}re, Florence and Hellings, Christoph and Wallraff, Andreas and Renner, Renato},
	date = {2026/05/01},
	doi = {10.1038/s41586-026-10521-8},
	id = {Kulikov2026},
	isbn = {1476-4687},
	journal = {Nature},
	number = {8116},
	pages = {1033--1038},
	title = {Experimental randomness amplification},
	url = {https://doi.org/10.1038/s41586-026-10521-8},
	volume = {653},
	year = {2026}}

@article{Liu_LongLivedHeraledEnt_2026,
  title = {Long-lived remote ion-ion entanglement for scalable quantum repeaters},
  author = {Liu, Wen-Zhao and Zhou, Ya-Bin and Chen, Jiu-Peng and Wang, Bin and Teng, Ao and Han, Xiao-Wen and Liu, Guang-Cheng and Zhang, Zhi-Jiong and Yang, Yi and Liu, Feng-Guang and Xue, Chao-Hui and Yang, Bo-Wen and Yang, Jin and Zeng, Chao and Pan, Du-Ruo and Zheng, Ming-Yang and Zhang, Xing-Jian and Shen, Cao and Zhen, Yi-Zheng and Xiao, You and Li, Hao and You, Li-Xing and Ma, Xiong-Feng and Zhao, Qi and Xu, Feihu and Wang, Ye and Wan, Yong and Zhang, Qiang and Pan, Jian-Wei},
  journal = {Nature},
  volume = {652},
  pages = {51--57},
  year = {2026},
  doi = {10.1038/s41586-026-10177-4},
  url = {https://doi.org/10.1038/s41586-026-10177-4}
}

@article{lu_deviceindependent_2026,
  title = {Device-Independent Quantum Key Distribution over 100 Km with Single Atoms},
  author = {Lu, Bo-Wei and Yang, Chao-Wei and Wang, Run-Qi and Gao, Bo-Feng and Zhen, Yi-Zheng and Wang, Zhen-Gang and Shi, Jia-Kai and Ren, Zhong-Qi and Hahn, Thomas A. and Tan, Ernest Y.-Z. and Xie, Xiu-Ping and Zheng, Ming-Yang and Jiang, Xiao and Zhang, Jun and Xu, Feihu and Zhang, Qiang and Bao, Xiao-Hui and Pan, Jian-Wei},
  year = 2026,
  month = feb,
  journal = {Science},
  volume = {391},
  number = {6785},
  pages = {592--597},
  issn = {0036-8075, 1095-9203},
  doi = {10.1126/science.aec6243},
  urldate = {2026-05-17},
  langid = {english}
}

@article{bluhm2022single,
  title={A single-qubit position verification protocol that is secure against multi-qubit attacks},
  author={Bluhm, Andreas and Christandl, Matthias and Speelman, Florian},
  journal={Nature Physics},
  volume={18},
  number={6},
  pages={623--626},
  year={2022},
  publisher={Nature Publishing Group UK London},
  url={https://www.nature.com/articles/s41567-022-01577-0}
}

@article{zhang_y:qc2011a,
  author = "Y. Zhang and S. Glancy and E. Knill",
  title = "Asymptotically Optimal Data Analysis for Rejecting Local Realism",
  journal = "Phys. Rev. A",
  volume = 84,
  pages = "062118/1--10",
  xxxep = "arXiv:1108.2468",
  year = 2011,
  note = "For supporting code, see arXiv:1108.2468."
}

@article{zhang_y:qc2013a,
  author = "Y. Zhang and S. Glancy and E. Knill",
  title = "Efficient Quantification of Experimental Evidence Against Local Realism",
  journal = "Phys. Rev. A",
  volume = 88,
  pages = "052119/1--8",
  xxxep = "arXiv:1303.7464",
  year = 2013
}

@article{rudolph2005further,
  title={Further results on the cross norm criterion for separability},
  author={Rudolph, Oliver},
  journal={Quantum Information Processing},
  volume={4},
  pages={219--239},
  year={2005},
  publisher={Springer},
  url={https://arxiv.org/abs/quant-ph/0202121}
}

@article{zhang2020experimental,
  title={Experimental low-latency device-independent quantum randomness},
  author={Zhang, Yanbao and Shalm, Lynden K and Bienfang, Joshua C and Stevens, Martin J and Mazurek, Michael D and Nam, Sae Woo and Abell{\'a}n, Carlos and Amaya, Waldimar and Mitchell, Morgan W and Fu, Honghao and others},
  journal={Physical review letters},
  volume={124},
  number={1},
  pages={010505},
  year={2020},
  publisher={APS},
  doi={10.1103/PhysRevLett.124.010505},
  note={arXiv:1812.07786}
}

@inproceedings{chandran2009position,
  title={Position based cryptography},
  author={Chandran, Nishanth and Goyal, Vipul and Moriarty, Ryan and Ostrovsky, Rafail},
  booktitle={Annual International Cryptology Conference},
  pages={391--407},
  year={2009},
  organization={Springer},
  url={https://ia.cr/2009/364}
}

@article{buhrman2014position,
  title={Position-based quantum cryptography: Impossibility and constructions},
  author={Buhrman, Harry and Chandran, Nishanth and Fehr, Serge and Gelles, Ran and Goyal, Vipul and Ostrovsky, Rafail and Schaffner, Christian},
  journal={SIAM Journal on Computing},
  volume={43},
  number={1},
  pages={150--178},
  year={2014},
  publisher={SIAM},
  note={arXiv:1009.2490}
}

@inproceedings{unruh2014quantum,
  title={Quantum position verification in the random oracle model},
  author={Unruh, Dominique},
  booktitle={Advances in Cryptology--CRYPTO 2014: 34th Annual Cryptology Conference, Santa Barbara, CA, USA, August 17-21, 2014, Proceedings, Part II 34},
  pages={1--18},
  year={2014},
  organization={Springer}
}

@article{beigi2011simplified,
  title={Simplified instantaneous non-local quantum computation with applications to position-based cryptography},
  author={Beigi, Salman and K{\"o}nig, Robert},
  journal={New Journal of Physics},
  volume={13},
  number={9},
  pages={093036},
  year={2011},
  doi={10.1088/1367-2630/13/9/093036}
}

@article{shafer:qc2009a,
  author = "G. Shafer and A. Shen and N. Vereshchagin and V. Vovk",
  title = "Test Martingales, {Bayes} Factors and $p$-Values",
  journal = "Statistical Science",
  volume = 26,
  pages = "84--101",
  year = 2011,
  doi = "10.1214/10-STS347"
}

@article{Masanes2006,
  title = {General properties of nonsignaling theories},
  author = {Masanes, Ll. and Acin, A. and Gisin, N.},
  journal = {Phys. Rev. A},
  volume = {73},
  issue = {1},
  pages = {012112},
  numpages = {9},
  year = {2006},
  month = {01},
  publisher = {American Physical Society},
  doi = {10.1103/PhysRevA.73.012112}
}

@article{brunner2007detection,
  title={Detection loophole in asymmetric {B}ell experiments},
  author={Brunner, Nicolas and Gisin, Nicolas and Scarani, Valerio and Simon, Christoph},
  journal={Physical Review Letters},
  volume={98},
  number={22},
  pages={220403},
  year={2007},
  publisher={APS},
  doi={10.1103/PhysRevLett.98.220403}
}

@article{brunner2014bell,
  title={Bell nonlocality},
  author={Brunner, Nicolas and Cavalcanti, Daniel and Pironio, Stefano
and Scarani, Valerio and Wehner, Stephanie},
  journal={Reviews of Modern Physics},
  volume={86},
  number={2},
  pages={419--478},
  year={2014},
  publisher={APS},
  doi="10.1103/RevModPhys.86.419"
}

@article{eberhard1993background,
  title={Background level and counter efficiencies required for a loophole-free Einstein-Podolsky-Rosen experiment},
  author={Eberhard, Philippe H},
  journal={Physical Review A},
  volume={47},
  number={2},
  pages={R747},
  year={1993},
  publisher={APS},
  doi={10.1103/PhysRevA.47.R747}
}

@article{eddins2022doubling,
  title={Doubling the size of quantum simulators by entanglement forging},
  author={Eddins, Andrew and Motta, Mario and Gujarati, Tanvi P and Bravyi, Sergey and Mezzacapo, Antonio and Hadfield, Charles and Sheldon, Sarah},
  journal={PRX Quantum},
  volume={3},
  number={1},
  pages={010309},
  year={2022},
  publisher={APS},
  doi={10.1103/PRXQuantum.3.010309}
}

@unpublished{knill:qc2017a,
  author = "E. Knill and Y. Zhang and P. Bierhorst",
  title = "Quantum Randomness Generation by Probability Estimation with Classical Side Information",
  xxxep = "arXiv:1709.06159",
  note = "arXiv:1709.06159",
  year = 2017
}

@article{Pironio_2011,
doi = {10.1088/1751-8113/44/6/065303},
year = {2011},
month = {01},
publisher = {},
volume = {44},
number = {6},
pages = {065303},
author = {Pironio, Stefano and Bancal, Jean-Daniel and Scarani, Valerio},
title = {Extremal correlations of the tripartite no-signaling polytope},
journal = {Journal of Physics A: Mathematical and Theoretical}
}

@book{wald1984generalrelativity,
  author = {Wald, Robert M.},
  title = {General Relativity},
  publisher = {University of Chicago Press},
  year = {1984},
  address = {Chicago},
  isbn = {978-0226870335}
}

@article{vidal1999robustness,
  title={Robustness of entanglement},
  author={Vidal, Guifr{\'e} and Tarrach, Rolf},
  journal={Physical Review A},
  volume={59},
  number={1},
  pages={141},
  year={1999},
  publisher={APS},
  doi={10.1103/PhysRevA.59.141}
}

@article{zhang_certifying_2018,
  title = {Certifying {{Quantum Randomness}} by {{Probability Estimation}}},
  author = {Zhang, Yanbao and Knill, Emanuel and Bierhorst, Peter},
  year = {2018},
  month = oct,
  journal = {Physical Review A},
  volume = {98},
  number = {4},
  eprint = {1811.11928},
  primaryclass = {quant-ph},
  pages = {040304},
  issn = {2469-9926, 2469-9934},
  doi = {10.1103/PhysRevA.98.040304},
  urldate = {2024-02-01},
  archiveprefix = {arXiv}
}

@article{allerstorfer2025making,
  title={Making existing quantum position verification protocols secure against arbitrary transmission loss},
  author={Allerstorfer, Rene and Bluhm, Andreas and Buhrman, Harry and Christandl, Matthias and Escol{\`a}-Farr{\`a}s, Lloren{\c{c}} and Speelman, Florian and Verduyn Lunel, Philip},
  journal={Physical Review Letters},
  volume={135},
  number={26},
  pages={260801},
  year={2025},
  publisher={APS},
  doi={10.1103/szwj-s7r6}
}

@article{duan2001long,
  title={Long-distance quantum communication with atomic ensembles and linear optics},
  author={Duan, L-M and Lukin, Mikhail D and Cirac, J Ignacio and Zoller, Peter},
  journal={Nature},
  volume={414},
  number={6862},
  pages={413--418},
  year={2001},
  publisher={Nature Publishing Group UK London},
  doi={10.1038/35106500}
}

@article{tomamichel2013monogamy,
  title={A monogamy-of-entanglement game with applications to device-independent quantum cryptography},
  author={Tomamichel, Marco and Fehr, Serge and Kaniewski, Jedrzej and Wehner, Stephanie},
  journal={New Journal of Physics},
  volume={15},
  number={10},
  pages={103002},
  year={2013},
  publisher={IOP Publishing},
  doi={10.1088/1367-2630/15/10/103002}
}

@inproceedings{linial2007lower,
  title={Lower bounds in communication complexity based on factorization norms},
  author={Linial, Nati and Shraibman, Adi},
  booktitle={Proceedings of the thirty-ninth annual ACM symposium on Theory of computing},
  pages={699--708},
  year={2007},
  doi={10.1145/1250790.1250892},
  mnotes={Not self-contained, hard to interpret, no examples, but
                  supposedly shows that prior entanglement doesn't
                  help much most of the time for communication
                  complexity. It states a quantum communication with
                  entanglement lower bound in terms of factorization
                  norms. Forster's theorem form "A linear lower
                  bound..." should help with the specific bounds for
                  the inner product function. }
                  }

@book{vershynin2020high,
  title={High-dimensional probability},
  author={Vershynin, Roman},
  publisher={Cambridge University Press},
  edition={2nd},
  year={2026},
  doi={10.1017/9781009490672},
url={https://users.math.msu.edu/users/iwenmark/Teaching/MTH994/Fall2022/HDP-book.pdf}
}

@incollection{carlet2010boolean,
  title={Boolean Functions for Cryptography and Error-Correcting Codes},
  author={Carlet, Claude},
  year={2010},
  booktitle={Boolean models and methods in mathematics, computer science, and engineering},
  series={Encyclopedia of Mathematics and its Applications},
  pages={257-397},
  publisher={Cambridge University Press}
}

@article{rothaus1976bent,
  title={On “bent” functions},
  author={Rothaus, Oscar S},
  journal={Journal of Combinatorial Theory, Series A},
  volume={20},
  number={3},
  pages={300--305},
  year={1976},
  publisher={Elsevier},
  doi={10.1016/0097-3165(76)90024-8}
}

@article{zhang2009communication,
  title={Communication complexities of symmetric XOR functions},
  author={Zhang, Zhiqiang and Shi, Yaoyun},
  journal={Quantum information and computation},
  volume={9},
  number={3\&4},
  pages={255--263},
  year={2009},
  doi={10.5555/2011781.2011786}
}

@article{cleve2013quantum,
  title={Quantum entanglement and the communication complexity of the inner product function},
  author={Cleve, Richard and Van Dam, Wim and Nielsen, Michael and Tapp, Alain},
  journal={Theoretical Computer Science},
  volume={486},
  pages={11--19},
  year={2013},
  publisher={Elsevier},
  doi={10.1016/j.tcs.2012.12.012}
}

@unpublished{montanaro2009communication,
  title={On the communication complexity of XOR functions},
  author={Montanaro, Ashley and Osborne, Tobias},
  note={arXiv preprint arXiv:0909.3392},
  year={2009}
}

@unpublished{miller2024perfect,
  title={Perfect cheating is impossible for single-qubit position verification},
  author={Miller, Carl A and Alnawakhtha, Yusuf},
  note={arXiv preprint arXiv:2406.20022},
  year={2024},
  doi={10.48550/arXiv.2406.20022}
}

@book{bhatia2013matrix,
  title={Matrix analysis},
  author={Bhatia, Rajendra},
  year={2013},
  publisher={Springer Science \& Business Media}
}

@inproceedings{chee1994semi,
  title={Semi-bent Functions},
  author={Chee, Seongtaek and Lee, Sangjin and Kim, Kwangjo},
  booktitle={Proceedings of the 4th International Conference on the Theory and Applications of Cryptology: Advances in Cryptology},
  pages={107--118},
  year={1994},
  doi={10.5555/647092.714074}
}

@article{asadi2025linear,
  title={Linear gate bounds against natural functions for position-verification},
  author={Asadi, Vahid and Cleve, Richard and Culf, Eric and May, Alex},
  journal={Quantum},
  volume={9},
  pages={1604},
  year={2025},
  doi={10.22331/q-2025-01-21-1604}
}

@inproceedings{buhrman2013garden,
  title={The garden-hose model},
  author={Buhrman, Harry and Fehr, Serge and Schaffner, Christian and Speelman, Florian},
  booktitle={Proceedings of the 4th conference on Innovations in Theoretical Computer Science},
  pages={145--158},
  year={2013},
  doi={10.1145/2422436.2422455}
}

@article{escola2023single,
  title={Single-qubit loss-tolerant quantum position verification protocol secure against entangled attackers},
  author={Escol{\`a}-Farr{\`a}s, Lloren{\c{c}} and Speelman, Florian},
  journal={Physical Review Letters},
  volume={131},
  number={14},
  pages={140802},
  year={2023},
  publisher={APS},
  doi={10.1103/PhysRevLett.131.140802}
}

@phdthesis{verduyn2024quantum,
  title={Quantum Position Verification: Loss-tolerant Protocols and Fundamental Limits},
  author={Verduyn Lunel, Philip},
  year={2024},
  school={Universiteit van Amsterdam},
  url={https://eprints.illc.uva.nl/id/eprint/2320/1/DS-2024-08.text.pdf}
}

@article{vovk2021values,
  title={E-values: Calibration, combination and applications},
  author={Vovk, Vladimir and Wang, Ruodu},
  journal={The Annals of Statistics},
  volume={49},
  number={3},
  pages={1736--1754},
  year={2021},
  publisher={Institute of Mathematical Statistics},
  doi={10.1214/20-AOS2020}
}

@article{grunwald2024safe,
  title={Safe testing},
  author={Gr{\"u}nwald, Peter and de Heide, Rianne and Koolen, Wouter},
  journal={Journal of the Royal Statistical Society Series B: Statistical Methodology},
  volume={86},
  number={5},
  pages={1091--1128},
  year={2024},
  publisher={Oxford University Press UK},
  doi={10.1093/jrsssb/qkae011}
}

\section*{Supplementary Information}

\addtocontents{toc}{\string\tocdepth@restore}

\tableofcontents

\section{Preliminaries}
\label{sec:preliminaries}

The Bell-test quantum position verification (QPV) protocol involves
three verifiers \(\Vap\), \(\Va\) and \(\Vb\), and a source of
entangled pairs of arbitrary quantum systems.  One member of each
entangled pair is sent to \(\Va\), who performs measurements on this
member. The other member of the pair is sent to a target region
\(\tR\).  Verifiers \(\Vap\) and \(\Vb\) send challenges and receive
prover responses. In our experimental implementation, \(\Vap\) and
\(\Va\) are co-located in the same lab, and we sometimes call \(\Va\)
a ``measurement station''. The entangled quantum systems are optical
wavepacket modes designed to contain a polarization-entangled pair of
photons.  We employ the label \(\Vs\) for the source of entangled
photons.  In the experiment, the source is co-located with \(\Va\) and
\(\Vap\). The goal of the Bell-test protocol is to verify that a
prover is in the target region \(\tR\).  The protocol consists of a
sequence of \(n\) trials, and success of the verification and the
protocol is determined by analyzing the statistics of the trial
records. For provers whose movements are unrestricted, success of the
protocol certifies that they are in the target region during at least
one trial of the protocol. If the protocol succeeds for a prover, we
say that the prover passes the protocol.  Our analysis shows that to
succeed, the prover must perform some non-trivial quantum operation in
the target region. The honest prover \(\Vp\) is in the target region
during every trial of the protocol.  The protocol is designed so that
such an honest prover can ensure that the protocol succeeds with a
reasonably high probability, while for any cooperating collection of
provers who are never in the target region the protocol fails with
high probability. Provers attempting to pass the Bell-test
protocol without being in the target region are called
adversaries. We demonstrate the honest prover strategy experimentally
by implementing a measurement station \(\Vp\) in a lab physically
separated from the verifier \(\Vap\) and \(\Vb\) labs.

In each trial of the Bell-test protocol, the source prepares two
quantum systems \(\Qa\) and \(\Qp\) and sends \(\Qa\) to \(\Va\) and
\(\Qp\) to the target region.  In spacetime, the target region \(\tR\)
consists of a disjoint union of target regions \(\tR_{k}\), where
\(\tR_{k}\) is contained inside the time interval for the \(k\)'th
trial.  For the \(k\)'th trial, \(\Vap\) and \(\Vb\) choose challenges
\(\cba_{k}\) and \(\cbb_{k}\) from finite sets \(I_{\cba}\) and
\(I_{\cbb}\) according to a probability distribution specified by the
protocol. In the experiment, the two challenges are bits and the
distribution is uniform.  When it is necessary to distinguish random
variables (RVs) from their values, we use standard capitalization
conventions. For example, \(\Cba_{k}\) and \(\Cbb_{k}\) are the random
variables whose instances have values \(\cba_{k}\) and \(\cbb_{k}\),
respectively. We use \(\Prob(\ldots)\) to denote probabilities of
events and \(\Exp(\ldots)\) to denote expectations of real-valued RVs.
We assume that the values of random variables belong to finite sets of
at least two elements. In the experiment, \(\Cba_{k}\) and
\(\Cbb_{k}\) are binary, independent and uniformly random.  Unless
explicitly stated, we do not assume that the trials are independent or
identically distributed.  Random variables without trial-index
subscripts refer to the random variables of a generic trial and are
used when the specific trial does not matter. In particular, for the
remainder of this section, we consider a generic trial of the
Bell-test protocol and drop the trial-index subscripts.

The honest prover receives \(\cba\) and \(\cbb\) and computes an
agreed-upon and publicly known prover settings function
\(\sfun(\cba,\cbb)\) with range \(\{1,\ldots,k_{\Vp}\}\). In parallel,
\(\Va\) chooses a measurement setting
\(\mqa\in \{1,\ldots,k_{\Va}\}\).  \(\Va\) applies the measurement
according to the setting \(\mqa\) and obtains measurement outcome
\(\oqa\in\{1,\ldots,c_{\Va}\}\).  The honest prover chooses their
measurement setting \(\mqp\) according to \(\mqp=\sfun(\cba,\cbb)\),
applies the corresponding measurement and obtains measurement outcome
\(\oqp\in\{1,\ldots,c_{\Vp}\}\).  The honest prover then sends
\(\oqp\) to each of \(\Vap\) and \(\Vb\).  In general, \(\Vap\) and
\(\Vb\) may receive responses from imperfect provers or adversaries,
in which case the two responses may not agree. The responses actually
received by \(\Vap\) and \(\Vb\) are denoted by \(\zqa\) and \(\zqb\),
where these values may belong to a larger set than \(\oqp\), for
example, to account for absence of responses and invalid responses.

In the experiment \(\sfun(\cba,\cbb)=(\cba -1)\oplus(\cbb-1)+ 1\),
where \(\cba\) and \(\cbb\) have values \(\cba, \cbb \in \{1,2\}\) and
\(\oplus\) denotes the exclusive OR operation. This is just the
regular exclusive OR with inputs and outputs in \(\{1,2\}\).  The
measurement setting and outcomes of \(\Va\) also have values
\(\mqa \in\{1, 2\}\) and \(\oqa\in\{1,2\}\).  The prover outcome
\(\oqp\) is binary, and \(\zqa\) and \(\zqb\) are binned to be binary
as well.

For each trial, the trial record \(\Trec\) consists of the
verifier-visible random variables, which are
\(\Cba,\Cbb,\Mqa,\Oqa,\Mqp, \Zqa,\Zqb\). These are the challenges from
\(\Vap\) and \(\Vb\), the measurement setting and outcome at \(\Va\),
the computed measurement setting at the prover \(\Vp\), and the
responses to the verifiers \(\Vap\) and \(\Vb\), respectively.  All
random variables in play, including those involving multiple trials,
are jointly distributed.  To determine whether the prover passed the
protocol successfully (``pass'') or failed (``fail''), the verifiers
perform a statistical test on the sequence of trial records.  The
security of the protocol for a given class of adversaries is
quantified by its soundness and completeness parameters.  For
\(\delta\in[0,1)\), the protocol is \(\delta\)-sound if for
adversaries in the given class, before the protocol starts, the
probability that the protocol will succeed is at most \(\delta\).  To
ensure good security, the soundness parameter \(\delta\) should be
extremely small. For \(\epsilon\in (0,1]\), the protocol is
\(\epsilon\)-complete for the actual, honest prover if this prover
succeeds with probability at least \(\epsilon\).  The soundness
parameter is established by proof.  The completeness parameter for the
actual, honest prover can only be estimated, but usefulness of the
protocol requires that there exist honest provers with completeness
much larger than the soundness.  Since we cannot be sure of the value
of the completeness parameter for the actual, honest prover, to
interpret the security of the protocol, it is desirable to be
conservative and assume that \(\epsilon\) itself is very small. On the
other hand, to ensure that the honest prover is securely
distinguishable from adversaries, it is necessary that
\(\epsilon/\delta\) is very large, requiring \(\delta\) to be much
smaller than the conservative value of \(\epsilon\). In our
experimental implementation, there are non-trivial tradeoffs between
the soundness, completeness, and the time required to run one instance
of the protocol or the number of trials per instance.  See
Sect.~\ref{sect:xanalysis}.

The target region is determined by the positions of the verifiers and
the timing of verifier events, see Sect.~\ref{subsect:spacetimeana}
for details.  In the experiment, the verifiers are stationary in a
common inertial frame of a flat (Minkowski) spacetime \(\cM\). The
space positions of verifiers \(\Va\), \(\Vap\) and \(\Vb\) are
\(\bm{x}_{\Va}\), \(\bm{x}_{\Vap}\) and \(\bm{x}_{\Vb}\).  Our
spacetime and security analysis does not depend on stationary
verifiers, but it simplifies the visualization of target regions and
represents the situation implemented in the experiment. In the
experiment, the honest prover \(\Vp\) is at rest at position
\(\bm{x}=0\).  For simplicity, the theoretical analysis assumes that
the relative timing of events in each trial is the same. The
  timing measurements are described in
  Sect.~\ref{supp:class-tim-res} and summarized in
  Table~\ref{tab-supp:timing_parameters_numbered}.  Let \(t\) be the
time at which the trial starts. We let \(t+s_{\Vap}\) and
\(t+s_{\Vb}\) be the times at which \(\Vap\) and \(\Vb\) begin to emit
their challenges \(\cba\) and \(\cbb\) toward the prover.  The final
times at which the verifiers accept the prover's responses are
\(t+r_{\Vap}\) and \(t+r_{\Vb}\).  The verifiers determine a response
value after these times based on the responses or their absence. The
trial, including recording of the trial record, is complete at time
\(t+\ttrial\). We require \(0\leq s_{\Vap} \leq r_{\Vap}\leq\ttrial\)
and \(0\leq s_{\Vb}\leq r_{\Vb} \leq \ttrial\).  In principle, we can
also specify the position of the source of the entangled pair of
quantum systems and the time at which it sends the two systems. For
our analysis this is subsumed by the constraints on prior entanglement
of adversaries as explained below.

We use statistical tests based on the test-factor
strategy~\cite{shafer:qc2009a,zhang_y:qc2011a,zhang_y:qc2013a} for
testing the hypothesis that the responses are from adversaries in a
given class. Given a general composite hypothesis, a test factor
for a trial is a non-negative function of the trial record whose
expectation is at most \(1\) under any prediction of that
hypothesis. The use of test factors simplifies the security proofs and
is required to enable use of the Bell-test protocol with
experimentally available loophole-free Bell tests. Test factors can be
productively used in most other QPV protocols to simplify security
proofs and protocol operation while supporting otherwise
difficult-to-implement features such as early stopping and heralding.
See Sect.~\ref{sect:heraldetc}.  For our purposes, a trial's test
factor \(W\) for detecting a class of adversaries is a non-negative
function of the trial record whose expectation is guaranteed to be at
most \(1\) for all adversaries in the given class. To have
completeness much larger than soundness, it is necessary that
implementations of an honest prover strategy can achieve an
expectation value \(\Exp(W)>1\), as we experimentally demonstrate. In
this work, we require that \(W\) is a function of
\(\Mqa,\Oqa,\Mqp, \Zqa,\Zqb\) only, so that it depends on
\(\Cba,\Cbb\) only through the settings function
\(\Mqp = f(\Cba,\Cbb)\).  The adversarial bound of \(1\) on the
expectation of \(W\) must hold conditional on the past of the
trial. The past includes all classical and quantum information
available before the trial begins, as specified in the spacetime
analysis below. In a generic trial, adversaries may either have no
prior entanglement or pre-share bounded entanglement. We construct
test factors for both scenarios. Test factors can be chained over
multiple trials by multiplication. According to the theory of test
factors, the product \(\cT\) of test factors over the trials of the
protocol has expectation at most \(1\) for all adversary strategies in
the given class. A key property of such test-factor products is that
the expectation of \(\cT\) remains upper-bounded by 1 even when
adversaries adapt their actions based on past information. By Markov's
inequality, \(\cT^{-1}\) is a \(p\)-value for adversaries of the given
class. For an experimental implementation of the protocol, an honest
prover strategy that ensures \(\Exp(\cT) \gg 1 \) is desirable. We use
the convention that a non-negative random variable \(R\) is a
\(p\)-value against a composite null hypothesis if the probability
that \(R\leq p\) given a model in the null hypothesis is at most
\(p\). We do not require the null hypothesis to contain models that
achieve this bound, nor do we restrict non-negative $p$-values to be
in the interval $[0,1]$.  We define the \(\log(p)\)-value for the
protocol as \(\log(\cT)\). Note that the ``$\log(p)$-value'' is the
negative of the logarithm of the $p$-value. The higher the
\(\log(p)\)-value, the stronger the evidence against the allowed
adversaries. Given the desired soundness parameter \(\delta\), the
protocol passes if the \(\log(p)\)-value is at least
\(\log(1/\delta)\) and fails otherwise. Thus, the protocol passes only
if the null hypothesis that the responses are from adversaries in the
given class can be rejected according to the \(p\)-value \(\cT^{-1}\)
at significance level \(\delta\). Unless specified otherwise, the base
of the logarithm is \(e\).

Test factors can be chosen to maximize the trial gain for a specific
honest prover \(\Vp\). The trial gain \(g\) is defined as the
expectation \(g=\Exp(\log(W))\) for \(\Vp\) of the logarithm of the
test factor.  The gain is the expected contribution per trial to the
\(\log(p)\)-value. The test factor is chosen by considering the honest
prover that is implemented in the experiment.  Since the implemented
prover is designed to produce independent and identically distributed
(i.i.d.) trial records, the value of \(\log(\cT)\) after \(n\) trials
has expectation \(n g\) with variance \(O(n)\).  Consequently, the
number of trials required for \(\Vp\) to succeed with reasonable
probability can be estimated as \(\log(1/\delta)/g\). Specific
estimates based on asymptotic normality are used to determine protocol
parameters, see Sect.~\ref{sect:xanalysis}.

\vfill
\pagebreak

\section{Bell-test protocol and assumptions}
\label{subsect:protocol_assumptions}

The Bell-test QPV protocol is detailed in Protocol~\ref{prot:pv}
below. A version of this protocol was implemented in the
experiment. The protocol is general and can be applied to arbitrary
configurations involving pairs of entangled quantum systems for which
choices of measurement settings exist, and for which the verifiers and
honest prover yield trial-record distributions that are outside a
convex set induced by the convex set of three-party quantum
non-signaling distributions as described in
Sect.~\ref{sect:reduction}. It is proven secure against unentangled or
weakly entangled adversaries with no restrictions on quantum
communication during a trial, provided that the settings function
satisfies the assumptions given below. To address the issue of adversaries with
significant amounts of prior entanglement, we apply an enhanced
version of the \(\epsilon\)-net construction pioneered in
Ref.~\cite{bluhm2022single} to the Bell-test protocol.  Given an an
initial test factor against unentangled adversaries, we show in
Sect.~\ref{sect:entangled_adversaries} how to modify this test factor
for scenarios where the adversaries pre-share significant bounded entanglement and
utilize arbitrary classical communication but no quantum communication
during trials. The completeness and soundness results are summarized
in Sect.~\ref{sec:results statements}.

\vfill
\pagebreak

\SetKwInOut{Given}{Given}
\SetKwInOut{Input}{Input}
\SetKwInOut{Output}{Output}
\SetInd{0.5em}{1em}

\SetAlgorithmName{Protocol}{}{}
\RestyleAlgo{boxruled}
\vspace*{\baselineskip}
\begin{algorithm}[H]
\caption{Bell-Test Quantum Position Verification (QPV) Protocol}\label{prot:pv}
\begin{minipage}{\dimexpr\textwidth-0.5in\relax}
\tcp{See Sects.~\textrm{\ref{sec:test_factors}},~\textrm{\ref{sec:smallentanglement}} and~\textrm{\ref{sect:parameter_choices}} for the required preparatory work for running this protocol. }
\textbf{Given:}
\begin{itemize}
  \item[] \(\delta\) --- desired soundness parameter. 
  \item[] \(n\) --- number of trials to be executed.
  \item[] \(\cnu\) --- joint challenge distribution.
  \item[] \(\sfun(\cba, \cbb)\) --- settings function. 
  \item[] \(\mu_{\Va}\) --- measurement settings distribution of verifier \(\Va\).
  \item[] \(W\) --- trial-wise test factor to be used. \(W\)
      depends on the class of adversaries and the Bell test used.
  \item[]  \(t_0\) --- initiation time.
  \item[] \(\ttrial\) --- trial duration.
  \item[] \(s_{\Vap},s_{\Vb},r_{\Vap},r_{\Vb}\) --- verifier send and receive times.
\end{itemize}

Let \(t_{1}\geq t_{0}\) be the first trial's start time\;
\For{\(k = 1\) \KwTo \(n\)}{ 
  Perform a QPV trial starting at time \(t_k \): 
  \begin{enumerate}[leftmargin=2em] 
  \item The source prepares a joint state of \(\Qa\) and \(\Qp\) (event \#4 in~\cref{tab-supp:timing_parameters_numbered}), and sends \(\Qa\)
    to \(\Va\) and \(\Qp\) to the target region. 
    This state preparation can be done before \(t_{k}\), provided that
    Assumption~\ref{ass:bell} is satisfied.
  \item \(\Vap\) and \(\Vb\) generate challenges \(\cba\) and \(\cbb\)
    according to the joint  distribution \(\cnu\) and 
    \(\Va\) chooses measurement setting \(\mqa\) (event \#5 in~\cref{tab-supp:timing_parameters_numbered}) according to probability distribution
    \(\mmua\). The challenges and settings for this trial may be generated 
    before \(t_{k}\) provided that Assumption~\ref{ass:distrib} is satisfied. 
  \item \(\Vap\) and \(\Vb\) send the challenges \(\cba\) and
    \(\cbb\) at times \(t_{k}+s_{\Vap}\) and \(t_{k}+s_{\Vb}\) from
    specified locations. See Assumption~\ref{ass:lab}. These are events \#1 \& \#2 in~\cref{tab-supp:timing_parameters_numbered}.
  \item \(\Va\) measures \(\Qa\) according to setting \(\mqa\),
    producing outcome \(\oqa\) (event \#6 in~\cref{tab-supp:timing_parameters_numbered}). For device-independence based on
    causality, the period between applying the setting to recording
    the measurement outcome is space-like separated from the trial's
    target region.
  \item \(\Vap\) and \(\Vb\) receive response values \(\zqa\) and
    \(\zqb\) from the provers at times \(t_{k}+r_{\Vap}\) and
    \(t_{k}+r_{\Vb}\) and at specified locations (events \#13 \& \#14 in~\cref{tab-supp:timing_parameters_numbered}). See
    Assumption~\ref{ass:lab}. 
  \item The verifiers save the reduced trial record
    \(\trec_{k}=(\mqa,\oqa,\sfun(\cba,\cbb),\zqa,\zqb)\) and compute
    \(w_{k}=W(\trec_{k})\).
    In the experiment \(w_{k}\) was
      computed later, after unblinding the data.
  \end{enumerate}
  Let \(t_{k+1} \geq t_{k}+\ttrial\) be the next trial's start time\;
}

Let \(p = \left( \prod_{k=1}^{n} w_k \right)^{-1}\)\tcp*{This is the \(p\)-value against adversaries}

\uIf{\(p \leq \delta\)} { 
  \Return{success}\tcp*{Protocol succeeded}
  }
\Else{
  \Return{fail}\tcp*{Protocol failed}
}
\end{minipage}
\end{algorithm}
\vspace*{\baselineskip}

\pagebreak

The assumptions required for the security of the protocol and to
ensure that soundness is satisfied for the given soundness parameter
\(\delta\) are listed next.

\vspace*{\baselineskip}
\noindent\textbf{Assumptions} (see
  Sect.~\ref{subsect:spacetimeana} for definitions and terms):
\begin{enuma}[label=\arabic*]
\item\label{ass:ccc}
  \textbf{Classical computation and communication:} All
  classical computation is error-free and satisfies the protocol
  specifications. This is a standard assumption in device-independent
  quantum protocols~\cite{nadlinger_experimental_2022, zhang_deviceindependent_2022}. Classical communication between the verifiers
  is also error-free. Most classical communication between the verifiers can
    be deferred until after the protocol completed and can be implemented by
    authenticated classical channels.
\item\label{ass:st} \textbf{Classical spacetime and worldlines:} The
  background spacetime is Minkowski. Worldlines are classical. This
  means that the spacetime circuit of the relevant entities is
  determined in each trial given the past.  In principle, the
  computations at the vertices could make choices of where to send the
  outputs.  We assume that there is a finite, bounded number of such
  choices, in which case we can arrange for the fixed spacetime
  circuit to have edges and vertices for every possible combination of
  such choices.  For this purpose, the systems on the edges may need
  to be enlarged by adding a ``vacuum'' state that indicates that the
  edge was not used. Vertex operations are extended accordingly.
\item\label{ass:timing} \textbf{Classical timing and location:} For
  the theoretical analysis of the protocol we assume that trial timing
  and location measurements are error-free. The timing and location
  measurements determine the target region.  In the experiment, timing
  and location measurements have uncertainties as described in
  Sect.~\ref{sect:experiment}, which affect the boundaries of the
  target region. The effect of these uncertainties on localization is
  illustrated in Fig.~\ref{supp:localization-advantage-comparison}.
\item\label{ass:lab} \textbf{Verifier lab security:} The labs of the
  three verifiers are secured to disallow interaction with
  non-verifier entities except for those that are part of the
  protocol. In particular, the only outgoing edges from the labs of
  verifiers \(\Vap\) and \(\Vb\) to non-verifiers are those that
  jointly carry the classical challenges and are causally consistent
  with the challenge timing. In practice, this means that the
  classical challenges do not ``leak'' out of the verifier labs apart
  from the established exit points at the appropriate challenge exit
  time, see~\cref{fig-supp:exit-point}. In addition, there are no
  outgoing edges from \(\Va\) to non-verifiers, which prevents
  information about the settings and outcomes at \(\Va\) from leaking
  to adversaries during a trial.  To ensure this condition, we assume
  that outgoing communications from \(\Va\) other than to the
  verifiers are blocked during a trial.  We further require that
  within a trial, there are no incoming edges from a non-verifier that
  causally precede edges from the verifiers to a non-verifier. For
  \(\Vap\) and \(\Vb\), this can be enforced by blocking incoming
  communications before the challenges are sent. \(\Va\) receives a
  quantum system from the source, which is treated as untrusted.  We
  require that the verifiers' processing of challenges and responses
  is completely classical.  In particular, challenges are classical
  and the responses are immediately decohered in the computational
  basis at the moment they enter the lab. Such assumptions of lab
  security are standard in device-independent and other cryptographic
  protocols~\cite{xu_deviceindependent_2022,nadlinger_experimental_2022,
    zhang_deviceindependent_2022}.
\item\label{ass:crevents} \textbf{Challenge and response events:}
  We identify specific events in spacetime where the challenges
  leave the labs and the responses enter the labs.  The only requirement
  on these events is that the target region as defined in
  Sect.~\ref{subsect:spacetimeana} is non-empty.  In the protocol and
  the experiment, these events have the specified space positions of
  \(\Vap\) and \(\Vb\) (see~\cref{fig-supp:exit-point}), which are stationary with respect to the earth's surface.
  The challenges leave the lab at or after times \(t_{k}+s_{\Vap}\)
  and \(t_{k}+s_{\Vb}\).  The responses enter the labs at or before
  times \(t_{k}+r_{\Vap}\) and \(t_{k}+r_{\Vb}\). All relevant experimentally measured timings are listed in~\cref{tab-supp:timing_parameters_numbered}.
\item\label{ass:acom}\label{ass:ent} \textbf{Adversary communication
    and entanglement:} We consider three complementary sets of
  assumptions. The sets have different requirements for the bit-length
  of the challenges \(\cba\) and \(\cbb\).  For the first,
  length-\(1\) bit strings suffice.  For the second and third,
  \(\cba\) and \(\cbb\) are bit strings of sufficient length \(l\) and
  the settings function has sufficient complexity, see
  Sect.~\ref{sect:entangled_adversaries}. In all cases, the settings
  function and challenge distributions must satisfy
  Assumptions~\ref{ass:funct} and~\ref{ass:distrib} below, and prior
  entanglement of adversaries for each trial is defined on a specific
  time slice determined by the trial timing as described in the
  spacetime analysis of Sect.~\ref{subsect:spacetimeana}.  This time
  slice defines the trial boundary for the purpose of quantifying the
  prior adversary entanglement. The protocol assumes that this prior
  entanglement is zero or satisfies bounds described in
  Sect.~\ref{sec:smallentanglement} or in
  Sect.~\ref{sect:entangled_adversaries}.  The test factors and the
  pass/fail decision depend on which assumption is used. The
  constraint on prior entanglement includes the source and the emitted
  or to be emitted entangled pair of quantum systems on the trial
  boundary.
  
  \noindent\emph{First set of assumptions:} Except for spacetime
  causality, there are no restrictions on adversary communication
  capabilities during a trial. The adversaries can communicate both
  classically and quantumly.  Quantum communication and memory between
  trials are indirectly restricted by the constraints on adversary
  entanglement. Here we assume that prior adversary entanglement is zero
  or satisfies bounds described in Sect.~\ref{sec:smallentanglement}.

  \noindent\emph{Second set of assumptions:} There is no
  constraint on classical communication between the adversaries, but
  quantum communication and prior entanglement is constrained.  By
  shifting quantum communication to prior entanglement and classical
  communication, the constraint is expressed in terms of a bound on
  the rank of the reduced density matrix of the prior adversary
  entanglement at the adversary nearer \(\Vb\).  As shown in
  Sec.~\ref{sect:entangled_adversaries}, the bound on this dimension
  is proportional to \(\sqrt{l}\).

  \noindent\emph{Third set of assumptions:}
  As in the second set of assumptions, classical communication is
  not constrained and quantum communication is shifted to prior
  entanglement.  The constraint is on the number of qubits, \(q\), and the number of 
  quantum gates, \(G\), used by the adversary nearer \(\Vb\). In
  Sec.~\ref{sect:entangled_adversaries}, the bound is of the
  form \(\log(q)G\leq c l\) for some constant \(c\).

\item\label{ass:bell} \textbf{Quantum source:} We treat the source as
  an untrusted component that may be under the control of
  adversaries. This implies that the source cannot be in the target
  region. In addition, for the first set of assumptions in
  Assumption~\ref{ass:ent}, the restriction on prior entanglement of
  the adversaries implicitly constrains the source and the entangled
  system prepared and communicated by the source. To prevent the
  adversaries from exploiting the source for shared entanglement it is
  necessary to secure the location of source and the timing of the
  source emissions.  Because the quantum system \(\Qa\) is treated as
  part of the adversary closer to \(\Vap\), the source needs to be
  located in the region closer to \(\Vap\).  To prevent the source
  from contributing entanglement to the adversaries, it is desirable
  for the source emission of \(\Qp\) to the prover location to be
  constrained so that the spacetime worldline of \(\Qp\) does not
  cross the trial boundary adjacent to the part of the adversary
  nearer \(\Vb\).  If the protocol is secure against adversaries with
  prior entanglement greater than that available from the source, then
  there is no need to secure the source location or emission. We show
  in Sect.~\ref{sect:entangled_adversaries} that such security is
  satisfied with sufficiently complex challenges and settings
  functions.
\item\label{ass:meas} \textbf{Quantum measurement:} \(\Va\) uses a
  quantum apparatus in their lab to measure \(\Qa\).  The apparatus is
  untrusted. Security of the protocol requires that the setting and
  the challenges are not known to the apparatus beforehand.  For the
  analysis here, we assume that the only incoming message during a
  trial is the quantum system \(\Qa\) from the source. We assume that
  outgoing messages are blocked, except for the eventual message of
  the measurement outcomes to the other verifiers. We assume and
  ensure in the experiment (see~\cref{fig-supp:spacetime_diagram})
  that the interval between applying a measurement setting and
  recording the measurement outcome is in the region where the
  challenge from \(\Vb\) is not visible and from which no signal can
  reach \(\Vb\) at or before the time point \(t_{k}+r_{\Vb}\). In
  particular this interval is space-like separated from the target
  region of a trial. In~\cref{fig-supp:spacetime_diagram} and using
  the conventions introduced before Eq.~\ref{eq:advcirc_defs}, the
  verifier measurement region consists of the subregions labeled
  \(*0*1\), where \(*\) denotes either \(0\) or \(1\).  An
  illustration of these subregions is provided
  in~\cref{fig:1+1d_regions}.
\item\label{ass:funct} \textbf{Settings function:} The main
  constraints on the settings function are for those implied by the
  assumptions on the settings and challenges in
  Assumption~\ref{ass:distrib} below. For adversaries with significant
  quantum resources, the bit length of the challenges must be large
  enough and the settings function must have sufficient complexity.
  
  In QPV, increasing the complexity of the challenges and the settings
  function leads to higher bounds on the minimum prior entanglement
  required for adversaries to successfully pass the
  protocol~\cite{das_practically_2021,bluhm2022single}. It can also be
  used to protect against computationally bounded adversaries without
  prior entanglement constraints~\cite{unruh2014quantum}.  Our
  inclusion in the experiment of a challenge from \(\Vap\) and use of
  the exclusive OR function anticipates the implementation of such
  strategies. 
\item\label{ass:distrib} \textbf{Verifier settings and challenge
    distributions:} For our analysis, we assume that verifier settings
  and challenge distributions are fixed and independent of the state
  coming into the trial.  In the spacetime circuit analysis given in
  Sect.~\ref{subsect:spacetimeana} below, this means that the
  probability distributions are independent of the state of the
  adversary including the source and the quantum systems that the
  source emitted on the time slice on which prior adversary
  entanglement is determined. For the reduction from general
  adversaries to two adversaries whose quantum messages to each other
  do not depend on the challenges, we require that 1) \(\Mqa\) and
  \(\Mqp\) are independent of \(\Cba\) so that
  \(\Prob(\mqa,\mqp,\cba) = \Prob(\mqa,\mqp)\Prob(\cba)\), and 2)
  \(\Mqa\) is conditionally independent of \(\Cba\) and \(\Cbb\) given
  \(\Mqp\) so that
  \(\Prob(\mqa,\cba,\cbb|\mqp)
  =\Prob(\mqa|\mqp)\Prob(\cba,\cbb|\mqp)\). See
  Sect.~\ref{sect:reduction} for how these independence assumptions
  are used.  The assumptions are satisfied if \(\Cba\) and \(\Cbb\)
  are uniformly random, independent of \(\Mqa\), and the cardinality
  \(|\{\cbb|f(\cba,\cbb)=\mqp\}|\) does not depend on \(\cba\) or
  \(\mqp\).  For simplicity of test-factor construction we choose
  uniform verifier and prover settings distributions for the
  experiment.
  
  It is possible to relax the assumption that the joint distribution of
  the verifier settings \(\Mqa\) and the prover settings \(\Mqp\) is
  fixed and independent of the past. This can be done by allowing
  mixtures of products of incoming states and joint distributions of
  \(\Mqa\) and \(\Mqp\), where the latter come from a fixed convex set
  of distributions near a target distribution.  The chosen test factors
  then need to be designed to have worst-case expectation of \(1\) for
  adversaries over the convex set of distributions. See
  Ref.~\cite{zhang2020experimental} for realizations of this strategy in
  the context of device-independent quantum randomness generation.\

  It is noteworthy that for rejecting unentangled adversaries, the
  assumptions on the settings and challenges can be satisfied with a
  settings function that does not depend on \(\cba\), in which case
  the challenge \(\cba\) can be omitted.  In this case, the role of
  \(s_{\Vap}\) is to define the trial boundary and the target
  region. Because \(\Qa\) is attributed to the adversary nearer
  \(\Vap\), the restriction on entanglement requires that \(\Qp\)
  cannot reach the adversary nearer \(\Vb\) before the trial boundary
  as defined by \(s_{\Vap}\). In the experiment this is satisfied by
  transmitting \(\Qp\) in fiber and accounting for the speed of light
  in the fiber. 
\end{enuma}


\section{Experiment}
\label{sect:experiment}
\subsection{Overview}

\begin{figure}[t!]
    \centering
    \includegraphics[width=\linewidth]{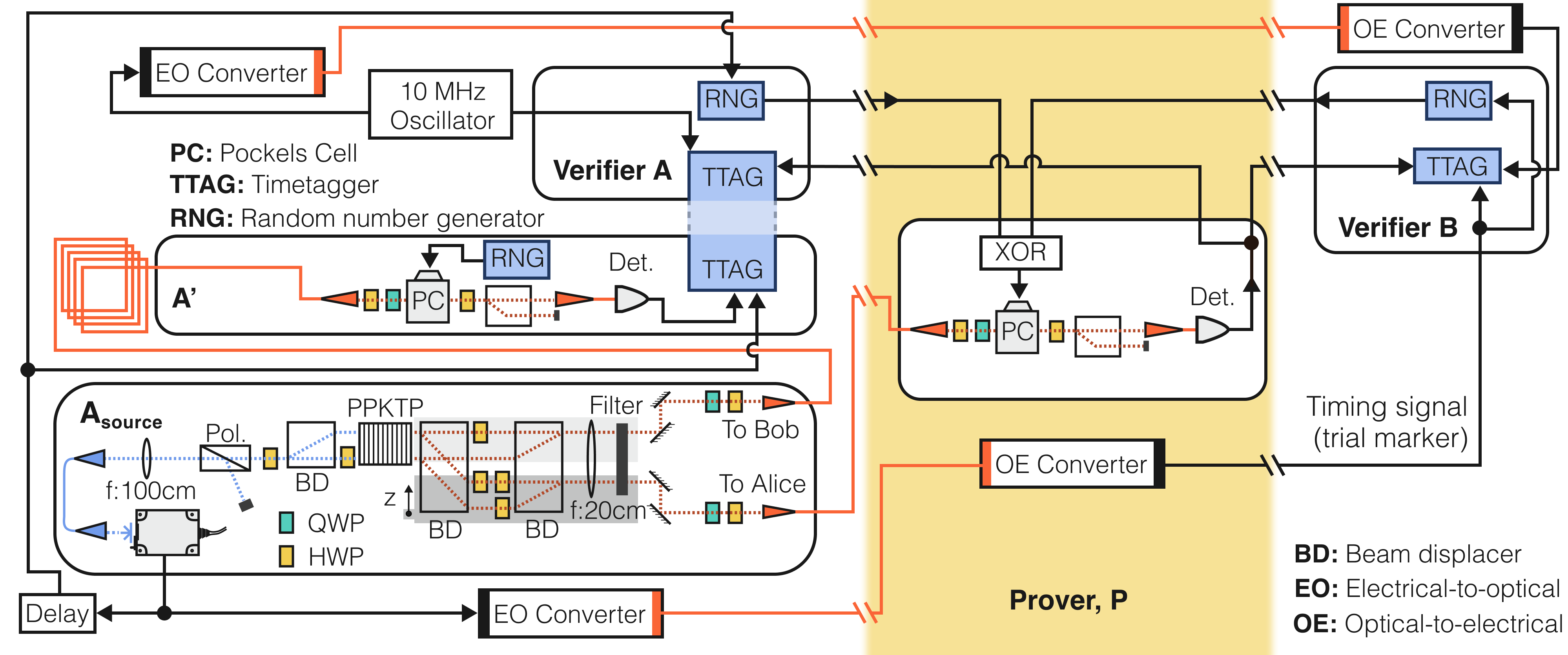}
    \caption{Experimental diagram for the position verification demonstration. Verifiers \(\Vap\), \(\Va\), and \(\Vb\), the honest prover \(\Vp\) as implemented, and the entangled photon source \(\Vs\) are indicated. The yellow shading around the honest prover indicates the region prover hardware must be in to pass the protocol (target region \(\tR\)). The RNGs and timetaggers are associated with the verifiers. Electrical and optical cables carry information to and from the various experimental stations. The entangled photons are distributed via optical cables, and the measurement choice bits are transmitted via high-speed coaxial cables. Electrical cables are indicated in black, and optical cables in orange. Electrical-to-optical (OE) and optical-to-electrical (EO) converters convert information between electrical and optical encoding. These are used in the lines that transmit synchronization signals among the verifiers. $\Vs$ is the source of entangled photons, which produces entangled photons via spontaneous parametric downconversion with a periodically poled potassium titanyl phosphate (ppKTP) crystal placed inside a dual Mach-Zehnder interferometer formed with beam displacers (BD) \cite{kavuri_traceable_2025}. For the measurements at \(\Va\) and \(\Vp\), the pair of measurement bases are implemented with quarter-waveplates (QWP) and half-waveplates (HWP), and fast switching between the bases is performed with a Pockels cell.}
    \label{fig-supp:expt-diagram}
\end{figure}

\cref{fig-supp:expt-diagram} shows a schematic of the components of the experiment. In the experiment, verifier $\Vap$ that sends challenge bits and receives responses, verifier $\Va$ that measures one of the entangled photons, and the entangled photon source $\Vs$ are in the same room. Verifier \(\Vb\) is in another room of the National Institute of Standards and Technology (NIST) building that is about \qty{200}{\meter} away.

The entangled photons \(\Qa\) and \(\Qp\) generated at $\Vs$ are
distributed through single-mode optical fiber to measurement station
\(\Va\) and the honest prover, \(\Vp\). For the successful protocol
runs presented here, the honest prover hardware is stationary and
located in a room about \qty{90}{\meter} from verifier
\(\Vap\). Subsequently, verifier \(\Va\) and the honest prover \(\Vp\)
measure the photons in one of two measurement bases, performing a Bell
test. We label the measurement settings of \(\Va\) and \(\Vp\)
determined by their measurement bases by \(\mqa\) and \(\mqp\).
Much of the hardware for this experiment builds on
loophole-free Bell tests and device-independent random number
generation (DIRNG) setups~\cite{shalm_deviceindependent_2021,
  zhang_certifying_2018}. For example, the measurements performed at
\(\Va\) are the same as at ``Bob'' in
Ref. \cite{kavuri_traceable_2025}, with the measurement basis \(\mqa\)
chosen using the same source of local randomness as in that work. The
resulting measurement outcomes are recorded to a timetagger. The
measurement basis at the honest prover \(\Vp\), however is decided by
the output of an exclusive OR (XOR) of random challenge bits 
\(\mqp = (\cba-1) \oplus (\cbb-1)+1\) released by the verifiers \(\Vap\) and
\(\Vb\). The random bits are sent to the honest prover at high speed
over specialized coaxial cables. The honest prover measures the photon
and replies to verifiers \(\Vap\) and \(\Vb\) with the measurement
outcome over fast coaxial cables. Verifiers \(\Vap\) and \(\Vb\) then
record these signals \(\zqa, \zqb\) on their timetaggers.

Provided that the recorded signals violate a Bell inequality, the time difference between when the random inputs were provided and the measurement outcome was received by $\Vap$ and $\Vb$ allows the verifiers to perform a secure ranging based on round-trip communication times. Any excess latencies during transmission (i.e., transmission slower than the speed of light), or in the measurement process at the honest prover implies a larger target region $\tR$ for localizing the prover. This highlights the importance of low-latency measurements at the implemented honest prover station and high-speed classical communication. In our experiment, this is enabled with the use of Pockels cells, custom high-speed electronics, and low-latency coaxial cables. The experimental realization of a QPV protocol requires careful design and optimization to minimize latency in electronics, measurement, and transmission.

With the Bell-test protocol, much of the ``quantum'' hardware (such as the source and the measurement stations) need not be trusted, or even be within the secure verifier labs, allowing for more flexibility in configuration choices. A careful account of the assumptions involved in our demonstration can be found in \cref{subsect:protocol_assumptions}.

\subsection{Protocol Implementation}
\label{subsec:exp_prot_imp}
The labs containing verifiers \(\Vap\) and \(\Vb\) are assumed to be trusted, in the sense that the equipment inside them is not under the influence of any adversaries. This is related to Assumptions \ref{ass:lab} and \ref{ass:distrib} in \cref{subsect:protocol_assumptions}. Physically, this means the signals are assumed to be shielded from adversaries (in that they cannot be altered or read by adversaries) until they physically exit the labs, and that any signals coming in from the outside are similarly shielded once they enter the verifier labs.

\begin{figure}[t!]
    \centering
    \includegraphics[width=\linewidth]{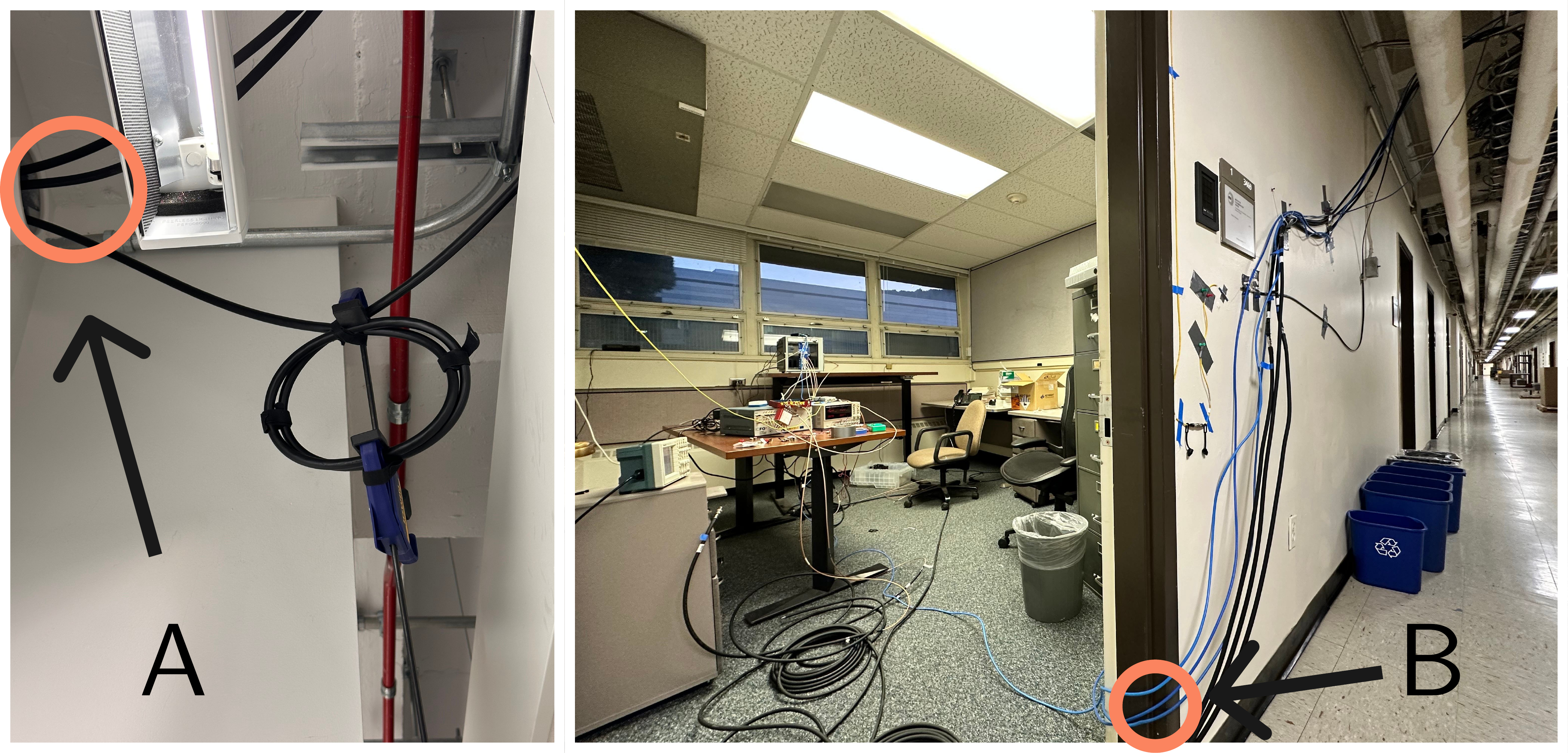}
    \caption[The physical exit points for the verifier signals]{Photographs of the physical entry/exit points of the classical and quantum signals that serve as the verifiers \(\Vap\) and \(\Vb\) in our demonstration. Any hardware and cabling inside the verifier \(\Vb\) lab is trusted. We do not need to trust the quantum hardware/measurement device, at Verifier \(\Vap\). See \cref{fig-supp:expt-diagram} for a full account of the devices at verifier \(\Vap\). At verifier \(\Vb\), the lab houses a random number generator for deciding the challenge \(u_{\Vb}\), and a timetagger that records classical timing information and the response from provers \(z_{\Vb}\). All information exiting and entering the Verifier \(\Vb\) lab is classical.}
    \label{fig-supp:exit-point}
\end{figure}

All signals exit and enter the verifier labs via access points in the walls through which cables are laid. In all our analyses, we rely on the trusted lab assumption stated to use these physical access points to define the verifier \(\Vap\) and \(\Vb\)'s physical locations. Here, we are modeling a situation where the verifier labs are carefully shielded from external signals, while an opening allows transmission of signals in and out of the labs. The times when challenge (response) signals exit (enter) these openings are the earliest (latest) time at which they can be influenced or read by adversaries. From the adversarial point of view, it thus suffices to reduce the verifiers to the spatial points corresponding to the signal entry and exit. See \cref{fig-supp:exit-point} for photographs of the entry/exit points. Signals traveling to or from the verifier labs and crossing these physical thresholds at time `$t$' are considered to have been generated or detected at `$t$'.

The protocol proceeds as a series of trials, with the number of trials performed in a single run of the protocol determined by previous calibration data and a desired threshold for soundness error (see~\cref{sec:preliminaries}). Each trial starts with the creation of an entangled pair of photons at $\Vs$. The photons are generated via spontaneous parametric downconversion. A solid-state pulsed laser at \qty{775}{\nano\meter} pumps an entangled pair generation source that probabilistically down-converts the pump to generate entangled pairs of photons at \qty{1550}{\nano\meter}. The laser operates at a repetition rate of \qty{80}{\mega\hertz}, meaning we attempt to produce entangled photons every \qty{12.5}{\nano \second}.

The source of entangled photons is largely similar to the one employed
in Ref.~\cite{kavuri_traceable_2025}, and is based on a twin
Mach-Zehnder interferometer constructed with beam displacers. The
entangled photon source is tuned to produce a non-maximally entangled
state, nominally
$0.383 |\mathrm{HH}\rangle + 0.924 |\mathrm{VV}\rangle$, conditional
on a pair of photons being spontaneously generated, which happens with
a probability of approximately $1/350$. This state (along with the
measurement angles at the measurement stations) was chosen based on a
numerical optimization maximizing the Bell violation. The optimization
takes experimental losses, dark counts, and imperfect state fidelity
into account.  For comparison to bounds on prior entanglement by
  adversaries, the robustness of entanglement of the nominal state is
  \((0.383+0.924)^{2}-1=0.71\). Accounting for the probability of
  photon pair emission, this gives a robustness of entanglement per
  trial of \(2\times 10^{-3}\).  The source sends one photon (\(\Qp\)) from
each of the entangled photon pairs to \(\Vp\) via optical fiber, and
the measurement station \(\Va\) makes measurements locally on its
photon (\(\Qa\)). The \(\Qa\) photon reaches the measurement station
at \(\Va\) after being delayed in a fiber spool. This delay ensures
that the measurement basis choice at this station is space-like
separated from the trial region, see Assumption \ref{ass:meas}
in~\cref{subsect:protocol_assumptions}.

Following the release of entangled photons from $\Vs$, the verifier
\(\Vap\) transmits a random bit (\(\cba\)). This bit is freshly
generated via a hardware random number generator at \(\Vap\), and its
value is immediately recorded on a timetagger located at \(\Vap\). The
bit is then encoded as an electronic non-return-to-zero (NRZ) signal
with about a \qty{1}{\volt} amplitude and transmitted via a
specialized coaxial fiber that has a central conductor surrounded by a
foam dielectric. The air in the foam reduces the relative permittivity
of the surrounding insulator and allows these cables to support a
signal propagation velocity that is between 85-87\% of the speed of
light~\cite{cooke_coaxial_2021}. In traditional low-density
polyethylene (LDPE) coaxial cables, the velocity of propagation is
closer to 66\%. This classical signal travels towards the prover
faster than the \qty{1550}{\nano\meter} photon \(\Qp\), which travels
through an optical fiber of roughly comparable length at approximately
66\% of the speed of light.  Some time after \(\cba\) is released,
verifier \(\Vb\) releases its bit (\(\cbb\)), which is recorded on a
timetagger at verifier \(\Vb\)). In our configuration, with verifier
\(\Vb\) situated closer to the honest prover \(\Vp\) than verifier
\(\Vap\), this delay is set so that the bits \(\cba\) and \(\cbb\)
reach the (known) location of the honest prover at approximately the
same time. The classical challenge bits reaching the honest prover at
the same time is optimal because any discrepancy in the arrival times
would mean that the honest prover would have to wait for either
\(\cba\) or \(\cbb\) before computing their XOR. Waiting in this way
would increase protocol latency at the honest prover hardware station
and the size of the realizable target region of the Bell-test
protocol. In our realization, the synchronization is slightly
imperfect and \(\cbb\) arrives at the prover \qty{9\pm2}{\nano\second}
after \(\cba\).  Once the bits \(\cba\) and \(\cbb\) reach the prover,
the prover computes the XOR of the bits
\(\mqp = (\cba-1) \oplus (\cbb-1)+1\) using an application-specific
integrated circuit (ASIC). This computation takes only a few
nanoseconds and does not contribute significantly to the latency at
the honest prover \(\Vp\). The honest prover then sets the basis of
the Pockels cell (see \cref{fig-supp:expt-diagram}) so that incoming
photons will be projectively measured in basis
$a1 = \qty{6.7}{\degree}$ or $a2 = \qty{-29.26}{\degree}$, depending
on whether \(\mqp =1\) or \(\mqp = 2\) respectively. Here, the angles
represent the projective bases corresponding to rotations of a linear
polarizer relative to the horizontal. The angles are the same as in
Ref.~\cite{kavuri_traceable_2025}. An incoming photon \(\Qp\) arrives
shortly after the Pockels cell is set. The photon is then measured in
the appropriate basis with a superconducting nanowire single photon
detector (SNSPD). The output is duplicated with a power splitter into
\(\zqa\) and \(\zqb\), which are sent to the verifiers over separate
specialized coaxial cables. These are similar to those used for the
transmission of \(\cba\) and \(\cbb\) to the honest prover \(\Vp\).

Approximately at the same time as the photon \(\Qp\) is measured at the honest prover, its partner photon \(\Qa\) is measured at the measurement station \(\Va\) using a local random bit produced at the station \(\mqa\). The photon is measured in the basis $b1 = \qty{-6.7}{\degree}$ or $b2 = \qty{29.26}{\degree}$ depending on if the value of \(\mqa\) is 1 or 2. The result of this local measurement, \(\oqa\), is recorded on the timetagger at verifier \(\Vap\).
Finally, the returning outputs from the honest prover \(\zqa\) to verifier \(\Vap\) are recorded on verifier \(\Vap\)'s timetagger, and \(\zqb\) on verifier \(\Vb\)'s timetagger. This concludes one trial in the experiment. Trials are repeated sequentially at a \qty{250}{\kilo\hertz} rate. The data recorded on the timetaggers, including the input random bits from the verifiers and the output bits back from any provers, constitutes the raw trial record. This raw trial record is then processed based on classical timing markers that are exchanged between the two verifiers via a trusted classical channel. A trusted classical channel is also used by the verifiers to exchange information used to align and process the collected data.

\subsection{Hardware and Timing}
\label{subsec:hardware_timing}

As discussed above, the experiment has three physically distinct locations, shown in, \cref{fig-supp:expt-diagram}. One lab, which we call ``$\Vap$'s lab'', contains $\Vap$, $\Va$, and $\Vs$; one contains $\Vb$; and one contains $\Vp$. 

A number of optical and electrical cables carry information back and forth between the labs in our demonstration. Because the timetaggers that record this information are separated from each other, and recorded trial information must be reconciled to a common timebase, the timetaggers must exchange timing information between themselves. In our demonstration, timing information is derived from an electronic ``sync'' signal, which is divided down from the \qty{775}{\nano\meter} pump laser pulsed at \qty{80}{\mega\hertz}, and serves as the master clock for pulse generation. This signal is first converted to optical pulses, which are sent to the honest prover's lab over optical fiber. The optical signal is then converted back to an electronic signal and propagated to verifier \(\Vb\). The sync signal is also used by the prover's hardware to sample the output of the XOR circuit correctly. In principle, the challenges can serve as sync signals, in which case sending separate sync signals is not required.

Another timing signal that the experiment uses is a \qty{10}{\mega\hertz} clock that serves as a timebase for the remote timetaggers. The ``sync'' signal serves as a trial marker, but it is not at the correct frequency to synchronize our remote timetaggers. The \qty{10}{\mega\hertz} clock is generated at verifier \(\Vap\), split off and supplied to the local timetagger, converted to an optical signal, and transmitted via optical fibers all the way to verifier \(\Vb\), where it is supplied to \(\Vb\)'s timetagger.

 The sync signal is divided down to \qty{250}{\kilo\hertz} from the electronic output monitor of the \qty{80}{\mega\hertz} pump laser such that the laser emits 320 pulses during every sync period. The laser pulses then propagate through a set of beam displacers and optics to reach a periodically poled potassium titanyl phosphate crystal in two spatial modes to produce the entangled state, consisting of the \(\Qp\) and \(\Qa\) portions of the entangled state. The time of emission of the quantum state \(\Qp\) from the verifier \(\Vap\) lab is much earlier than the emission of the classical bit \(\cba\). This is to ensure that the photon traveling slowly through fiber arrives at the honest prover in time to be measured. Because the measurement station \(\Va\) is in the same room as the verifier \(\Vap\), the fiber is looped up into a spool to delay measurement by \(\qty{791\pm 1}{\nano \second}\), as mentioned in a previous section. These single mode optical fibers are each of length \(\approx \qty{160}{\meter}\) and have a transmission efficiency of over 99\%.

Other processes are also clocked off of the electrical sync signal. In addition to the sync signal that is propagated to verifier \(\Vb\), it is also sent to verifier \(\Vap\) after passing through a digital delay circuit. The sync signal triggers hardware random number generators at $\Vap$ and $\Vb$ to in-part determine when they each emit a random bit (\(\cba\) and \(\cbb\)) to the prover \(\Vp\) over custom coaxial cables. The delay at verifier \(\Vap\) is then set so that the random bits \(\cba\) and \(\cbb\) arrive at the prover at approximately the same time. The relative timing of the bits is decided so that there is sufficient time for the XOR of the bits to be computed and the Pockels cell to be set to either a high or low voltage based on the result of the XOR by the time the photon \(\Qp\) arrives.

The electronic sync signal is also provided to the measurement station \(\Va\), which measures its photon \(\Qa\) based on a random number generator that is triggered again by this sync signal. The output of this measurement is immediately recorded on a local timetagger.
Finally, the prover \(\Vp\) transmits the measurement outcome from its measurement of photon \(\Qp\) to $\Vap$ and $\Vb$ over electrical coax cables. These signals containing the measurement outcome are recorded on timetaggers. Due to electronic errors, in about $2\times 10^{-6}$ of trials, $\Vap$'s and $\Vb$'s records of the prover's measurement outcome disagree. Besides the outcomes, a local record of the RNG bits \(\cba\) and \(\cbb\) is recorded on the timetaggers every trial, along with the timetag of the sync signal to serve as a reference. Because the entire experiment is clocked on this sync signal, the timetagger records can be reconciled with one another using the trusted delay measurements. This allows the creation of a trial-by-trial timing record of all relevant events.

\section{Classical Spacetime Measurements}
\label{supp:class-tim-res}
Determining the localization regions achieved by the Bell-test protocol requires making measurements on the timing signals and the physical distances separating the verifiers. Here, we provide a summary of our spacetime measurement techniques and results.

We define time $t=0$ as when the synchronization signal exits the divider circuit, which electronically divides the monitor output of the \qty{775}{\nano \meter} pump laser. Several techniques are used to estimate the timing of various events relative to this reference during a trial. For instance, to establish the time when the sync signal reaches the prover \(\Vp\) and the far verifier station \(\Vb\), we measure some electrical latencies directly with an oscilloscope, measure certain electrical cable latencies using electrical reflectometry, measure optical cable latencies using optical time domain reflectometry (OTDR), and perform relative delay measurements with timetaggers.

\paragraph{Direct Oscilloscope Measurements}
Where appropriate, we measure the electrical latency of components in a single-pass configuration. We split pulses produced by a pulse generator with a $\approx \qty{1}{\nano \second}$ rise time, and measure relative delays on a $\qty{1}{\giga\hertz}$ oscilloscope employing length-matched coaxial cables.
For single-pass optical measurements (always in free-space), we employ physical measurements of the path lengths. 

\paragraph{Time Domain Reflectometry}
When single-pass measurements were not appropriate, we use electrical and OTDR to measure component latencies. Here, a pulsed source is directed through one end of an electrical or optical path such that the reflections off the far end can be timed and divided by two to compute the component latency. 

\begin{figure}
    \centering
    \includegraphics[width=0.7\linewidth]{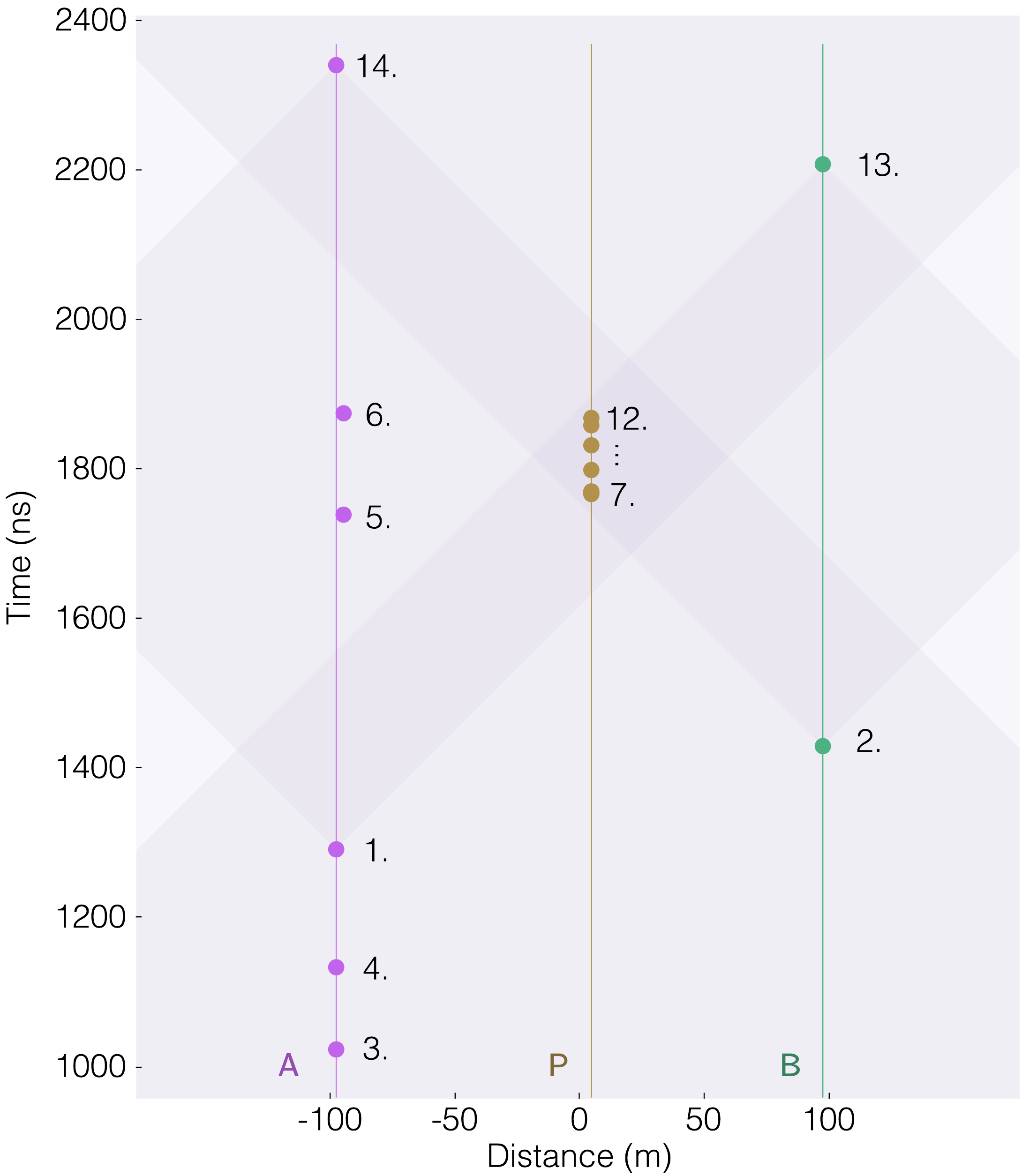}
    \caption{Spacetime diagram of the position verification protocol. Important events during one trial of the protocol (listed in~\cref{tab-supp:timing_parameters_numbered}) are marked. The purple lightcones are forward and backward lightcones from the times $s_{\Vap}$, $s_{\Vb}$, $r_{\Vap}$, and $r_{\Vb}$ (labeled 1, 2, 14 and 13 respectively) when the verifiers send the challenge bits and receive the response bits (see~\cref{subsec:target_region}). The target region is the darker diamond that contains points 7 through 12. The events labeled 5 and 6 occur at \(\Va\), which is slightly offset from \(\Vap\).  The origin (time $t=0$) of both this figure and~\cref{tab-supp:timing_parameters_numbered} is when the sync signal corresponding to a particular trial exits the divider board. (see~\cref{subsec:hardware_timing}).}
    \label{fig-supp:spacetime_diagram}
\end{figure}

The results for key events in the experiment are listed in \cref{tab-supp:timing_parameters_numbered} and depicted in~\cref{fig-supp:spacetime_diagram}. While provided here for completeness, not all of these times need to be characterized or measured for the execution of the position verification protocol by the verifiers. The only times that must remain constant during a protocol run are the times when the verifiers send the challenge bits and receive the response bits (numbered 1, 2, 13 and 14). The constancy of these times can be checked with the verifiers' timetaggers during data runs. The timetaggers are assumed to be functioning correctly and outside adversarial influence (see assumptions \ref{ass:timing} and \ref{ass:lab}). For our demonstration, we verify the timings on a few data sets, and assume that the timings remain constant during the actual data runs.

\begin{table}[h!]
\centering
\renewcommand{\arraystretch}{1.2}

\begin{tabular}{@{}r l r@{}} 
\toprule
\multicolumn{1}{c}{\textbf{Index}} & \textbf{Parameter}                     & \multicolumn{1}{c}{\textbf{Value}} \\ 
\midrule
1. & Verifier \(\Vap\): random \(\cba\) out           & \timeCbaOut \\
2. & Verifier \(\Vb\): random \(\cbb\) out           & \timeCbbOut \\
3. & Verifier \(\Vap\): photon \(\Qp\) exits laser       & \timeQpLas \\
4. & Verifier \(\Vap\): photon \(\Qp\) emitted    & \timeQpEm \\
5. & Verifier \(\Va\): setting \(\mqa\) created               & \timeSetMqa \\
6. & Verifier \(\Va\): outcome \(\oqa\) recorded                 & \timeOutOqa \\
7. & Prover \(\Vp\): \(\cba\) in from \(\Vap\)             & \timeCbaIn \\
8. & Prover \(\Vp\): \(\cbb\) in from \(\Vb\)             & \timeCbbIn \\
9. & Prover \(\Vp\): signal into Pockels cell                        & \timeSigPc \\
10. & Prover \(\Vp\): photon \(\Qp\) arrives at Pockels cell                & \timePhotPc \\
11. & Prover \(\Vp\): photon \(\Qp\) arrives SNSPD             & \timePhotSnspd \\
12. & Prover \(\Vp\): signals \(\zqa, \zqb\) created          & \timeZCreate \\
13. & Verifier \(\Vb\): response \(\zqb\) in            & \timeZqbIn \\
14. & Verifier \(\Vap\): response \(\zqa\) in              & \timeZqaIn \\
\bottomrule
\end{tabular}
\caption[Timings for position verification]{A summary of the times events occur in a position verification trial, referenced to the time the synchronization signal exits the divider circuit as time $t=\qty{0}{\nano\second}$. Relevant events in the trial span about \qty{1050}{\nano\second}, if we consider the classical signal exiting the verifier \(\Vap\) lab as the start time at verifier \(\Vap\). The protocol run time would increase with larger separations between the verifiers \(\Vap\) and \(\Vb\). The trial rate is \qty{4}{\micro \second}. See~\cref{fig-supp:spacetime_diagram} for a visualization of these key events.}
\label{tab-supp:timing_parameters_numbered}
\end{table}

\subsection{Physical Distances}
The physical distance between the verifiers is assumed to be known and constant during the protocol, see \cref{subsect:protocol_assumptions}, assumption 5. We determine this distance with measuring tape by hand-measuring a set of orthogonal spans. The location of the prover hardware that we install for demonstrating successful runs of the protocol has a bearing on the choices of time delays used for the classical position verification. We measure these distances with measuring tape also. In a real-world implementation of this protocol, these distances must first be established using classical geomapping techniques before one could implement the position verification protocol. 

The major contributors to uncertainty in the physical measurements is the uncertainty in the orthogonality of the various spans in the building, and the size of the exit points of the verifier labs. To estimate the effect of various uncertainties, we perform Monte-Carlo sampling based on the uncertainties of the individual spans, and the uncertainty in the solid angles connecting the various spans. Uncertainty in the solid angles shortens the best estimate for the separation between the verifiers because deviations from \qty{90}{\degree} for the angles between the orthogonal spans reduce the verifier separation. We estimate that the building is true to \qty{1}{\degree}, and employ a projected normal distribution to model angular uncertainty. The largest source of uncertainty is the \qty{200}{\milli\meter} uncertainty in the size of the verifier locations (cable exit points) \(\Vap\) and \(\Vb\), pictured in~\cref{fig-supp:exit-point}. A summary of the distance measurements is presented in \cref{supp:distance_measurements}.

\begin{table}[t!] 
    \centering 
    \renewcommand{\arraystretch}{1.2}
    \begin{tabular}{@{}l r@{}} 
        \toprule
        \textbf{Distance}                          & \multicolumn{1}{c}{\textbf{Value}} \\
        \midrule
        Verifier \(\Vap\) - Verifier \(\Vb\)                             & \valVerifierSep \\
        Verifier \(\Vap\) - Prover \(\Vp\) (proj.)                & \valDistVtwoP \\
        Verifier \(\Vb\) - Verifer \(\Va\) TTAG (proj.)                  & \valDistVtwoBttag \\
        Verifier \(\Vb\) - Verifer \(\Va\) RNG (proj.)                   & \valDistVtwoBrng \\
        \bottomrule
    \end{tabular}
    \caption[Measured distances for position verification]{This table summarizes key measured distances and their associated uncertainties for our experimental layout. Because there are only two classical verifiers transmitting challenges in our demonstration, the only relevant distances are those projected onto the line joining the verifiers \(\Vap\) and \(\Vb\), indicated by ``(proj.)''.}
    \label{supp:distance_measurements}
\end{table}

\subsection{Target Regions in the Spatial Dimensions}\label{subsec:target_region}

Using the distance and timing measurements, the verifiers can establish the target region that effectively localizes the successful prover. A formal treatment of the target region is presented in~\cref{subsect:spacetimeana}. In this section, we outline how the target region realized by the experimentally implemented protocol instance is calculated.

\subsubsection{Experimental Parameters for Determining Target Regions}
Our experiment is a series of trials, and times for trial $k$ are described relative to the start time \(t_k\) for that trial. To compute the extent of the target region \(\tR\) in 3+1 dimensions, and lower-dimensional projections thereof, as in Fig. 3  of the main text, only four times are relevant. These are the times when the verifiers send classical challenge bits and the latest times they accept responses. Following the notation introduced in~\cref{sec:preliminaries}, \(t_{k}+s_{\Vap}\) and \(t_{k}+s_{\Vb}\) are the times at
which \(\Vap\) and \(\Vb\) emit their challenges \(\cba\) and
\(\cbb\) toward the prover. The verifiers agree beforehand that responses must be
received by times \(t_k+r_{\Vap}\) and
\(t_k+r_{\Vb}\) at \(\Vap\) and \(\Vb\) respectively. In our implementation, the verifiers time-tag when challenges are emitted, and when responses are received.

A successful prover must simultaneously satisfy the following constraints related to the timing of signals in every trial (in addition to producing a Bell violation at the verifiers). 
\begin{enumerate}
    \item Receive challenge \(\cba\) (released at \(t_k+s_{\Vap}\)) from \(\Vap\) and reply with \(\zqa\) to \(\Vap\) (by \(t_k+r_{\Vap}\)). 
    \label{expt_supp:constraint_1}
    \item Receive challenge \(\cbb\) (released at \(t_k+s_{\Vb}\)) from \(\Vb\) and reply with \(\zqb\) to \(\Vb\) (by \(t_k+r_{\Vb}\)).
    \label{expt_supp:constraint_2}
    \item Receive challenge \(\cba\) (released at \(t_k+s_{\Vap}\)) from \(\Vap\) and reply with \(\zqb\) to \(\Vb\) (by \(t_k+r_{\Vb}\)).
    \label{expt_supp:constraint_3}
    \item Receive challenge \(\cbb\) (released at \(t_k+s_{\Vb}\)) from \(\Vb\) and reply with \(\zqa\) to \(\Vap\) (by \(t_k+r_{\Vap}\)).
    \label{expt_supp:constraint_4}
\end{enumerate}

These constraints require that the prover measures \(\Qp\) with low enough latency to be able to respond in time and that the messages are transmitted to and from a prover over fast enough classical channels. As explained in the main text,
\begin{itemize}
    \item Constraint~\ref{expt_supp:constraint_1} enforces a sphere of radius \(\frac{c\ (r_{\Vap} - s_{\Vap})}{2}\) centered around \(\Vap,\ \Sp{\Vap}{c(r_{\Vap} - s_{\Vap})}\).
    \item Constraint~\ref{expt_supp:constraint_2} enforces a sphere of radius \(\frac{c\ (r_{\Vb} - s_{\Vb})}{2}\) centered around \(\Vb,\ \Sp{\Vb}{c(r_{\Vb} - s_{\Vb})}\).
    \item Constraint~\ref{expt_supp:constraint_3} enforces an ellipsoid of rotation with foci at the verifiers and a major axis $c(r_{\Vb} - s_{\Vap}), \El{c(r_{\Vb} - s_{\Vap})}$.
    \item Constraint~\ref{expt_supp:constraint_4} enforces an ellipsoid of rotation with foci at the verifiers and a major axis $c(r_{\Vap} - s_{\Vb}),\ \El{c(r_{\Vap} - s_{\Vb})}$.
\end{itemize} 

The dimensions of these shapes are indicated in~\cref{supp:timing-const-regions}. The target region for each trial is the intersection of these spheres and ellipsoids, which are all rotationally symmetric about the axis joining the verifiers.

Because of the periodic nature of the trials, there is also a periodic constraint in the times that a prover must be located in these regions, which is not captured in the spatial projections.

\begin{table}[t!] 
    \centering 
    \renewcommand{\arraystretch}{1.2} 
    \begin{tabular}{@{}l r@{}} 
        \toprule
        \textbf{Parameter Description}       & \multicolumn{1}{c}{\textbf{Value}} \\
        \midrule
        Sphere \(\Sp{\Vap}{c(r_{\Vap} - s_{\Vap})}\), radius             & \valCircleOneRad \\ 
        Sphere \(\Sp{\Vb}{c(r_{\Vb} - s_{\Vb})}\), radius             & \valCircleTwoRad \\
        Ellipsoid \(\El{c(r_{\Vb} - s_{\Vap})}\), major axis        & \valEllipseTwo \\
        Ellipsoid \(\El{c(r_{\Vap} - s_{\Vb})}\), major axis        & \valEllipseOne \\
        Focal distance for ellipsoids (verifier separation)           & \valVerifierSep \\ 
        \bottomrule
    \end{tabular}
    \caption[Localization regions from the timing constraints]{Extents of the geometric localization regions derived from the timing and distance measurements in the experimental setup. All values are presented with their associated uncertainties, rounded to two decimal places. The units for these measurements are meters.}
    \label{supp:timing-const-regions}
\end{table}

\subsubsection{Computing Target Regions}
When projecting the target regions to the spatial dimensions (to perform position localization), the aforementioned two spheres and two ellipsoids are the only relevant geometries that verifiers need to estimate. Estimating these geometries requires measurement of the physical distance between the verifiers determined via the procedures outlined in~\cref{supp:distance_measurements}, and the start and end transmission times determined by the verifiers' timetagger measurements (see~\cref{tab-supp:timing_parameters_numbered}). 

For the QPV protocol, the target region $\tR$ (where an honest prover must be located) projected to spatial dimensions is the intersection of the two spheres and two ellipses from Constraints~\ref{expt_supp:constraint_1}-\ref{expt_supp:constraint_4} above, \(\tR = \Sp{\Vap}{c\Delta\tau_1} \cap \Sp{\Vb}{c\Delta\tau_2} \cap \El{c\Delta\tau_3} \cap  \El{c\Delta\tau_4}\). This is shown in~\cref{subsect:spacetimeana}.

The target regions can also be projected into lower spatial dimensions
(one and two dimensions).  For a one-dimensional projection, the
projection of interest is along the line joining the verifiers, and
for two dimensions, it is the plane roughly parallel to the ground and
containing the line joining the verifiers. Because of the rotational
symmetry of the regions, projection onto any plane rotated along the
line joining the verifiers yields an identical target region.

When projecting onto lower-dimensional regions, the spheres and ellipsoids reduce to lines in one dimension and circles and ellipses in two dimensions. The target region $\tR$ can also be computed as a projection into these lower-dimensional regions. To perform these calculations, we use the known distances between the verifiers from~\cref{tab:calib_dist}, and the timings from~\cref{tab-supp:timing_parameters_numbered} to numerically compute the lengths, areas and volumes of the target regions of the quantum protocol (quantum target region) in one, two and three dimensions. We also use the uncertainties in these quantities to run a Monte Carlo simulation to yield an estimate of the uncertainty in the length/area/volume of the target region.

\subsubsection{Comparing Classical and Quantum Target Regions}

An alternative protocol where the verifiers only require an honest prover to compute a classical function based on inputs \(u_{\Vap}\) and \(u_{\Vb}\) can be considered, see~\cref{subsect:spacetimeana}. Because such a protocol requires no quantum resources on the part of the prover or the verifiers, it serves as a useful comparison to the Bell-test protocol. If the target region achievable with the classical position verification protocol is comparable in size to the quantum target region \(\tR\), we gain little by adding quantum resources to the Bell-test protocol. On the other hand, if the quantum target region is significantly smaller than the classically achievable target region, we take the Bell-test protocol to have demonstrated a quantum advantage. We note that the security of such a classical protocol is based on the impossibility of faster-than-light signaling and thus does not require restrictions on the entanglement shared between the adversaries that the Bell-test protocol requires (detailed in~\ref{sec:smallentanglement}, also see Assumption~\ref{ass:ent}). 

We employ the following metric to quantify the advantage of using quantum resources,
\begin{align}
    {\rm Quantum\ Advantage} &= \frac{\rm size\ of\ classical\ target\ region}{\rm size\ of\ quantum\ target\ region\ (\tR)}
\label{supp_eq:quantum_advantage}
\end{align}
The ``size'' in the metrics may be a length, area, or volume when considering one, two or three-dimensional position verification.
An ideal quantum protocol (without experimental latencies in transmission and processing) can achieve an infinite Quantum Advantage, as defined above. In general, the goal of a good quantum protocol is to maximize the Quantum Advantage.

In this work, we compare our quantum experiment to two classical protocols:

\noindent\textbf{The ideal classical protocol} involves the verifiers sending challenges at the speed of light, and the implemented prover replying back with zero latency, also at the speed of light. The target region is the point exactly at the midpoint of the line joining the verifier. This represents the classical protocol with the smallest possible target region.

\noindent\textbf{The comparable classical protocol} involves the verifiers sending challenges and receiving responses at the same times as our quantum protocol (\(t_k + r_{\Vap}, t_k + r_{\Vb}, t_k + s_{\Vap}, t_k + s_{\Vb}\)). The motivation here is to consider a practically implementable classical protocol. For any prover implementation, computing classical functions has lower latency than making quantum measurements. So it is reasonable that a classical protocol with the latencies we demonstrate here for the quantum protocol can be implemented.

For the classical protocols, a pair of provers can collude to pass the protocol as long as one of them is in \(\Big(\Sp{\Vap}{c\Delta\tau_1} \cap\El{c\Delta\tau_3}\Big)\), and the other in \( \Big( \Sp{\Vb}{c\Delta\tau_2} \cap  \El{c\Delta\tau_4}\Big)\). This is shown in~\cref{subsect:spacetimeana}. We take this to mean that the classical target region is the union of these two regions for all our comparisons.

For \textbf{the ideal classical protocol}, the ellipsoids from Constraints~\ref{expt_supp:constraint_3} and \ref{expt_supp:constraint_4} reduce to degenerate line segments. This is because the major axis is equal to the distance between the verifiers ($c(r_{\Vap} - s_{\Vb}) = c(r_{\Vb} - s_{\Vap}) = d_{\Vap \Vb}$). The classical target region is the line segment joining the verifiers. Because this is an \textit{ideal} protocol, the target region is one-dimensional. The only uncertainty associated with the size of this region is in the distance between the verifiers. The quantum advantage for the ideal classical protocol is the ratio of its target region to the quantum target region projected along the line joining the verifiers. We performed Monte Carlo simulations to estimate the uncertainty in this ratio. The results are presented in~\cref{supp:localization-advantage-comparison}.

\begin{figure}[!ht]
    \centering
\includegraphics[width=\linewidth]{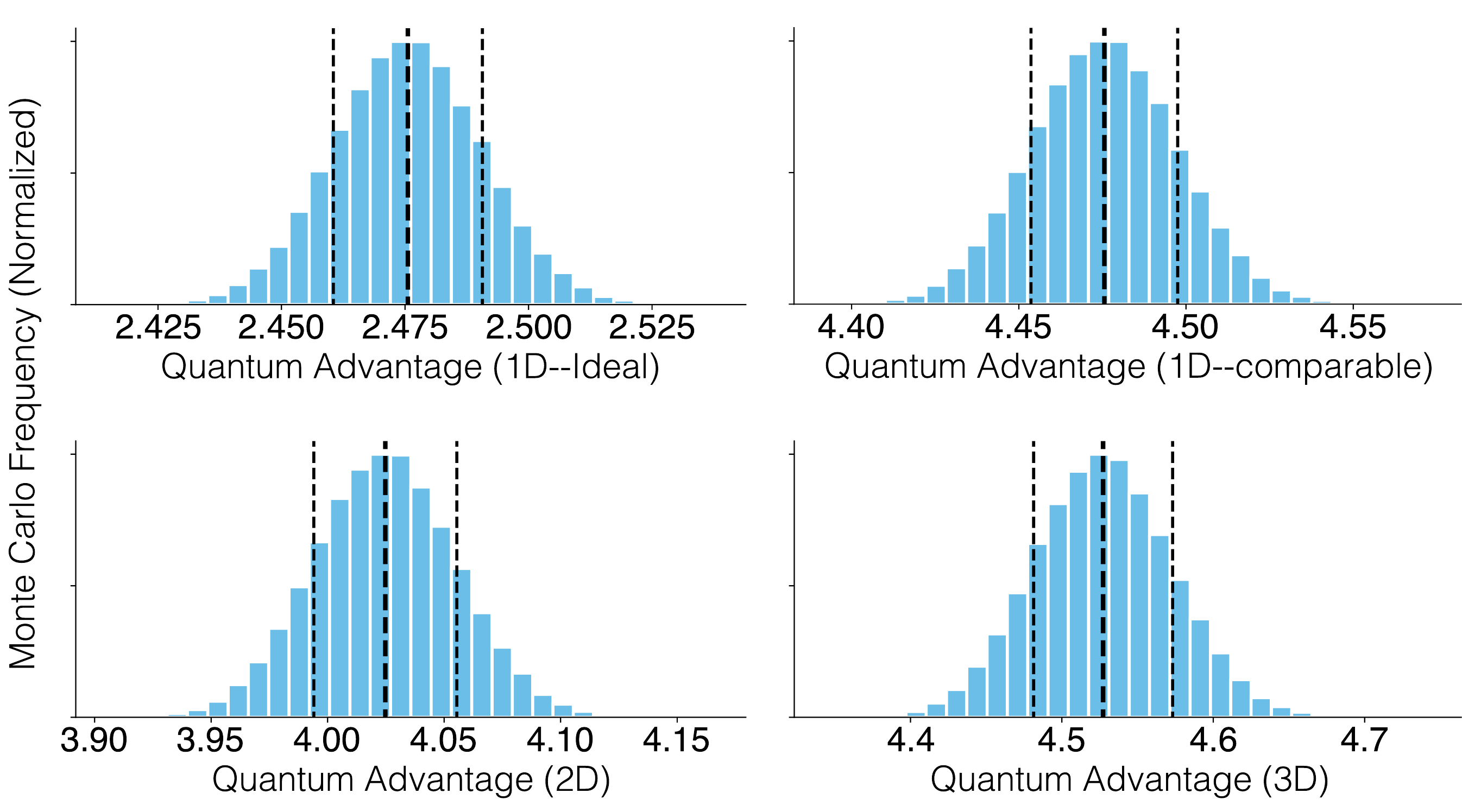}
    \caption{Histograms showing the relative frequencies of the quantum advantage from \(10^5\) Monte Carlo simulations. The simulations sample the measured distances and timings based on estimated uncertainties to arrive at the quantum advantage distributions. The uncertainties on individual timing and distance measurements are modeled as Gaussian (with the exception of angular uncertainties, which use projected Gaussians). The largest source of uncertainty is the physical ambiguity in the definition of the verifier \(\Vap\) and \(\Vb\) positions, see~\cref{fig-supp:exit-point}. The mean and \(\pm 1\sigma\) for each distribution are indicated with vertical black dashed lines. The top row plots the quantum advantage against ideal and comparable quantum protocols in 1 dimension, while the bottom row plots the advantage against the comparable protocol in 2 and 3 dimensions.}
    \label{supp-fig:mc-hist}
\end{figure}

For \textbf{the comparable classical protocol}, the target regions in one, two and three dimensions all have non-zero sizes, and we compare these to the sizes of the quantum target region. This computation is similar to that done for the ideal classical protocol. The results of these comparisons are presented in~\cref{supp:localization-advantage-comparison}, \cref{supp-fig:mc-hist} and in the main text.

\begin{table}[!ht] 
    \centering
    \renewcommand{\arraystretch}{1.2} 
    \begin{tabular}{@{}l c c@{}}
        \toprule
        \textbf{Dimension}                                & \textbf{Quantum Advantage} \\
        \midrule
        1D (vs Ideal Classical)                 & \valMeanInvIdealLine \\
        1D (vs Comparable Classical)              & \valMeanInvLineInt \\
        \addlinespace 
        2D (vs Comparable Classical)    & \valMeanInvArea \\
        \addlinespace
        3D (vs Comparable Classical)   & \valMeanInvVol \\
        \bottomrule
    \end{tabular}
    \caption[Localization Ratios and Quantum Advantages]{Comparison of localization ratios and quantum advantages for QPV protocols in different dimensions. The quantum protocol achieves zero quantum advantage compared to the ideal classical protocol in 2D and 3D because the ideal classical size is zero in 2D and 3D. Reported uncertainties are one standard deviation.}
    \label{supp:localization-advantage-comparison}
\end{table}

\pagebreak


\section{Theory}\label{sec:theory}

\subsection{Statements of results}
\label{sec:results statements}

We prove soundness and completeness of the Bell-test protocol in
Sect.~\ref{subsect:protocol_assumptions} for versions of the
assumptions in Sect.~\ref{subsect:protocol_assumptions}, specifically
versions of Assumption~\ref{ass:ent}.  
The same configuration of verifiers and honest prover
witnesses completeness regardless of which version of the
assumptions is used. In all cases, for soundness \(0<\delta<1\) and
completeness \(0<\epsilon<1\), completeness can be achieved with a
number of trials of order
\(O(\log(1/\delta)) + O(\log(1/(1-\epsilon)))\), where the suppressed
constants depend on the quality of the Bell-test on which protocol
completeness is based and the class of adversaries under
consideration. 
\begin{result}\label{res:completeness} Protocol completeness:
  For all \(\epsilon\in (0,1)\), the protocol of
  Sect.~\ref{subsect:protocol_assumptions} is
  \(\epsilon\)-complete. To witness completeness it suffices for the
  verifiers and honest prover to violate local realism for some
  Bell-test configuration with two measurement settings for the
  prover. For every class of adversaries under consideration and a
  given violation of local realism, the number of trials required
  for completeness \(\epsilon\) and soundness \(\delta\) is
  \(O(\log(1/\delta))+ O(\log(1/(1-\epsilon)))\). 
\end{result}
The proof of this result is in Sect.~\ref{sect:reduction}. Because
there is no restriction on how the entangled state for the Bell-test protocol 
is obtained by the verifiers and prover before each trial,
completeness is preserved if a slow quantum network is used to
prepare these states. The quantum link may be heralded and can
support prover commitment strategies~\cite{allerstorfer2025making}.
Therefore, loss in photonic links is well tolerated. See
Sect.~\ref{sect:heraldetc} for how the protocol supports heralded sources 
and commitment.

The soundness results depend on the class of adversaries under
consideration. In each case, soundness holds for all space
dimensions, and the source is untrusted and may be controlled by one
of the adversaries. Classical communication by the adversaries is
always unconstrained.  The first soundness result involves the class of
adversaries with no prior entanglement.
\begin{result}\label{res:soundness1}
  Protocol soundness against adversaries with no prior entanglement:
  For all \(\delta>0\), the protocol of
  Sect.~\ref{subsect:protocol_assumptions} is \(\delta\)-sound
  against adversaries who have no prior entanglement for each
  trial. The adversaries can freely use quantum computation and
  communication during each trial.
\end{result}
The proof of Result~\ref{res:soundness1} is contained in
Sects.~\ref{subsect:spacetimeana},~\ref{sect:reduction}
and~\ref{sec:test_factors}. The first step of the proof follows the traditional path
of reducing the adversary attacks to those involving two adversaries
on opposite sides of the target region~\cite{buhrman2014position}.
The Bell-test protocol requires an analysis that is independent of space
dimension, takes into account the time required by the honest prover
to perform its measurement task, and supports device-independence
with test factors.  For this we adapt the spacetime circuit analysis
introduced in Ref.~\cite{unruh2014quantum} and fully formalize the
reduction to two adversaries. The subsequent analysis of adversary attacks
deviates from the traditional path in order to reduce the statistics
of adversarial trial records to a three-party non-signaling scenario
for which we can construct test factors.

The absence of prior entanglement for the above soundness result
implies a constraint on the location of the entanglement source at the
trial boundary: The source must be located on the side closer to
verifier \(\Va\), which requires that we prevent the quantum system
destined for the prover from leaking to the other side. This
constraint on the location of the source is mitigated by the next
soundness result and absent in the last two soundness results.

The second soundness result allows for small amounts of prior entanglement.
\begin{result}\label{res:soundness2} Protocol soundness against
  adversaries with small amounts of prior entanglement: For all
  \(\delta>0\), the Bell-test protocol of
  Sect.~\ref{subsect:protocol_assumptions} is \(\delta\)-sound
  against adversaries with small amounts of prior entanglement when
  quantified in terms of robustness of entanglement.  The
  adversaries can freely use classical and quantum computation and
  communication during each trial.
\end{result}
This is proven in Sect.~\ref{sec:smallentanglement}. The amount of prior
entanglement allowed depends on the quality of the Bell test used by
the verifiers and prover. The amount of prior entanglement is
quantified either as maximum robustness of entanglement over trials,
or as average robustness of entanglement over trials.  For the
experiment, the upper bounds on allowed prior entanglement are smaller than the
entanglement created by the source, limiting the strength of the
result.  The technique used here can be viewed as a quantified
strengthening of Lemma~5.3 of Ref.~\cite{beigi2011simplified}.  It
is also related to the entanglement forging method of
Ref.~\cite{eddins2022doubling}.  Neither our above bounds nor those based
on Ref.~\cite{beigi2011simplified} have so far led to bounds on
prior entanglement of adversaries that are higher than the resources
used by the verifiers and honest prover when quantum computation and
communication during a trial is unrestricted.

To prove soundness against adversaries with significant amounts of
prior entanglement requires increasing the complexity of the settings
function \(f(\cba,\cbb)\).  For the next two results, \(l\) is the
bit-length of \(\cba\) and \(\cbb\). The results hold with \(l=2l'+1\)
and \(f(\cba,\cbb)=g_{0}(\cba\oplus\cbb)\) where
\(g_{0}(x_{0},\ldots x_{2l'})=x_{0}+\sum_{i=1}^{l'}x_{i}x_{i+l'}\mod 2\),
where for this purpose we use \(\{0,1\}\) for the range of \(f\) and
the set of settings choices for the prover.
\begin{result}\label{res:soundness3} Protocol soundness against adversaries with
  logarithmic prior entanglement: There are efficiently
  implementable settings functions \(f(\cba,\cbb)\)
  such that for all \(\delta>0\), the Bell-test protocol of
  Sect.~\ref{subsect:protocol_assumptions} is
  \(\delta\)-sound against adversaries with prior entanglement
  on at most \(\log(l)/2-O(1)\) qubits, no quantum communication but arbitrary quantum
  computation and classical communication during each trial.
\end{result}
The proof is in Sect.~\ref{sect:entangled_adversaries}. Since
quantum communication can be replaced by prior entanglement and
classical communication, the soundness can be interpreted as being
against adversaries with at most \(\log(l)/2-O(1)\) qubits of prior
entanglement and quantum communication.  To support practical
applications, we fully quantify the allowed prior-entanglement bound. To maintain
a given soundness and completeness, it suffices to increase the
number of trials by a constant factor. We choose a factor of \(6\) in
the examples given in Sect.~\ref{sect:entangled_adversaries}.  The
suppressed constant in the bound \(\log(l)/2-O(1)\) depends on the test factor used and
its gain.  See the paragraphs starting at Eq.~\ref{eq:epsilon_n_d}
in Sect.~\ref{sect:entangled_adversaries}.  The proof of
Result~\ref{res:soundness3} is inspired by the proof of Prop. 6.1 in
Ref.~\cite{bluhm2022single}, but requires a different strategy for
applying an \(\epsilon\)-net and a different
communication-complexity bound. A specific analysis of the
relationship between adversarial test-factor expectations and guessing
probabilities for the settings function is required to efficiently
preserve completeness under all circumstances.

Instead of bounding prior entanglement and quantum communication,
we can bound qubits and gates used by the adversary nearer \(\Vb\).
\begin{result}\label{res:soundness4} Protocol soundness against adversaries with
  linear quantum operations: There are efficiently implementable
  settings functions \(f(\cba,\cbb)\) such that for all
  \(\delta>0\), the Bell-test protocol of
  Sect.~\ref{subsect:protocol_assumptions} is \(\delta\)-sound
  against adversaries with arbitrary classical communication but no
  quantum communication, and where the adversary nearer \(\Vap\) can
  use arbitrary local quantum resources, but the adversary nearer
  \(\Vb\) uses \(O(l/(\log(q)+1))\) quantum operations in a basic
  universal set, where \(q\) is the number of qubits used by this
  adversary.
\end{result}
The proof of this result is based on the proof of
Result~\ref{res:soundness3} and given at the end of
Sect.~\ref{sect:entangled_adversaries}.  Again, to support practical
applications, we fully quantify the bound on quantum resources allowed by adversaries.  The
proof of Result~\ref{res:soundness4} is motivated by the
argument in Ref.~\cite{asadi2025linear}.
The suppressed constant in the bound
\(O(l/(\log(q)+1))\) depends on the size of the basic universal set of
quantum operations, which includes qubit measurements.

In Refs.~\cite{bluhm2022single,asadi2025linear}, the preferred
settings functions are either random or the inner product function
\(f(\cba,\cbb)= \sum_{i}\cba_{i}\cbb_{i}\mod 2\). Random settings
functions are not scalably implementable. The inner product function
does not satisfy the assumptions of the Bell-test protocol, see
Assumptions~\ref{ass:funct} and ~\ref{ass:distrib}. In particular,
if one verifier's challenge is all zeros, the value of the inner
product function is zero, which determines the prover setting.  The
inner product function is also a poor choice for establishing
adequate lower bounds on the communication complexity of the
communication scenario that plays a key role in the proof of
Result~\ref{res:soundness3}.

\subsection{Spacetime analysis}\label{subsect:spacetimeana}

For defining the target region and the restrictions on prior
entanglement of adversaries, we use a version of the spacetime circuit
formalism and adversary analysis introduced in
Ref.~\cite{unruh2014quantum} for QPV protocols with parties that have
access to so-called random oracles for defining the settings function.
The analysis of Ref.~\cite{unruh2014quantum} applies to any number of
verifiers and needs no restriction on the amount of prior entanglement
of adversaries. However, the protocol involves one round consisting of
a number of parallel trials, depends on quantum random oracles and
does not determine a trial boundary on which to define prior
entanglement.  We require a different version of this spacetime
analysis to identify the adversary capabilities and the trial boundary
on which entanglement is restricted.  For this purpose, we provide a
self-contained description of the formalism and analysis.  We first
define the target region and the trial boundary on which adversaries'
entanglement is restricted. We then reduce the problem of analyzing
adversary capabilities without prior entanglement to the case of
having two separate, initially unentangled adversaries that can each
communicate to the other only once before committing to responses to
the verifiers.

\noindent{\textbf{Spacetime circuits:}}
We assume familiarity with the fundamental concepts of relativistic
spacetime. Our theory applies to any time-oriented globally-hyperbolic
Lorenzian manifold \(\cM\). Minkowski space is the most familiar
example of such a manifold. For a comprehensive treatment of the
relevant concepts see, for example,
Ref.~\cite{wald1984generalrelativity}. Here we give a brief overview
of the concepts used here. An event or point in spacetime is
identified by its position in the manifold.  A worldline is a
forward-directed light- or time-like path in spacetime. For our
purposes, a worldline may be identified with the image of a
differentiable function \(\bm{r}(x)\) from an interval \(I\) on the
reals to spacetime, where \(I\) may be open or closed on either end,
and \(d\bm{r}(x)/dx\) is forward-directed, light-like or time-like on
the interior of \(I\). The worldline is inextendible if it is not a
proper subset of another worldline.  Event \(\bm{r}'\) causally follows
event \(\bm{r}\) if there is a worldline that contains both \(\bm{r}\)
and \(\bm{r}'\), and \(\bm{r}'\) follows \(\bm{r}\) on the
worldline. Because we assume global hyperbolicity, this defines the
causal partial order on spacetime, and we write
\(\bm{r}\preceq \bm{r}'\) if \(\bm{r}'\) causally follows
\(\bm{r}\). The closed forward lightcone \(C_{\bm{r},+}\) at
\(\bm{r}\) consists of the events that causally follow \(\bm{r}\),
including \(\bm{r}\). Similarly, the closed backward lightcone
\(C_{\bm{r},-}\) at \(\bm{r}\) consists of the events that causally
precede \(\bm{r}\), including \(\bm{r}\).  Lightcones have the
property that no worldline can exit a forward lightcone or enter a
backward lightcone.  This means that the intersection of an
inextendible worldline with a lightcone is either empty or a closed
half-line.  Backward lightcones are instances of past sets, which are
defined as subsets \(J\) of spacetime such that if \(\bm{r}\in J\) and
\(\bm{r}'\preceq \bm{r}\), than \(\bm{r}'\in J\).  Boundaries of past
sets are instances of time slices, which we define to be closed
space-like subsets of spacetime whose spacetime complement has two
connected components, one consisting of the events that causally
precede the time slice, and the other consisting of the events that
causally follow the time slice. Space-like subsets satisfy that for
every pair \(\bm{r}\) and \(\bm{r}'\) in the subset, neither
\(\bm{r}\preceq\bm{r}'\) nor \(\bm{r}'\preceq\bm{r}\).  Global
hyperbolicity is equivalent to the existence of a time slices
satisfying the property that every inextendible worldline intersects
the time slice. In Minkowski space, the boundaries of light cones are
examples of time slices that do not satisfy this property.  A subset
\(D\) of spacetime is causally convex if for every worldline that
starts and ends in \(D\), the worldline is contained in
\(D\). Equivalently, if \(\bm{r}\preceq \bm{r}'\) are events in \(D\)
and \(\bm{r}''\) satisfies \(\bm{r}\preceq \bm{r}''\preceq\bm{r}'\),
then \(\bm{r}''\in D\). A special case of a causally convex subset is
a light diamond from one event \(\bm{r}\) to another \(\bm{r}'\),
which is defined as the intersection of the forward light cone
\(C_{\bm{r},+}\) at \(\bm{r}\) and the backward light cone
\(C_{\bm{r}',-}\) at \(\bm{r}'\).  Light diamonds are empty if
\(\bm{r}'\) does not causally follow \(\bm{r}\).  The intersection of
a worldline with a light diamond forms a time-bounded interval, which
may be empty.

The following is a conceptual description of the behavior of entities
such as verifiers and provers in spacetime: All entities exist on
worldlines.  The entities can interact with communications.
Communications travel on worldlines starting on one entity's worldline
and ending at another's. We refer to the events where an entity sends
or receives a communication as interaction vertices and assume that
there are finitely many such vertices in every compact region of
spacetime.

Spacetime circuits abstract the conceptional description
above. Spacetime circuits do not distinguish between worldlines of
communications and entities connecting vertices, and confine the
dynamics of systems to the vertices. In the conceptual picture,
entities may perform operations along their worldlines. In
the spacetime circuit, such operations are moved forward to the next
interaction vertex along the line and combined into one operation.

Spacetime circuits \(\cS\) are defined by the following: (1) A set of
interaction vertices \(\cV=\cV(\cS)\), which are events in
spacetime. (2) A set of causally directed edges \(\cE=\cE(\cS)\)
representing worldlines connecting two events in spacetime, at least
one of which is an interaction vertex in \(\cV\). The two events
connected by an edge \(E\) are referred to as the source vertex
\(V_{\Src}(E)\) and the target vertex \(V_{\Tgt}(E)\), where
\(V_{\Tgt}(E)\) causally follows \(V_{\Src}(E)\).  An edge \(E\) is
internal if both endpoints are in the set \(\cV\). It is external if
exactly one of the endpoints is in \(\cV\).  (3) For each edge
\(E\in\cE\) an assignment of a quantum system \(Q(E)\) carried by
\(E\). (4) For each vertex \(V\in\cV\) a quantum operation
\(\cO(V)\) that jointly transforms incoming quantum systems to
outgoing quantum systems at \(V\). For a vertex \(V\in\cV\), the
incoming quantum systems are those associated with edges \(E\) whose
target vertex is \(V\), that is, \(V_{\Tgt}(E)=V\).  The outgoing
quantum systems are those associated with edges \(E\) whose source
vertex is \(V\).  The quantum operation at a vertex without incoming
edges is a state preparation on the joint quantum systems of the
outgoing edges. The operation at a vertex with no outgoing edges
discards the quantum systems of the incoming edges.  We refer to the
state of quantum systems of edges simply as the state of the
edges. 

Spacetime circuits describe the dynamics of the edge
systems. Intuitively, for any time slice, the systems of edges
crossing this time slice have a joint state, and the quantum
operations on the vertices determine the joint state on a later time
slice given the joint state on an earlier one. To treat this dynamics
formally, we introduce a number of causal concepts for spacetime
circuits.  The vertices and edges of a spacetime circuit form a
directed acyclic graph, which induces a partial order over the union
of the vertex and edge sets.  A directed path in the circuit is a
sequence of alternating vertices and edges \((A_{k})_{k=0}^{m}\) with
\(m\geq 0\) that starts with either a vertex or an edge, where the
vertices and edges along the sequence are incident on each other and
in causal order. Specifically, if \(A_{k}\) is an edge and \(A_{k+1}\)
is a vertex, then \(A_{k+1}=V_{\Tgt}(A_{k})\), and if \(A_{k}\) is a
vertex and \(A_{k+1}\) is an edge, then \(A_{k}=V_{\Src}(A_{k+1})\).
Topologically, we can associate to each edge its open worldline
between its source vertex and its target vertex.  Then the union of
the vertices and edges on a directed path is a worldline. We use
directed paths to define the circuit partial order.  For two vertices
or edges \(A\) and \(B\), \(A\) precedes \(B\) in the circuit partial
order, written \(A\leq B\), iff there is a directed path starts with
\(A\) and ends with \(B\).  This partial order is weaker than the
partial order induced by spacetime causality, in the sense that we can
have \(A\) causally precede \(B\) in spacetime but not \(A\leq B\)
in the circuit partial order. 

A vertex subset \(\cV'\) of \(\cV\) is causally convex if for every
directed path starting and ending in \(\cV'\), all vertices on the
path are in \(\cV'\).  We define the sets of in-edges and out-edges of
\(\cV'\) as
\(\In(\cV')=\{E\in\cE: V_{\Src}(E) \notin\cV', V_{\Tgt}(E)\in\cV'\}\)
and
\(\Out(\cV') = \{E\in\cE: V_{\Src}(E) \in\cV',
V_{\Tgt}(E)\notin\cV'\}\).  That is, \(\In(\cV')\) includes all edges
entering \(\cV'\) from outside, while \(\Out(\cV')\) includes all
edges exiting \(\cV'\) to the outside.  A causally convex set of
vertices \(\cV'\) defines a spacetime subcircuit \(\cS(\cV')\) with
vertex set \(\cV'\) and edge set \(\cE'\) consisting of the edges in
\(\cE\) with at least one endpoint in \(\cV\).  The external edges of
\(\cS(\cV')\) consist of the union of \(\In(\cV')\) and
\(\Out(\cV')\).  A causally convex set of vertices \(\cV'\) and its
associated spacetime subcircuit \(\cS(\cV')\) define a quantum
operation \(\cO(\cV')\) that maps the quantum systems of its in-edges
to those of its out-edges. This quantum operation can be constructed
recursively as follows: 1) Every set \(\cV'\) consisting of a single
vertex \(V\) is causally convex, and its quantum operation is given by
\(\cO(\cV')=\cO(V)\).  2) If \(\cV'\) consists of more than one
vertex, choose a vertex \(Z\in \cV'\) that is maximal in the circuit
partial order restricted to \(\cV'\). That is, no other vertex in
\(\cV'\) lies after \(Z\) along a directed path.  Then the reduced set
\(\cV'' = \cV'\setminus \{Z\}\) is causally
convex. Suppose that we have constructed the
quantum operation \(\cO(\cV'')\) of \(\cV''\). The in-edges of \(Z\)
partition into \(\cZ_{1}=\In(\{Z\})\cap\Out(\cV'')\) consisting of
the in-edges of \(Z\) that are out-edges of \(\cV''\), and
\(\cZ_{2}=\In(\{Z\})\cap\In(\cV')\) consisting of the in-edges of
\(Z\) that are in-edges of \(\cV'\). The in-edges of \(\cV'\)
consist of the in-edges of \(\cV''\) and the edges of \(\cZ_{2}\).
The quantum operation of \(\cV'\) is constructed by adjoining the
quantum systems of edges in \(\cZ_{2}\) to those of the edges \(\In(\cV'')\), applying
\(\cO(\cV'')\) to the latter, and then applying the \(\cO(Z)\) to both
the systems of edges in \(\cZ_{2}\) and the systems of edges in
\(\cZ_{1}\) that appeared at the output of \(\cO(\cV'')\).

The spacetime-circuit notion of a time slice and its state is captured
by an antichain of edges and the joint state of their systems.  A set
of edges \(\cA\) is an antichain if no pair of edges in \(\cA\) is
comparable in the circuit partial order, that is, there is no directed
path connecting two edges in \(\cA\). For an antichain of edges
\(\cA\), consider the set \(\cV'\) of vertices \(V\) for which there
exists an edge \(E\) in \(\cA\) such that \(V\leq E\). Intuitively,
\(\cV'\) consists of the vertices that lie in the causal past of the
antichain \(\cA\). Then the set \(\cV'\) is causally convex.  To see
this, consider a directed path from vertex \(V_{1}\in\cV'\) to vertex
\(V_{2}\in\cV'\), and let \(V_{3}\) be a vertex on the path.  By
definition of the circuit partial order, \(V_{3}\leq V_{2}\), and by
construction of \(\cV'\), \(V_{2}\leq E\) for some edge \(E\in\cA\).
Therefore, \(V_{3}\leq E\), which implies that \(V_{3}\in\cV'\). The
state of the systems of the antichain is determined by the joint state
of the quantum systems on the in-edges of \(\cV'\).  The quantum
operation \(\cO(\cV')\) transforms this joint state into a joint state
of the quantum systems of its out-edges. Because the source vertices
of the edges in \(\cA\) are in \(\cV'\), the edges in \(\cA\) are
among the out-edges of \(\cV'\). The state of \(\cA\) is obtained by
tracing out the systems on the out-edges of \(\cV'\) that are not in
\(\cA\).

For a spacetime circuit \(\cS\) with vertex set \(\cV\) and a subset
\(D\) of spacetime, the intersection \(\cV\cap D\) consists of all
vertices in \(\cV\) that lie within the region \(D\). If \(D\) is
causally convex in spacetime, then \(\cV\cap D\) is causally convex in
the spacetime circuit \(\cS\). This is because the circuit partial
order on the vertices of \(\cV\) is weaker than the partial order
induced by spacetime causality. Consequently, If \(V\)
and \(V'\) are vertices in \(\cV\cap D\) and \(V''\) is a vertex of
\(\cV\) satisfying \(V\leq V''\leq V'\) in the circuit partial order
for \(\cS\), then \(V\preceq V''\preceq V'\) in the causal partial
order, so \(V''\in D\).

\begin{figure}[ht!]
  \centering

  \pgfmathsetmacro{\xunit}{5.5/16}
  \pgfmathsetmacro{\yunit}{3.5/12}
  \begin{tikzpicture}[x=\xunit in, y=\yunit in]
    \path[use as bounding box] (0,0) rectangle (16.5,13);
    \clip (0,0) rectangle (16.5,13);

    \def\xA{6}
    \def\xB{10}
    \coordinate (sA) at (\xA, 2);
    \coordinate (rA) at (\xA, 8.16);   
    \coordinate (sB) at (\xB, 2.83);
    \coordinate (rB) at (\xB, 7.39); 

    
    \begin{scope}
      \fill[gray!20] (rA) -- ($(rA)+(-12,-12)$) -- ($(rA)+(12,-12)$) -- cycle;
      \fill[gray!20] (rB) -- ($(rB)+(-12,-12)$) -- ($(rB)+(12,-12)$) -- cycle;
      \fill[white] (sA) -- ($(sA)+(-12,12)$) -- ($(sA)+(12,12)$) -- cycle;
      \fill[white] (sB) -- ($(sB)+(-12,12)$) -- ($(sB)+(12,12)$) -- cycle;
    \end{scope}

    
    \draw[->, thick] (0.5, 0) -- (0.5, 12) node[above] {$t$};
    \draw[->, thick] (0, 0.5) -- (16, 0.5) node[right] {$x$};

    \draw (\xA, 0) -- (\xA, 12) node[above=2mm] {$\Vap$};
    \draw (\xB, 0) -- (\xB, 12) node[above=2mm] {$\Vb$};
    
    \begin{scope}[gray, dashed]
      \draw (sA) -- ($(sA)+(-10,10)$); \draw (sA) -- ($(sA)+(10,10)$);
      \draw (sB) -- ($(sB)+(-10,10)$); \draw (sB) -- ($(sB)+(10,10)$);
      \draw (rA) -- ($(rA)+(-10,-10)$); \draw (rA) -- ($(rA)+(10,-10)$);
      \draw (rB) -- ($(rB)+(-10,-10)$); \draw (rB) -- ($(rB)+(10,-10)$);
    \end{scope}

    \fill (sA) circle (2.5pt) node[left=2mm] {$t_{k}+s_{\Vap}$};
    \fill (rA) circle (2.5pt) node[left=2mm] {$t_{k}+r_{\Vap}$};
    \fill (sB) circle (2.5pt) node[right=2mm] {$t_{k}+s_{\Vb}$};
    \fill (rB) circle (2.5pt) node[right=2mm] {$t_{k}+r_{\Vb}$};

    \node at (1.5,3.2) {\(T\)};
    \node at (8, 1) {\(0000\)};
    \node at (7, 3.6) {\(1000\)};
    \node at (9.4, 4) {\(0100\)};
    \node at (8.4, 5.1) {\(1100\)};
    \node at (13, 2.8) {\(0010\)};
    \node at (10.9, 5.4) {\(0110\)};
    \node at (9.4, 6.1) {\(1110\)};
    \node at (3, 2.5) {\(0001\)};
    \node at (5,5.6) {\(1001\)};
    \node at (7.4,6.1) {\(1101\)};
    \node at (2.5,8) {\(1011\)};
    \node at (13.5,7.5) {\(0111\)};
    \node at (8.4,8.5) {\(1111\)};
    \node at (1.8,5) {\(0011\)};
    \node at (15,5) {\(0011\)};

  \end{tikzpicture}
  \caption[Schematic \(1+1\)-dimensional spacetime cross-section of a trial.]{Schematic \(1+1\)-dimensional spacetime cross-section of a trial.
    The trial boundary \(T\) is the boundary of the gray-shaded region, which is
    the \(1+1\)-dimensional cross-section of the region \(\cM_{\In}\) defined in the text.
    The \(k\)\textsuperscript{th} trial is shown and \(t_{k}\) is the trial's
    start time according
    to the protocol. The regions are labeled by 4-bit strings according to the convention
    introduced before Eq.~\eqref{eq:advcirc_defs}, where
    bits \(1\) and \(2\) determine whether
    the points are causally after \(t_{k}+s_{\Vap}\) and \(t_{k}+s_{\Vb}\), respectively;
    bits \(3\) and \(4\) determine whether the points are not causally before
    \(t_{k}+r_{\Vap}\) and \(t_{k}+r_{\Vb}\), respectively.
    The region labeled with \(0011\) appears
    in two places, and regions labeled with 
    \(1010\) and \(0101\) do not appear in this cross-section.
    \label{fig:1+1d_regions} }
\end{figure}
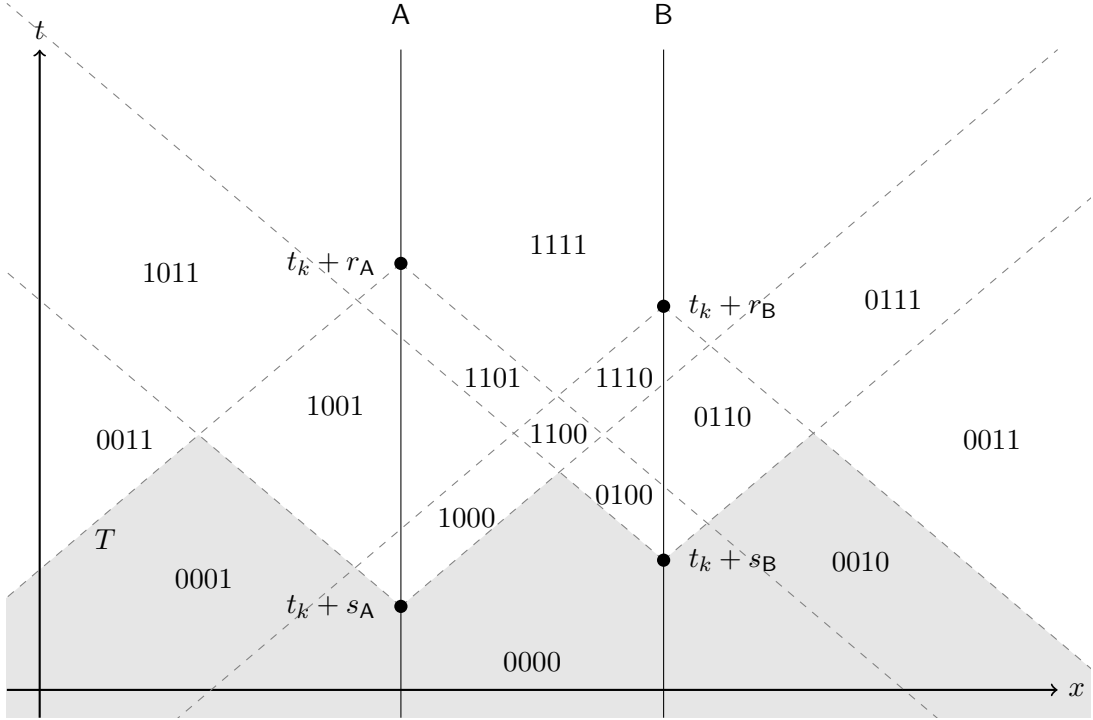

\noindent{\textbf{Spacetime circuit for verifiers and provers:}}
We model verifiers and provers as a spacetime circuit \(\cS\) that
spans all trials of the Bell-test QPV protocol.  We consider a generic trial
with start time $t$. We specify an event with the notation
$\bm{r}=(\bm{x},s)$, where $\bm{x}$ is the space location, and $s$
is the time. For \(\Ef=\Vap,\Vb\), let \(C_{\Ef,+}\) be the forward
lightcone from \(\bm{r}_{\Ef,s}=(\bm{x}_{\Ef},t+s_{\Ef})\) and
\(C_{\Ef,-}\) the backward lightcone from
\(\bm{r}_{\Ef,r}=(\bm{x}_{\Ef},t+r_{\Ef})\), where \(t+s_{\Ef}\) and
\(t+r_{\Ef}\) are the times at which the verifier \(\Ef\) sends a
challenge and receives a response, respectively. \(C_{\Ef,+}\) is the
region within which the challenge sent by \(\Ef\) is visible to
provers, and \(C_{\Ef,-}\) is the region within which it is possible
for a prover to send a response to \(\Ef\).  Let
\(D_{\Ef}=C_{\Ef,+}\cap C_{\Ef,-}\) be the light diamond from
\(\bm{r}_{\Ef,s}\) to \(\bm{r}_{\Ef,r}\), representing the region in
which a valid challenge–response interaction with the verifier \(\Ef\)
could occur.  The target region for the trial is defined as
\(E=D_{\Vap}\cap D_{\Vb}\), which we assume to be non-empty.  The
target region for the Bell-test protocol is the union of the target regions
for each trial.  We analyze spacetime circuits in \(d+1\)-dimensional
spacetime, where \(d\geq 1\).  For an illustration of the relevant
regions in a \(1+1\)-dimensional Minkowski cross-section containing
the verifiers in the experiment, see Fig.~\ref{fig:1+1d_regions}.  The
labeling of the regions is introduced before
Eq.~\eqref{eq:advcirc_defs}, when we reduce the available adversary
strategies to those achievable by two adversaries.

For each trial, the trial boundary on which entanglement of
adversaries is restricted is defined as the time slice that is formed
by the boundary \(T\) of the set
\(\cM_{\In}=(C_{\Vap,-}\cup C_{\Vb,-}) \cap (\cM\setminus
(C_{\Vap,+}\cup C_{\Vb,+})\), where \(\cM\) is the spacetime
manifold. See Fig.~\ref{fig:1+1d_regions} for a visualization of \(T\)
in the \(1+1\)-dimensional cross-section containing the verifiers in
the experiment.  We refer to \(T\) as the trial boundary.  The set
\(\cM_{\In}\) is a past set. \(\cM_{\In}\) can be described as the set
of events at which the trial's challenges are not yet visible, but
from which it is possible to send a response to at least one of the
verifiers \(\Vap\) or \(\Vb\). All events relevant to the trial occur
in the union \(C_{\Vap,-}\cup C_{\Vb,-}\).  If \(\cM_{\In}'\) is the
region defined in the same way as \(\cM_{\In}\) but for the next
trial, then \(\cM_{\In}'\) contains \(C_{\Vap,-}\cup C_{\Vb,-}\) of
the current trial. This ensures that all events of the current trial
occur before the trial boundary of the next one, and, similarly, all
events of previous trials occur before the current boundary \(T\). The
restriction on adversarial entanglement is formulated in terms of the
spacetime circuit's edges that exit \(\cM_{\In}\) and is defined in
terms of the systems listed at~\eqref{ent:systems} after introducing
the adversary spacetime circuit.

We now determine the adversary spacetime circuit, which by definition
satisfies that there are no vertices in the trial's target region
\(E\).  Later we generalize the analysis by allowing a limited class
of vertices in the target region \(E\) to prove that honest provers
must employ quantum devices in the target region to successfully pass
the protocol.  The trial circuit \(\cC\) is defined as the spacetime
circuit associated with the vertices in the causally convex set
\((C_{\Vap,-}\cup C_{\Vb,-})\cap (C_{\Vap,+}\cup C_{\Vb,+})\).  The
trial record is determined by the trial circuit \(\cC\) and the state
on the in-edges of \(\cC\). The vertices of \(\cC\) partition into
verifier vertices and adversary vertices, where verifier vertices
include any operations that are performed by verifier-owned devices
and needed to implement the verifiers' protocol actions.  The edges
connecting verifiers and adversaries are restricted by
Assumption~\ref{ass:lab}. In particular, the only edges from
verifiers to adversaries are those carrying the challenges.  In
addition, there are no edges
from adversaries to verifiers that precede an edge from verifiers to
adversaries. This restriction implies that no directed path starting
and ending at adversary vertices can pass through a verifier
vertex. Therefore, the set of adversary vertices is causally
convex. Let \(\cA\) be the spacetime subcircuit associated with the
adversary vertices.

\noindent{\textbf{Reduction to two adversaries and one round of
    communication:}} Such a reduction is given for a protocol with
idealized timing in Ref.~\cite{buhrman2014position} for one space
dimension. Here we provide the reduction for any number of space
dimensions.  To analyze the capabilities of the adversary spacetime
circuit \(\cA\), we partition the vertices of the circuit into
causally convex subsets determined by the locations of vertices
relative to the verifiers' sending and receiving vertices. For every
event \(\bm{r}\) in spacetime, we assign a \(4\)-bit string
\(l(\bm{r}) = b_{1}b_{2}b_{3}b_{4}\) defined as follows: \(b_{1}=1\)
if \(\bm{r}\in C_{\Vap,+}\), \(b_{2}=1\) if \(\bm{r}\in C_{\Vb,+}\),
\(b_{3}=1\) if \(\bm{r}\notin C_{\Vap,-}\), and \(b_{4}=1\) if
\(\bm{r}\notin C_{\Vb,-}\), with \(b_{j}=0\) if the condition for
\(b_{j}=1\) is not satisfied. Each bit encodes whether the event
\(\bm{r}\) lies inside or outside an associated forward or backward
lightcone. We form bit strings by concatenating the bit values.  For
two such bit strings \(\bm{b}\) and \(\bm{b}'\), we define a partial
order on bitstrings by bit-wise comparison and write
\(\bm{b}\leq \bm{b}'\) if each bit of \(\bm{b}\) is less than or equal
to the corresponding bit of \(\bm{b}'\). This partial order reflects
the causal ordering of events: If an event \(\bm{r}\) causally
precedes another event \(\bm{r}'\), then \(l(\bm{r})\leq l(\bm{r}')\).

For each \(4\)-bit string \(\bm{b}\), the set
\(\cM_{\bm{b}}=\{\bm{r}:l_{}(\bm{r})=\bm{b}\}\) is causally
convex. This is because \(\cM_{\bm{b}}\) is the intersection of four
causally convex sets specified by the values of the bits of
\(\bm{b}\). Since the intersection of causally convex sets is also
causally convex, \(\cM_{\bm{b}}\) inherits this property.  The set
\(\cM_{1100}\) is the target region \(E\). This region is the
intersection of the four closed lightcones \(C_{\Vap,\pm}\) and
\(C_{\Vb,\pm}\), and is therefore also closed.  For a subset \(B\) of
\(4\)-bit strings, let \(\cM_{B}=\cup_{\bm{b}\in B}\cM_{\bm{b}}\).  In
general, a subset \(I\) of a partially ordered set \(P\) is called
order-convex if for every \(x,y\in I\) with \(x\leq y\) and \(z\in P\)
with \(x\leq z\leq y\), we have \(z\in I\). In particular, causally
convex sets of interaction vertices in a spacetime circuit are
order-convex for the circuit partial order.  If \(B\) is order-convex,
then the set \(\cM_{B}\) is causally convex. To see this, consider
arbitrary events \(\bm{r},\bm{r}'\in\cM_{B}\) and an event
\(\bm{r}''\) in \(\cM\) with \(\bm{r}\preceq\bm{r}''\preceq\bm{r}'\).
We have to show that \(\bm{r}''\in\cM_{B}\).  Be the relationship
between the causal ordering and the partial order of bit strings,
\(l(\bm{r})\leq l(\bm{r}'')\leq l(\bm{r}')\). Since \(B\) is
order-convex, \(l(\bm{r}'')\in B\) which implies that
\(\bm{r}''\in\cM_{B}\).  We adopt the convention that the bit symbol
``\(*\)'' stands for either \(0\) or \(1\), and a bit string that has
\(*\)'s denotes the set of bit strings obtained by replacing the
\(*\)'s with \(0\) or \(1\). With this convention, the strings
\(\bm{b}\) with alphabet \(\{0,1,*\}\) denote order-convex subsets of
bitstrings and consequently, the corresponding sets
\(\cM_{\bm{b}}\) are causally convex.  The regions that appear in the
cross-section shown in Fig.~\ref{fig:1+1d_regions} are labeled by
bitstrings according to this convention.  The regions \(\cM_{1010}\)
and \(\cM_{0101}\) do not appear in the figure.

Let \(\cV\) be the set of vertices of \(\cA\).  For an order-convex
set $B$ of \(4\)-bit strings, \(\cM_{B}\) is causally convex in
spacetime, from which it follows that \(\cV\cap\cM_{B}\) is causally
convex in \(\cA\).  Since the set of vertices \(\cV\) of \(\cA\) are
causally convex in the trial circuit \(\cC\), \(\cV\cap\cM_{B}\)  is also causally
convex in \(\cC\).  We define \(\cA_{B}=\cS(\cV\cap\cM_{B})\) to be
the spacetime subcircuit of \(\cA\) and of \(\cC\) determined by the
vertices in \(\cV\cap\cM_{B}\).  We partition \(\cA\) into the
following vertex-disjoint spacetime subcircuits:
\begin{align}
  \Ain &= \cA_{\{00*0,000*\}}
         ,\notag\\
  \Aastart &=\cA_{\{100*,1010\}}
         ,\notag\\
  \Abstart &=\cA_{\{01*0,0101\}}
         ,\notag\\
  \Aaend &= \cA_{1101}
         ,\notag\\
  \Abend &= \cA_{1110}
         ,\notag\\
  \Aout &= \cA_{**11}
          .
          \label{eq:advcirc_defs}
\end{align}

\begin{figure}
  \includegraphics[scale=0.6]{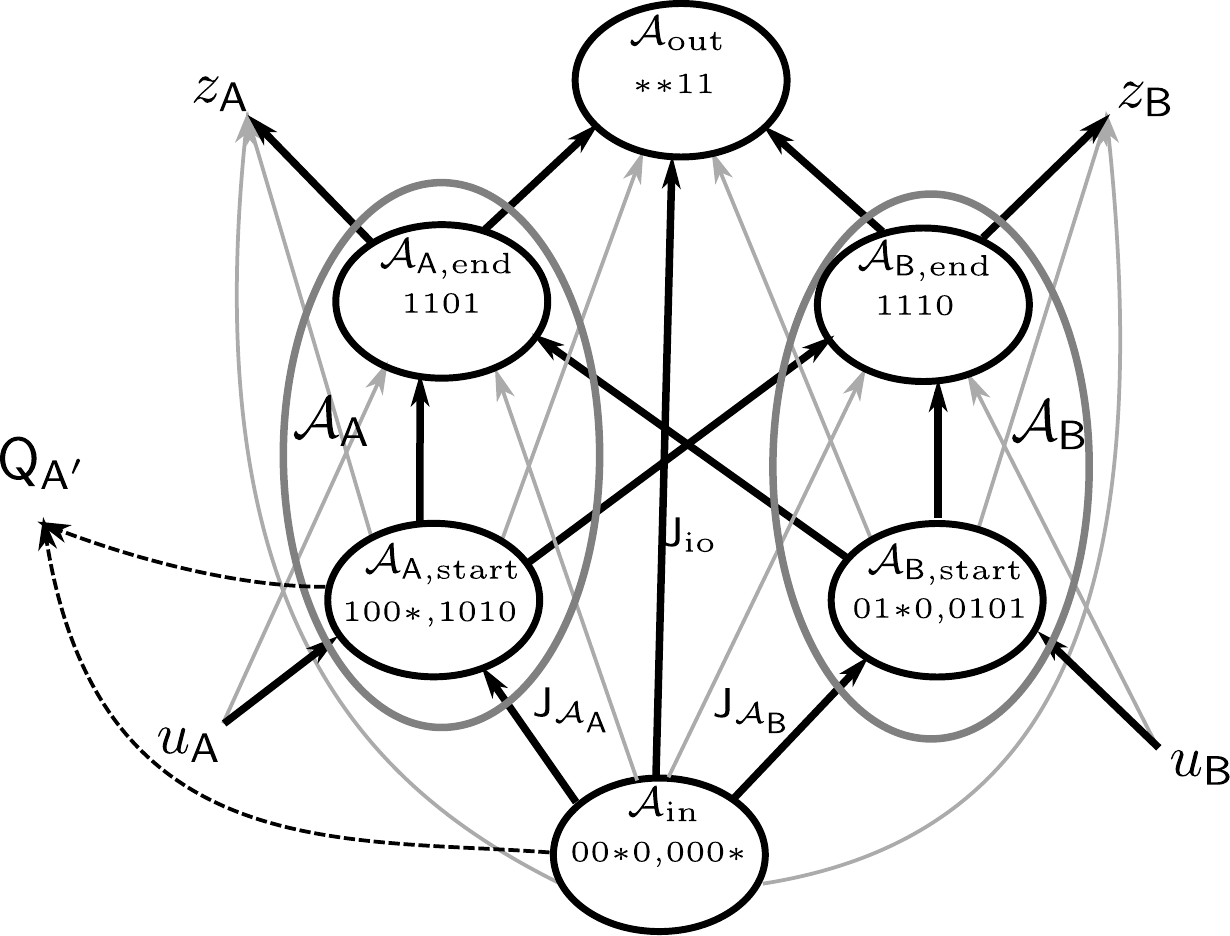}
  \caption[Adversary spacetime circuit.]{Adversary spacetime circuit.
    The diagram depicts the adversary circuits and their
    interconnections. The circuit receives challenges \(\cba\) and
    \(\cbb\) coming from the verifiers and delivers responses \(\zqa\)
    and \(\zqb\) to the verifiers. 
    The adversary circuit partitions into six subcircuits indicated by the black
    ellipses. Each subcircuit enacts a quantum operation from its
    in-edges to its out-edges.  The binary strings indicate the
    associated spacetime regions as described in the text.
    The entangled quantum systems
    prepared by the source are \(\Qa\) and \(\Qp\). Because the source
    is treated as part of the adversary, \(\Qp\) is carried on the internal
    edges of the adversary circuit, while \(\Qa\) may be sent to \(\Va\) from
    \(\Ain\) or from \(\Aastart\), where we show the edges that alternatively
    or jointly carry \(\Qa\) with dashed arrows.
    By Assumption~\ref{ass:meas},
    \(\Aaend\), \(\Abstart\) and \(\Abend\) cannot send anything to
    \(\Va\) before completion of \(\Va\)'s measurement process. We therefore
    do not need to include edges from these subcircuits for producing \(\Qa\).
    The part
    of the circuit directly relevant to the protocol consists of two
    groups indicated by gray ellipses and labeled by \(\Aa\) on the
    left and \(\Ab\) on the right. The black arrows depict the
    principal circuit connections. The light-gray edges may be present
    in the spacetime circuit but are removed by passing them through
    one of the bypassed adversary circuits in the abstract analysis of
    adversary capabilities. }
    \label{fig:cabs}
\end{figure}

We let \(\Aa\) be the part of the adversary that consists of \(\Aastart\)
and \(\Aaend\), and \(\Ab\) the part that consists of \(\Abstart\) and \(\Abend\):
\begin{align}
  \Aa &= \cA_{\{100*,1010,1101\}}
        ,\notag\\
  \Ab &= \cA_{\{01*0,0101,1110\}}
        .
        \label{eq:two_advcirc_defs}
\end{align}
Because \(\cA\) contains no vertices in the target region, the
bit-string suffixes in the above list do not include \(1100\).  The
partition of \(\cA\) is shown in Fig.~\ref{fig:cabs} with arrows
indicating where there can be edges connecting the parts.  \(\Ain\) is
the part of the adversary in \(\cM_{\In}\) and consists of the
vertices of \(\cA\) that do not yet have access to the challenges but
are in principle able to send a response to at least one of the
verifiers.  \(\Aout\) consists of the adversary vertices that can no
longer respond to either of the verifiers. The edges from \(\Aastart\)
to \(\Abend\) and from \(\Abstart\) to \(\Aaend\) allow for one-round
communication between \(\Aa\) and \(\Ab\). There is no edge
connecting \(\Aastart\) and \(\Abstart\) or connecting \(\Aaend\) and
\(\Abend\), as the two parts are space-like separated from each other.

\newcommand{\QAa}{{\sf J}_{\cA_{\Vap}}}
\newcommand{\QAb}{{\sf  J}_{\cA_{\Vb}}}
\newcommand{\Qio}{{\sf  J}_{{\rm io}}}

The restrictions on prior entanglement are expressed in terms of the
joint state of the systems carried by the outgoing edges of \(\Ain\),
where we exclude direct edges to \(\Aout\).  These direct edges do not
contribute to the current trial of the protocol.  The systems carried
by edges that may contribute relevant entanglement partition into
\begin{align}
  1)\;&\text{the system on the edge from \(\Ain\) that may contribute to
        \(\Qa\) and}\notag\\&\text{the system \(\QAa\) carried by
  edges from \(\Ain\) to \(\Aa\),}
  \notag\\
  2)\; &\text{ the system \(\QAb\) carried by edges from \(\Ain\) to \(\Ab\).}
        \label{ent:systems}
\end{align}
After we eliminate the indirect edges in the abstract treatment of the
circuit below, these edges go to \(\Va\) and \(\Aastart\), and
\(\Abstart\), respectively.  In our analysis, we choose to restrict
the entanglement between \(\Qa\QAa\) and \(\QAb\).  Except for the
untrusted measurement apparatus of \(\Va\), the verifiers may be
assumed to be classical on exiting \(\cM_{\In}\). Restrictions on
communication by the measurement apparatus are described in
Assumption~\ref{ass:meas}.

Because of the absence of adversary vertices in the target region
\(\cM_{1100}\), the vertices of \(\Aa\) and \(\Ab\) are causally
convex in the circuit partial order, making \(\Aa\) and \(\Ab\)
spacetime subcircuits of \(\cA\).  In particular, because it would
violate causality, there is no directed path that starts at a vertex
in \(\Aa\), enters \(\Ab\), and returns to \(\Aa\), or vice versa.
Consequently, in the absence of the light-grey arrows in
Fig.~\ref{fig:cabs}, the adversary actions in a trial can be viewed as
follows: 1) An initial state preparation by \(\Ain\), which is in the
causal past of the trial, followed by local processing by \(\Aastart\)
and \(\Abstart\). \(\Ain\) and \(\Aastart\) may contribute to 
the system \(\Qa\) to
be measured by \(\Va\). 2) A single round of communication
consisting of one message from \(\Aastart\) to \(\Abend\), and another
from \(\Abstart\) to \(\Aaend\). 3) Responses sent from \(\Aaend\) to
\(\Vap\) and from \(\Abend\) to \(\Vb\). After completing the trial,
the adversaries pass a quantum state to \(\Aout\) for use in the next
trial.  The actions of the adversaries can therefore be represented by
two adversaries \(\Aa\) and \(\Ab\) who prepare an initial joint state
in their past, receive challenges locally, and communicate once in one
round of communication before committing to responses to the
verifiers.  The light-grey arrows in Fig.~\ref{fig:cabs} will be
eliminated without weakening the capabilities of the adversary in the
abstract analysis of the circuit in Sect.~\ref{sect:reduction}.

\noindent\textbf{Projection of target region:}
A single trial's target region in spacetime is \(\cM_{1100}\).  For
visualization, it is helpful to project this region onto the
space dimension of an inertial frame.  In the experiment, both
verifiers are stationary relative to the ground, so we use the
inertial frame in which the verifiers are at rest.  To determine
the spatial projection of \(\cM_{1100}\), we find the set of 
space positions \(\bm{x}\) whose
stationary worldlines intersect \(\cM_{1100}\). The stationary
worldline for space position \(\bm{x}\) is given by \(P=\{(\bm{x},t)\}_{t\in\rls}\).
The distances from \(\bm{x}\) to the verifiers \(\Vap\) and \(\Vb\) are
denoted by \(l_{\Vap}\) and \(l_{\Vb}\).  Let \(t_{1},t_{2},t_{3}\), and 
\(t_{4}\) be the times at which the worldline \(P\) enters
or leaves the lightcones \(C_{\Vap,+}\), \(C_{\Vb,+}\), \(C_{\Vap,-}\)
and \(C_{\Vb,-}\). Then, with units where the speed of light is \(c=1\), 
we have 
\begin{align}
  t_{1} &=t+s_{\Vap}+l_{\Vap}\nonumber\\
  t_{2} &=t+s_{\Vb}+l_{\Vb},\nonumber\\
  t_{3} &=t+ r_{\Vap}- l_{\Vap},\nonumber\\
  t_{4} &=t+ r_{\Vb}- l_{\Vb}.
\end{align}
To ensure that the stationary worldline \(P\) intersects the target
region \(\cM_{1100}\), the condition
\(\max(t_{1},t_{2}) \leq \min(t_{3},t_{4})\) must be satisfied.  From
the equivalent four inequalities \(t_{j}\leq t_{k}\) for
\(j\in\{1,2\}\) and \(k\in\{3,4\}\), we get
\begin{align}\label{eq:space_constraints}
  t_{1}\leq t_{3} &\leftrightarrow 2 l_{\Vap}\leq r_{\Vap}-s_{\Vap},\nonumber\\
  t_{2}\leq t_{4}&\leftrightarrow 2 l_{\Vb}\leq r_{\Vb}-s_{\Vb},\nonumber\\
  t_{1}\leq t_{4}&\leftrightarrow l_{\Vap}+l_{\Vb}\leq r_{\Vb}-s_{\Vap},\nonumber\\
  t_{2}\leq t_{3}&\leftrightarrow l_{\Vap}+l_{\Vb}\leq r_{\Vap}-s_{\Vb}.
\end{align}
The first two inequalities constrain \(\bm{x}\) to lie within spheres
centered at the verifiers \(\Vap\) and \(\Vb\), while the last two
constrain \(\bm{x}\) to ellipsoids with foci at the verifiers. Thus,
the projection of \(\cM_{1100}\) onto space in the verifiers' inertial
frame is the intersection of these spheres and ellipsoids.

\noindent\textbf{Classically achievable target region:}
Classical position verification with the two verifiers \(\Vap\) and
\(\Vb\) involves the verifiers' sending challenges \(\cba\) and
\(\cbb\) and the prover responding to both verifiers with the value of
\(g(\cba,\cbb)\) for some fixed function \(g\).  In the spacetime
circuit picture, to succeed with probability \(1\) it is necessary and
sufficient to have one prover vertex in the region where \(\cba\) and
\(\cbb\) are visible and there is sufficient time to send
\(g(\cba,\cbb)\) to \(\Vap\), and another prover vertex in the region
where \(\cba\) and \(\cbb\) are visible and there is sufficient time
to send \(g(\cba,\cbb)\) to \(\Vb\). These constraints are related to
the analysis of adversaries for classical position verification in
Ref.~\cite{chandran2009position}.  If either of these conditions is
not met, the prover has a nonzero probability of failure. The first
condition requires a prover vertex in
\(C_{\Vap,+}\cap C_{\Vb,+}\cap C_{\Vap,-}\), while the second requires
another prover vertex in \(C_{\Vap,+}\cap C_{\Vb,+}\cap C_{\Vb,-}\).
We identify the union of these two spacetime regions as the effective
target region \(E_{c}\) for classical position verification. Although
this target region includes the locations of the verifiers themselves,
it serves as a useful reference for comparing with the Bell-test
protocol's target region \(E\). The projection of \(E_{c}\) onto space
in the verifiers' inertial frame can be determined in the same way as
the projection of \(E\). To determine the projection of the prover
vertex in \(C_{\Vap,+}\cap C_{\Vb,+}\cap C_{\Vap,-}\), the
inequalities that need to be satisfied are
\begin{align}
  t_{1} \leq t_{3} 
  & \leftrightarrow 2 l_{\Vap}\leq r_{\Vap}-s_{\Vap},
    \notag\\
  t_{2}\leq t_{3} 
  &\leftrightarrow l_{\Vb}+l_{\Vap}\leq r_{\Vap}-s_{\Vb}
                   .
\end{align}
To determine the projection of the prover vertex in
\(C_{\Vap,+}\cap C_{\Vb,+}\cap C_{\Vb,-}\), 
the inequalities to be satisfied are
\begin{align}
  t_{1} \leq t_{4}
  & \leftrightarrow l_{\Vb}+l_{\Vap}\leq r_{\Vb}-s_{\Vap},
    \notag\\
  t_{2}\leq t_{4}
  &\leftrightarrow 2 l_{\Vb}\leq r_{\Vb}-s_{\Vb}
                   .
\end{align}
Both of these regions are intersections of an ellipsoid and a sphere. The projection of the target region \(E_{c}\) is the union of the two.

\subsection{Reduction to three-party non-signaling and connection to two-party local realism}
\label{sect:reduction}

In the previous section, we reduced the adversary capabilities to two
spacetime circuits \(\Aa\) and \(\Ab\) with limited inputs and
restricted communication. We now assume that there is no entanglement
between \(\Qa\QAa\) and \(\QAb\), which implies that \(\Aa\) and
\(\Ab\) share no prior entanglement.  We relax this assumption first
for weak entanglement in Sect.~\ref{sec:smallentanglement} and then
for significant amounts of entanglement in
Sect.~\ref{sect:entangled_adversaries}.  The goal of this section is
to characterize the probability distributions of the trial record
achievable by the adversary in terms of a three-party non-signaling
scenario. Our use of test factors dictates that we characterize these
probability distributions for any joint state of all entities' edges
crossing the trial boundary \(T\) defining the start of the
trial. This joint state includes the trial records of previous trials
as well as all other information that can influence the trial. In the
spacetime analysis, this accounts for all possible dependencies on
the past.  To check that a test factor \(W\) has expectation at most
\(1\) for all probability distributions achievable by the adversary,
it suffices to consider extremal joint states as expectations cannot
increase under convex combinations.

Before proceeding, we eliminate redundant edges in the high-level
description of the adversary spacetime circuit depicted in
Fig.~\ref{fig:cabs}.  These are the edges associated with connections
shown in light-gray. For this purpose, we detach the circuit from its
specific spacetime realization and consider it as an abstract quantum
circuit with processing vertices
\(\Ain,\Aastart,\Aaend,\Abstart,\Abend,\Aout\) that are directionally
connected with inputs and outputs as shown in Fig.~\ref{fig:cabs}. In
this abstract circuit, each of the redundant edges bypasses one or
more intervening vertices. We can re-route the redundant edges by
passing them through the intervening vertices, re-defining the process
at these intervening vertices to copy the system carried by the
redundant edge from the vertex input to the vertex output. The
modified circuit contains no bypassing edges, and it 
preserves the power of the adversary.

The connection in Fig.~\ref{fig:cabs} with system \(\Qio\) in
Fig.~\ref{fig:cabs} is not re-routed because it does not contribute to
the actions of the adversary during the trial. Entanglement between it
and the systems on other edges into \(\Aa\) and \(\Ab\) is allowed,
provided there is no entanglement between \(\Qa\QAa\) and \(\QAb\)
after tracing out \(\Qio\).  The test factor \(W\) does not depend on
\(\Qio\), since the trial record is independent of this system.  In
addition to \(\Qio\), we also need to consider the classical state of
the verifiers on edges exiting the causal past \(\cM_{\In}\) of the
trial.  The classical state of the verifiers may include shared
randomness that is kept secret from and therefore independent of other
entities before the trial and that can help generate measurement
settings and challenges. Such randomness is used to satisfy the
independence requirements of Assumption~\ref{ass:distrib}.
Accordingly, and because the reduced state on \(\Qa\QAa\) and
\(\QAb\) is separable, the adversary's expectation of the test factor
\(W\) is maximized by some pure product state over the systems
\(\Qa\QAa\) and \(\QAb\). We now fix such a product state. This
state may be assumed to be known by every component of the adversary.

\newcommand{\Rsys}{{\sf R}}

\begin{figure}
  \includegraphics[scale=0.6]{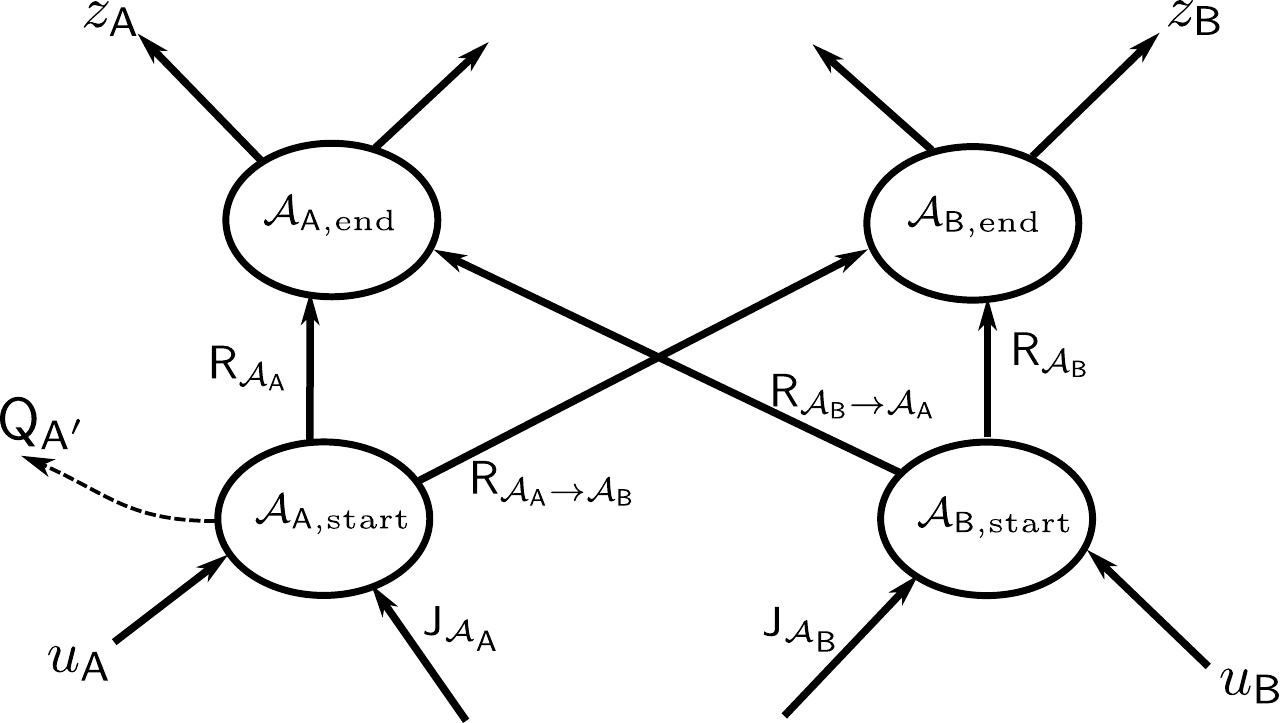}
  \caption{Adversary spacetime circuit with bypassing edges removed
    and communicated systems labeled. The dashed arrow may contribute
    to \(\Qa\), see Fig.~\ref{fig:cabs}.}
  \label{fig:cabs_simplified}
\end{figure}

The first half of the adversary strategy, executed by \(\Aastart\) and
\(\Abstart\), can be described as follows: \(\Aastart\) transforms its
input systems \(\QAa\) into output systems \(\Rsys_{\Aa}\) and
\(\Rsys_{\Aa\rightarrow\Ab}\), where \(\Rsys_{\Aa}\) includes
\(\Qa\). This transformation may depend on the classical challenge
\(\cba\), as it is an input to \(\Aastart\). The resulting system
\(\Rsys_{\Aa}\) is then passed on to \(\Aaend\) with the component
\(\Qa\) being passed on to \(\Va\), and \(\Rsys_{\Aa\rightarrow\Ab}\)
is passed on to \(\Abend\). Because the quantum inputs to \(\Aa\) and
\(\Ab\) are in a product state, the quantum system \(\QAb\),
originally transmitted from \(\Ain\) to \(\Abstart\), can instead be
treated as being internally prepared by \(\Abstart\) in a pure state.
The operation by \(\Abstart\) is equivalent to transforming \(\QAb\)
into quantum systems \(\Rsys_{\Ab}\) and
\(\Rsys_{\Ab\rightarrow\Aa}\). The joint state of these two systems
may depend on the classical challenge \(\cbb\), which is an input to
\(\Abstart\).  The systems \(\Rsys_{\Ab}\) and
\(\Rsys_{\Ab\rightarrow\Aa}\) are then passed on to \(\Abend\) and
\(\Aaend\), respectively. The communicated systems are shown in
Fig.~\ref{fig:cabs_simplified} with the bypassing edges removed.
Without loss of generality, we may assume that the challenge \(\cba\)
is separately passed on by \(\Aastart\) to both \(\Aaend\) and
\(\Abend\), and similarly, \(\cbb\) is passed on by \(\Abstart\). The
relevant second half of the strategy, which determines the classical
challenge responses \(\zqa\) and \(\zqb\) recorded by the verifiers,
can be treated as measurements performed by \(\Aaend\) and \(\Abend\).
Each of these measurements may depend on both challenges \(\cba\) and
\(\cbb\). In the following, we modify the adversary strategy while
improving the corresponding expectation of the test factor \(W\), and
show that for the modified strategy the trial record follows a
three-party non-signaling distribution.

We first show that the quantum messages
\(\Rsys_{\Ab}\Rsys_{\Ab\rightarrow\Aa}\) from \(\Abstart\) can be
eliminated without decreasing the expectation of the test factor \(W\)
for the adversary strategy.  Given that the initial state of \(\QAb\)
is a fixed pure state and considering the causal structure shown in
Fig.~\ref{fig:cabs_simplified}, the state of
\(\Rsys_{\Ab}\Rsys_{\Ab\rightarrow\Aa}\) depends on \(\cbb\) but is
independent of the other systems involved in the trial.  \(\Abstart\)
can therefore prepare it as a partial trace over an ancillary system
of a \(\cbb\)-dependent pure state \(\ket{\psi_{\cbb}}\). \(\Abstart\) can modify \(\Rsys_{\Ab}\) so
that it includes the ancillary system and let \(\Abend\) remove or
ignore it. We can therefore assume that
\(\Rsys_{\Ab}\Rsys_{\Ab\rightarrow\Aa}\) is in state
\(\ket{\psi_{\cbb}}\) when the challenge is \(\cbb\).  Instead of
\(\Abstart\) preparing and passing on \(\Rsys_{\Ab}\) and
\(\Rsys_{\Ab\rightarrow\Aa}\), we can have \(\Aastart\) prepare
independent systems
\(\Rsys_{\Ab,\cbb'}\Rsys_{\Ab\rightarrow\Aa,\cbb'}\) for each possible
value \(\cbb'\) of \(\Cbb\), where the state of
\(\Rsys_{\Ab,\cbb'}\Rsys_{\Ab\rightarrow\Aa,\cbb'}\) is
\(\ket{\psi_{\cbb'}}\). In addition to passing on the systems
\(\Rsys_{\Aa}\) and \(\Rsys_{\Aa\rightarrow\Ab}\), \(\Aastart\) passes
on all systems \(\Rsys_{\Ab\rightarrow\Aa,\cbb'}\) to \(\Aaend\) and
all systems \(\Rsys_{\Ab,\cbb'}\) to \(\Abend\).  Since both
\(\Aaend\) and \(\Abend\) know the actual challenge \(\cbb\) used in
the trial, they can discard all systems \(\Rsys_{\Ab,\cbb'}\) and
\(\Rsys_{\Ab\rightarrow\Aa,\cbb'}\) for \(\cbb'\ne\cbb\). After this
step, the joint state at \(\Aaend\) and \(\Abend\) and \(\Qa\)
relevant for the measurements producing the responses and the verifier
measurement outcomes is identical to that in the original strategy.
The trial-record probability distribution is therefore unchanged, and
the edges leaving \(\Abstart\) only carry the classical message
consisting of \(\cbb\) to \(\Aaend\) and \(\Abend\).

The action by which \(\Aastart\) creates the system
\(\Rsys_{\Aa}\Rsys_{\Aa\rightarrow\Ab}\) may depend on \(\cba\).  Next
we show that the adversary circuit can be modified so that this
creation of \(\Rsys_{\Aa}\Rsys_{\Aa\rightarrow\Ab}\) becomes
independent of \(\cba\), without reducing the adversary's expectation
\(\Exp(W)\) of the test factor \(W\).  Let
\(\bar W_{\cba}=\Exp(W|\Cba=\cba)\) be the expectation of \(W\)
conditional on the challenge \(\cba\). We have
\(\Exp(W)=\sum_{\cba}\Prob(\Cba=\cba) \bar W_{\cba}\), which is a
convex combination of the \(\bar W_{\cba}\). It follows that
\(\max(\bar W_{\cba})\geq \Exp(W)\).  Let
\(\cba_{0}= \argmax_{\cba:\Prob(\cba)>0}\bar W_{\cba}\). We now modify
the adversary circuit so that \(\Aastart\) always prepares
\(\Rsys_{\Aa}\Rsys_{\Aa\rightarrow\Ab}\) as if the challenge from
\(\Vap\) were \(\Cba=\cba_{0}\), regardless of the actual \(\cba\)
used in the trial. \(\Aaend\) and \(\Abend\) modify their measurements
according to those that would have been performed if the challenge
from \(\Vb\) were \(\cbb'\), where they jointly choose \(\cbb'\) from
the probability distribution of the challenges from \(\Vb\)
conditional on \(\Cba=\cba_{0}\) and \(\Mqp=\mqp=f(\cba,\cbb)\), with
\(\mqp\) being the actual prover setting and \(\cba,\cbb\) the actual
challenges in this trial. To see that this conditional probability
distribution is well-defined, we note that by
Assumption~\ref{ass:distrib}, \(\Mqp\) and \(\Cba\) are
independent. Since $\mqp$ is the actual prover setting,
$\Prob(\mqp) > 0$. Similarly, $\Prob(\Cba=\cba_{0}) > 0$ by
construction. Due to independence, the joint probability is simply the
product $\Prob(\mqp, \cba_{0}) = \Prob(\mqp)\Prob(\cba_{0})$, which is
strictly positive. This ensures that the conditioning event
\((\Cba=\cba_{0}, \Mqp=\mqp)\) has non-zero probability and the
conditional distribution for choosing $\cbb'$ is well-defined.  The
method for choosing \(\cbb'\) defines the random variable \(\Cbb'\)
and ensures that \(f(\cba_{0},\Cbb')=\Mqp=f(\Cba,\Cbb)\).  The joint
choice of \(\cbb'\) requires joint randomness, which can be shared by
\(\Aastart\) as part of their classical message to \(\Abend\).
Although \(\Aastart\) does not know \(\mqp\), \(\Aastart\) can
generate the needed shared randomness for each possible value of
\(\Mqp\) and pass on the shares to \(\Aaend\) and \(\Abend\).
\(\Aaend\) and \(\Abstart\) have both learned \(\mqp\) and can choose
from the shared randomness accordingly.

From now on, we abbreviate events and conditions of the form \(R=r\)
for RVs \(R\) by leaving out ``\(R=\)'' inside probability and
expectation expressions, provided that this does not introduce
ambiguity.  We claim that the expectation of \(W\) after the
modifications of \(\Aa\) and \(\Ab\) is \(\bar W_{\cba_{0}}\). To show
this, let \(\Oqa'\) be the verifier \(\Va\) outcome RV and \(\Zqa'\)
and \(\Zqb'\) the response RVs with the modifications. We compute the
expectation of \(W(\Mqa,\Oqa',\Mqp, \Zqa',\Zqb')\) below.  By construction, and
because \(\Mqa\) is conditionally independent of \(\Cba\) and \(\Cbb\)
given \(\Mqp\) (see Assumption~\ref{ass:distrib}), the state
being measured to obtain \(\Oqa'\), \(\Zqa'\) and \(\Zqb'\) and the
measurements being performed are the same as if \(\Cba=\cba_{0}\) and
\(\Cbb=\cbb'\), where \(\Mqp=f(\cba,\cbb)=f(\cba_{0},\cbb')\).  We
first verify this equivalence by calculating the probability distribution of
the settings and outcomes as follows. By assumption and then by construction,
\begin{align}
  \Prob(\mqa,\cbb'|\mqp)
  &=\Prob(\mqa|\mqp)\Prob(\cbb'|\mqp)
    \notag\\
  &=\Prob(\mqa|\Cba=\cba_{0},\mqp)\Prob(\Cbb=\cbb'|\Cba=\cba_{0},\mqp)
    \notag\\
  &=\Prob(\mqa,\Cbb=\cbb'|\Cba=\cba_{0},\mqp).
\end{align}
We apply this identity to determine
\begin{align}
  \Prob(\mqa,\oqa',\mqp,\zqa',\zqb')  
  &=\sum_{\cbb'} \Prob(\mqa,\oqa',\cbb',\mqp, \zqa',\zqb')  
  \notag \\
  &=
    \sum_{\cbb'}\Prob(\oqa',\zqa',\zqb'|\mqa,\cbb',\mqp)
    \Prob(\mqa,\cbb'|\mqp)\Prob(\mqp)
    \notag\\
  &=\sum_{\cbb'}
    \Prob(\Oqa=\oqa',\Zqa=\zqa',\Zqb=\zqb'|\mqa,\Cba=\cba_{0},\Cbb=\cbb',\mqp)
    \notag\\
  &\hphantom{=\sum_{\mqa,\cbb'}
    \Prob}\times
    \Prob(\mqa,\Cbb=\cbb'|\Cba=\cba_{0},\mqp)
    \Prob(\mqp)
    \notag\\
  &=
    \sum_{\cbb'}
    \Prob(\mqa,\Oqa=\oqa',\Zqa=\zqa',\Zqb=\zqb',\Cbb=\cbb'|\Cba=\cba_{0},\mqp)
    \notag\\
  &\hphantom{=\sum_{\mqa,\cbb'}
    \Prob}\times
    \Prob(\mqp)
    \notag\\
  &=
    \Prob(\mqa,\Oqa=\oqa',\Zqa=\zqa',\Zqb=\zqb'|\Cba=\cba_{0},\mqp)
    \Prob(\mqp)
    \notag\\
  &=\Prob(\mqa,\Oqa=\oqa',\mqp,\Zqa=\zqa',\Zqb=\zqb'| \Cba=\cba_{0}),
    \label{eq:barw-1}
\end{align}
where in the first two lines, we calculate probabilities involving the
RVs \(\Oqa',\Zqa',\Zqb'\) according to the modified adversary circuit,
and in the last step we used \(\Prob(\mqp)=\Prob(\mqp|\Cba=\cba_{0})\)
by independence of \(\Mqp\) and \(\Cba\) according to
Assumption~\ref{ass:distrib}.  We can now calculate the expectation of
\(W\) for the modified adversary circuit.
\begin{align}
 &\Exp(W(\Mqa,\Oqa',\Mqp, \Zqa',\Zqb'))\hspace*{-1.5in}
    \notag\\
  &=\sum_{\mqa,\oqa',\mqp,\zqa',\zqb'}
    W(\mqa,\oqa',\mqp,\zqa',\zqb') \Prob(\mqa,\oqa',\mqp,\zqa',\zqb')  
    \notag\\
  &=\sum_{\mqa,\oqa',\mqp,\zqa',\zqb'}
    W(\mqa,\oqa',\mqp,\zqa',\zqb')
    \Prob(\mqa,\Oqa=\oqa',\mqp,\Zqa=\zqa',\Zqb=\zqb'|\Cba=\cba_{0})  
    \notag\\
  &=\sum_{\mqa,\oqa,\mqp,\zqa,\zqb}
    W(\mqa,\oqa,\mqp,\zqa,\zqb)
    \Prob(\mqa,\oqa,\mqp,\zqa,\zqb|\Cba={\cba_{0}})
    \notag\\
  &= \bar W_{\cba_{0}},
\end{align}
where we used Eq.~\eqref{eq:barw-1} for the second identity and a
change of variables for the third.

Because of Assumption~\ref{ass:distrib}, the incoming adversary state
including that of the source and quantum systems being shared is
independent of the verifier settings and challenge distribution.
Consequently, with the above modifications to the adversary circuit,
we have arrived at a situation where \(\Aaend\), \(\Abend\) and
\(\Va\) receive a joint quantum state \(\rho\) that does not depend on
the challenges \(\cba\), \(\cbb\) or the measurement setting \(\mqa\)
used by \(\Va\) in the trial.  The challenges and the measurement
setting constitute classical inputs to \(\Aaend\), \(\Abend\) and
\(\Va\), and the measurements performed by \(\Aaend\) and \(\Abend\)
now depend only on \(\cbb'\).  We can further modify the measurements
so that they depend only on the actual prover setting \(\mqp\).  For
this additional modification, consider a particular \(\mqp\).  Then we
have, for the above modified adversary circuit and every \(\cbb'\) for
which \(f(\cba_{0},\cbb')=\mqp\)
\begin{align}
  \Exp(W|\mqp,\cbb')=\Exp(W|\cbb') 
  &\leq \max_{\cbb''} \{\Exp(W|\Cbb'=\cbb''): 
    f(\cba_{0},\cbb'')=\mqp\},
\end{align}
where we define conditional expectations and probabilities to be zero
whenever the conditioning event has zero probability.  The adversary's
expectation of \(W\) therefore improves if given
\(\mqp=f(\cba,\cbb)\), \(\Aaend\) and \(\Abend\) always perform the
measurements for an input \(\cba_{0},\cbb''\) that maximizes
\(\Exp(W|\cba_{0},\cbb'')\) subject to \(f(\cba_{0},\cbb'')=\mqp\).
With this additional modification, \(\cbb''\) is determined by
\(\mqp\), which implies that the joint randomness used for the first
modification can now be omitted.

Instead of providing both \(\cba\) and \(\cbb\) as
inputs, it suffices to provide just \(\mqp\). The local measurement
settings now depend only on the local inputs: \(\mqa\) for \(\Va\),
\(\mqp\) for \(\Aaend\), and \(\mqp\) for \(\Abend\). Because the
measurement settings are local to \(\Va\), \(\Aaend\) and \(\Abend\),
and the joint state \(\rho\) does not depend on the inputs, we can
define the conditional distribution of the three measurement outcomes,
namely, the outcome \(\Oqa\) of \(\Va\), the response \(\Zqa\) from
\(\Aaend\), and the response \(\Zqb\) from \(\Abend\). This
conditional distribution can be defined not only for the case where
the inputs for \(\Aaend\) and \(\Abend\) are identical, but also when
their inputs are different.  Consequently, we constructed a modified
adversary strategy that yields an expectation of \(W\) at least as
high as the original, and for which the probability distribution of
the trial record is derived from the full set of conditional
distributions according to a three-party, necessarily non-signaling,
scenario with a joint quantum state for the three parties, \(k_{\Va}\)
settings and \(c_{\Va}\) outcomes for the first party, \(\Va\), and
\(k_{\Vp}\) settings and \(c_{\Vp}\) outcomes for each of the second
and third parties, \(\Aaend\) and \(\Abend\).  The expectation of
\(W\) is evaluated according to a distribution of inputs where the
second and third party's inputs are always the same.  It follows that
to ensure that \(W\) is a test factor against adversaries without
prior entanglement, it suffices to verify that its maximum expected
value over all three-party non-signaling distributions does not exceed
\(1\), where the three inputs are \(\mqa\), \(\mqp\) and \(\mqp\) with
probability distribution \(\Prob(\mqa,\mqp)\).

\noindent\textbf{Existence of test factors for protocol completeness:}
The Bell-test protocol requires finding test factors
against adversaries that ensure completeness of the protocol for an
implementable honest prover.  It suffices to consider non-negative
candidate functions \(W(\mqa,\oqa,\mqp,\zqa,\zqb)\) that can be
computed from the trial record and do not depend directly on \(\cba\)
and \(\cbb\).  To take advantage of the reduction to the three-party
quantum scenario derived above, we write \(W\) in terms of a function
\(W'(\oqa, \mqa; \zqa, b; \zqb, b')\) according to
\begin{align}
  W(\mqa,\oqa,\mqp,\zqa,\zqb)
  &= W'(\oqa, \mqa; \zqa, \mqp; \zqb, \mqp).
    \label{eq:WfromW'}
\end{align}
The reduction to three parties implies that the conditional
distributions accessible by adversary strategies for the protocol can
be written as
\begin{align}
  \Prob(\oqa,\zqa,\zqb|\mqa,\mqp)
  &=\mu(\oqa,\zqa,\zqb|\mqa,b=\mqp, b'=\mqp), 
\end{align}
where the conditional distributions \(\mu(\oqa,\zqa,\zqb|\mqa,b, b')\)
satisfy both quantum and non-signaling constraints for three parties
according to the configuration described in the previous paragraph.
We use \(b\) and \(b'\) to refer to the inputs of parties \(\Aaend\)
and \(\Abend\) in the extended three party scenario where \(\Aaend\)
and \(\Abend\) may have different inputs determining their
measurements.  For a review of the quantum and non-signaling
constraints, see~Ref.~\cite{brunner2014bell}.  We do not take
advantage of the quantum constraints and instead consider all
non-signaling conditional distributions.  Let \(\nu(\mqa,\mqp)\) be
the probability distribution of \(\mqa, \mqp=f(\cba,\cbb)\) used in
the Bell-test protocol, where \(\nu(\mqa,\mqp)>0\) for
each \(\mqa,\mqp\). Both conditional and unconditional probability
distributions are denoted by \(\mu,\nu,\ldots\) and treated as
functions of their arguments, for which we avoid the use of RV
conventions. The candidate function \(W'\) is a test factor against
adversaries if for every three-party non-signaling conditional
distribution \(\mu(\oqa,\zqa,\zqb|\mqa,b, b')\), we have
\begin{align}
  \Exp(W') = \sum_{\mqa,b,\oqa,\zqa,\zqb}W'(\oqa, \mqa;\zqa, b;\zqb, b)\nu(\mqa,b)
  \mu(\oqa,\zqa,\zqb|\mqa,b,b) &\leq 1.
                                 \label{eq:3party_expectation}
\end{align}
The construction and optimization of functions \(W'\) satisfying this condition
is described in Sect.~\ref{sec:test_factors}.

The three-party non-signaling constraints are sufficient for
constructing test factors for the general Bell-test protocol.  In the
case where the prover has only two settings, that is, \(k_{\Vp}=2\)
which is relevant for the experimental implementation, a significant
simplification can be obtained by taking advantage of a connection to
two-party local realism and standard Bell tests.  We use this
simplification to prove completeness. For this, we consider only
test factors \(W'\) that are symmetric under the exchange of
parties \(\Aaend\) and \(\Abend\), that is,
\begin{align}
  W'(\oqa, \mqa;\zqa, b;\zqb, b') &=
                                 W'(\oqa,\mqa;\zqb,b';\zqa, b)
\end{align}
for all values of the arguments. As a result, the expectation of \(W'\)
is unchanged if we symmetrize the non-signaling conditional 
distribution \(\mu(\oqa,\zqa,\zqb|\mqa,b, b')\) under the exchange of parties \(\Aaend\) and \(\Abend\) according to
\begin{align}
  \mu'(\oqa,\zqa,\zqb|\mqa,b, b')=
  \frac{1}{2}\left(\mu(\oqa,\zqa,\zqb|\mqa,b, b')
  +\mu(\oqa,\zqb,\zqa|\mqa,b', b)\right).
\end{align}
Because the exchanged conditional distribution is also non-signaling,
\(\mu'\) is the mixture of two non-signaling conditional
distributions. Since mixtures of non-signaling conditional
distributions are non-signaling, \(\mu'\) is non-signaling.

For any three-party non-signaling distribution
\(\mu'(\oqa,\zqa,\zqb|\mqa,b, b')\) that is symmetric under the
exchange of parties \(\Aaend\) and \(\Abend\), the two marginal
conditional distributions \(\mu'(\oqa,\zqa|\mqa,b)\) for \(\Va\) and
\(\Aaend\), and \(\mu'(\oqa,\zqb|\mqa, b')\) for \(\Va\) and
\(\Abend\), are identical. In the terminology of
Ref.~\cite{Masanes2006}, this means that the marginal
\(\mu'(\oqa,\zqa|\mqa,b)\) is 2-shareable. According to ``Result 1''
in this reference, any such 2-shareable probability distribution with
\(k_{\Vp}=2\) is local realistic. For the marginal conditional
distribution \(\mu'(\oqa,\zqa|\mqa,b)\), the two parties are \(\Va\)
and \(\Aaend\) with inputs \(\mqa\) and \(b\), respectively.  In the
experiment, we directly sample according to
\(\mu(\oqa,\zqa,\zqb|\mqa,b,b)\) with probability \(\nu(\mqa,b)\).
From this we can in principle simulate samples from
\(\mu'(\oqa,\zqa,\zqb|\mqa,b,b)\) by swapping \(\zqa\) with \(\zqb\)
with probability \(1/2\) before recording the trial results. Because
of symmetry of \(W'\), it is not necessary to do this explicitly. That
is, the value of \(W'\) is not affected by explicitly performing the
probabilistic swap.

We do not directly observe different inputs \(b,b'\) to
\(\Aaend\) and \(\Abend\), but the analysis above shows that
there exists a three-party non-signaling distribution consistent with the observations. 
By non-signaling of the distribution \(\mu'\),
\(\mu'(\oqa,\zqa|\mqa,b,b')=\mu'(\oqa,\zqa|\mqa,b,b)\). That is, the
observed marginal for \(\Va\) and \(\Aaend\) conditional on their
inputs does not depend on which \(b'\) is given as input to
\(\Abend\).  Consequently, although we cannot directly observe input
combinations with \(b\ne b'\), for the three-party non-signaling
distribution implied by the best adversary strategies obtained above,
after symmetrization the observable conditional marginal for \(\Va\)
and \(\Aaend\) is the 2-shareable and hence local-realistic marginal
conditional distribution \(\mu'(\oqa,\zqa|\mqa,b)\) for all such
adversary strategies.

Ideally, and with high probability in the experiment, the honest
prover's responses satisfy \(\zqa=\zqb\).  Let
\(\sigma(\oqa,\zqa,\zqb|\mqa,\mqp)\) be the achieved conditional
distribution for the honest prover, and
\(\sigma'(\oqa,\zqa|\mqa,\mqp)=
(\sigma(\oqa,\zqa|\mqa,\mqp)+\sigma(\oqa,\zqb|\mqa,\mqp))/2\) its
symmetrized marginal. The honest prover behaves so that \(\sigma'(\oqa,\zqa|\mqa,\mqp)\)
violates some two-party Bell inequality with \(k_{\Vp}=2\) values for
input \(\mqp\).  Therefore, \(\sigma'\) is outside the
local-realistic polytope.  Given the conditional distribution
\(\sigma'(\oqa,\zqa|\mqa,\mqp)\) and the input distribution
\(\nu(\mqa,\mqp)\), a test factor \(W_{\LR}(\oqa,\mqa;\zqa,\mqp)\)
against local realism can be constructed that witnesses the Bell
violation and has positive gain~\cite{zhang_y:qc2011a}. 
Here, \(W_{\LR}(\oqa,\mqa;\zqa,\mqp)\) is non-negative and satisfies that its expectation is at most \(1\) under any local realistic predictions. If we set
\begin{align}
  W'(\oqa, \mqa;\zqa, b;\zqb, b')
  &= \frac{1}{2}(W_{\LR}(\oqa,\mqa;\zqa,b)+W_{\LR}(\oqa,\mqa;\zqb,b'))
  \label{eq:LRtf}
\end{align} 
then \(W'\) is symmetric under the exchange of parties \(\Aaend\) and
\(\Abend\). Therefore, the argument presented in the 
paragraphs above applies. Given this and in view of the
conditions satisfied by the test factor \(W_{\LR}\) against local
realism, we conclude that \(W'\) is a test factor against three-party
non-signaling conditional distributions, where the three inputs have
probability distribution \(\nu(\mqa,b)\delta_{b,b'}\). Here,
\(\delta_{b,b'}\) is the Kronecker delta. Consequently, \(W'\)
satisfies Eq.~\eqref{eq:3party_expectation} and so is a test factor
against adversary strategies.

Because \(W_{\LR}\) has positive gain for the symmetrized marginal
\(\sigma'(\oqa,\zqa|\mqa,\mqp)\) introduced above, \(W'\) has positive
gain \(g\) for the honest distribution in the experiment, but
expectation at most \(1\) for adversary strategies.  Therefore, the
completeness of the Bell-test protocol is ensured, provided that the
number of trials is large enough.  Since the gain \(g\) is the
expectation of \(\log(W')\) for the honest distribution and the
\(p\)-value against adversaries after a finite number of trials is
given by \(1/(\prod_{k} w'_k)\) with \(w'_k\) being the value taken by
\(W'\) in the \(k\)'th trial, an estimate for the number of trials
required for soundness \(\delta\) is \(\bar n=\log(1/\delta)/g\). If
\(n\geq\bar n\), then the average \(\log(p)\)-value for the protocol
is \(ng \geq \log(1/\delta)\). For completeness \(\epsilon\in (0,1)\),
the probability of passing needs to be at least \(1-\epsilon\). The
probability of passing is the probability that the \(\log(p)\)-value
for the protocol is at least \(\log(1/\delta)\). Since the honest
distribution is close to i.i.d. and the number of trials \(n\) is
large, the \(\log(p)\)-value is expected to be close to Gaussian with
variance \(n\) times the variance of \(\log(W')\). We use this as a
heuristic to choose the number of trials required to ensure soundness
and completeness given a reference honest distribution in the
experimental protocol, see Sect.~\ref{sect:xanalysis}.

To obtain a mathematical bound on the number of trials required for
completeness \(\epsilon\), we may conservatively assume that at each
trial, conditional on the past, the expected value of \(\log(W')\) is
at least some value \(g\) that can be estimated before the experiment
based on the expected honest distribution. Optimized test factors
\(W'\) generally satisfy that \(\log(W')\) is bounded. Whenever
necessary, this boundedness can be enforced by mixing \(W'\) with the
trivial test factor \(1\). Suppose that
\(\log(W')\in [\lambda_{l},\lambda_{u}]\) and let
\(c=|\lambda_{u}-\lambda_{l}|\).  Let \(W'_{k}\) be \(W'\) evaluated
on the \(k\)'th trial record and \(M_{k}=\sum_{j=1}^{k}\log(W'_{j})\).
According to the Azuma-Hoeffding inequality,
\begin{align}
  \Prob(M_{n}< ng-\Delta) \leq e^{-\Delta^{2}/(2 n c^{2})}.
\end{align}
For completeness \(\epsilon\), we set \(\Delta=c\sqrt{2n\log(1/(1-\epsilon))}\).
To determine the number of trials required, we solve for the minimum \(n\)
satisfying
\begin{align}
  \log(1/\delta)
  &\leq ng-\Delta
    \notag\\
  &=ng-c\sqrt{2n\log(1/(1-\epsilon))}
    \notag\\
  &=\sqrt{ng}(\sqrt{ng}-c\sqrt{2\log(1/(1-\epsilon))/g}).
\end{align}
This inequality is satisfied if
\(\sqrt{ng}=\sqrt{\log(1/\delta)}+c\sqrt{2\log(1/(1-\epsilon))/g}\),
because then both \(\sqrt{ng}\) and
\(\sqrt{ng}-c\sqrt{2\log(1/(1-\epsilon))/g}\) are greater than or equal to 
\(\sqrt{\log(1/\delta)}\). Consequently, it suffices to set
\begin{align}
  n &=2\pqty(\log(1/\delta)/g+2c^{2}\log(1/(1-\epsilon))/g^{2})
      ,
\end{align}
which shows that for completeness \(\epsilon\) with soundness
\(\delta\), \(O(\log(1/\delta))+O(\log(1/(1-\epsilon)))\) trials
suffice. The constants depend on the performance of the honest
prover and are finite provided that \(g\) is positive.

The connection to the hypothesis test against local realism implies
that the Bell-test protocol is robust against experimental imperfections such as
loss. In our experimental implementation, both the verifier \(\Va\)
and the prover \(\Vp\) have two measurement settings, where under each
setting there are two possible measurement outcomes. That is,
\(k_{\Va}=k_{\Vp}=2\) and \(c_{\Va}=c_{\Vp}=2\). According to
Ref.~\cite{eberhard1993background}, it is possible to violate a
two-party Bell inequality in this configuration for any photon loss
less than \(1/3\) for each party. In our experiment, the source is
  near \(\Va\)'s measurement apparatus, so \(\Va\) can have
  substantially less loss than the honest prover. If \(\Va\) has no
loss, then any photon loss between the source and the honest prover of
less than \(1/2\) can be tolerated~\cite{brunner2007detection}.

The test-factor construction described in Sect.~\ref{sec:test_factors}
below for position verification uses a test factor for two-party local
realism as a starting point but exploits the greater flexibility of
the three-party non-signaling constraints.  In general, the test
factors constructed directly for the non-signaling constraints without
taking advantage of the simplification based on local realism need not
be expressable in terms of Eq.~\eqref{eq:LRtf}.

\noindent \textbf{Requirement for quantum vertices in the target
  region:} The Bell-test QPV protocol fails if the honest prover uses
only classical devices.  Here, a device is considered classical if it
decoheres its inputs in a fixed product basis and produces separable
outputs for each product basis input. We claim that no prover, either
honest or adversarial, who uses only classical devices in the target
region can successfully pass the protocol.  Consider adversary
circuits that include vertices in the target region \(\cM_{1100}\).
According to our conventions, such vertices change the adversary into
an honest prover. Here we show that if the joint action of the
vertices in \(\cM_{1100}\) satisfies the definition of being removable
given below, then the adversary can not increase their expectation of
the test factors used in the protocol over an adversary with no such
vertices.  We also show that if the vertices act classically as
formally defined below, then they are removable.  This is the sense in
which an honest prover must have a quantum presence in the target
region to pass the protocol.

The \(\cA_{1100}\) subcircuit is now not empty and is part of the
adversary circuit \(\cA\) and needs to be added to the list of
components of \(\cA\) in Eq.~\eqref{eq:advcirc_defs}.  \(\cA_{1100}\)
receives inputs from \(\Ain\), \(\Aastart\) and \(\Abstart\) and
passes outputs to \(\Aout\), \(\Aaend\) and \(\Abend\). As in the
analysis of \(\cA\) given above, to simplify the circuit, without
loss of generality, we can pass the input from \(\Ain\) through
\(\Aastart\) or \(\Abstart\) and pass the output to \(\Aout\) through
\(\Aaend\) or \(\Abend\). With this simplification, the quantum
operation of \(\cA_{1100}\) can be viewed as an operation with two
inputs, from \(\Aastart\) and \(\Abstart\), and two outputs, to
\(\Aaend\) and \(\Abend\).

\begin{figure}
  \includegraphics[scale=0.4]{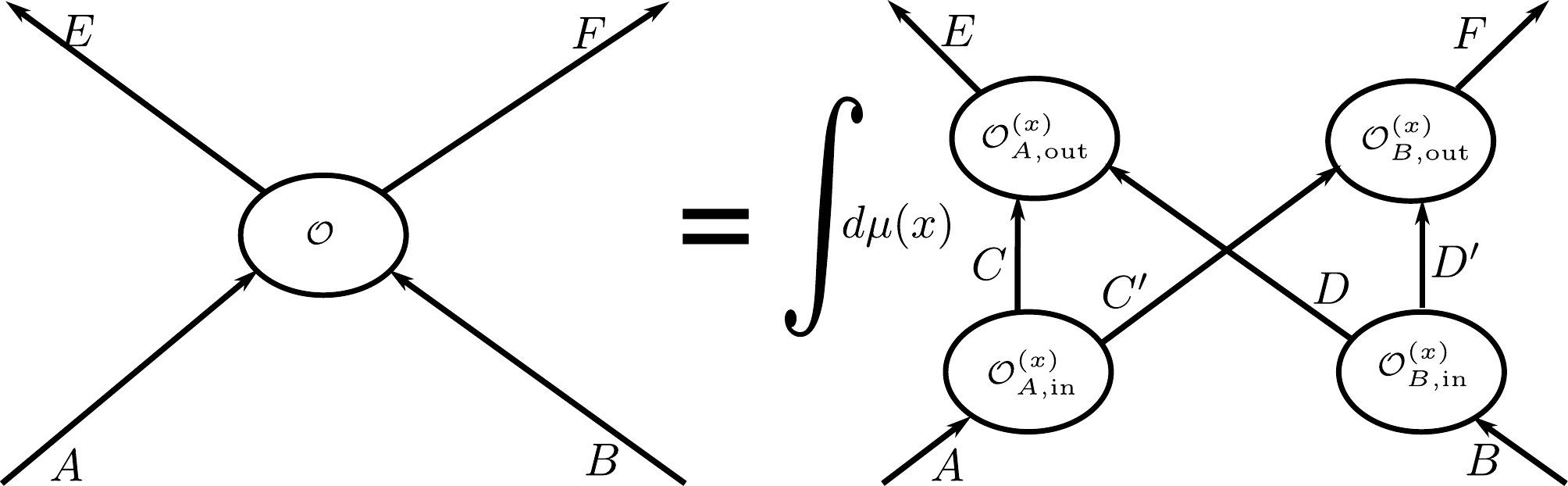}
  \caption{Removable quantum operations.
  \label{fig:removable}}
\end{figure}

Consider a general two-input two-output quantum operation
\(\cO: \rho_{AB}\mapsto \cO(\rho_{AB})_{EF}\), where \(A,B\) are the
input quantum systems and \(E,F\) the output quantum systems.  For
\(\cA_{1100}\), the operation is implemented by \(\cA_{1100}\), and
the inputs and outputs are the edges from \(\Aastart\) and
\(\Abstart\), and to \(\Aaend\) and \(\Abend\).  We call \(\cO\)
removable if 1) there exist quantum systems \(C,C',D,D'\), a
probability distribution \(\mu(x)\) over a classical variable \(x\),
and the following \(x\)-dependent quantum operations:
\(\cO_{A,\In}^{(x)}\) from \(A\) to \(C\) and \(C'\),
\(\cO_{B,\In}^{(x)}\) from \(B\) to \(D\) and \(D'\),
\(\cO_{A,\Out}^{(x)}\) from \(C\) and \(D\) to \(E\), and
\(\cO_{B,\Out}^{(x)}\) from \(C'\) and \(D'\) to \(F\). And 2) these
quantum operations satisfy
\begin{align}
  \cO &= \int d\mu(x) \left(\cO_{A,\Out}^{(x)}\otimes \cO_{B,\Out}^{(x)}\right)\circ
        \left(\cO_{A,\In}^{(x)}\otimes \cO_{B,\In}^{(x)}\right).
        \label{eq:removableops}
\end{align}
The right-hand side is the replacement for the removable operation \(\cO\).
The situation is illustrated in Fig.~\ref{fig:removable}.

To make the correspondence between \(\cA_{1100}\) and \(\cO\), let
\(A\) be the system coming in from \(\Aastart\), \(B\) the system
coming in from \(\Abstart\), \(E\) the system going to \(\Aaend\) and
\(F\) the system going to \(\Abend\).  To show that an adversary with
removable \(\cA_{1100}\) can do no better than an adversary with no
vertices in the target region, with \(\cO\) the quantum operation
implemented by \(\cA_{1100}\), let
\(\cO_{A,\In}^{(x)}, \cO_{A,\Out}^{(x)}, \cO_{B,\In}^{(x)}\) and
\(\cO_{B,\Out}^{(x)}\) be the operations for the replacement in
Eq.~\eqref{eq:removableops}. We modify the other components of the
adversary circuit to eliminate \(\cA_{1100}\). For this purpose, we
modify \(\Ain\) and its outgoing edges so that \(\Ain\) randomly
selects \(x\) with probability distribution \(\mu(x)\) and sends \(x\)
to both \(\Aastart\) and \(\Abstart\) by adding it to its outgoing
edges. After completing their existing quantum operations,
\(\Aastart\) and \(\Abstart\) apply \(\cO_{A,\In}^{(x)}\) to \(A\) and
\(\cO_{B,\In}^{(x)}\) to \(B\), respectively, producing systems
\(C,C'\) and \(D,D'\).  They then send \(C\) and \(D\) to \(\Aaend\),
and \(C'\) and \(D'\) to \(\Abend\), along with \(x\).  Upon receiving
these systems and \(x\), \(\Aaend\) and \(\Abend\) apply
\(\cO_{A,\Out}^{(x)}\) to \(C\) and \(D\), and \(\cO_{B,\Out}^{(x)}\)
to \(C'\) and \(D'\), which after averaging over \(x\), reproduces the
outputs they previously obtained from \(\cA_{1100}\) in the original
circuit.

Removable quantum operations include quantum routers that forward
subsystems from the inputs to the outputs according to a fixed routing
rule independent of the inputs, and semiclassical operations that
decohere the inputs in a product basis and map each basis element to a
separable state.  To implement such a semiclassical operation in the
form given on the right-hand side of Eq.~\eqref{eq:removableops},
decoherence is implemented by operations \(\cO_{A,\In}\) and
\(\cO_{B,\In}\) that do not depend on the classical variable \(x\)
introduced below. These operations implement the decoherence in their
respective bases and copy their decohered states to their outputs. Let
the basis elements of \(A\) and \(B\) be denoted by \(\ket{a}_{A}\)
and \(\ket{b}_{B}\), where \(a\) and \(b\) range over finite index
sets. The semiclassical operation produces, for each product state
\(\ket{a}_{A}\ket{b}_{B}\), a separable state.  We express this
separable state as an explicit mixture
\(\sum_{x_{a,b}}\mu_{a,b}(x_{a,b})\ket{\phi(x_{a,b})}_{E}\ket{\psi(x_{a,b})}_{F}\),
where for each \(a,b\), \(\mu_{a,b}(x_{a,b})\) is a probability
distribution on a finite set depending on \(a,b\).  We let the
variable \(x\) be a list \((x_{a,b})_{a,b}\), where each \(x_{a,b}\)
is independently distributed with probability distribution
\(\mu_{a,b}(x_{a,b})\).  The operations \(\cO_{A,\Out}^{(x)}\) and
\(\cO_{B,\Out}^{(x)}\) are then designed to independently prepare the
states \(\ket{\phi(x_{a,b})}_{E}\) and \(\ket{\psi(x_{a,b})}_{F}\),
given the inputs \(\ket{a}_{C}\ket{b}_{D}\) and
\(\ket{a}_{C'}\ket{b}_{D'}\), respectively. After integrating over the
probability distribution of \(x\), this reproduces the original
semiclassical operation.  Such semiclassical operations include
probabilistic classical computations and Markov processes that
decohere the inputs in a product basis and perform probabilistic
classical computation given the input basis elements to produce a
random output, which is then encoded in a fixed output product basis.

\subsection{Test-factor construction}
\label{sec:test_factors}

Test factors for detecting adversaries without prior entanglement are
constructed to maximize the gain per trial for the actual honest
prover.  This requires estimates of the conditional probabilities
\(\sigma(\oqa,\zqa,\zqb|\mqa,\mqp)\) for the protocol with the actual
prover. The quality of the estimates does not affect the validity of
the test factors constructed but may affect the probability of success
for the honest prover.  We obtain such estimates from calibration data
obtained before runs of the protocol. Ideally, the prover always
responds with \(\zqa=\zqb\).  However, in the experiment we observe
trials where \(\zqa\ne\zqb\). The method for estimating
\(\sigma(\oqa,\zqa,\zqb|\mqa,\mqp)\) is described in
Sect.~\ref{sect:xanalysis}.  Here, we assume that
\(\sigma(\oqa,\zqa,\zqb|\mqa,\mqp)\) is given.  The associated
trial-wise probability distribution is
\begin{align}
  \sigma(\oqa,\zqa,\zqb,\mqa,\mqp)
  &=\sigma(\oqa,\zqa,\zqb|\mqa,\mqp) \mmuap(\mqa,\mqp),
\end{align}
where \(\mmuap(\mqa,\mqp)\) is the probability of \(\mqa\) and
\(\mqp=f(\cba,\cbb)\), which is determined by the joint  distribution of settings 
and challenges. See Assumption~\ref{ass:distrib}. Let \(\cP\) be the
polytope of conditional distributions
\(\mu(\oqa,\zqa,\zqb|\mqa,b,b')\) that are three-party non-signaling. See~\cite{Pironio_2011} for a detailed treatment of the geometry and extremal structure of \(\cP\). For such a \(\mu\in\cP\), define
\begin{align}
  \mu(\oqa,\zqa,\zqb,\mqa,b,b')
  &=
  \mu(\oqa,\zqa,\zqb|\mqa,b,b') \mmuap(\mqa,b)\delta_{b,b'} .
\end{align}
We can maximize the gain by solving the following problem:
\begin{itemize}
\item[]\textbf{Gain optimization problem:}
  \begin{itemize}
  \item[] \textbf{Given:} Conditional probabilities
    \(\sigma(\oqa,\zqa,\zqb|\mqa,\mqp)\).
  \item[] \textbf{Maximize over $W$:}
    \begin{align}
      \Exp_{\sigma}(\log(W))={}
      \hspace{-.5in}&
                      \notag\\
                    &
                      \sum_{\oqa, \zqa, \zqb, \mqa,\mqp}
                      \log(W(\mqa,\oqa,\mqp,\zqa,\zqb))\sigma(\oqa,\zqa,\zqb,\mqa,\mqp).
    \end{align}
  \item[] \textbf{Constraints:}
    \begin{align}
      W   &\geq 0,
            \notag\\
      \Exp_{\mu}(W)
          & =
            \sum_{\oqa, \zqa, \zqb, \mqa,\mqp}
            W(\mqa,\oqa,\mqp,\zqa,\zqb)\mu(\oqa,\zqa,\zqb,\mqa,\mqp,\mqp)
            \notag\\
          &\leq 1
            \hphantom{{}={}}\textrm{for all \(\mu\in\cP\)}.
    \end{align}
  \end{itemize}
\end{itemize}
The gain optimization problem requires maximizing a concave function
over the convex set \(\cP\) of three-party non-signaling
distributions.  For the case of two settings per party and two
outcomes for each setting as in our experimental implementation, this
can be done directly. However, to simplify the calculations, we
restricted the test factor \(W\) to be symmetric under the exchange of
\(\zqa\) and \(\zqb\). In our experiment, the probability of
\(\zqa\ne\zqb\) is very low, on the order of \(10^{-6}\). If it is
zero, then the probability distribution of \((\oqa,\zqa,\mqa,\mqp)\)
conditional on \(\zqa=\zqb\) for three-party non-signaling is local
realistic, a consequence of the results of Ref.~\cite{Masanes2006}
discussed above.  With this as motivation, we took advantage of
existing strategies for optimizing the gain of test factors in
two-party Bell tests against local realism~\cite{zhang_y:qc2011a}.
From calibration data, we estimated the actual prover's probability
distribution of \((\oqa,\zqa,\mqa,\mqp)\) conditional on
\(\zqa=\zqb\), and then determined the gain-optimizing test factor
\(W_{\LR}(\oqa, \mqa;\zqa,\mqp)\) against local realism given this
probability distribution. We then considered test factors \(W\) of the
form
\begin{align}
  W(\mqa,\oqa,\mqp,\zqa,\zqb)
  &=\delta_{\zqa,\zqb}W_{\LR}(\oqa, \mqa;\zqa,\mqp)
       + \lambda (1-\delta_{\zqa,\zqb}),\label{eq:robust_tf}
\end{align}
where \(\lambda\) was chosen to have the largest value
for which the resulting \(W\) is
a test factor against three-party non-signaling, that is, for which \(W\) satisfies the
constraints of the gain optimization problem. Although these test
factors are not gain-optimal, they are robust against variations in the
small probability that \(\zqa\ne\zqb\). Empirical tests indicate that
for the probability distributions encountered in our experiments, near-optimal gain is achieved.

\subsection{Test factors against small amounts of prior entanglement}
\label{sec:smallentanglement}

We can exclude small amounts of prior entanglement shared by
adversaries, depending on the expectation of the chosen test factor
that can be achieved by the actual prover. For general protocols, such
bounds can be derived from Lemma~5.3 of
Ref.~\cite{beigi2011simplified}. This lemma obtains a lower bound of
order \(\delta'/\delta\) on the total dimension of the entangled
quantum systems required for adversaries to pass the protocol if the
protocol is \(\delta\)-sound when adversaries start with separable
states and the soundness is changed to \(\delta'>\delta\) to
  account for prior entanglement.  This
bound is obtained by replacing the adversary's entangled state before
the protocol with the completely mixed state and disallowing quantum
communication during the protocol. Here we develop quantified bounds
that apply trial-by-trial or on average over a protocol by taking
advantage of properties of test factors.  For bounds that
overcome this limitation by increasing the complexity of the settings
function, see Sect.~\ref{sect:entangled_adversaries}. We quantify
entanglement in terms of the robustness of
entanglement~\cite{vidal1999robustness}, see also
Ref.~\cite{rudolph2005further}.  For a density matrix \(\rho\), we
write \(R(\rho)\) for the robustness of entanglement.  \(R(\rho)\) is
an entanglement monotone.  For separable states, \(R(\rho)=0\), and
for entangled states, \(R(\rho)>0\). In particular, for a bipartite
pure entangled state \(\rho\) with Schmidt coefficients \(a_{j}\), we
have \(R(\rho)=\pbr(\sum_{j} a_{j})^{2}-1\).  The significance of
robustness of entanglement is that for every bipartite entangled state
\(\rho\) in finite dimensions, there exist separable states \(\sigma\)
and \(\tau\) such that \(\rho=(1+\xi)\sigma-\xi\tau\) with
\(\xi=R(\rho)\), and no such representation exists with
\(0\leq \xi< R(\rho)\). These properties of robustness of entanglement
can be found in Ref.~\cite{vidal1999robustness}. We also remark that
for a bipartite pure state \(\rho\), the R\'enyi entropy of order
\(1/2\) of its reduced density matrix is given by
\(S_{1/2}=\log(R(\rho)+1)\), which establishes a direct connection
between the robustness of entanglement and a conventional entropic
measure.

Consider a test factor \(W\) for the Bell-test protocol
against adversaries without prior entanglement.  Let \(\bar w_{\min}\)
be the average over settings and challenges of the minimum value of
\(W\) conditional on the settings and challenges, that is
\begin{equation}\label{eq:w_min_avg}
\bar{w}_{\min}=\sum_{\mqa,\mqp}\mmuap(\mqa,\mqp)\min_{\oqa,\zqa,\zqb}W(\mqa,\oqa,\mqp,\zqa,\zqb).
\end{equation}
Since the probability distribution of settings and challenges is
fixed, the expectation of \(W\) is at least \(\bar w_{\min}\).
Without loss of generality, we may assume that \(\bar w_{\min}<1\), as
the test factor is otherwise useless for the 
protocol.  If the entangled state \(\rho\) of the two subsystems
\(\Qa\QAa\) and \(\QAb\) on leaving the causal past \(\cM_{\In}\) of
the trial satisfies \(R(\rho)=\xi'\leq\xi\), then we represent the
state as an affine combination of separable states \(\sigma\) and
\(\tau\) as \((1+\xi')\sigma-\xi'\tau\).  By construction, the
expectation of \(W\) for each of \(\sigma\) and \(\tau\) is between
\(\bar w_{\min}\) and \(1\). By linearity over the affine combination
of the expectation of \(W\), the expectation of \(W\) for state
\(\rho\) is at most
\( (1+\xi') -\xi' \bar w_{\min} = 1+\xi'(1-\bar w_{\min}) \leq
1+\xi(1-\bar w_{\min})\).  Let \(\bar w_{u}=1+\xi(1-\bar w_{\min})\).
This implies that \(W/\bar w_{u}\) is a test factor against
adversaries whose prior entanglement has robustness bounded from above
by \(\xi\).  By using the test factor \(W/\bar w_{u}\) for each trial,
we can effectively reject such adversaries. 

We can also test the hypothesis that adversaries have prior
entanglement with average robustness below some threshold, averaged
over all trials.  For this, suppose that at trial \(k\), conditional
on the past, the adversaries have prior entanglement with robustness
\(\Xi_{k}\).  We treat \(\Xi_{k}\) as a random variable determined by
the past of trial \(k\).  Let \(W_{k}\) denote the instantiation of
the test factor \(W\) on the record of the \(k\)'th trial.
Conditional on the past, \(\Xi_{k}\) takes on a specific, determined
value \(\xi_{k}\), and the expectation of \(W_{k}\) given this past is
at most
\(1+\xi_{k}(1-\bar w_{\min})\leq e^{\xi_{k}(1-\bar w_{\min})}\).
Equivalently, the past-conditional expectation of
\(W_{k} e^{-\Xi_{k}(1-\bar w_{\min})}\) is at most \(1\). By the law
of total expectation, the product \(Z\) of
\(W_{k} e^{-\Xi_{k}(1-\bar w_{\min})}\) over \(n\) trials also has
expectation at most \(1\). By Markov's inequality, we have
\(\Prob(Z\geq 1/\delta )\leq \delta \Exp(Z)\), where \(\delta>0\)
is the desired soundness parameter. Rearranging the inequality in this
expression, we conclude that the probability that
\begin{align}
\frac{1}{n(1-\bar w_{\min})}\pbr(\sum_{k}\log(W_{k})-\log(1/\delta))
  &\geq \frac{1}{n}\sum_{k} \Xi_{k}
  \label{eq:robustness_ineq}
\end{align}
is at most \(\delta\).  We can therefore take the left-hand side of
this inequality as a lower bound on the average robustness of
entanglement that the adversaries must share prior to each trial in
order to pass the protocol with probability at least \(1-\delta\).  We
modify the protocol so that it succeeds when the lower bound on the
average robustness certified at soundness parameter \(\delta\) exceeds
a threshold chosen ahead of time. Let \(r_{\text{th}}\) be the chosen
threshold for average robustness of entanglement per trial.  Then, to
pass the protocol we require
\(\frac{1}{(1-\bar w_{\min})n}(\sum_{k}\log(W_{k})-\log(1/\delta))
\geq r_{\text{th}}\). We can rewrite this inequality as
\(\prod_{k}W_{k}e^{-r_{\text{th}}(1-\bar w_{\min})} \geq 1/\delta\).
Therefore, it suffices to replace the trial-independent test factor in
Protocol~\ref{prot:pv} by \(We^{-r_{\text{th}}(1-\bar w_{\min})}\) to test
against adversaries with prior entanglement.  

Before the run of the protocol, the verifiers determine the values of
the robustness threshold \(r_{\text{th}}\) for which the actual prover has
a high probability of passing the protocol given enough
trials. Suppose that the test factor \(W\) against adversaries without
prior entanglement has gain \(g=\Exp(\log(W))\) for the actual prover,
and define \(\bar w_{\min}\) as above.  For \(n\) trials with the
actual prover, the expectation of the left-hand side in
Eq.~\ref{eq:robustness_ineq} is
\(g/(1-\bar w_{\min})-\log(1/\delta)/\pbr(n(1-\bar w_{\min}))\).  The
term multiplying \(\log(1/\delta)\) goes to zero as
\(n\rightarrow\infty\), so given sufficiently many trials, it is
possible to reject adversaries with average robustness of entanglement
per trial up to \(g/(1-\bar w_{\min})\).

Any test factor $W$ for QPV against adversaries without prior
entanglement can be used in the above expressions to also test against
adversaries with prior entanglement. For the experiment, we modify $W$
to minimize the number of trials required to ensure that the actual
prover can pass the protocol with chosen threshold \(r_{\text{th}}\)
and with sufficiently high probability.  The modification
consists of replacing $W$ with \(W'=\lambda W+(1-\lambda)\) and
minimizing the required number of trials over
\(\lambda\in[0,1/(1-w_{\min})]\), where $w_{\min}$ is the minimum
value that $W$ can take.  See Sect.~\ref{sect:parameter_choices} for
details.

\subsection{Heralded sources and commitment to overcome transmission
  loss}
\label{sect:heraldetc}

Significant advantages of test factors include that they permit early
stopping if thresholds on the product of trial-wise test factors are
reached, and that they can be adapted based on past trials if
desired. The latter property enables setting the next trial's test
factor to \(1\), if conditions for performing the next trial are not
satisfied.  A test factor of \(1\) is equivalent to ignoring the
contribution of the next trial. A common situation where it is
desirable to not perform the next trial is if heralded sources are
used. In this case, provided the herald is committed early enough,
such as before the trial boundary or at \(\Va\) before the measurement
is applied, a test factor of \(1\) can be used for trials where the
herald indicates that no entanglement has been generated.  Heralded
sources can be used to overcome the problem of loss in transmission.
For example, the protocol of Ref.~\cite{duan2001long}, known as the
DLCZ protocol, can extend entanglement links to arbitrary distances
with the help of local quantum memories, and a recent experiment
employing high-fidelity entanglement over \qty{100}{\kilo \meter}
using a single-photon-interference
protocol~\cite{lu_deviceindependent_2026} demonstrates that heralded
entanglement can already be established over large distances between
atomic memories with fairly long storage times of around
$\qty{300}{\milli \second}$.

For bounded distances, the commitment method of
Refs.~\cite{allerstorfer2025making,verduyn2024quantum} can be used to
avoid the need for quantum memory and allow for transmission loss
\(\gamma\) arbitrarily close to \(1\) if the detectors have
  efficiency \(1\).  The probability of a successful trial with
commitment is effectively \(1-\gamma\).  In terms of distance \(l\),
we usually have \(1-\gamma = e^{-\lambda l}\) where \(\lambda\) is a
loss rate. Consequently, commitment can enable a significant increase
in distance without the use of quantum repeaters, albeit at the cost
of an exponential increase in the number of trials required for the
protocol.  When using commitment, the prover, before entering the
target region, decides whether they are ready to commit to the
upcoming trial.  Commitment is determined by whether or not they
detected the presence of the photon. They send a commitment bit with
value \(1\) for ``commit'' to verifier \(\Va\), before they enter the
target region, ensuring that it will be received by verifier \(\Va\)
or \(\Vap\) early enough.  The verifiers must certify that commitment
happened before the target region, for which it suffices if they
receive the commitment in the complement of the set that causally
depends on the target region. For simplicity, we assume that verifier
\(\Vap\) and \(\Va\) are co-located and receive the commitment before
they exit region \(1001\) in Fig.~\ref{fig:1+1d_regions}. The
verifiers set the test factor to \(1\) in post-processing if the
prover indicated not being ready. To avoid the requirement for local
memory in the case that the source is near \(\Va\),
the verifier measurement is performed immediately after
their photon arrives. This requires replacing the causal separation
assumption of Assumption~\ref{ass:meas} with physical security. In
particular, \(\Va\) must prevent the measurement apparatus from
communicating measurement results beyond \(\Va\)'s laboratory during
the period from when the setting is applied to the end of the trial.
Alternatively, \(\Va\) can make their own commitment measurement
before generating and applying their measurement setting. We remark
that the quantum measurement apparatus used for deciding whether to
commit does not need to be trusted.

Refs.~\cite{allerstorfer2025making,verduyn2024quantum} prove that the
commitment method works in some generality for QPV protocols that use
simple correct/incorrect decisions for each trial.  To verify security
of the commitment method for Bell-test QPV, it is necessary to extend
the analysis of adversary capabilities to account for the fact that
adversary \(\Aastart\), but not \(\Aaend\), \(\Abstart\) or
\(\Abend\), can control the commitment bit received by \(\Va\).  The
reduction to three-party non-signaling in Sect.~\ref{sect:reduction}
needs no modification, because the commitment bit can be added to
\(\Qa\) as an additional classical bit, and \(\Qa\) is included in
\(\Rsys_{\Aa}\).  The resulting three-party state including the
commitment bit is independent of the challenges and the verifier
setting. Conditioning on the commitment bit having value \(1\) results
in a conditional three-party state, which again results in a
three-party quantum non-signaling scenario. In particular the test
factor has expectation bounded above by \(1\) conditional on the
commitment bit. Setting the test factor to \(1\) if the commitment bit
is \(0\) preserves this bound. In general, any conditioning performed
by \(\Va\) before applying the measurement setting and independent of
this setting performs similarly.

\subsection{Security against adversaries with significant quantum resources}
\label{sect:entangled_adversaries}

In order to obtain security against adversaries who possess larger
amounts of entanglement than can be handled by the results of
Sect.~\ref{sec:smallentanglement}, we use quantum position
verification protocols with more complex settings functions.
Ref.~\cite{bluhm2022single} considers settings functions which accept
two length-$l$ bit strings $x$ and $y$ as input, which are transmitted
to the prover by the left and right verifiers, respectively.
Ref.~\cite{bluhm2022single} Prop.~6.1 establishes a lower bound of
\(l^{1/2}/8\) on the dimension of the memory registers needed for
adversaries to defeat the soundness of their protocols for concrete
settings functions and sufficiently small noise levels. The memory
registers include classical registers such as those needed for
classical communication. They also show non-constructively the
existence of settings functions requiring dimension at least
\(2^{l/2-5}\) under the same assumptions.  See also
Ref.~\cite{escola2023single}. The proofs are based on $\epsilon$-net
arguments and on results in communication complexity. Note that in
this section, \(\epsilon\) denotes the parameter associated with the
\(\epsilon\)-net technique, not the completeness parameter of the
protocol.

Here, we use a similar approach to analyze the Bell-test protocol for
the case where provers have two setting choices and the verifier
settings and challenges are independent and uniformly distributed. We
show that there are simple and concrete settings functions such that
the adversary closer to verifier \(\Vb\) needs quantum systems of
dimension \(d=\Omega(l^{1/2})\) for prior  entanglement to
  defeat soundness of the protocol. See Eq.~\eqref{eq:entdim_bound}
for an explicit inequality relating \(l\) and \(d\). Whereas
Ref.~\cite{bluhm2022single} only provided a lower bound on the total
size of both classical and quantum registers that cheating adversaries
must use, we provide a lower bound specifically on the size of the
adversaries' quantum registers, thus allowing for free classical
computation and communication. Crucially, our results require no
restriction on the noise level in the quantum systems used by the
verifiers and honest prover.

In our security analysis, we assume that the adversaries possess an
initial entangled quantum state. We do not directly consider the case
in which the adversaries use quantum communication with one another,
since in this setting quantum communication can be simulated with
teleportation.  To achieve this, it suffices to extend the initial quantum state
 with one Bell pair for each qubit to be communicated.  We let $d$
denote the local dimension of the entangled quantum state.  Following
Ref.~\cite{bluhm2022single}, we approximate the states obtained after
the initial measurements made by the adversary nearer verifier \(\Vb\)
with the states of an \(\epsilon\)-net. The approximation
  demonstrates that the adversary nearer verifier \(\Vb\) can reduce
  the size of the classical messages to the other adversary to the
  size of the \(\epsilon\)-net, and conditional on these messages, we
  recover a three-party quantum non-signaling scenario. 
  We then take advantage of the properties of test factors to deduce
  the existence of a solution to a communication-complexity problem.
  By applying bounds on such solutions, we obtain a lower bound on
  \(d\).

An alternative approach to proving bounds on adversary resource
requirements is given in~Ref.~\cite{asadi2025linear}. In this approach
the number of quantum gates and measurements required by successful
adversaries is bounded from below.  Since the sequence of gates,
measurements and measurement outcomes can be efficiently encoded, it
is possible to avoid the use of \(\epsilon\)-net
approximations. Ref.~\cite{asadi2025linear} effectively obtains a
linear lower bounds on the number of gates and measurements in terms
of \(l\).  Unlike Ref.~\cite{bluhm2022single}, there is no restriction
on classical communication. In this section, we also extend these
bounds to Bell-test QPV. For this purpose, in the proof of the bounds
on dimension of the entangled quantum states below, we can directly
substitute gate sequences for \(\epsilon\)-nets to obtain a linear
bound on quantum gates and measurements.  This bound is established at
the end of the section for the adversary closer to verifier \(\Vb\)
for the Bell-test protocol. The other adversary's local quantum
resources are unconstrained, and classical communication is free. See
Eq.~\eqref{eq:gatequbit_bound} for an explicit bound relating \(l\) to
the number of gates and qubits used for a standard universal gate set.

In this section, we use simplified notation to refer to the parties
and states.  We refer to verifier \(\Va\)'s quantum measurement
apparatus as \(\cV\) and to the adversaries as \(\cB,\cC\), where
  \(\cB\) is the adversary nearer \(\cV\). The adversary strategy
consists of two phases, where in the first phase they receive the
challenges and measure and manipulate their shared entangled state,
and in the second phase they receive the other adversary's classical
message and send their responses to the verifiers.  The quantum system
destined for \(\cV\) is \(V\) and the quantum systems of \(\cB\) and
\(\cC\) are \(B\) and \(C\).  Without loss of generality, the initial
state is pure on systems \(VBC\), where \(VB\) is in initially in the
possession of \(\cB\), and \(C\) is in the possession of
\(\cC\). \(V\) must be sent to \(\cV\) by \(\cB\) because the
measurement station must produce its output before \(\cC\) has a
chance to send anything to it.  Let \(\ket{\psi}_{VBC}\) be the
initial state. We let \(d\) be the dimension of the support of the
reduced state on system \(C\), which is also the dimension of the 
support of system \(VB\). During the first phase of the trial, \(\cB\)
and \(\cC\) receive their challenges \(x\) and \(y\).  Conditional on
challenge \(x\), \(\cB\) applies a quantum instrument to \(VB\) with
outcome \(b\), and \(\cC\) applies another quantum instrument to \(C\)
and obtains outcome \(c\).  System \(V\) goes to the verifier to be
measured by the verifier's measurement device, which is not trusted
and therefore may be assumed to receive both \(x\) and \(b\) along
with system \(V\).  Systems \(B\) and \(C\) are stored for the second
phase of the trial during which \(\cB\) and \(\cC\) make measurements
to determine the values to be sent to the respective verifiers.  (See
Figure~\ref{fig:qpvcheatingmodel}.) We consider the transmission of
subsystem \(V\) as an unrestricted quantum communication by adversary
\(\cB\). By increasing the instrument outcome spaces, we can, without
loss of generality, assume that the operations conditional on outcomes
are pure, meaning that as operations on density matrices they have
just one Kraus operator \(X\) and act as
\(\ket{\phi}\mapsto X\ket{\phi}\) on the Hilbert space. The
  probability of the corresponding instrument outcome is
  \(\bra{\phi}X^{\dagger}X\ket{\phi}\).

\begin{figure}
  \centering
  \begin{tabular}{c}
    \begin{tikzpicture}
      \draw (2,0) node[draw, circle] {\(V\)};
      \draw (3,0) node[draw, circle] {\(B\)};
      \draw (1.5,-0.5) rectangle (3.5,0.5);
      \draw (2.5,1.75) node {\(x\)};
      \draw (2.5,-1.75) node {\(b\)};
      \draw (3.5,-0.75) node {\(\cB\)};
      \draw [<-] (2.5,-1.55) -- (2.5,-0.55) node[midway, left] {};
      \draw [<-] (2.5,0.55) -- (2.5,1.55) node[midway, left] {};
      
      \draw (8,0) node[draw, circle] {\(C\)};
      \draw (7.5,-0.5) rectangle (8.5,0.5);
      \draw (8,1.75) node {\(y\)};
      \draw (8,-1.75) node {\(c\)};
      \draw (8.5,-0.75) node {\(\cC\)};
      \draw [<-] (8,-1.55) -- (8,-0.55) node[midway, left] {};
      \draw [<-] (8,0.55) -- (8,1.55) node[midway, left] {};
      \draw (-4,0.25) node[draw,circle] {\large 1};
    \end{tikzpicture}  \\
    \hline \\
    \begin{tikzpicture}
      \draw (0,0) node[draw, circle] {\(V\)};
      \draw (-0.5,-0.5) rectangle (0.5,0.5);
      \draw (0.5,-0.75) node {\(\cV\)};
      \draw [<-] (0,-1.55) -- (0,-0.55);
      \draw [<-] (0,0.55) -- (0,1.55);
      \draw (0,1.75) node {\(z\)};
      \draw (0,-1.75) node {\(m_v\)};
      \draw (4,0) node[draw, circle] {\(B\)};
      \draw (3.5,-0.5) rectangle (4.5,0.5);
      \draw (4.5,-0.75) node {\(\cB\)};
      \draw (4,1.75) node {\(y,c\)};
      \draw (4,-1.75) node {\(m_b\)};
      \draw [<-] (4,-1.55) -- (4,-0.55) node[midway, left] {};
      \draw [<-] (4,0.55) -- (4,1.55) node[midway, left] {};

      \draw (8,0) node[draw, circle] {\(C\)};
      \draw (7.5,-0.5) rectangle (8.5,0.5);
      \draw (8.5,-0.75) node {\(\cC\)};
      \draw [<-] (8,-1.55) -- (8,-0.55) node[midway, left] {};
      \draw [<-] (8,0.55) -- (8,1.55) node[midway, left] {};
      \draw (8,1.75) node {\(x,b\)};
      \draw (8,-1.75) node {\(m_c\)};
      \draw (-4,0.25) node[draw,circle] {\large 2};
    \end{tikzpicture}     
  \end{tabular}
  \caption{A theoretical cheating model.  In the first phase, the systems \(V,B\) and \(C\) are measured by the adversarial parties \(\cB\) and \(\cC\).  In the second phase, the same systems are re-measured by the first verifier's measurement apparatus \(\cV\) and the parties \(\cB\) and \(\cC\).}
  \label{fig:qpvcheatingmodel}
\end{figure}

Before proceeding, we show that if \(d=2^{2^{l}}\), then the
adversaries can perfectly simulate any pair of verifier measurement
station and honest prover sharing a pair of qubits from a
source. For some protocols with trusted devices, such perfect
attacks are impossible, as shown in
Ref.~\cite{miller2024perfect}. The following is a simple instance of
a ``garden-hose'' attack~\cite{buhrman2013garden} adapted to our
scenario.  In our case, the attack is enabled by adversary \(\cB\)'s
control of the quantum state that is measured by \(\cV\).  The
attack strategy requires that \(\cB\) and \(\cC\) share \(2^{l}\)
pairs of qubits \(\{P_{x}\}_{x}\), each in the state that is
intended to be shared between the verifier measurement station and
the honest prover. In the first phase of the trial, when \(\cB\)
receives challenge \(x\), \(\cB\) sends their share of \(P_{x}\) to
\(\cV\) and \(x\) to \(\cC\).  When \(\cC\) receives challenge
\(y\), for each possible \(x'\), \(\cC\) applies measurement
\(f(x',y)\) to their share of \(P_{x'}\), using the same positive
operator valued measure (POVM) as would be used by the honest prover
being simulated.  The measurement outcome is recorded as
\(r(x',y)\). \(\cC\) sends to \(\cB\) the challenge $y$ along with a
string of length \(2^{l}\) whose symbol in position \(x'\) has value
\(r(x',y)\).  In the second phase, \(\cC\), having learned \(x\),
responds with \(r(x,y)\) and \(\cB\) does the same, using the
information received from the first phase.

Returning to the analysis of adversaries with constrained entanglement
dimension, let \(B_{x,b}\) and \(C_{y,c}\) be the operators applied by
the quantum instruments of \(\cB\) and \(\cC\) upon challenges \(x\)
and \(y\) and outcomes \(b\) and \(c\), respectively.  We have
\(\sum_{b}B^{\dagger}_{x,b}B_{x,b}=\Id\) and similarly for
\(C_{y,c}\). To simplify notation, we treat states \(\ket{\phi}\) of
system \(VBC\) as \(d'\times d\) dimensional matrices \(\phi\), where
\(d'\geq d\), \(VB\) is represented by the row space, and \(C\) by the
column space.  In this representation, \(B_{x,b}\) are matrices of
size \(d'\times d'\) acting on the left, and \(C_{y,c}\) are matrices
of size \(d\times d\) whose transpose is acting on the right. That is,
conditional on \(x,b,y,c\), the normalized state obtained after the
instruments act is
\(\psi_{x,b,y,c}=B_{x,b}\psi C_{y,c}^{T}/|B_{x,b}\psi C_{y,c}^{T}|\),
where \(\psi\) is the initial entangled state, which is known to
\(\cB\) and \(\cC\). Since the support of the reduced density matrix
of the initial state on \(VB\) has dimension at most \(d\), we assume
that the matrix \(\psi\) is non-zero only on the first \(d\) rows.
Here \(|\cdot |\) denotes the Hilbert-Schmidt norm
\(|M|=\sqrt{\tr(M^{\dagger}M)}=\sqrt{\tr(MM^{\dagger})}\), which is
the usual Hilbert-space norm for states represented by matrices.  In
principle, the quantum instruments of \(\cC\) could map the
\(d\)-dimensional support \(\cH\) of \(\psi\)'s reduced density matrix
on system \(C\) to arbitrary places in a higher dimensional
space. However, for a given \(y,c\) the reduced density matrix of
state \(\psi C_{y,c}^{T}\) still has support of at most \(d\)
dimensions. We can therefore ensure that the output of \(C_{y,c}\) is
in \(\cH\) by applying an isometry \(V_{y,c}\) to the output that
returns the output to \(\cH\). A standard such isometry is obtained
from the polar decomposition
\(C_{y,c}=V_{y,c}^{\dagger}\tilde C_{y,c}\), where \(\tilde C_{y,c}\)
is positive semidefinite Hermitian on \(H\).  \(\cC\) knows which
isometry was applied and can reverse it in the second phase before
performing their final measurement, or simply integrate it into the
final measurement.  In order to reduce the real dimension of the
matrices \(\psi C_{y,c}^{T}\) to \(d^{2}\), we further modify
\(C_{y,c}\) by post-processing with a unitary depending on \(y,c\)
applied to system \(C\) such that \(\psi C_{y,c}^{T}\) restricted to
the first \(d\) rows is a Hermitian matrix, where \(\psi C_{y,c}^{T}\)
is by construction zero on the other rows.  This can be done by
replacing \(C_{y,c}\) with \(U^{\dagger T}V^{T}C_{y,c}\), where
\( \psi C_{y,c}^{T}\) has singular value decomposition
\(UDV^{\dagger}\). Here \(U\) may be considered as a \(d\times d\)
unitary matrix acting on the first \(d\) rows when multiplied on the
left.

Let \(W\) be a test factor used for Bell-test QPV that has expectation
at most \(1\) for the three-party quantum non-signaling scenario
derived in Sect.~\ref{sect:reduction}. The goal is to show that if the
entanglement dimension of the entangled state shared by the
adversaries at the trial boundary is at most \(d\), then the
expectation of the test factor \(W\) for the adversaries exceeds \(1\)
by at most a factor \(1/\kappa\), where \(\kappa\in (0,1)\) and
\(\kappa\) can be made close to \(1\) by increasing \(l\). To protect
against such adversaries it then suffices to replace the test factor
\(W\) by \(W'=\kappa W\).  The basic idea of the
proof is that the communication from adversary \(\cC\) to \(\cB\) can
be compressed to consist of only one of \(L\) symbols, where \(L\) is
determined by \(d\). At the same time, \(\cB\) is able to guess the
value of the prover setting \(f(x,y)\) knowing only their challenge
and the compressed symbol from \(\cC\), with a guessing advantage
linearly related to \(1/\kappa-1\). If \(l\) is large enough and
the function \(f(x,y)\) has sufficient complexity, a
communication-complexity analysis shows that \(1/\kappa-1\) must
be small.   The proof is obtained in a series of steps.  We begin by
fixing \(d\) and the excess expectation \(\epsilon'=\Exp(W)-1\) of
\(W\) achieved by the adversaries.  The steps are as follows:
\begin{enuma}
\item \textbf{Construction of an \(\epsilon\)-net:} We
  construct an \(\epsilon\)-net \(\cN\) of \(L\) normalized states
  such that for each state \(\psi C_{y,c}^{T}\) there is a member
  \(\phi_{y,c}\) of the \(\epsilon\)-net within distance
  \(\epsilon\) of \(\psi C_{y,c}^{T}/|\psi C_{y,c}^{T}|\). Later, we
  express \(\epsilon\) as a positive multiple of \(\epsilon'\) with
  coefficient determined by the values of \(W\). The number \(L\) of
  states in the \(\epsilon\)-net is monotonically increasing in
  \(d\) and \(1/\epsilon\).
\item \textbf{New adversary strategy:} For each \(x,b\), we replace
  the states \(\psi_{x,b,y,c}\) by the states
  \(B_{x,b}\phi_{y,c}/|B_{x,b}\phi_{y,c}|\) to obtain a new, albeit
  artificial strategy for the adversaries.  With this replacement
  the new probability distribution \(\mu'\) of the trial record and
  \(x,b,y,c\) is within total-variation distance \(2\epsilon\) of
  the original probability distribution \(\mu\).
\item \textbf{Limited decrease of test-factor expectation:} We show
  that the decrease in the expectation of the test factor for the
  new strategy compared to the original one is bounded by a linear
  multiple of \(\epsilon\). By choosing \(\epsilon\) small enough,
  we can ensure that the test factor expectation for the new
  strategy still exceeds \(1\) by a positive multiple of
  \(\epsilon\).
\item \textbf{Guessing probability and communication scenario:} For
  each \(x,b\) and each relevant \(\phi\in\cN\), we reduce the new
  strategy to a three-party quantum non-signaling scenario
  corresponding to the events with \(\phi_{y,c}=\phi\).  We show that
  if, conditional on \(x,b\) and \(\phi\), the test factor expectation
  exceeds \(1\) by a given amount \(\delta\), then the settings
  distribution for \(\cB\) and \(\cC\) in the non-signaling scenario
  must be biased away from \(1/2\), making it possible for \(\cB\) to
  guess \(f(x,y)\) knowing only \(x,b\) and \(\phi\).  The guessing
  probability then also applies to the original strategy's state and
  probability distribution, with a controlled decrease due to the
  small total-variation distance previously established.  We find that
  by averaging over all \(x,b\) and \(\phi\in\cN\), adversary \(\cB\)
  can guess \(f(x,y)\) with guessing probability at least
  \(1/2+\epsilon\), being assisted only by initial shared entanglement
  with \(\cC\) and a message from an \(L\)-letter alphabet from
  \(\cC\). The alphabet consists of the members of \(\cN\) encoded
    classically, and the message replaces the original message to
  \(\cB\) from \(\cC\). 
\item \textbf{Communication-complexity bound:} We identifying the
  strategy for \(\cB\) to guess \(f(x,y)\) with a simultaneous
  message passing protocol for \(f(x,y)\) with non-trivial guessing
  probability. We show that the guessing probability for this
  protocol can be bounded in terms of \(L\), \(l\) and the maximum
  entry of the binary Fourier transform of the communication matrix
  of \(f(x,y)\). For semi-bent functions \(f(x,y)=g_{0}(x\oplus y)\),
  the maximum entry is small, leading to a favorable bound.
\item \textbf{Test-factor modification:} From the bounds obtained,
  we determine explicit factors \(\kappa\) depending on \(l\)
  satisfying that \(W'=\kappa W\) is a test factor against
  adversaries with \(d=O(l^{1/2})\).  Furthermore, we show how to
  modify the strategy to protect against adversaries whose average
  over the trials of the logarithm of the entanglement dimension
  is bounded by a bound that can be made arbitrarily large by increasing \(l\).
\item \textbf{Gate-complexity bound:} Finally, we bypass the
  \(\epsilon\)-net construction to show that if adversary \(\cB\) uses
  fewer than \(q\) qubits and \(G=O(l/(\log_{2}(q)+1)\) gates, then
  the adversaries pass the protocol with probability at most
  the soundness parameter. 
\end{enuma}

\vspace*{\baselineskip}
\noindent\textbf{Construction of the \(\epsilon\)-net:}
Consider \(\epsilon>0\). 
We make use of an \(\epsilon\)-net \(\cN\) on Hermitian matrices of
dimension \(d\times d\). The \(\epsilon\)-net serves to
  approximate the Hermitian matrices \(\psi C^{T}_{y,c}\). The real dimension of the space of
such matrices is \(d^{2}\). The space of Hermitian matrices is a real
inner product space under the trace inner product that induces the
Hilbert-Schmidt norm. We therefore identify this space with
\(\rls^{d^{2}}\). The distance between two \(d\times d\) Hermitian
  matrices \(\phi\) and \(\phi'\) may be expressed explicitly as
  \(|\phi-\phi'|=\sqrt{\tr((\phi-\phi')^\dagger(\phi-\phi'))}\) and
  can be written as a standard Euclidean distance in the basis
  constructed from the Gell-Mann matrices. There exists a set
\(\cN\) consisting of \(L\leq(1+2/\epsilon)^{d^{2}}\) Hermitian
matrices \(\phi\) with \(|\phi|=1\) such that for every \(y,c\) with
\(|\psi C_{y,c}^{T}|>0\) there is a member \(\phi\) of \(\cN\) within
distance \(\epsilon\) of the normalized state
\(\psi C_{y,c}^{T}/|\psi C_{y,c}^{T}|\). This follows from
Ref.~\cite{vershynin2020high}, Proposition 4.2.10.  This proposition
states that for every subset \(K\) of \(\rls^{d^{2}}\), there is an
\(\epsilon\)-net \(\cN\) for \(K\) of size
\(\textrm{vol}(K+(\epsilon/2)B_{d^{2}})/\textrm{vol}((\epsilon/2)B_{d^{2}})\). The
\(\epsilon\)-net is a set of points in \(K\) such that every member of
\(K\) is within \(\epsilon\) of one of these points.  Here,
\(\textrm{vol}(\cdot)\) denotes volume, and \(B_{d^2}\) is the unit
ball in \(\rls^{d^2}\). For our application, we let \(K\) be the unit
sphere in \(\rls^{d^{2}}\), and observe that
\(K+(\epsilon/2)B_{d^{2}}\) is contained in the ball of radius
\(1+\epsilon/2\). The ratio of volumes in the proposition is then the
ratio of the volume of a ball of radius \(1+\epsilon/2\) to the volume
of a ball of radius \(\epsilon/2\).  Since the normalized states
\(\psi C_{y,c}^{T}/|\psi C_{y,c}^{T}|\) are in the unit sphere,
\(\cN\) satisfies the claimed properties.

\vspace*{\baselineskip}
\noindent\textbf{New adversary strategy:}
Next, we reduce the space consisting of the input \(y\) and the
instrument outcomes \(c\) for \(\cC\) to \(L\) possible outcomes
without significantly decreasing the verifiers' test factor
expectation after the second phase of the trial. 
For each \(y,c\) such that \(|\psi C_{y,c}^{T}|>0\), choose
\(\phi_{y,c}\in\cN\) such that
\(\pqty|{\phi_{y,c}-\psi C_{y,c}^{T}/|\psi C_{y,c}^{T}|}|\leq \epsilon\). If \(|\psi C_{y,c}^{T}|=0\), we
set \(\phi_{y,c}=0\).

We consider \(\phi_{y,c}\in\cN\) both as a state and as a symbol from
an alphabet of \(L\) letters. The cases where \(\phi_{y,c}=0\) have
zero probability of occurring and therefore do not contribute.  Let
\(\mu(b,c|x,y)=|B_{x,b}\psi C_{y,c}^{T}|^{2}\) be the probability of
outcomes \(b,c\) conditional on \(x\) and \(y\).  When
\(\mu(b,c|x,y)>0\), the resulting normalized state is
\(\psi_{x,b,y,c}=B_{x,b}\psi C_{y,c}^{T}/\sqrt{\mu(b,c|x,y)}\).  We
set \(\psi_{x,b,y,c}=0\) if \(\mu(b,c|x,y)=0\).  We modify the first
phase of the trial by artificially replacing \(\psi_{x,b,y,c}\) with
\begin{align}
  \phi_{x,b,y,c}=B_{x,b}\phi_{y,c}/|B_{x,b}\phi_{y,c}|,
\end{align}
where we set \(\phi_{x,b,y,c}=0\) if \(\mu(b,c|x,y)=0\).
It is not necessary for this replacement to be physically realizable, but it enables the
reduction to a communication-complexity problem.

We claim that for each pair of \(x\) and \(y\), the expectation over
\(b,c\) of \(|\phi_{x,b,y,c}-\psi_{x,b,y,c}|^{2}\) is bounded by
\(4\epsilon^{2}\). The claim is established by the following
sequence of inequalities.
{\allowdisplaybreaks
  \begin{align}
    \sum_{b,c}\mu(b,c|x,y)|\phi_{x,b,y,c}-\psi_{x,b,y,c}|^{2}\hspace*{-2in}
    &
      \notag\\
    &=\sum_{b,c}\pqty|
      \frac{|B_{x,b}\psi C_{y,c}^{T}|}{|B_{x,b}\phi_{y,c}|}B_{x,b}\phi_{y,c}
      - B_{x,b}\psi C_{y,c}^{T}
      |^{2}
      \notag\\
    &=\sum_{b,c}\pqty|{
      \frac{|B_{x,b}\psi C_{y,c}^{T}|-|\psi C_{y,c}^{T}||B_{x,b}\phi_{y,c}|}{|B_{x,b}\phi_{y,c}|}B_{x,b}\phi_{y,c}
      + |\psi C_{y,c}^{T}|B_{x,b}\phi_{y,c}
      - B_{x,b}\psi C_{y,c}^{T}
      }|^{2}
      \notag\\
    &\leq
      \sum_{b,c}2\pqty|
      \frac{|B_{x,b}\psi C_{y,c}^{T}|-|\psi C_{y,c}^{T}||B_{x,b}\phi_{y,c}|}{|B_{x,b}\phi_{y,c}|}B_{x,b}\phi_{y,c}|^{2}
      +2\pqty|{
      |\psi C_{y,c}^{T}|B_{x,b}\phi_{y,c}
      - B_{x,b}\psi C_{y,c}^{T}
      }|^{2}
      \label{eq:ent_rep1_1}\\
    &=      2\sum_{b,c}\pqty|
      {|B_{x,b}\psi C_{y,c}^{T}|-|\psi C_{y,c}^{T}||B_{x,b}\phi_{y,c}|}
      |^{2}
      +2\sum_{b,c}\pqty|{
      |\psi C_{y,c}^{T}|B_{x,b}\phi_{y,c}
      - B_{x,b}\psi C_{y,c}^{T}
      }|^{2}
      \notag\\
    &\leq 
      4\sum_{b,c}\pqty|{
      |\psi C_{y,c}^{T}|B_{x,b}\phi_{y,c}
      - B_{x,b}\psi C_{y,c}^{T}
      }|^{2}
      \label{eq:ent_rep1_2}\\
    &=
      4\sum_{b,c}
      \tr(B_{x,b}\pqty(|\psi C_{y,c}^{T}|\phi_{y,c}-\psi C_{y,c}^{T})
      \pqty(|\psi C_{y,c}^{T}|\phi_{y,c}-\psi C_{y,c}^{T})^{\dagger}B_{x,b}^{\dagger})
      \notag\\
    &=
      4\sum_{b,c}
      \tr(B_{x,b}^{\dagger}B_{x,b}\pqty(|\psi C_{y,c}^{T}|\phi_{y,c}-\psi C_{y,c}^{T})
      \pqty(|\psi C_{y,c}^{T}|\phi_{y,c}-\psi C_{y,c}^{T})^{\dagger})
      \notag\\
    &=
      4\sum_{c}
      \tr(\pqty(|\psi C_{y,c}^{T}|\phi_{y,c}-\psi C_{y,c}^{T})
      \pqty(|\psi C_{y,c}^{T}|\phi_{y,c}-\psi C_{y,c}^{T})^{\dagger})
      \notag\\
    &=
      4\sum_{c}
      |\psi C_{y,c}^{T}|^{2}\pqty|{\phi_{y,c}-\psi C_{y,c}^{T}/|\psi C_{y,c}^{T}|}|^{2}
      \notag\\
    &\leq 4\epsilon^{2}\sum_{c}\tr(\psi (C_{y,c}^{\dagger} C_{y,c})^{T}\psi^{\dagger})
      \notag\\
    &= 4\epsilon^{2}.
      \label{eq:ent_rep1}
  \end{align}
}
To obtain line~\eqref{eq:ent_rep1_1}, we used
\(|u+v|^{2}\leq 2|u|^{2}+2|v|^{2}\).  For line~\eqref{eq:ent_rep1_2}
we applied the triangle inequality in the form
\(||u|-|v||\leq|u-v|\) to the first summand's terms.  \Ac{Instead of
  using \(\epsilon\)-nets, one can pick \(L\) states \(\phi\)
  uniformly randomly from the unit sphere and estimate the
  expectation of the third-to-last line in Eq.~\eqref{eq:ent_rep1}
  with respect to the closest of the \(L\) states for each term.
  When this is done, to obtain an expectation of at most
  \(4\epsilon^{2}\) requires a significantly smaller value of \(L\)
  compared to the \(\epsilon\)-net. However, the impact on the
  minimum \(l\) required against adversaries with entanglement
  dimension \(d\) is not substantial.  The advantage of using random
  states is that it can be adapted to the spectrum of the reduced
  density matrix \(\rho\) of the entanglement at \(\cC\).  To do so,
  one can modify \(\phi\) to
  \(\phi'=\rho^{1/4}\phi\rho^{1/4}/|\rho^{1/4}\phi\rho^{1/4}|\) and
  use \(\phi'\) instead of \(\phi\). With \(\phi\) in this
  expression still uniformly random, the distribution of \(\phi'\)
  is an angular central Gaussian distribution. The problem then can
  be reduced to a problem of order statistics of a difference
  between a \(\chi^{2}\) random variable and a weighted positive sum
  of independent \(\chi^{2}\) random variables. It may be challenging
  to obtain suitable analytic lower bounds on the order statistic in
  order to connect \(L\) to an entanglement measure.}

Next we claim that the probability distribution of the trial record
according to the adversary strategy changes by at most \(2\epsilon\)
in total-variation distance after we make the state replacement of the
previous paragraph.  For the purpose of this proof, we change the
notation for verifier measurement settings and measurement outcomes
used in other sections. Let \(z\) be the verifier measurement
setting. The probability distribution of \(z,x,b,y,c\) is not changed
by the replacement. Let \(m_{v},m_{b},m_{c}\) be the measurement
outcomes obtained by \(\cV,\cB,\cC\) in the second phase of the trial,
where the same measurement POVMs are used before and after the
replacements. Let \(\mu, \mu'\) the distributions of
\(z,x,b,y,c,m_{v},m_{b},m_{c}\) before and after the replacement.  By
construction, we have \(\mu'(x,b,y,c)=\mu(x,b,y,c)\).  Since
total-variation distance satisfies the data-processing inequality, it
suffices to bound the total-variation distance of the trial record by
the total-variation distance between \(\mu\) and \(\mu'\). We have
\begin{align}
  2\textrm{TV}(\mu,\mu') &=
                           \sum_{z,x,b,y,c,m_{v},m_{b},m_{c}}
                           |\mu(z,x,b,y,c,m_{v},m_{b},m_{c})-\mu'(z,x,b,y,c,m_{v},m_{b},m_{c})|
                           \notag\\
                         &=
                           \sum_{z,x,b,y,c}\mu(z,x,b,y,c)
                           \notag\\
                         &\hphantom{\sum_{z,x,b,y,c}}\times
                           \sum_{m_{v},m_{b},m_{c}}|\mu(m_{v},m_{b},m_{c}|z,x,b,y,c)-\mu'(m_{v},m_{b},m_{c}|z,x,b,y,c)|.
                           \label{eq:tv_mu_mu'}
\end{align}
Consider a general POVM with POVM
operators \(P_{i}\) for \(i\) in an index set \(I\) and normalized
states \(\ket{\sigma}\) and \(\ket{\sigma'}\). Write
\(\ket{\Delta}=\ket{\sigma}-\ket{\sigma'}\).  We can estimate
\begin{align}
  \sum_{i}|\bra{\sigma}P_{i}\ket{\sigma}-\bra{\sigma'}P_{i}\ket{\sigma'}|
  &=\sum_{i}|(\bra{\sigma}-\bra{\sigma'})P_{i}\ket{\sigma}
    -\bra{\sigma'}P_{i}(\ket{\sigma'}-\ket{\sigma})|
    \notag\\
  &\leq \sum_{i}(|\bra{\Delta}P_{i}\ket{\sigma}| + |\bra{\sigma'}P_{i}\ket{\Delta}|)
    \notag\\
  &\leq 2|\ket{\Delta}|,
    \label{eq:genpovmdiff}
\end{align}
where, for the last line we applied the following inequality
twice: For every normalized state \(\ket{\rho}\),
\begin{align}
  \sum_{i}|\bra{\Delta}P_{i}\ket{\rho}|
  &=    \sum_{i}|\bra{\Delta}P_{i}^{1/2}P_{i}^{1/2}\ket{\rho}|
    \notag\\
  &\leq \sum_{i}\bra{\Delta}P_{i}\ket{\Delta}^{1/2}\bra{\rho}P_{i}\ket{\rho}^{1/2}
    \notag\\
  & \leq \pqty(\sum_{i}\bra{\Delta}P_{i}\ket{\Delta})^{1/2}
    \pqty(\sum_{k}\bra{\rho}P_{k}\ket{\rho})^{1/2}
    \notag\\
  &= \braket{\Delta}^{1/2}\braket{\rho}^{1/2}= |\ket{\Delta}|.
\end{align}
Here we applied the Cauchy-Schwarz inequality twice, first to the
inner product of states in the first line, and then to the sum in
the second line after recognizing the sum as an inner product.
Returning to Eq.~\eqref{eq:tv_mu_mu'} and substituting the
inequality of Eq.~\eqref{eq:genpovmdiff} we obtain
\begin{align}
  2\textrm{TV}(\mu,\mu')
  &\leq
    2\sum_{z,x,b,y,c}\mu(z,x,b,y,c)
    |\phi_{x,b,y,c}-\psi_{x,b,y,c}|
    \notag\\
  &     = 2\sum_{x,b,y,c}\mu(x,b,y,c)|\phi_{x,b,y,c}-\psi_{x,b,y,c}|
    \notag\\
  &
    \leq 2\pqty(\sum_{x,b,y,c}\mu(x,b,y,c))^{1/2}
    \pqty(\sum_{x,b,y,c}\mu(x,b,y,c)|\phi_{x,b,y,c}-\psi_{x,b,y,c}|^{2})^{1/2}
    \notag\\
  &\leq 4\epsilon,
\end{align}
where we used the Cauchy-Schwarz inequality to obtain the third line and Eq.~\eqref{eq:ent_rep1} to obtain the last line.

\vspace*{\baselineskip}
\noindent\textbf{Limited decrease of test-factor expectation:}  We can now obtain a bound on how much the expectation of the test
factor \(W\) can be reduced by the above artificial state
replacement.  The test factor \(W\) depends only on \(z\),
\(f(x,y)\) and \(m_{v},m_{b},m_{c}\).
Let \(W_{\max}(z)=\max_{f,m_{v},m_{b},m_{c}}W(z,f,m_{v},m_{b},m_{c})\)
and \(W_{\min}(z)=\min_{f,m_{v},m_{b},m_{c}}W(z,f,m_{v},m_{b},m_{c})\).
Let \(\Delta_{\max}=\max_{z}(W_{\max}(z)-W_{\min}(z))\).
Since
\(\mu(z)=\mu'(z)\), and the total-variation distance between \(\mu\)
and \(\mu'\) is at most \(2\epsilon\), it follows that
\begin{align}
  \Exp_{\mu'}(W) &\geq \Exp_{\mu}(W)-2\epsilon\Delta_{\max}.
                   \label{eq:exptfbound}
\end{align}
We provide the explicit calculation to establish this inequality. In
general, let \(w(z,r)\) be a real-valued function of variables \(z\)
and \(r\). Let \(w_{\max}(z)=\max_{r}w(z,r)\) and
\(w_{\min}(z)=\min_{r}w(z,r)\). Let
\(\delta_{\max}=\max_{z}(w_{\max}(z)-w_{\min}(z))\). Let \(\nu\) and
\(\nu'\) be probability distributions such that \(\nu(z)=\nu'(z)\)
and \(\TV(\nu,\nu')\leq\epsilon'\).  Because the distribution of \(z\)
is the same for \(\nu\) and \(\nu'\), we have
\(\TV(\nu,\nu')=\Exp_{\nu}(\TV(\nu(\cdot|z),\nu'(\cdot|z)))\).  Let
\(R_{z}=\{r:\nu'(r|z)\geq \nu(r|z) \} \) and \(\bar R_{z}\) the
complement of \(R_{z}\) in the domain for the variable \(r\).  Then
\begin{align}
  \Exp_{\nu'}(w) - \Exp_{\nu} (w)
  &= \sum_{z} \nu(z) \sum_{r}(\nu'(r|z) - \nu(r|z))w(z,r)
    \notag\\
  &= \sum_{z}\nu(z) \pqty(\sum_{r\in R_{z}}(\nu'(r|z)-\nu(r|z))w(z,r)
    -\sum_{r\in\bar R_{z}}(\nu(r|z)-\nu'(r|z)) w(z,r))
    \notag\\
  &\geq\sum_{z}\nu(z) \pqty(\sum_{r\in R_{z}}(\nu'(r|z)-\nu(r|z))w_{\min}(z)
    -\sum_{r\in\bar R_{z}}(\nu(r|z)-\nu'(r|z)) w_{\max}(z))
    \notag\\
  &=\sum_{z}\nu(z) \pqty(\TV(\nu(\cdot|z),\nu'(\cdot|z))w_{\min}(z)
    -\sum_{r\in\bar R_{z}}\TV(\nu(\cdot|z),\nu'(\cdot|z)) w_{\max}(z))
    \notag\\
  &= -\sum_{z}\nu(z) \TV(\nu(\cdot|z),\nu'(\cdot|z)) (w_{\max}(z)-w_{\min}(z))
    \notag\\
  &\geq
    -\sum_{z}\nu(z) \TV(\nu(\cdot|z),\nu'(\cdot|z)) \delta_{\max}
    \notag\\
  &= -\TV(\nu,\nu')\delta_{\max} \notag \\
  & \geq -\epsilon'\delta_{\max}.
\end{align}

\vspace*{\baselineskip}
\noindent\textbf{Guessing probability and communication scenario:}
Let \(\bar\Delta=\sum_{z}\mu(z)(W_{\max}(z)-W_{\min}(z))\) be the
average over \(z\) of \((W_{\max}(z)-W_{\min}(z))\).  Suppose that
\(\Exp_{\mu}(W) \geq 1+\epsilon(3\bar\Delta+2\Delta_{\max})\).  Then,
in view of Eq.~\eqref{eq:exptfbound},
\(\Exp_{\mu'}(W) \geq 1+3\epsilon\bar\Delta\). This expectation can be
expressed as the average over \(\phi\in\cN\) and \(x,b\) of the
expectation of \(W\) with respect to \(\mu'\) conditional on
\(\phi_{y,c}=\phi\) and \(x,b\).  Consider a fixed \(\phi\in\cN\) and
suppose that conditional on \(\phi_{y,c}=\phi\) and \(x,b\), the
expectation with respect to \(\mu'\) of \(W\) is
\(1+\delta(\phi,x,b)\). Fix \(\phi,x,b\) and consider the case
\(\delta(\phi,x,b)\geq 0\). Conditional on \(\phi_{y,c}=\phi\) and
\(x,b\), the situation in the modified second phase of the trial is
that \(\cV,\cB,\cC\) share a fixed quantum state
\(\phi_{x,b,y,c}=B_{x,b}\phi /|B_{x,b}\phi|\).  This is a three-party
scenario with a fixed shared quantum state, and measurements are made
based on additionally provided classical inputs \(z,y,c\) with a
conditional probability distribution that is the same as the original
conditional distribution \(\mu(z,y,c|\phi_{y,c}=\phi,x,b)\).
\(\cV\)'s quantum measurement device receives \(z\) and \(\cB\) and
\(\cC\) both get \(y,c\), where in this scenario, \(\cV,\cB,\cC\) may
be assumed to know the given values \(\phi,x,b\) before receiving the
additional input \(z\) and \(y,c\), respectively.  The test factor
\(W\) depends only on \(z\), \(f(x,y)\) and the measurement outcomes.
We claim that with the addition of shared classical randomness with a
specific distribution for \(\cB\) and \(\cC\), this scenario is
quantum non-signaling with respect to the setting choice \(z\) at
\(\cV\) and identical settings choices \(f=f(x,y)\) for \(\cB\) and
\(\cC\).  Here we take \(\{0,1\}\) as the set of values for
\(f=f(x,y)\).  The value of \(z\) received by \(\cV\) is uniform and
independent of \(y,c\), and \(\cV\) applies a \(z\)-conditional POVM
producing output \(m_{v}\), where the POVM does not depend on
\(y,c\). Consequently there is no signaling from \(\cV\) to \(\cB\cC\)
in the sense that the probability distributions of \(m_{b},m_{c}\) and
\(y,c\) observed by \(\cB\cC\) do not depend on the value of \(z\).
Similarly, the probability distribution of \(z\) and \(m_{v}\) do not
depend on the value of \(y,c\).  To obtain the correct non-signaling
conditions for \(\cB\) and \(\cC\) separately, we express \(f(x,y)\)
in terms of additional shared randomness that does not depend on the
value of \(f(x,y)\). The shared randomness consists of auxilliary
variables \(y_{0},c_{0}, y_{1},c_{1}\) that are independent of \(y,c\)
and all other pre-existing variables, and shared between \(\cB\) and
\(\cC\). The probability of \(y_{0},c_{0}\) is defined as the
probability of \(y=y_{0},c=c_{0}\) conditional on \(f(x,y)=0\), which
is given by \(\mu(y=y_{0},c=c_{0}|f(x,y)=0,\phi_{y,c}=\phi,x,b)\). The
variables \(y_{1},c_{1}\) are independent of \(y_{0},c_{0}\) and have
the distribution of \(y=y_{1},c=c_{1}\) conditional on \(f(x,y)=1\).
The auxiliary variables can be added to the quantum state \(\phi\) in
the form of an extra, independent mixed state \(\sigma\) shared
between \(\cB\) and \(\cC\) that is perfectly correlated between
\(\cB\) and \(\cC\).  \(\cB\) and \(\cC\) on input \(f=f(x,y)\) can
simulate the original scenario by each of them assigning
\(y=y_{f},c=c_{f}\) and using these values as before. This can be
interpreted as \(\cB\) and \(\cC\) each locally applying an
\(f\)-conditional POVM to the extended shared quantum state consisting
of \(\phi\) and \(\sigma\), producing outcomes \(m_{b}\) and
\(m_{c}\).  The scenario is now defined consistently also for
different inputs \(f\) and \(f'\) for \(\cB\) and \(\cC\).  If the
joint distribution of \(z\) and \(f=f(x,y)\) is uniform conditional on
\(\phi_{y,c}=\phi\) and \(x,b\), the conditional expectation of \(W\)
is at most \(1\) as \(W\) is a test factor for such non-signaling
scenarios with uniform settings \(z\) and \(f\). Since, in this
section, we are assuming that verifier measurement settings and
challenges are uniform and independent, the distribution of \(z\)
alone is uniform and independent of \(f(x,y)\), even conditional on
\(\phi_{y,c}=\phi\) and \(x,b\). That is
\(\mu'(z|f(x,y)=f,\phi_{y,c}=\phi,x,b)=\mu'(z)=\mu(z)\).  We can
therefore obtain a lower bound on the guessing probability of the
value of \(f=f(x,y)\) by \(\cB\) given only the symbol \(\phi\) in
addition to \(x,b\) and the conditional expectation of \(W\).  To
obtain the lower bound, we instead show that if the guessing
probability for \(f\) is less than \(1/2+\epsilon'\), then the
expectation of \(W\) can exceed \(1\) by at most
\(\epsilon'\bar\Delta\).  Suppose that the guessing probability is
less than \(1/2+\epsilon'\), that is, for each \(f\in\{0,1\}\), we
have \(\mu'(f|\phi_{y,c}=\phi,x,b)<1/2+\epsilon'\).  Let \(\nu\) be
the distribution \(\mu'\) conditioned on \(\phi_{y,c}=\phi,x,b\) that
governs the constructed non-signaling scenario.  Since
\(W_{\min}(z)\leq \Exp_{\nu}(W|z,f)\leq W_{\max}(z)\) and by
independence of \(z\), we obtain
\begin{align}
  \Exp_{\nu}(W|z)
  &= \sum_{f}\nu(f|z) \Exp_{\nu}(W|z,f)
    \notag\\
  &=
    \sum_{f}\nu(f) \Exp_{\nu}(W|z,f)
    \notag\\
  &= \sum_{f}\frac{1}{2} \Exp_{\nu}(W|z,f)
    +\sum_{f}\pqty(\nu(f)-\frac{1}{2})
    \Exp_{\nu}(W|z,f)
    \notag\\
  &<\sum_{f}\frac{1}{2} \Exp_{\nu}(W|z,f)
    + \epsilon'(W_{\max}(z)-W_{\min}(z)).
    \label{eq:exp(w|z)}
\end{align}
The average over \(z\) of the first summand on the last line is bounded by \(1\)
since it is the average of \(W\) with respect to the uniform
settings distribution.  Averaging Eq.~\eqref{eq:exp(w|z)}
over \(z\) gives
\begin{align}\label{eq:exp(w)_ub}
  \Exp_{\nu}(W) &< 1+\epsilon'\bar\Delta.  
\end{align}
It follows that if \(\delta(\phi,x,b)\geq 0\) and
  \(\Exp_{\nu}(W)=1+\delta(\phi,x,b)\), then there is a value of
\(f\) for which
\(\nu(f)=\mu'(f=f(x,y)|\phi_{y,c},x,b) \geq
1/2+\delta(\phi,x,b)/\bar\Delta\), that is, the guessing probability
for \(f\) is at least \(1/2+\delta(\phi,x,b)/\bar \Delta\).  Let
\(\bar\delta\) be the average over \(\phi\) and \(x,b\) of
\(\delta(\phi,x,b)\).  Since the guessing probability is always at
least \(1/2\), the average of the guessing probability over
\(\phi,x,b\) with distribution \(\mu'\) is lower bounded by
\begin{align}
  \Exp_{\mu'}(1/2+\max(\delta(\phi,x,b)/\bar \Delta, 0))
  &\geq \Exp_{\mu'}(1/2+\delta(\phi,x,b)/\bar \Delta)
    \notag\\
  &=\Exp_{\mu'}(1/2+(W-1)/\bar\Delta)
    \notag\\
  &\geq 1/2 + 3\epsilon,
\end{align}
where we applied \(\Exp_{\mu'}(W)\geq 1+ 3\epsilon\bar\Delta\).
Because the total-variation distance between \(\mu\) and \(\mu'\) is
bounded by \(2\epsilon\), the average guessing probability for \(f\)
for \(\cB\) when the variables have their original distribution
\(\mu\) and \(\cB\) learns only \(\phi=\phi_{y,c}\) from \(\cC\) is
reduced by at most \(2\epsilon\) to at least \(1/2+\epsilon\). This
holds because the guessing probability is the expectation of the
binary random variable describing whether the guess is correct or not,
and the total-variation distance is the maximum change in expectation
of such binary random variables. 

\vspace*{\baselineskip}
\noindent\textbf{Communication-complexity bound:}
We have shown that if the adversarial expectation of \(W\)
exceeds \(1\) by at least \(\epsilon(3\bar\Delta+2\Delta_{\max})\) and
the adversary \(\cC\), instead of communicating \(y,c\) to
\(\cB\), sends only \(\phi=\phi_{y,c}\) from the \(L\) members of the
\(\epsilon\)-net \(\cN\), then \(\cB\), knowing \(x,b\), can guess the
value of \(f=f(x,y)\) with an average probability of at least
\(1/2+\epsilon\). Thus \(\epsilon\) is the guessing advantage of
\(\cB\) for \(f\) in this modified adversary protocol. To make the
guess, \(\cB\) takes advantage of complete knowledge of the shared
entangled state and the quantum instruments used to compute the
necessary conditional probabilities according to \(\mu\).  This
adversary protocol is equivalent to an entanglement-assisted protocol
in the simultaneous message passing (SMP) model, where on uniformly
random inputs \(x\) and \(y\), \(\cB\) and \(\cC\) send messages to a
third party \(\cD\), sharing with each other ahead of time only a
pure quantum state on a bipartite Hilbert space of total
dimension \(d\times d\). \(\cB\)'s message, obtained before
transmitting the subsystem destined for \(\cV\), is from an alphabet
of arbitrary size and consists of \(x,b\), and \(\cC\)'s message is
from an alphabet of size \(L\) and consists of
\(\phi=\phi_{y,c}\in\cN\).   \(\cD\) uses the two messages to
guess the value of \(f(x,y)\), succeeding with probability at least
\(1/2+\epsilon\).  The dimension of the entangled state in this SMP
protocol is \(d\times d\) because \(\psi\)'s reduced density matrix on
\(C\) has dimension \(d\).  We consider functions \(f(x,y)\) of the
form \(g(x\oplus y)\) with \(x\) and \(y\) bit strings of length
\(l\). For such functions, we show below how to determine lower bounds
on \(L\) for such protocols. The conventional communication complexity
for such protocols is \(\log_{2}(L)\).  The best bounds for functions
satisfying our assumptions on the settings function are obtained with
balanced, semi-bent
functions~\cite{rothaus1976bent,carlet2010boolean,chee1994semi}.
Balance ensures that the settings function satisfies the assumptions
stated in Sect.~\ref{subsect:protocol_assumptions} when the verifier setting \(z\) and challenges \(x,y\) are
uniform and independent.  A simple example of an efficiently
implementable, balanced, semi-bent function for \(l=2k+1\) is the
exclusive OR of one bit with the split inner product on \(2k\)
additional bits defined as
\(g_{0}(u)=u_{0}+ \sum_{j=1}^{k} u_{j}u_{j+k} \mod 2\). 

The analysis of the lower bound on \(L\) given \(\epsilon\) and \(d\)
for the SMP protocol described in the previous paragraph consists of
three parts.  In the first part the problem is reduced to estimating a
trace of the sandwiched product of the communication matrix defined
below with a factorization of a guessing
matrix with entries given in Eq.~\eqref{eq:guessing matrix entries},
where the factors' Hilbert-Schmidt norm can be bounded.
In the second part the guessing probability is related to
the maximum Fourier coefficient of the communication matrix and the
Hilbert-Schmidt norms of the factors of the guessing matrix. In the
third part, we instantiate the communication matrix with a semi-bent
function of \(x\oplus y\) taking advantage of the small value of the
maximum Fourier coefficient of such functions' communication
matrix. The techniques used are from
Refs.~\cite{zhang2009communication,cleve2013quantum,montanaro2009communication}
and inspired by the factorization norm analyses of
Ref.~\cite{linial2007lower}.  \Ac{The argument used here is based on
  an AI generated outline of a proof for a simple SMP result with
  two-sided bounds on alphabet sizes. The AI and MK were not able to
  find a good match of the argument or the result in the cited papers,
  but it is fairly described as a synthesis of what can be found
  there.}

We switch to column vector notation for quantum states and
transition from guessing bits to guessing values in \(\{-1,1\}\).
The communication matrix \(M\) for the SMP protocol is defined by its entries
\(M_{x,y}=(-1)^{f(x,y)} \in\{-1,1\}\).  The steps of the
SMP protocol are as follows:  
\begin{enumerate}
\item \(\cB\) and \(\cC\) share a pure quantum state \(\psi\) in a
  bipartite Hilbert space \(\cH_{B}\otimes\cH_{C}\). To establish the
  general result below, the dimension of this Hilbert space is
  unconstrained.
\item \(\cB\) receives a uniformly random input \(x\) and \(\cC\) a uniformly random input \(y\). 
\item \(\cB\) applies POVM \((B_{x,i})_{i=1}^{L'}\) to \(\cH_{B}\) and
  sends outcome \(i\) to \(\cD\), while \(\cC\) applies POVM
  \((C_{y,j})_{j=1}^{L}\) to \(\cH_{C}\) and sends outcome \(j\) to
  \(\cD\). Here \(\cB\) can include classical information about \(x\)
  in the POVM labels so that it can be extracted from the outcomes,
  but \(\cC\) is limited to \(L\) outcomes, which restricts \(\cC\)'s
  ability to convey information about \(y\). Here we overloaded
    notation to denote the POVM operators, which are obtained from the
    instrument operators in the QPV protocol according to
    \(B_{x,i}=B_{x,b_{i}}^{\dagger}B_{x,b_{i}}\) for a labeling of the
    outcomes \(b\) for \(\cB\), and
    \(C_{y,j}=\sum_{c:\phi_{y,c}=\phi_{j}}{C_{y,c}}^{\dagger}C_{y,c}\)
    for some indexing \(\phi_{j}\) of the members of \(\cN\).
\item  \(\cD\) produces the guess
\(R_{ij}\in \{-1,1\}\) of \(M_{x,y}\). 
\end{enumerate}
The probability of outcome
\(i,j\) given \(x,y\) is
\begin{align}
  \nu(i,j|x,y)=\psi^{\dagger}B_{x,i}\otimes C_{y,j}\psi
  = ((B_{x,i}\otimes\Id) \psi)^{\dagger} ((\Id \otimes C_{y,j})\psi),
\end{align}
where we used \(B_{x,i}=B_{x,i}^{\dagger}\), since POVM elements are
Hermitian.  Let \(1/2+s_{x,y}\) be the
probability that \(\cD\) correctly guesses \(M_{x,y}\) on input
\(x,y\) and \(s=2^{-2l}\sum_{x,y}s_{x,y}\) its expectation over
\(x,y\).  We have
\begin{align}
  2s_{x,y} = \sum_{i,j} \nu(i,j|x,y)M_{x,y}R_{i,j}.
\end{align}
Write vectors in \(\cH_{B}\otimes\cH_{C}\) in a fixed orthogonormal basis
\((e_{k})_{k=1}^{d'}\), where \(d'\) is the dimension of
\(\cH_{B}\otimes\cH_{C}\). 
For \(\varphi\in\cH_{B}\otimes\cH_{C}\), \(\varphi_{k}\) is the \(k\)'th coefficient of this vector
in this basis. 
Let \(\bf{B}\) be the matrix of dimension
\(2^{l}\times (L'd')\) whose entries are
\begin{align}
  \bm{B}_{x,ik}=((B_{x,i}\otimes\Id)\psi)_{k}^{*},
\end{align}
where \(ik\) denotes the pair \((i,k)\) and indexes the columns with
\(i\in\{1,\ldots, L'\}\) and \(k\in\{1,\ldots, d'\}\).
Let \(\bm{C}\) be the matrix of dimension \(Ld'\times 2^{l}\) whose entries are
\begin{align}
  \bm{C}_{jk',y}=((\Id\otimes C_{y,j})\psi)_{k'}.
\end{align}
Intuitively, $\bm{B}$ is the
matrix of row vectors that catalogs the action of $\cB$'s
measurements on the shared state for every input $x$, while $\bm{C}$
is the matrix of column vectors
cataloging $\cC$'s measurements for every input
$y$. Their form enables the use of matrix multiplication to
  express the guessing advantages.  Let \(\bm{R}\)
be the \(L'd'\times Ld'\) matrix whose entries are
\begin{align}
  \bm{R}_{ik,jk'}= R_{ij}\delta_{k,k'}.
\end{align}
We have
\begin{align}
  \sum_{i,j}\nu(i,j|x,y) R_{i,j}
  &= \sum_{i,j} R_{i,j}\sum_{k}
    ((B_{x,i}\otimes\Id) \psi)_{k}^{*} ((\Id \otimes C_{y,j})\psi)_{k}
    \notag\\
  &=
    \sum_{ik,jk'} 
    ((B_{x,i}\otimes\Id) \psi)_{k}^{*} R_{i,j}\delta_{k,k'}((\Id \otimes C_{y,j})\psi)_{k'}
    \notag\\
  &= \pqty(\bm{B}\bm{R}\bm{C})_{x,y}.
    \label{eq:guessing matrix entries}
\end{align}
Therefore
\begin{align}
  2s=2^{-2l}\tr(M^{T}(\bm{B} \bm{R}\bm{C})) = 2^{-2l}
  \tr(\bm{C}M^{T} (\bm{B}\bm{R})),
  \label{eq:2s=}
\end{align}
where \(M^{T}=M^{\dagger}\) since \(M\) is real.  It follows that \(2|s|\) is upper-bounded by $2^{-2l}$
times the product of the Hilbert-Schmidt norms \(|\bm{C}|\) and
\(|\bm{B}\bm{R}|\) of \(\bm{C}\) and \(\bm{B}\bm{R}\) and the maximum
singular value \(\|M\|\) of \(M\). This is a consequence of the
Cauchy-Schwarz inequality for the Hilbert-Schmidt inner product, the
properties of Schatten \(p\)-norms, and the fact that the
Hilbert-Schmidt norm is the Schatten \(2\)-norm.  Schatten \(p\)-norms
are unitarily invariant, submultiplicative, and non-increasing under
multiplication by a matrix of maximum singular value less than \(1\). For example, see Ref.~\cite{bhatia2013matrix},
Chapter IV.  \Ac{The reference covers unitarily invariant norms in
  general and the results need to be instantiated by hand. See the
  Wikipedia entry on Schatten norm for a quick summary of relevant
  properties.}  By the Cauchy-Schwarz inequality,
  \(|\tr(\bm{C}M^{T} (\bm{B}\bm{R}))| \leq
  |\bm{C}||M^{T}(\bm{B}\bm{R})|\), and by the last property,
  \(|M^{T}(\bm{B}\bm{R})|\leq \|M\||\bm{B}\bm{R}|\). We have
\begin{align}
  |\bm{C}|^{2}
  &= \tr(\bm{C}^{\dagger}\bm{C})
    \notag\\
  &= \sum_{y=1}^{2^l} \sum_{j=1}^{L} \sum_{k'=1}^{d'} |\bm{C}_{jk',y}|^{2}
    \notag\\  
  &= \sum_{y=1}^{2^l} \sum_{j=1}^{L}
    \psi^{\dagger}(\Id\otimes C_{y,j}^{2})\psi
    \notag\\
  &\leq
    \sum_{y=1}^{2^l} \sum_{j=1}^{L}
    \psi^{\dagger}(\Id\otimes C_{y,j})\psi
    \notag\\
  &=
    \sum_{y=1}^{2^l} 
    \psi^{\dagger}(\Id\otimes \Id)\psi
    \notag\\
  &= 2^{l},
\end{align}
where we used the fact that the POVM element \(C_{y,j}\) is Hermitian and,
because \(0\leq C_{y,j}\leq \one\), satisfies \(C_{y,j}^{2}\leq C_{y,j}\).
For \(|\bm{B}\bm{R}|\) we get 
\begin{align}
  |\bm{B}\bm{R}|^{2}
  &= \tr((\bm{B}\bm{R})^{\dagger}\bm{B}\bm{R})
    \notag\\
  &= \sum_{x=1}^{2^l}\sum_{j=1}^{L}\sum_{k'=1}^{d'}\pqty|\sum_{i=1}^{L'}\sum_{k=1}^{d'}
    \bm{R}_{ik,jk'}((B_{x,i}\otimes \Id)\psi)_{k}^{*}|^{2}
    \notag\\
   &=\sum_{x=1}^{2^l}\sum_{j=1}^{L}\sum_{k'=1}^{d'}\pqty|\sum_{i=1}^{L'}
    R_{i,j}((B_{x,i}\otimes \Id)\psi)_{k'}^{*}|^{2}
    \notag\\  
  &=
    \sum_{x=1}^{2^l} \sum_{j=1}^{L}\sum_{k'=1}^{d'}
    \sum_{i=1}^{L'}\sum_{i'=1}^{L'}
    R_{i,j}R_{i',j}\pqty(\psi^{\dagger}(B_{x,i}\otimes\Id))_{k'}((B_{x,i'}\otimes\Id)\psi)_{k'}
    \notag\\
  &= \sum_{x=1}^{2^l}\sum_{j=1}^{L}
    \psi^{\dagger}\pqty(\sum_{i=1}^{L'}R_{ij}(B_{x,i}\otimes \Id))^{2}\psi
    \notag\\
  &=\sum_{x=1}^{2^l}\sum_{j=1}^{L} \pqty|\sum_{i=1}^{L'}R_{ij}(B_{x,i}\otimes\Id)\psi|^{2}, 
    \label{eq:BRchain}
\end{align}
where we used that the matrix \(\bm{A}=\bm{B}\bm{R}\) has entries \(\bm{A}_{x, jk'} = \sum_{i=1}^{L'} \sum_{k=1}^{d'} \bm{R}_{ik,jk'} ((B_{x,i}\otimes\Id)\psi)_{k}^{*}\) to obtain the second line.
Let \(V_{j}=\sum_{i=1}^{L'}R_{ij}B_{x,i}\).
Let \(I\) be the set of indeces \(i\) such that \(R_{ij}=1\) and let \(\bar I\) be
the complement of \(I\) in \(\{1,\ldots,L'\}\). Define
\(V_{j+}=\sum_{i\in I} R_{ij}B_{x,i}= \sum_{i\in I}B_{x,i}\) and
\(V_{j-}=-\sum_{i\in \bar I}R_{ij}B_{x,i}=\sum_{i\in \bar I}B_{x,i}\).
Then \(V_{j+}\) and \(V_{j-}\) are positive semidefinite,
\(V_{j+}-V_{j-}=V_{j}\) and \(V_{j+}+V_{j-}=\Id\). Therefore the maximum eigenvalue
of \(V_{j}\) is at most \(1\), and the minimum eigenvalue is at least \(-1\).
It follows that \(\sum_{i=1}^{L'}R_{ij}(B_{x,i}\otimes\Id) = V_{j}\otimes\Id\) has
maximum singular value at most \(1\). Continuing Eq.~\eqref{eq:BRchain} gives
\begin{align}
  |\bm{B}\bm{R}|^{2}
  &= \sum_{x=1}^{2^l}\sum_{j=1}^{L}|(V_{j}\otimes\Id)\psi|^{2}
    \notag\\
  &\leq
    \sum_{x=1}^{2^l}\sum_{j=1}^{L}1 = 2^{l}L.
\end{align}
Applying the bounds obtained to Eq.~\eqref{eq:2s=} gives  
\begin{align}
  2s \leq \|M\| 2^{-2l}\sqrt{2^{l}}\sqrt{2^{l}L } = \|M\| 2^{-l}\sqrt{L}.
\end{align}

For functions \(f(x,y)\) of the form \(f(x,y)=g(x\oplus y)\), the
matrix \(M\) can be diagonalized with the Fourier transform. For
odd \(l\), the function \(f(x,y)=g_{0}(x\oplus y)\) satisfies that its
communication matrix \(M_{g_{0}}\) has maximum absolute eigenvalue
\(2^{(l+1)/2}\), see Ref.~\cite{chee1994semi}.  Because \(M_{g_{0}}\) is symmetric,
the maximum absolute eigenvalue directly corresponds to its maximum
singular value, yielding \(\|M_{g_{0}}\| = 2^{(l+1)/2}\). \Ac{The
  Wikipedia entry on bent function defines semi-bent functions and
  summarizes the relevant properties.}  Therefore, for odd \(l\)
  and \(g_{0}\), 
\begin{align}\label{eq:bound_communicomplexity}
  2s \leq 2^{-(l-1)/2}\sqrt{L}, 
\end{align}
which demonstrates an exponential suppression of the guessing
advantage as a function of $l$, regardless of the state shared between
\(\cB\) and \(\cC\) in the SMP protocol.   Together with the
  previous analysis of the relationship between test-factor
  expectation and guessing advantage for adversary protocols in
  Bell-test QPV, this implies that given an upper bound on the dimension of prior
  entanglement shared by adversaries, it suffices to increase \(l\) to
  reduce the excess test-factor expectation achievable by
  adversaries. Once the excess is made small enough, the modification
  in the next step can be used to adjust the test factor to preserve
  protocol soundness.

\vspace*{\baselineskip}
\noindent\textbf{Test-factor modification:} Returning to the test-factor estimates, the \(\epsilon\)-net
  analysis of adversary strategies showed that if the test-factor
  expectation according to the original adversarial distribution
  \(\mu\) satisfies 
\(\Exp_{\mu}(W)\geq 1+\epsilon(3\bar{\Delta}+2\Delta_{\max})\), then
the adversary's guessing advantage \(s\) is at least
  \(\epsilon\). On the other hand, through the
communication-complexity analysis, we obtained the upper bound of
  Eq.~\eqref{eq:bound_communicomplexity} on \(s\) given \(l\) and
  \(L\).  Combining these bounds and substituting the size of the
\(\epsilon\)-net for \(L\), we find that the
following inequality must be satisfied:
\begin{align}
  \epsilon
  \leq 2^{-(l-3)/2}(1+2/\epsilon)^{d^{2}/2}.
  \label{eq:epsilon_n_d}
\end{align}
This imposes a lower bound on \(d\) given \(l\) and
  \(\epsilon\). For asymptotic bounds, we rewrite and simplify the
bound.  It is necessary to satisfy
\(\epsilon 2^{(l-3)/2} \leq (3/\epsilon)^{d^{2}/2}\). Taking
logarithms gives
\begin{align}
  (l-3)-2\log_{2}(1/\epsilon) \leq d^{2} \log_{2}(3/\epsilon) .
  \label{eq:entdim_bound}
\end{align}
In the next two paragraphs, we show that for a given test
factor and actual prover, \(\epsilon\) and the number of trials
required for a given soundness can be chosen to be constant
independent of \(l\) and \(d\). Therefore, \(d\) must be
\(\Omega(l^{1/2})\) for the inequality to be satisfied. We can
  interpret this as requiring that the adversaries share at least
  \(\log_{2}(l)/2-O(1)\) qubits to defeat soundness of the protocol.

Suppose we wish for soundness against adversaries with support of
dimension at most \(d\) for \(\cC\)'s density matrix of the prior
entanglement.  This requires choosing \(l\) large enough and adjusting
the test factor by multiplying it by a factor of \(\kappa<1\). Let
\(W\) be a test factor against unentangled adversaries with positive
gain \(\Exp(\log(W))=g\) for the actual prover.  The expectation of
\(W\) for this prover is therefore at least \(e^{g}\).  In the
derivation above, we assumed that the adversarial expectation
satisfies \(\Exp_{\mu}(W)\geq 1+\epsilon(3\bar\Delta+2\Delta_{\max})\)
to obtain Eq.~\eqref{eq:epsilon_n_d}, which determines a lower bound
\(d_{0}\) on \(d\) as a function of \(l\) and \(\epsilon\), where
  \(d_{0}\) is the smallest integer \(d\) for which
  Eq.~\eqref{eq:epsilon_n_d} is satisfied.  By inspection, 
\(d_{0}\) is monotonically increasing with both \(l\) and
\(\epsilon\). This indicates that a higher adversarial expectation
\(\Exp_{\mu}(W)\) requires a strictly larger prior-entanglement
  dimension \(d\). For a fixed expectation
\(\Exp_{\mu}(W)\), tolerating a higher dimension of prior entanglement
necessitates increasing the number of challenge bits
\(l\).  Let
\(\Delta_{5}=3\bar\Delta+2\Delta_{\max}\).  If we fix \(\epsilon\) and
multiply \(W\) by \(\kappa=1/(1+\epsilon\Delta_{5})\) to obtain
\(W'=\kappa W\), then \(W'\) is a test factor against adversaries with
pre-shared entangled systems having dimension \(d\leq
d_{0}\).  The value of \(\epsilon\) is constrained by the need for
the actual prover to still have positive gain for \(W'\), which is
  satisfied by ensuring \(e^g > (1+\epsilon\Delta_{5})\). We
therefore first choose \(\epsilon\) to ensure sufficient gain for
\(W'\), then, given a threshold 
  \(d_{0}\), choose \(l\) large enough to ensure soundness against
adversaries with prior-entanglement dimension \(d\leq d_{0}\). 

The gain of the actual prover for \(W'\) is
\(-\log(1+\epsilon\Delta_{5})+g\).  To ensure that the actual
prover retains a gain of  \(g/6\), we choose \(\epsilon\) so
that \(\log(1+\epsilon\Delta_{5})=5g/6\). It then suffices
to increase the number of trials by a factor of \(6\) for the same
soundness when using \(W'\). We now have 
\begin{align} \label{eq:epsilon_choice}
  \epsilon = (e^{5g/6}-1)/\Delta_{5}.
\end{align}
Given the chosen \(\epsilon\) and for any fixed dimension \(d\), there
exists a threshold \(l_{0}\) such that the inequality in
Eq.~\eqref{eq:epsilon_n_d} is violated for all \(l>l_{0}\). Therefore,
by choosing \(l\) large enough, adversaries with arbitrarily large
\(d\) and no limit on classical communication can be rejected with
high probability.

For example, with \(d=2\) and a typical test factor in our experiment,
the gain is \(g=\gexp\) and \(\Delta_{5}= \Deltafiveexp\).  This
determines \(\epsilon=\Epsilonexp\) according to
Eq.~\eqref{eq:epsilon_choice}.  The minimum odd \(l\) for which
Eq.~\eqref{eq:epsilon_n_d} is not satisfied with \(d=2\) is then
\(l=\lexptwo\). For \(d=3\) and \(d=4\) this requires \(l=\lexpthree\)
and \(l=\lexpfour\), respectively.  For a test factor suitable for an
extremely effective combination of verifiers and prover who can
achieve maximal CHSH violation with a Bell state, we found
\(g=0.032075\) and \(\Delta_{5}= \Deltafivechsh\). This requires
\(l=\lchshtwo\) for \(d=2\) and \(l=\lchshfour\) for \(d=4\).  \Ac{ A
  typical test factor in our experiment has \(g=\gexp\),
  \(\Delta_{\max}=\Deltamaxexp\), \(\bar\Delta=\Deltabarexp\).  A test
  factor for a protocol with maximal CHSH violation has \(g=\gchsh\),
  \(\Delta_{\max}=\Deltamaxchsh\), \(\bar\Delta=\Deltabarchsh\).  See
  \texttt{definitions.tex} for numbers and octave code.}

We can also use a version of the strategy used in
Sect.~\ref{sec:smallentanglement} to protect against adversaries with
bounded average prior entanglement over the trials, with entanglement
quantified as \(\log(d)\). For a proof of principle, we use the
following conservative version of Eq.~\eqref{eq:epsilon_n_d}:
\begin{align}
  \epsilon
  \leq 2^{-(l-3)/2}(1+2/\epsilon)^{d^{2}/2}d,
  \label{eq:epsilon_n_dw}
\end{align}
for integers \(d\geq 1\).  Extending the definition of the
  right-hand side to all real \(d\geq 0\), let \(d'_{\epsilon}\) be
the real solution for \(d\) satisfying equality in
Eq.~\eqref{eq:epsilon_n_dw}, where we keep \(l\) fixed but can vary
\(\epsilon>0\).  For \(\epsilon>0\), let
\(d_{\epsilon}=\max(2,d'_{\epsilon})\), and for \(\epsilon\leq 0\), we
set \(d_{\epsilon}=1\).  Then \(d_{\epsilon}\) is a lower bound on the
dimension required by the adversaries to achieve the assumed
expectation of \(W\). The maximum \(\max(2,d'_{\epsilon}) \) takes
into account that by the results for adversaries without prior
entanglement, the dimension \(d\) is at least \(2\) to achieve an
expectation \(\Exp(W)\geq 1+\epsilon\Delta_{5}\) with \(\epsilon>0\)
in the above derivation.  Suppose that at trial \(i\), conditional on
the past, \(\Exp(W_{k})= 1+\epsilon'_{k}\Delta_{5}\), where \(W_{k}\)
is the instantiation of \(W\) at the \(k\)'th trial. Let
\(\epsilon_{k}=\max(0,\epsilon'_{k})\).  If \(d_{k}\) is the
  entanglement dimension of the entanglement shared by the adversaries
  before the \(k\)'th trial,  then
\(d_{k}\geq d_{\epsilon_{k}}\).  Because the expectation of \(W\)
cannot exceed \(\bar W_{\max}=\Exp(W_{\max}(z))\), the relevant range
for \(\epsilon_{k}\) is between \(0\) and
\(\epsilon_{\max}=(\bar W_{\max}-1)/\Delta_{5}\).

To apply the method of Sect.~\ref{sec:smallentanglement}, we need a
lower bound of the form \(\log(d_{\epsilon}) \geq \lambda\epsilon\)
for some \(\lambda>0\). We have \(\log(d_{0})=0\). If we can
  determine \(\gamma>0\) such that
  \(d_{\epsilon}\geq 1+\gamma\epsilon\), then by concavity of the logarithm
\begin{align}
  \log(d_{\epsilon}) \geq \log(1+\gamma\epsilon)
  \geq \big(\log(1+\gamma\epsilon_{\max})/\epsilon_{\max}\big)\epsilon
\end{align}
for \(\epsilon\in [0,\epsilon_{\max}]\).  We can then set
\(\lambda=\log(1+\gamma\epsilon_{\max})/\epsilon_{\max}\) to satisfy
\(\log(d_{\epsilon})\geq \lambda\epsilon\).  To determine such a
\(\gamma\), we first show \(\tilde d = d'_{\epsilon}/\epsilon\) is
decreasing as a function of \(\epsilon\).  Equivalently, on the
interval \([0,\epsilon']\), the graph of
\(\epsilon\mapsto d'_{\epsilon}\) sits above the straight line from
\((0,0)\) to \((\epsilon',d'_{\epsilon'})\),
yielding the lower bound
\begin{align}
  d'_{\epsilon} \geq (d'_{\epsilon_{\max}}/\epsilon_{\max})\epsilon
\end{align}
for the relevant range \([0,\epsilon_{\max}]\) of \(\epsilon\).  
To show the monotonicity of
\(\tilde d\), from the bound in Eq.~\eqref{eq:epsilon_n_dw}, we have
\begin{align} \label{eq:epsilon_tilde_d}
  1=2^{-(l-3)/2}(1+2/\epsilon)^{\epsilon^{2}\tilde d^{2}/2}\tilde d. 
\end{align}
Consider the function \(g(\epsilon)=(1+2/\epsilon)^{\epsilon^{2}}\),
which converges to \(1\) as \(\epsilon \to 0\). To determine its
behavior for all \(\epsilon > 0\), we analyze its logarithmic
derivative by defining
\(h(\epsilon) = \epsilon^{-1}\frac{d}{d\epsilon}\log(g(\epsilon)) =
2\log(1+2/\epsilon) - 2/(\epsilon+2)\). Setting
\(y=2/\epsilon\in (0,\infty)\), we have
\(h(\epsilon)=2\log(1+y)-y/(1+y)\). We have
\(\log(1+y)=\int_{0}^{y}dz/(1+z)\geq \int_{0}^{y}dz/(1+y)=y/(1+y)\).  Therefore \(h(\epsilon)>0\), which guarantees that
\(g(\epsilon)\) is monotonically increasing for all positive
\(\epsilon\).  Because the right-hand side of
Eq.~\eqref{eq:epsilon_tilde_d} is fixed to be \(1\), the growth of
\(g(\epsilon)\) requires that \(\tilde d\) strictly decrease as
\(\epsilon\) increases, as promised.

To determine the desired linear lower bound on \(d_{\epsilon}\), 
let \(\epsilon_{2}\) satisfy \(d'_{\epsilon_{2}}=2\). If
\(\epsilon_{2}>\epsilon_{\max}\), let \(\epsilon_{2}=\epsilon_{\max}\).
The definition of \(d_{\epsilon}\) implies that
\(d_{\epsilon}\geq 1+\gamma\epsilon\) for
  \(\epsilon\in[0,\epsilon_{\max}]\), where \(\gamma\) is the minimum
of \((d'_{\epsilon_{\max}}-1)/\epsilon_{\max}\) and
\((d'_{\epsilon_{2}}-1)/\epsilon_{2}=1/\epsilon_{2}\).

   We now return to the resulting inequality
  \(\log(d_{\epsilon}) \geq \lambda \epsilon\) with
  \(\lambda=\log(1+\gamma\epsilon_{\max})/\epsilon_{\max}\) for \(\epsilon\leq \epsilon_{\max}\).  We claim
  that \(\lambda\) is increasing in \(l\). For this it suffices to
  show that \(\gamma\) is increasing in \(l\).  To satisfy
  Eq.~\eqref{eq:epsilon_tilde_d} at a given \(\epsilon\), \(\tilde d\)
  and therefore \(d'_{\epsilon}\) must increase as \(l\) increases. If
  we fix \(d'_{\epsilon_{2}}=2\) and increase \(l\), then
  \(\epsilon_{2}\) must decrease.  Thus, both of the terms whose
  minimum determines \(\gamma\) increase with \(l\), and therefore so
  does \(\gamma\).

The expectation of the test-factor product over \(n\) trials
is bounded above by
\begin{align}
  \prod_{k=1}^{n}(1+\epsilon_{k}\Delta_{5})
  &\leq \prod_{k=1}^{n} e^{\epsilon_{k}\Delta_{5}}
    \notag\\
  &=\prod_{k=1}^{n} e^{(\Delta_{5}/\lambda)\lambda\epsilon_{k}}
    \notag\\
  &\leq\prod_{k=1}^{n} e^{(\Delta_{5}/\lambda)\log(d_{k})}
\end{align}
Let \(\bar q = \sum_{k}\log(d_{k})/n\) be the average of
\(\log(d_{k})\) over the trials. As in
Sect.~\ref{sec:smallentanglement}, to reject adversaries with
\(\bar q \leq r_{\text{th}}\), it suffices to use the modified test
factor \(W'=e^{-(\Delta_{5}/\lambda)r_{\text{th}}}W\), where \(W\) be
a test factor against unentangled adversaries.  For a given target
average \(r_{\text{th}}\), we can increase \(l\) to increase
\(\lambda\) until the gain of the actual prover for \(W'\) is a
constant fraction of the original gain \(g\) for \(W\).   For maintaining a gain of at
least \(g/6\), it suffices to choose \(l\) large enough to satisfy
\(r_{\text{th}}(\Delta_{5}/\lambda) \leq 5 g/6\).
Sect.~\ref{sec:smallentanglement}. 

\vspace*{\baselineskip}
\noindent\textbf{Gate-complexity bound:}
For the last result of this section, we obtain a bound on the number
of gates and measurements that adversary \(\cC\) needs to perform to
increase the expectation of the test factor \(W\) used for excluding
unentangled adversaries by a specified amount.  We assume that the
\(\cC\)'s quantum system \(C\) consists of \(q=\log_{2}(d)\)
qubits, which may carry arbitrary prior entanglement with
  \(\cB\)'s quantum system \(B\).  For simplicity of expressions, we
  assume that \(q\) is a power of \(2\).  Upon receiving input \(y\),
\(\cC\) executes one- and two-qubit unitary and measurement operations
from a finite set of such operations where the measurements have
finitely many outcomes. The outcomes determine the value \(c\) in the
protocol as described in Fig.~\ref{fig:qpvcheatingmodel}.

We can assign to each operation and its outcome (if any) a binary code
of length bounded by \(E\log(q)\). The constant \(E\) depends on the
gate set used. If the gate set consists of the controlled-NOT,
Hadamard, \(T\)-gate, and one-qubit logical-basis measurements, the
binary code for each operation and its outcome can be shorter than
\(2\log_{2}(q)+2\). One such encoding is to use the first
  two bits to indicate which of the four gates in the set is applied.
  If it is a two qubit gate, we use \(2\log_{2}(q)\) bits to encode
  which two of the \(q\) qubits are acted on. If it is a one-qubit
  gate, we use  \(\log_{2}(q)\) bits to indicate which one.  If it
  is a measurement, we use \(1\) more bit to record the outcome. 

Let \(\log_{2}(L_{g})\) be the maximum length of the encoded sequence
of operations with outcomes for adversary \(\cC\) in the adversary
  protocol. Assuming that the gate set used is the specific one as illustrated above with a fixed-length encoding where each operation uses exactly \(2\log_2(q)+2\) bits, the maximum number of gates that \(\cC\) uses in the protocol
is then \(\log_{2}(L_{g})/(2\log_{2}(q)+2)\).   The state \(\psi C_{y,c}^{T}\) is
determined by the encoded sequence, so the number of such states that
can occur is at most \(L_{g}\). In the \(\epsilon\)-net argument
above, we can let \(\phi_{y,c}=\psi C_{y,c}^{T}/|\psi C_{y,c}^{T}|\),
which is now one of \(L_{g}\) many possible states. Since
\(\phi_{y,c}\) does not involve an approximation, the chain of
distances used in the \(\epsilon\)-net argument can be bypassed. In
effect, we now have \(\mu=\mu'\), and then, following the statement
right after Eq~\eqref{eq:exp(w)_ub}, we can conclude that if
\(\Exp_{\mu}(W) \geq 1 + \epsilon\bar\Delta\), the above described SMP
protocol has guessing probability at least \(1/2+\epsilon\).
Combining this with Eq.~\eqref{eq:bound_communicomplexity}, we obtain
the bound
\begin{align}\label{eq:bound_commuicomplexity2}
  \epsilon\leq 2^{-(l-3)/2}\sqrt{L_{g}}.
\end{align}
Given a choice of \(\epsilon>0\), for soundness against 
adversaries where \(\cC\) uses less than
\(G=\log_{2}(L_{g})/(2\log_{2}(q)+2)\) gates on at most \(q\)
  qubits,  it suffices to choose \(l\) large enough to ensure
that
\begin{align}
  \epsilon\geq 2^{-(l-3)/2}2^{G(\log_{2}(q)+1)}
  \label{eq:bound_Gepsilon}
\end{align}
and adjust the test factor by a scaling factor of \(\kappa=1/(1+\epsilon\bar\Delta)\). 
Equivalently, we choose \(l\) such that 
\begin{align}
  l\geq 2G(\log_{2}(q)+1)+2\log_{2}(1/\epsilon) + 3,
  \label{eq:gatequbit_bound}
\end{align}
which is linear in the number \(G\) of quantum gates  used by adversary \(\cC\) and the
logarithm of the number of qubits \(q\). 

To preserve positive gain for the actual prover, we choose
\(\epsilon\) small enough. We repeat the construction used against
adversaries with bounded prior entanglement and no constraints on
local quantum computation.  Given the test factor \(W\) against
adversaries with no prior entanglement and its positive gain \(g\)
for the actual prover, the expectation of \(W\) for this prover is
at least \(e^{g}\). As indicated above, we set \(W'=\kappa W\) with
\(\kappa=1/(1+\epsilon \bar\Delta)\) for a test factor against
adversaries where \(\cC\) uses at most \(G\) gates on \(q\)
qubits. To ensure that the actual prover retains a gain of \(g/6\)
in this setting, we again choose \(\epsilon\) such that
\(\log(1+\epsilon\bar\Delta)=5g/6\). Accordingly, we set
\(\epsilon = (e^{5g/6}-1)/\bar\Delta\).  For the typical test factor
and gain in our experiment given above, we have
\(\epsilon=\Epsilonexpg\). According to
Eq.~\eqref{eq:gatequbit_bound}, we can protect against adversaries
with up to \(G=\lexpgbnd\) gates and measurements on
\(q\leq \lexpgq\) qubits with \(l=\lexpgl\).

\section{Analysis of experimental results}
\label{sect:xanalysis}

The data analysis consists of three main steps: parameter
determination, calibration prior to each protocol instance, and
execution of each instance. For the first step, parameter
determination, we used data collected on September 19, 2024 and use
this data to estimate the distribution of the trial records for the
implemented prover. We refer to a prover for whom the trial records
are i.i.d. according to the estimated distribution as the reference
prover. This prover is honest. The second and third
steps---calibration and analysis---were performed using data collected
on September 20 and October 7, 2024.  On each day, the data was saved
across a varying number of files, with each file containing one minute
of data. The trial rate in our experiment is approximately \(250,000\)
trials per second.

\subsection{Parameter Determination}
\label{sect:parameter_choices}

The first step of data analysis is to determine key parameters 
necessary for evaluating the protocol. These parameters include:
\begin{enumerate}
    \item The \emph{soundness error} \(\delta\),
    \item The \emph{target success probability} \(\epsilon\) for the reference prover,
    \item The \emph{number of trials} $n$ (the amount of data) to be used per protocol instance, and
    \item The \emph{success threshold} \(r_{\text{th}}\) on the
      certified average robustness of entanglement pre-shared by
      adversaries for implementing a successful attack.
\end{enumerate}
We fixed the soundness error to \(\delta = \valSound\), which is considered 
sufficiently small for practical security purposes. In addition, we set the 
target success probability to \(\epsilon = \valComplete\), corresponding to a 
two-sigma confidence level.

The procedure for determining the remaining parameters is as
follows. In the case where the adversaries under consideration do not
share prior entanglement, we examine the trade-off between the number
of trials \(n\) and the soundness error \(\delta\) (that is, the
\(p\)-value threshold). We choose \(n\) such that for the reference
prover, the observed \(p\)-value falls below \(\delta\) with
probability at least \(\epsilon = \valComplete\). To consider
adversaries with bounded prior entanglement, we analyze how \(n\)
affects the achievable success threshold \(r_{\text{th}}\), which
serves as a lower bound on the entanglement per trial required for the
adversaries to succeed. In this case, \(n\) and \(r_{\text{th}}\) are
selected jointly to ensure that for the reference prover, the
resulting \(p\)-value falls below \(\delta\) with probability at least
\(\epsilon = \valComplete\). To estimate these probabilities, we
proceed as follows: First, we use data collected on September 19, 2024
to estimate the conditional distribution of measurement outcomes given
the settings. We refer to this data as the
calibration data for parameter determination. Second, we determine a
high-gain test factor for the estimated distribution.  And finally, we
analyze the statistical behavior of the logarithm of the test
factor for the reference prover.  

We begin by estimating the conditional distribution
\(\sigma(\oqa,\zqa,\zqb|\mqa,\mqp)\) from calibration data. The
calibration data can be summarized by counts
\(n(\oqa,\zqa,\zqb; \mqa,\mqp)\). These counts are the number of
trials for which the settings \((\mqa,\mqp)\) and outcomes
\((\oqa,\zqa,\zqb)\) were recorded. A representative example of the
counts for each such combination of values for the trials obtained in
one \(5\)-minute run is shown in Tab.~\ref{tab:calibration_counts},
and this example is used as the calibration data for parameter
determination.  Ideally, we always have \(\zqa = \zqb\), but in
practice, due to technical imperfections, \(\zqa\) and \(\zqb\)
occasionally differ.  Such differences can also arise from adversarial
manipulations, where an adversary may sacrifice the probability of
returning identical responses to improve performance conditional on a
match.  In the experiment, the observed frequency of such differences
was normally less than \(d=2\times 10^{-6}\). Because differences
occur rarely, we regularize the four conditional probabilities
\(\sigma(\oqa,\zqa,\zqb|\mqa,\mqp)\) given settings \((\mqa,\mqp)\)
with \(\zqa\ne\zqb\) by setting all of them to the fixed value
\(d/4 = 0.5\times 10^{-6}\) for the purpose of constructing test
factors and choosing protocol parameters. When constructing test
factors, we constrain the test factor to take a constant value for all
settings and outcomes with \(\zqa\ne\zqb\). Consequently, all such
events contribute identically to the logarithm of the test factor, and
so the choice of a uniform distribution over these events in the
regularization does not affect the gain calculation.

\begin{table}[!htbp]
\centering
\caption[Calibration counts]{Calibration counts \(n(\oqa, \zqa, \zqb; \mqa, \mqp)\) for settings \((\mqa, \mqp)\) and outcomes \((\oqa, \zqa, \zqb)\).
All settings and outcomes are binary with two
    possible values denoted by \(1,2\). The first four columns correspond to cases where
  \(\zqa = \zqb\), while the last column aggregates all cases where
  \(\zqa \ne \zqb\). These counts are based on \(5\) minutes of data
  collected on September 19, 2024. Random resampling suggests that
  this amount of data is sufficient for a reliable estimation of the
  underlying conditional distribution. }
\begin{tabular}{w{c}{1cm} w{c}{1.5cm} *{5}{w{c}{2cm}}}

\hline\hline
 & & \multicolumn{5}{c}{\((\oqa, \zqa, \zqb)\)} \\
\cline{2-7}
 & & \((1,1,1)\) & \((2,1,1)\) & \((1,2,2)\) & \((2,2,2)\) & Others \\
\cline{2-7}
\multirow{4}{*}{\rotatebox[origin=c]{90}{\((\mqa, \mqp)\)}} & \((1, 1)\) & 18,764,031 & 2,339 & 2,390 & 4,794 & 16 \\
 & \((2, 1)\) & 18,751,159 & 9,481 & 1,647 & 5,474 & 29 \\
 & \((1, 2)\) & 18,752,299 & 1,527 & 6,879 & 5,730 & 32 \\
 & \((2, 2)\) & 18,745,211 & 14,655 & 12,333 & 364 & 35 \\
\hline\hline
\end{tabular}
\label{tab:calibration_counts}
\end{table}

Instead of directly determining the conditional distribution
\(\sigma(\oqa,\zqa,\zqb|\mqa,\mqp)\) by fitting the observed counts,
we first restrict the counts to the events where \(\zqa=\zqb\).  We
define the restricted counts as
\(\tilde{n}(\oqa,\oqp; \mqa,\mqp)=n(\oqa,\oqp,\oqp;\mqa,\mqp)\) for
each combination of settings \((\mqa,\mqp)\) and outcomes
\((\oqa,\oqp)\).  The counts \(\tilde{n}(\oqa,\oqp; \mqa,\mqp)\)
determine empirical conditional frequencies
\begin{align}
  \tilde{f}(\oqa,\oqp|\mqa,\mqp)=
  \tilde{n}(\oqa,\oqp; \mqa,\mqp)/\sum_{\oqa,\oqp}\tilde{n}(\oqa,\oqp;\mqa,\mqp).
\end{align}
If the probability that \(\zqa\ne\zqb\) is zero, 
the counts and frequencies are sampled according to some conditional
distribution \(\tilde{\sigma}(\oqa,\oqp|\mqa,\mqp)\), which must be in the set
\(\cQ_{\Oqa,\Oqp\lvert\Mqa,\Mqp}\) of two-party non-signaling
distributions that satisfy Tsirelson's bounds. Here we do not consider
other quantum constraints.  The set
\(\cQ_{\Oqa,\Oqp\lvert\Mqa,\Mqp}\), as characterized in Sect.~IX.~A
of Ref.~\cite{knill:qc2017a}, forms a convex polytope.  We fit
\(\tilde{\sigma}(\oqa,\oqp|\mqa,\mqp)\in\cQ_{\Oqa,\oqp\lvert\Mqa,\Mqp}\) to
the frequencies \(\tilde{f}(\oqa,\oqp|\mqa,\mqp)\) by maximum likelihood.
That is,
we obtain the estimate \(\tilde{\sigma}(\oqa,\oqp\lvert\mqa,\mqp)\) as the unique 
optimal solution of the following constrained maximum-likelihood problem:
\begin{align}
  \underset{\mu(\oqa,\oqp\lvert\mqa,\mqp)}{\argmax}\quad &\sum_{\oqa,\oqp,\mqa,\mqp}\tilde{n}(\oqa,\oqp;\mqa,\mqp)\log\mu(\oqa,\oqp\lvert\mqa,\mqp)\\
  \text{subject to}\quad &\mu(\oqa,\oqp\lvert\mqa,\mqp)\in \cQ_{\Oqa,\Oqp\lvert\Mqa,\Mqp}\label{eq:maximum_likelihood}
\end{align}
The estimate obtained is shown in Tab.~\ref{tab:calib_dist}.
We then determine  the conditional probabilities 
\(\sigma(\oqa,\zqa,\zqb|\mqa,\mqp)\) as follows:
\begin{equation}
\sigma(\oqa,\zqa,\zqb|\mqa,\mqp)=\delta_{\zqa,\zqb}(1-d)\tilde{\sigma}(\oqa,\Oqp=\zqa|\mqa,\mqp)+\left(1-\delta_{\zqa,\zqb}\right)d/4. 
\label{eq:calibrated_discrepant_distribution}
\end{equation}

\begin{table}[!htbp]
\centering
\caption[Settings-conditional distribution]{The settings-conditional distribution of outcomes
  \(\tilde{\sigma}(\oqa,\oqp|\mqa,\mqp)\) obtained by maximum likelihood using
  the calibration data in Tab.~\ref{tab:calibration_counts}. This
  probability distribution is used solely for determining key
  parameters in the position verification protocol and not intended to
  make a statement about the actual probability distribution during
  the calibration or analysis phase of each instance of the
  protocol. All displayed values are rounded to \(10\) digits after
  the decimal point.}\label{tab:calib_dist}
\begin{tabular}{w{c}{1cm} w{c}{1.5cm} *{4}{w{c}{3cm}}}
  \hline\hline
 & & \multicolumn{4}{c}{\((\oqa, \oqp)\)} \\
\cline{2-6}
 & & \((1,1)\) & \((2,1)\) & \((1,2)\) & \((2,2)\) \\
\cline{2-6}
\multirow{4}{*}{\rotatebox[origin=c]{90}{\((\mqa, \mqp)\)}} & \((1, 1)\) &  0.9994906521 & 0.0001264110 & 0.0001261772 & 0.0002567597 \\
 & \((2, 1)\) & 0.9991117479 & 0.0005053152 & 0.0000885493 & 0.0002943876 \\
 & \((1, 2)\) & 0.9992478451 & 0.0000802128 & 0.0003689843 & 0.0003029579 \\
 & \((2, 2)\) & 0.9985476132 & 0.0007804446 & 0.0006526840 & 0.0000192581 \\
\hline\hline
\end{tabular}
\end{table}

Next, following the discussion in Sect.~\ref{sec:test_factors}, we construct a test 
factor for the two-party Bell test using the strategy developed in 
Ref.~\cite{zhang_y:qc2011a}. This test factor is optimal for rejecting local realism~\cite{zhang_y:qc2011a} with respect to 
the settings-conditional distribution of outcomes \(\tilde{\sigma}(\oqa,\oqp | \mqa,\mqp)\) 
shown in Tab.~\ref{tab:calib_dist}, given a uniform settings distribution 
\(\mmuap(\mqa,\mqp)\) defined as \(\mmuap(\mqa,\mqp)=1/4\) for all \(\mqa,\mqp\in\{1, 2\}\). Building on this, we construct a family 
of test factors of the form given in Eq.~\eqref{eq:robust_tf}, parameterized
by \(\lambda\). By maximizing \(\lambda\), we obtain a well-performing 
test factor, as presented in Tab.~\ref{tab:pv_testfactor}, for the task of position verification against adversaries without prior entanglement. 
From now on, we refer to this task as basic position verification
and to the corresponding protocol as the basic protocol.
In addition to its role in parameter determination below, this test
factor could also be used to analyze future data for basic position verification. 

\begin{table}[!htbp]
\centering
\caption[Trial-wise test factor]{Trial-wise test factor \(W(\mqa,\oqa, \mqp, \zqa, \zqb)\) for
  basic position verification, constructed using the calibration data
  in Tab.~\ref{tab:calibration_counts}. The four columns shown
  correspond to cases where \(\zqa = \zqb\). For all other cases with
  \(\zqa \ne \zqb\), which are not displayed in the table, the test
  factor takes a constant value of \(0.9118409194\), independent of
  the settings and outcomes. All displayed values
  are rounded to \(10\) digits after the decimal point.}
\begin{tabular}{w{c}{1cm} w{c}{1.5cm} *{4}{w{c}{3cm}}}
\hline\hline
 & & \multicolumn{4}{c}{\((\oqa, \zqa, \zqb)\)} \\
\cline{2-6}
 & & \((1,1,1)\) & \((2,1,1)\) & \((1,2,2)\) & \((2,2,2)\) \\ 
\cline{2-6}
\multirow{4}{*}{\rotatebox[origin=c]{90}{\((\mqa, \mqp)\)}} & \((1, 1)\) &  1.0000133425 & 0.8853069445 & 0.8825655759 & 1.0836976498 \\ 
 & \((2, 1)\) & 1.0000098326 & 0.9751727669 & 0.8016191273 & 1.0926205335 \\ 
 & \((1, 2)\) &  1.0000079803 & 0.7988759064 & 0.9699591897 & 1.0846655877 \\ 
 & \((2, 2)\) &  0.9999688446 & 1.0248059103 & 1.0300176352 & 0.7390162290 \\ 
\hline\hline
\end{tabular}
\label{tab:pv_testfactor}
\end{table}

\begin{figure}
  \includegraphics[scale=0.5]{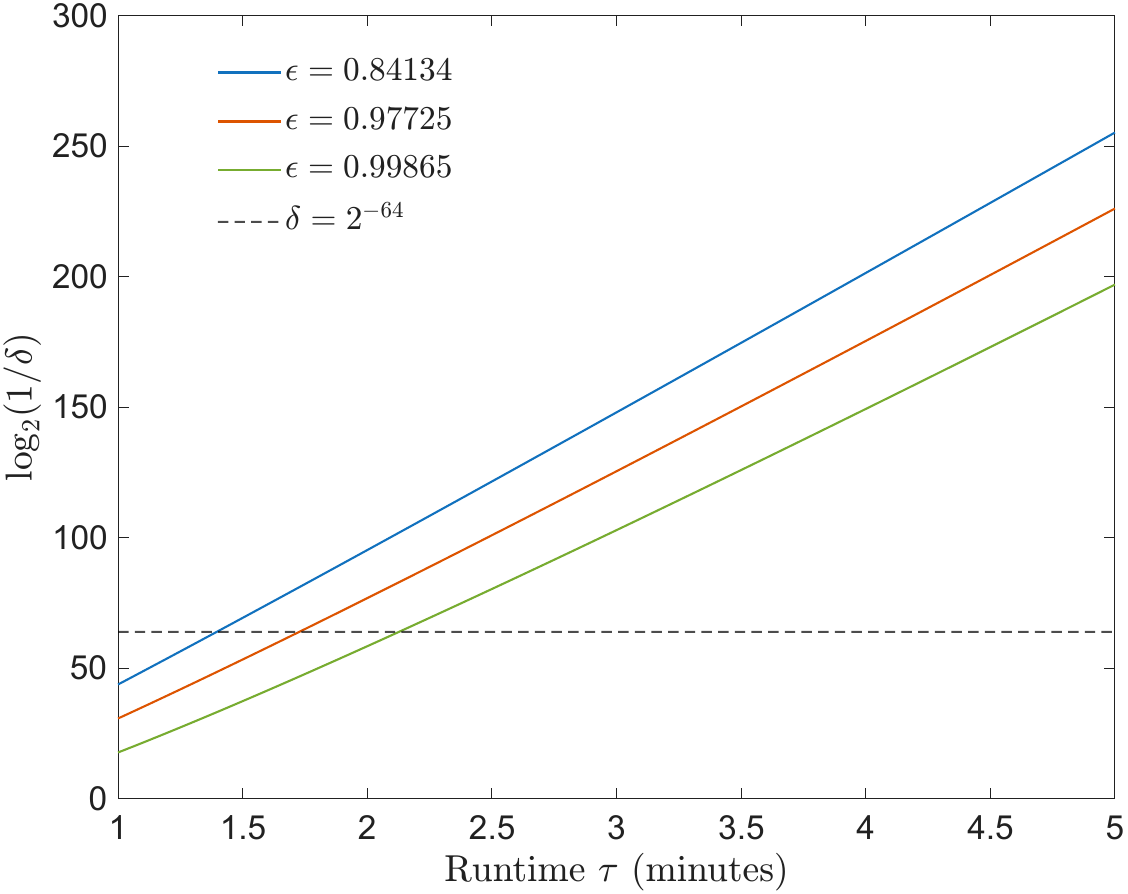}
  \caption[Required runtime]{Required runtime of the basic protocol (\(x\)-axis) to achieve a target soundness
    error \(\delta\) (\(y\)-axis) with high probability. For a desired
    soundness error \(\delta\) to be achieved with success probability
    at least \(\epsilon\), the experiment must run for at least the
    duration \(\tau\) indicated on the \(x\)-axis. The three curves
    correspond to confidence levels of one-sigma
    (\(\epsilon=0.84134\)), two-sigma (\(\epsilon=0.97725\)), and
    three-sigma (\(\epsilon=0.99865\)), respectively.  In our
    experiment, we generate approximately \(250,000\) trials per
    second.  For parameter determination, we assume that the trial
    generation rate is exact.  }
    \label{fig:tradeoff_basic}
\end{figure}

Finally, we analyze the statistical behavior of the logarithm of the
obtained test factor.  According to our estimates of
\(\sigma(\oqa,\zqa,\zqb|\mqa,\mqp)\) in
Eq.~\eqref{eq:calibrated_discrepant_distribution},
\(\tilde{\sigma}(\oqa,\oqp | \mqa,\mqp)\) in Tab.~\ref{tab:calib_dist}, 
and the uniform settings distribution used in our
implementation, the base-\(2\) logarithm of the trial-wise test factor
shown in Tab.~\ref{tab:pv_testfactor}, \(\log_2(W)\), has a mean of
\(g_{w}=3.79135\times 10^{-6}\) and a variance of
\(v_{w}=1.13029 \times 10^{-5}\). For the completeness analysis, we
assume the reference prover so that the trials are i.i.d.  and
distributed according to the above estimates. We consider the
case where the test factor \(W_{k}\) used in each trial \(k\) is
identical to the one presented in Tab. \ref{tab:pv_testfactor}.  In
view of the central limit theorem, we approximate the sum
\(\sum_{k=1}^{n}\log_2(W_k)\) over \(n\) i.i.d. trials as normally
distributed with mean $n g_w$ and variance $n v_{w}$. For a given
soundness error \(\delta\), the condition
\(\sum_{k=1}^{n} \log_2(W_k) \geq \log_2(1/\delta)\) must be satisfied
to ensure a \(p\)-value below \(\delta\). Accordingly, the success
probability of obtaining a \(p\)-value less than \(\delta\) after
\(n\) trials is heuristically estimated to be
\(p_{\mathrm{succ}}=Q\left(-\left(n
    g_{w}-\log_2(1/\delta)\right)/\sqrt{nv_{w}}\right)\), where the
function $Q$ is the tail distribution function of the standard normal
distribution. To ensure \(p_{\mathrm{succ}}\geq \epsilon\), the number
of trials \(n\) must be above a certain threshold. By varying the
values of \(\delta\) and \(\epsilon\), we obtain the trade-off curves
in Fig.~\ref{fig:tradeoff_basic}. The trade-off curves show the
  value of \(\delta\) that can be achieved at three different success
  probabilities as a function of the running time of the basic protocol.
  For the chosen values \(\delta=\valSound\) and \(\epsilon=\valComplete\),
  the figure shows that \(2\) minutes of data is sufficient for basic
position verification.

\begin{figure}
  \includegraphics[scale=0.5]{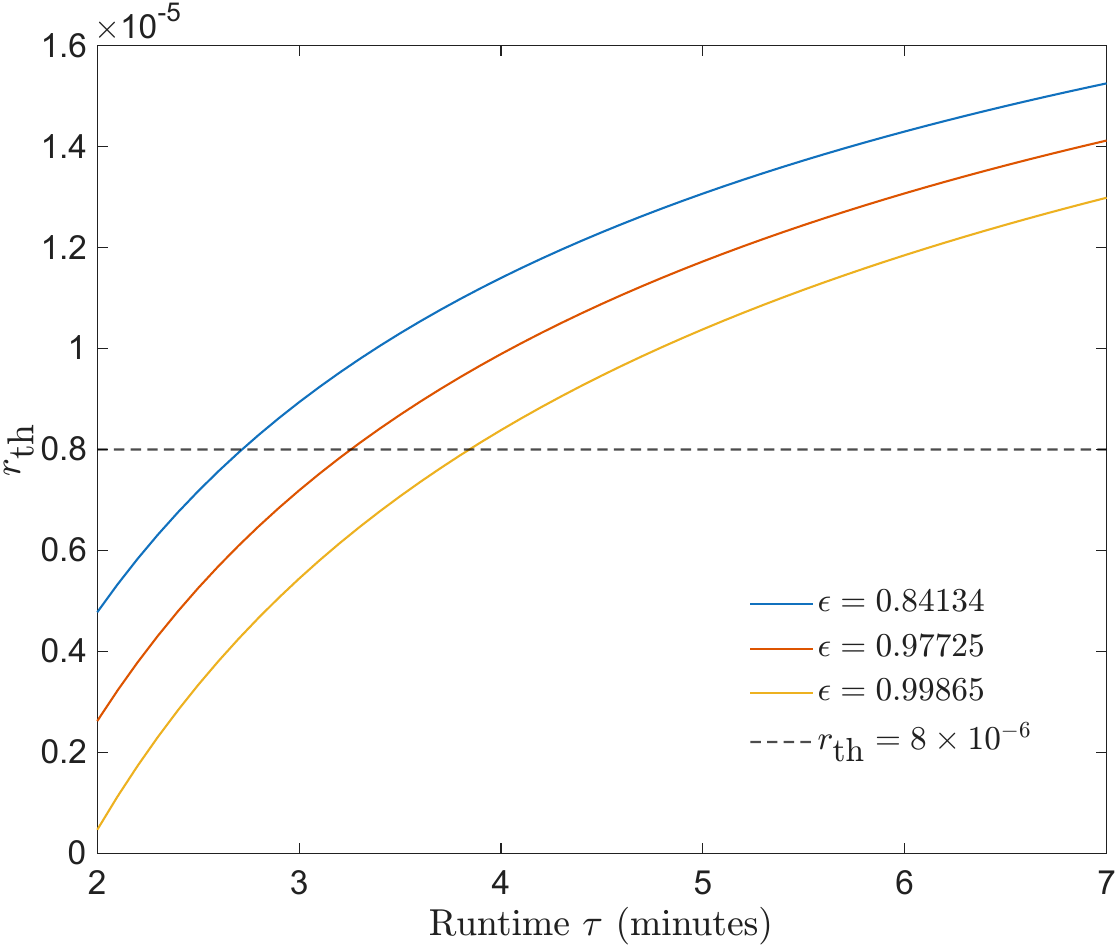}
  \caption[Required runtime with entanglement]{
    Required runtime (\(x\)-axis) to ensure, with
  high probability, that the certified average robustness of 
  entanglement pre-shared by adversaries reaches the target success threshold \(r_{\text{th}}\) (\(y\)-axis). The certification is performed with soundness error \(\delta=\valSound\). 
  For a desired success threshold \(r_{\text{th}}\) to be reached with success 
  probability at least \(\epsilon\), the experiment must run for at least the 
  duration \(\tau\) indicated on the \(x\)-axis. The three curves 
  correspond to confidence levels of one-sigma (\(\epsilon=0.84134\)), two-sigma 
  (\(\epsilon=0.97725\)), and three-sigma (\(\epsilon=0.99865\)), respectively. 
  }
    \label{fig:tradeoff_ent}
\end{figure}

Similarly, we analyze the behavior of the test factor adapted for position 
verification in the presence of adversaries with bounded prior entanglement. 
Following Sect.~\ref{sec:smallentanglement}, we consider trial-wise test factors 
of the form \(W'=\lambda W+(1-\lambda)\),  where \(\lambda\in[0,1/(1-w_{\min})]\) 
and \(w_{\min}\) is the minimum value of the test factor \(W\) for basic 
position verification. Let \(\bar{w}'_{\min}\), defined analogously to \(\bar{w}_{\min}\) in~\eqref{eq:w_min_avg}, denote 
the average of the minimum values of the test factor \(W'\) conditioned on the 
measurement settings and challenges. Let \(W'_k\) represent the instantiation of the test 
factor \(W'\) used in the \(k\)'th trial. Given \(n\) trials and a 
soundness error \(\delta\), the certified lower bound on the average 
robustness of prior entanglement
is given by 
\begin{equation}
  \frac{1}{n(1 - \bar{w}'_{\min})} \left( \sum_{k} \log(W'_k) - \log\left(1/\delta\right)\right).
  \label{eq:elb}
\end{equation}
Thus, to ensure that the certified lower bound exceeds the threshold 
\(r_{\text{th}}\), we must have 
\begin{equation} \label{eq:thres_entPV}
\sum_{k}\log(W'_{k}) \geq n r_{\text{th}} (1-\bar w'_{\min})+\log(1/\delta).
\end{equation}
To characterize the statistical behavior of this condition, we compute
the mean and variance of the random variable \(\log(W')\).  As before,
we resort to the central limit theorem to estimate the success
probability for the reference prover. For each desired success
probability \(\epsilon\) and each fixed value of the threshold
\(r_{\text{th}}\), we optimize over the parameter \(\lambda\) that
characterizes the test factor \(W'\). This optimization allows us to
determine the minimum value of \(n\) required to satisfy the condition
in Eq.~\eqref{eq:thres_entPV} with success probability \(\epsilon\).
This process also yields the corresponding optimized test factor
\(W'\), which could be used to analyze future data for
position verification against adversaries with prior entanglement.
Varying the values of \(\epsilon\) and \(r_{\text{th}}\) yields the
requirement curve illustrated in Fig.~\ref{fig:tradeoff_ent}.
Based on this figure, we set \(r_{\text{th}}=\valThresh\).
  For the chosen values of \(\delta=\valSound\) and
  \(\epsilon=\valComplete\), the figure shows that \(4\) minutes of data
is sufficient for position verification against adversaries with
average robustness of prior entanglement at most \(r_{\text{th}}\).

\subsection{Calibration and Analysis}
\label{sect:calib_analysis}


With the protocol parameters obtained above, we executed multiple
instances of the protocol using data collected on September 20, 2024,
for the case of adversaries without prior entanglement, and data
collected on October 7, 2024, for the case involving adversaries with
limited prior entanglement. On each day, trials were performed and
recorded in successive \(1\)-minute intervals, resulting in separate
\(1\)-minute data files with about \(15\) million trials each. The
data was blinded until after the protocol parameters and analysis
strategies were fixed. The protocol implementation and analysis are
summarized in Protocol~\ref{prot:pv_direct_implt}. Because the
performance of the Bell test underlying the protocol varied in time,
we optimized the test factor to be used for each instance of the
protocol based on the \(10\) \(1\)-minute files immediately preceding the
instance that are free of detection errors as described in the next
paragraph.  We refer to this data as the calibration data for the
instance. For each instance, we used the calibration data to estimate
the settings-conditional distribution of outcomes and to construct the
corresponding test factor, following the approach described in
Sect.~\ref{sect:parameter_choices}.

In our experiment, some \(1\)-minute recording intervals experienced
detector errors, an event that happened at most a few times per day.
Such errors were recorded with the corresponding data files. At the
beginning of each day, the first \(10\) \(1\)-minute files free
of detector errors were used for calibration of the first protocol
instance of the day. The next \(2\) or \(4\) files, regardless of
whether or not detector errors happened, made up the first protocol
instance, depending on whether the adversaries had no prior
entanglement or bounded prior entanglement, respectively.  Subsequent
instances were processed sequentially using the next \(2\) or \(4\)
files if available; otherwise the instance used all the remaining data
files. For each such instance, calibration data consisted of the
previous \(10\) files without detector errors.

Nominally, each \(2\)-minute interval consists of \(30\) million
trials, and each \(4\)-minute interval consists of \(60\) million
trials.  However, due to clock synchronization and other uncertainties
in our experiment, the actual number of recorded trials deviates from
these nominal values.  To ensure the validity of the analysis, we
fixed the number of trials \(n\) processed to exactly \(30\) million
for each 2-minute interval and \(60\) million for each 4-minute
interval. If the actual number of trials exceeded \(n\), the data were
truncated and the excess trials were discarded. If it fell short,
artificial trials were appended to reach \(n\) trials. In either
case---whether testing for adversaries with prior entanglement or
without---the appended artificial trials were analyzed using the
trivial test factor that assigns a value of \(1\) to all possible
measurement settings and outcomes, which is a valid
choice. Consequently, the specific settings and outcomes assigned to
each artificial trial are irrelevant. Since the logarithm of \(1\) is
\(0\), these artificial trials did not contribute to the accumulated
sum of the logarithms of the trial-wise test factors and thus did not
influence the final analysis result. Therefore, in practice, it is not
necessary to physically append artificial trials if the actual number
of trials is less than \(n\). This approach ensures the correctness of
the analysis regardless of fluctuations in the actual number of trials
available for processing. 

\SetAlgorithmName{Protocol}{}{}
\RestyleAlgo{boxruled}
\vspace*{\baselineskip}
\begin{algorithm}[H]
  \caption{Protocol Implementation and Analysis.}\label{prot:pv_direct_implt}
  \begin{minipage}{\dimexpr\textwidth-0.5in\relax}
  \Input{
    Sequence of 1-minute data files collected in a day.
  }

  \Given{
  \begin{itemize}
    \item[] \(\delta = \valSound\) --- soundness error (i.e., \(p\)-value threshold).
    \item[] \(\epsilon = \valComplete\) --- target success probability
      for the reference prover.
    \item[] \(n\) --- number of analysis trials: \(3\times10^7\) (nominally, 2-minute data) or \(6\times10^7\) \\
    (nominally, 4-minute data), depending on the adversary model.
  \item[] \(r_{\text{th}} = \valThresh\) --- threshold on the certified average robustness of entanglement \\ for adversaries with bounded prior entanglement.
  \end{itemize}
  }

  \Output{
    \(P\) --- binary success flag: \(P=1\) (success) or \(P=0\) (failure).
  }

  \BlankLine

  Identify the \(10\)\textsuperscript{th} file in the sequence that is
  free of detector errors, and set the current file to the one
  immediately following it\;
  
  \ForEach{protocol instance}{
    Set \(P \gets 0\)\;
    Initialize the running number of calibration files: \(K_{\text{calib}} \gets 0\)\;
    Initialize the running calibration counts for all settings \(z=(\mqa, \mqp)\) and outcomes \(c=(\oqa, \zqa, \zqb)\):
    \(\{\text{Data}_{\text{calib}}(cz)\}_{cz} \gets 0\)\;

    \While{a previous \(1\)-minute file exists}{
      \If{file has no detector error}{
        Load file; increment \(K_{\text{calib}} \gets K_{\text{calib}} + 1\)\;
        Increment $\{\text{Data}_{\text{calib}}(cz)\}_{cz}$ with counts from the file\;
      }
      \If{\(K_{\mathrm{calib}} = 10\)} 
      {\textbf{break}\tcp*{Load up to 10 files for calibration}
      }
    }

    Use $\{\text{Data}_{\text{calib}}(cz)\}_{cz}$ to estimate the settings-conditional 
    distribution of outcomes for the purpose of test-factor construction (see Sect.~\ref{sect:parameter_choices})\;
    
    Construct test factor \(W\) (for adversaries without prior entanglement) or \(W'\) (for adversaries with bounded prior entanglement) as in Sect.~\ref{sect:parameter_choices}\;
    
    Load the trials from the next \(2\) or \(4\) \(1\)-minute files,
    if available. Otherwise, load from all remaining files. Truncate
    to \(n\) trials if necessary\;
    
    Compute the accumulated sum of the logarithms of test factors
    \(S_{\text{run}}\). For adversaries with prior entanglement,
    return the certified lower bound \(r_{\text{lb}}\) on the average
    robustness of entanglement according to Eq.~\eqref{eq:elb} with
    the replacement of \(\sum_{k} \log(W'_k)\) with
    \(S_{\text{run}}\)\; 
    
    Calculate the success threshold \(S_{\text{th}}=\log(1/\delta)\)
    for adversaries without prior entanglement and
    \(S_{\text{th}} = \log(1/\delta)+nr_{\text{th}}(1-\bar{w}'_{\min})
    \) for adversaries with bounded prior entanglement\;
    
    \If{\(S_{\text{run}} \geq S_{\text{th}}\)}{
      Set \(P \gets 1\)\tcp*{Protocol succeeded}
    }

    Update the current file to the file immediately following the ones
    used for this protocol instance, if this is not the last
    instance\; }
  \end{minipage}
\end{algorithm}

\begin{figure}
  \includegraphics[scale=0.47]{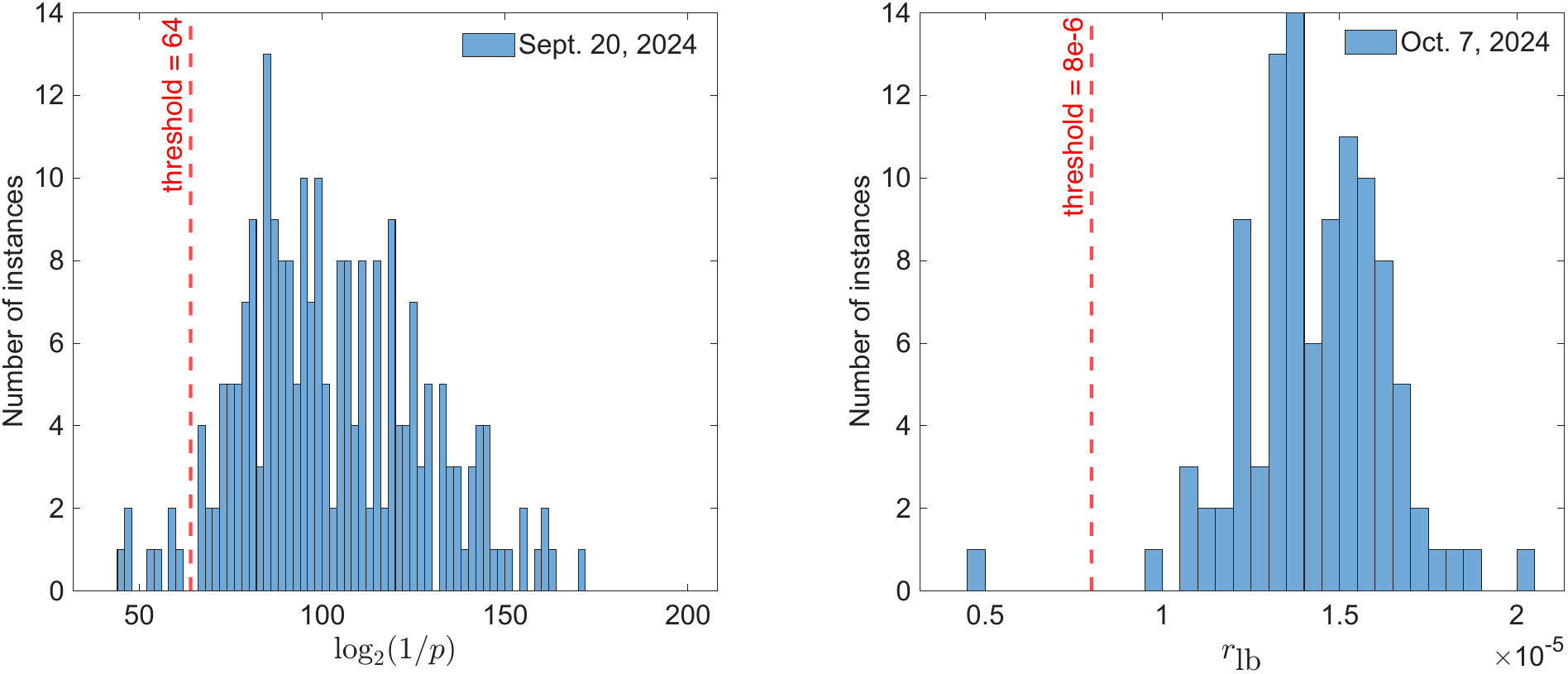}
  \caption[Histograms of the final analysis results.]
   {Histograms of the final analysis results. The left panel
    shows the results for basic position verification, while the right
    is for position verification against adversaries with bounded prior
    entanglement. The histograms show the number of instances
    binned according to the \(\log(p)\)-value and 
    the certified lower bound \(r_{\text{lb}}\) on the average 
    robustness of prior entanglement, respectively. These values are determined
    by the accumulated sum of 
    the logarithms of the trial-wise test factors used in each 
    instance. The value of \(r_{\text{lb}}\) is given by 
    Eq.~\eqref{eq:elb} for each instance.
    }
    \label{fig:hist_final}
\end{figure}

\begin{figure}
  \includegraphics[scale=0.47]{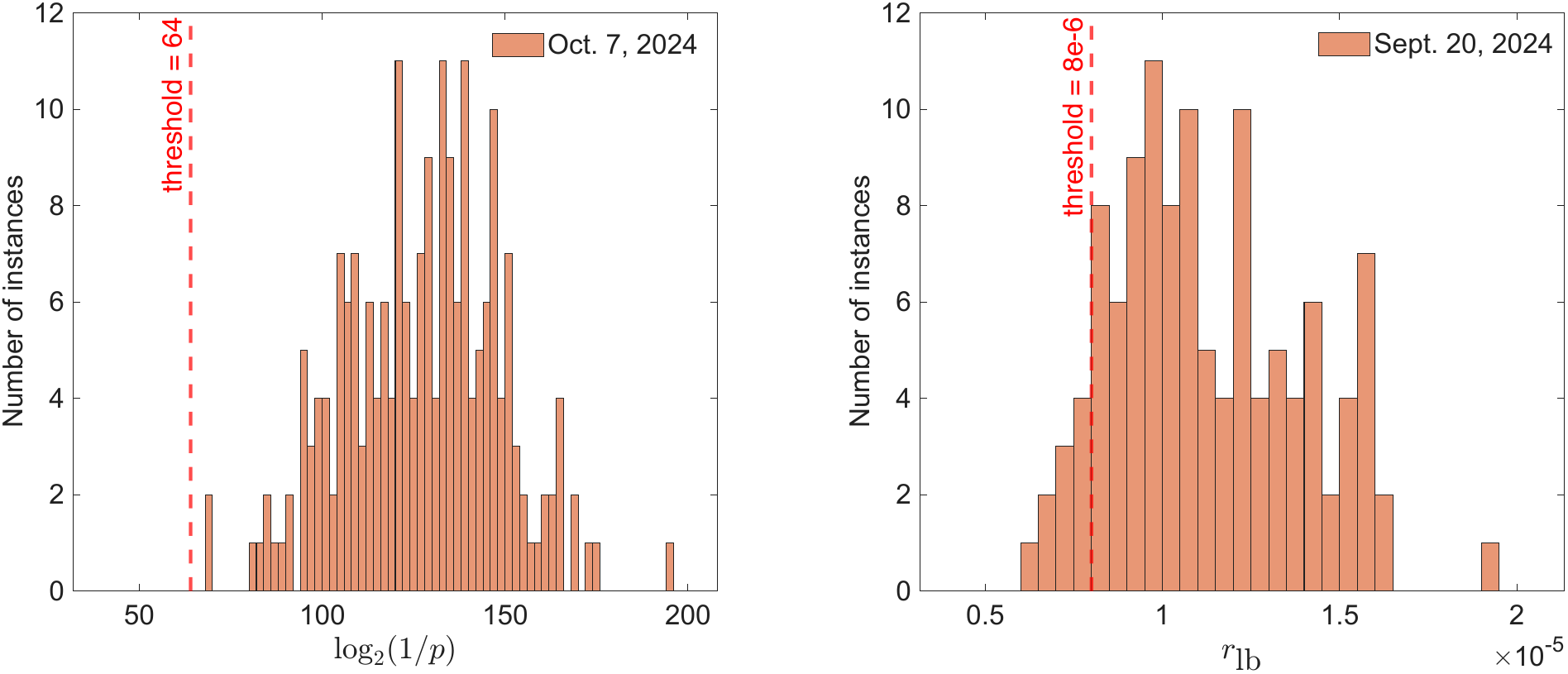}
  \caption[Histograms of the reversed analysis results]{Histograms of
    the reversed analysis results. The left panel shows the results
    for basic position verification, while the right is for position
    verification against adversaries with bounded prior
    entanglement. The plotted quantities are defined in the caption of
    Fig.~\ref{fig:hist_final}.  }
  \label{fig:hist_revers}
\end{figure}

The analysis procedure follows the implementation outlined in
Protocol~\ref{prot:pv_direct_implt}. Before unblinding the
protocol instance data and conducting the final analysis, we
validated that the protocol including the parameter determination
functioned correctly using data collected on September 19, 2024. We
then proceeded with the final analysis.  Using \(474\) \(1\)-minute
data files collected on September 20, 2024, we performed basic
position verification, resulting in \(\valBasicInstances\) instances, 
\(\valBasicFail\) of which
failed.  For position verification against adversaries with
bounded prior entanglement, we used the \(420\) \(1\)-minute data
files collected on October 7, 2024 to perform \(\valEntInstances\) instances. 
All but the final instance succeeded; the last one failed due to
insufficient remaining data. Histograms of the analysis results are
presented in Fig.~\ref{fig:hist_final}. After completing the
analysis of the protocol instances on the unblinded data, for
diagnostic purposes, we reversed the data assignment for the two
types of analyses. In the reversed analysis, using the data from
September 20, 2024, we obtained \(116\) instances of position
verification against adversaries with bounded prior entanglement, with
\(10\) instances failing.  Using the data from October 7, 2024, we
obtained \(205\) instances of basic position verification, all of which
succeeded. The results of the reversed analysis are shown in
Fig.~\ref{fig:hist_revers}.

In the analysis described above, we certified a lower bound
$r_{\text{lb}}$ on the average robustness of prior entanglement to
quantify the resources required for a successful adversarial
attack. Alternatively, we can formulate this as a hypothesis-testing
task: we aim to reject the null hypothesis that the average robustness
of entanglement pre-shared by adversaries is bounded above by a
specific threshold $r_{\text{th}}$. For this purpose, we introduce a
modified test factor $W''=W'e^{-r_{\text{th}}(1-\bar{w}'_{\min})}$,
where $W'$ is the test factor used for certifying the average
robustness, and $\bar{w}'_{\min}$ is the corresponding averaged
minimum value (see Sect.~\ref{sect:parameter_choices} for details on
determining $W'$ and $\bar{w}'_{\min}$). Let $W''_{k}$ denote the
instantiation of the modified test factor for the $k$'th
trial. Following the discussion below Eq.~\eqref{eq:robustness_ineq},
we compare the accumulated product over all trials,
$\prod_{k} W''_{k}$, against $1/\delta$ to determine whether the
protocol instance successfully rejects the null hypothesis. When
  the number of available trials meets or exceeds the required number of trials
  $n$ for the instance,  we analyze exactly \(n\) trials. In this case, the
rejection condition $\prod_{k} W''_{k}\geq 1/\delta$ is equivalent to
satisfying Eq.~\eqref{eq:thres_entPV}. If the number of available trials 
is less than \(n\), we virtually append artificial trials for which we set 
\(W'=1\).  For each such artificial trial, \(\Exp(W'')=\Exp(W'e^{-r_{\text{th}}(1-\bar{w}'_{\min})})\leq 1\) for adversaries with bounded average robustness, 
  so the analysis according to the discussion below 
  Eq.~\eqref{eq:robustness_ineq} in Sect.~\ref{sec:smallentanglement}
  remains valid.  Among the \(103\) instances using the data from
  October 7, 2024, there were \(4\) instances where the available
  number of analysis trial was less than \(n\). The logarithm of the product
$\prod_{k} W''_{k}$ for each instance using the data from October 7,
2024 is shown in the bottom panel of Fig.~4 of the main text.

\end{document}